UNIVERSITY OF CAMBRIDGE

# Quantum Chromodynamics and the Precision Phenomenology of Heavy Quarks

Matthew Alexander Lim
of Pembroke College

This dissertation was submitted to the University of Cambridge for the degree of Doctor of Philosophy in August 2018.

# Quantum Chromodynamics and the Precision Phenomenology of Heavy Quarks

**Matthew Alexander Lim**

In this thesis we consider the phenomenology of the theory of strong interactions, Quantum Chromodynamics (QCD), with particular reference to the ongoing experimental program at the Large Hadron Collider in CERN. The current progress in precision measurement of Standard Model processes at the LHC experiments must be matched with corresponding precision in theoretical predictions, and to this end we present calculations at next-to-next-to-leading order in perturbation theory of observable quantities involving quarks and gluons, the strongly interacting particles of the SM. Such calculations form the most important class of corrections to observables and are vital if we are to untangle signals of New Physics from LHC data.

We consider in particular the amplitudes for five parton interactions at 1- and 2-loop order and present full (in the 1-loop case) and partial (in the 2-loop case) analytic results in terms of rational functions of kinematic invariants multiplying a basis of master integrals. We address the problem of the solution of a system of integration-by-parts identities for Feynman integrals and demonstrate how some current difficulties may be overcome.

We consider also the properties of the top quark, and present the NNLO, real-virtual contributions to the calculation of its decay rate. The results are presented as helicity amplitudes so that the full behaviour of the top spin is retained. These amplitudes constitute a necessary ingredient in the complete calculation of top quark pair production and decay at NNLO which will be an important theoretical input to many experimental analyses.

Turning to a more phenomenological study, we consider the extraction of two important SM parameters, the top mass and the strong coupling constant, from measurements of top pair production at the ATLAS and CMS experiments. We compare with NNLO theory predictions and use a least-squares method to extract the values of the parameters simultaneously. We find best fit values of the parameters which are compatible with previous extractions performed using top data with the current world averages published by the Particle Data Group.

We consider the issue of PDF choice and the circumstances in which a heavy quark can be considered a constituent of the proton. In particular, we look at the production of a Higgs boson in association with bottom quarks in four and five flavour schemes, in which the b may or may not be included in the initial state. We show that theoretical predictions in both schemes are well-motivated and appropriate in different scenarios, and moreover that results in the schemes are consistent provided a judicious choice of the renormalisation and factorisation scales is made. We suggest a typical scale choice motivated by considerations of consistency and find it to be somewhat lower than the typical hard scale of the process.

# Declaration

This dissertation is the result of my own work and includes nothing which is the outcome of work done in collaboration except as declared in the Preface and specified in the text. It is not substantially the same as any that I have submitted, or, is being concurrently submitted for a degree or diploma or other qualification at the University of Cambridge or any other University or similar institution except as declared in the Preface and specified in the text. This dissertation does not exceed the word limit for the Degree Committee of Physics and Chemistry.

Matthew Alexander Lim

# Acknowledgements

Firstly, I would like to thank my supervisor Dr Alexander Mitov for his help and guidance in completing this thesis. His patience, attention to detail and general approach to physics as a discipline have helped instil in me better habits and taught me much about how to think about and approach research problems. He has also provided me with many opportunities to develop as a scientist. For all of this, I am very grateful.

I would also like to thank several other members of the Cavendish High Energy Physics group. I am grateful to Dr Maria Ubiali for her advice and support over the course of two degrees, as well as for proofreading this manuscript. My thanks go also to Dr Andrew Papanastasiou for many useful discussions during our work together.

A large part of what I have learnt about physics has come from interesting discussions with my peers. Conversations with David Sutherland, Thomas Cridge and Benjamin Brunt have been particularly helpful and I thank them all, especially the last for his assistance with various computing issues. Members of the SUSY Working Group have also been supportive and made insightful suggestions during meetings. Thanks also go to Emma Slade for proof-reading this manuscript.

I am grateful to the Science and Technologies Facilities Council and the Cambridge Philosophical Society for the support they have provided over the course of this work.

Finally, I would like to thank my friends and family for their unwavering support.

# Preface

This work is based in part on published material. Chapter 6 is taken almost verbatim from the paper [1]. Parts of that work were also submitted towards an MSci degree at the University of Cambridge as a Part III project—however, no figure which appears in that project is included in this thesis and while the focus of the paper is broadly the same, the scope has been significantly expanded and there is a considerable amount of new material in the final work. Section 3.3 of Chapter 3 is based on the paper [2], while Chapter 5 is based on a paper set to appear soon.

*To my grandparents*

# Contents

“*Wir müssen wissen—wir werden wissen!*”

— David Hilbert

# Chapter 1.

# Introduction

*"Among thousands of men hardly one strives after perfections; among those who strive hardly one knows Me in truth."*
— Krishna; Chapter 7, verse 3; the Bhagavad Gita

The 20$^{\text{th}}$ century saw perhaps the greatest progress of any period of history in our understanding of the fundamental laws of Nature. Through many efforts, both theoretical and experimental, our forerunners developed and perfected the theory we now know as the Standard Model (SM), a mathematical description of the elementary particles of matter and the forces which govern their behaviour. If the fact that Man should be able to describe any natural phenomemon in terms mathematical were not astounding enough[3], the accuracy to which the Standard Model has been verified ought to be cause for marvel. The magnetic moment of the electron, for example, has been measured to 0.28 parts per trillion and implies agreement with the Standard Model prediction to a few parts per billion[4]. Our most rigorous experiments have, as yet, failed to find any signficant discrepancy with the model in a series of tests far too long to enumerate here. This places it in a position supreme as the best tested theory of all time.

## 1.1. The Standard Model of particle physics

The Standard Model is a quantum field theory built on the gauge group $SU(3) \times SU(2)_L \times U(1)_Y$ and describes three of the four fundamental forces known to us—the electromagnetic, weak and strong interactions. It is comprised of 16 spin-1/2 fermionic fields, 4 types of spin-1 gauge boson and a single scalar, the Higgs boson, discovered in

2012[5,6]. Schematically, the Lagrangian is written as

$$\mathcal{L}_{\rm SM} = -\frac{1}{4}F_{\mu\nu}F^{\mu\nu} + i\bar{\psi}\not{D}\psi + \psi_i y_{ij}\psi_j\phi + \text{h.c.} + |D_\mu\phi|^2 - V(\phi) \tag{1.1}$$

where the gauge bosons are encoded in the field strength tensor $F^{\mu\nu}$, the fermionic fields are the $\psi_i$, the Higgs field and potential are given by $\phi$ and $V(\phi)$ and interactions between fields enter in the covariant derivative $D_\mu$. The $SU(2)_L \times U(1)_Y$ component of the gauge group describes the unified electroweak force, a theory due to Weinberg and Salam, and is associated with two neutral gauge bosons, the photon $\gamma$ and the $Z$, and two charged, the $W^\pm$. Of these, the photon corresponds to the sub-group of electromagnetism $U(1)_{\rm EM}$ and is strictly massless while the other three obtain masses after electroweak symmetry breaking (EWSB) via the Higgs mechanism. This mechanism also accounts for the masses of the fermions through Yukawa couplings $y_{ij}$ of the fermionic and Higgs fields. The remnant of EWSB, corresponding to the last of four degrees of freedom left in the Higgs complex doublet, is identified as the Higgs boson, a neutral spin-0 particle. The other $SU(3)$ component describes the strong interaction, which couples the fermionic quark fields to massless gauge bosons themselves charged under the group, the gluons. We shall have much to say about this sector of the SM in later chapters.

The rich phenomenology of the SM is determined by the variety and strength of its couplings. Each gauge sector is associated with a coupling constant, $g$ and $g'$ being the electroweak couplings ($SU(2)_L \times U(1)_Y$) and $g_s$ the strong coupling (the remaining $SU(3)$). These are usually translated into three numbers which quantify the relative strength of each force, $\alpha = e^2/4\pi \approx 1/137$, the fine structure constant for the electromagnetic force, $G_F = \sqrt{2}g^2/8M_W^2 \approx 1.17 \times 10^{-5}\,\text{GeV}^{-2}$, the Fermi constant for the weak force and $\alpha_s = g_s^2/4\pi \approx 0.118$ for the strong force. The comparative sizes of these couplings mean that quantum corrections due to strong interactions usually dominate over electroweak corrections. As will later be discussed, the couplings also vary with the energy scale at which they are measured. While the electroweak couplings increase with the energy, the nature of the self-interaction of the gluon fields causes the strong coupling to decrease at higher energies—this property is known as asymptotic freedom.

A peculiar feature of the Standard Model is the existence of multiple generations of the fermionic fields, differing only in their mass. Each charged lepton, neutrino and quark appears in triplicate—in order of increasing mass we call the charged leptons and their associated neutrinos electron, muon and tau, the $e = +2/3$ quarks up, charm and top and the $e = -1/3$ quarks down, strange and bottom. It is not known at present why there are only 3 generations or indeed why there should be multiple generations at all. While it is believed that the coupling of the electroweak force to the leptons is universal[1], couplings to different generations of quarks are modified by a unitary transformation known as the Cabbibo-Kobayashi-Masakawa matrix. Interestingly, this matrix allows the introduction of charge-parity violation within the Standard Model when at least 3 generations are present, though measurements of the entries indicate that this alone cannot suffice to explain the observed amount of violation in the Universe.

### 1.1.1. Failings of the Standard Model

Although in many ways an almost incredibly successful theory, the SM is not without its limitations. We have mentioned some of these already, namely the 'generation problem' and its inability to account for the observed matter-antimatter asymmetry (a consequence of CP violation). Perhaps the most signficant absence is a description of gravity at energies near the Planck scale—while general relativity functions perfectly well as a quantum theory of gravity up to $M_P \sim \mathcal{O}(10^{18})$ GeV, beyond this an adequate theory of quantum effects is needed. Although the size of gravitational effects is believed to be sufficiently small that at current collider experiments they should play no rôle[2], this in itself presents a further question, namely the reason for the hierarchy between the electroweak scale and the Planck scale. A naïve calculation of quantum corrections to the Higgs mass, for example, would suggest a value $\sim \Lambda_P$ when in fact the measured mass is $\sim 125$ GeV—this can only be achieved via a fine-tuning of the radiative corrections in order to make the mass light. This is known as the electroweak hierarchy problem, to which many solutions have been proposed (most prominently, supersymmetry) and yet none experimentally validated.

Many other problems remain. The lack of a particle candidate for dark matter, a possible explanation for large scale astrophysical anomalies, the difference between the measured value of the cosmological constant and that calculated from the zero-point energy of the Universe, the structure of the CKM matrix, the near but imperfect unification of the gauge couplings at high energy scales, the observation of neutrino masses, the lack of observation of CP violation in the strong sector—all these defy a Standard Model explanation. It is the aim of modern particle physics to account for as many of these phenomena as possible and provide a unified description of Nature. Currently, the largest experimental program operational is the Large Hadron Collider (LHC) at CERN in Geneva, where four large experiments (and several smaller ones) are now taking data from proton-proton collisions to test the SM and search for evidence of New Physics. Later parts of this thesis will be concerned with the ATLAS and CMS experiments in particular, which are general purpose detectors designed to measure a wide range of SM processes and detect any exotic phenomena. There are however many other experiments worldwide which address equally important issues: to name but a few, the measurement of neutrino flavour oscillations, the detection of double neutrinoless beta decay and the properties of $B$ mesons. Together they push towards the ultimate goal of finding a more complete description of Nature.

[1] Recent data from the LHCb experiment at the Large Hadron Collider suggest that in fact this may not be the case, and a hierarchical structure may exist within the leptonic sector. Anomalies in the rates of decay of $B$ mesons point towards a difference in the rates to electrons and to muons—specifically, a measurement of $BR(B \to K^*\mu^+\mu^-)/BR(B \to K^*e^+e^-)$ seems to differ by a few $\sigma$ from the SM predicted value of 1. At the time of writing, however, there is insufficient evidence to claim any statistically significant discrepancy.

[2] This is not strictly true, as there exist theories of New Physics which posit situations where gravity enters at the TeV scale[7,8]. However, current experimental data does not favour such a scenario.

### 1.1.2. The rôle of perturbative corrections

Any realistic quantum field theory which purports to describe Nature is inherently non-linear due to the interaction terms in the Lagrangian. It is well known that non-linear models are harder to solve than linear models, and as a result no complete solution to a physical interacting field theory is known. The SM is no exception, and therefore in order to calculate quantities of interest (such as rates of scattering of the fields) we are forced to make a perturbative approximation and truncate at some fixed order. The leading order approximation often gives little better than a rough estimate of the order of magnitude of the quantity one wishes to compute, and in order to attain any precision at all one usually needs at least a next-to-leading order (NLO) calculation. At the moment, the state of the art is at next-to-next-to-leading order (NNLO) for most processes of interest, though in a few special cases even higher order corrections have been calculated[9,10]. Although in principle we should consider corrections from all sectors of the Standard Model to a theoretical prediction, as we argued earlier the relative size of $\alpha_s$ means that it is corrections due to the strong interaction which are of highest importance and it is this type that we shall examine in this thesis.

It is clear that in order to detect the effects of something new, one needs to know precisely what one should expect to see based on known physics. Precision in experimental measurement is rapidly reaching percent level, and disentangling New Physics therefore requires theoretical predictions of matching precision. One might demarcate crudely two approaches to the hunt for New Physics, one concerned with searching for exotic resonances and one concerned with detecting deviations from Standard Model predictions. Precision theory is relevant to both of these, whether it be in calculating the expected backgrounds to a signal or obtaining the SM value for a parameter which might be modified by the presence of an exotic state.

## 1.2. Quantum Chromodynamics

We now cease further discussion of the electroweak theory and physics Beyond the Standard Model and turn instead to the $SU(3)$ sector of the Standard Model describing the strong interaction, which will be the subject of the remainder of this thesis.

The development of a new kind of particle detector, the bubble chamber, in the 1950s led to the advent of the field of strange particle spectroscopy in the 1960s. A number of new resonances were observed, beginning with the $\Sigma(1385)$ in late 1960[11] followed by the $K^-(892)$ and $\Lambda(1405)$ in 1961[12,13] and the $\Xi(1530)$ in 1962[14,15]. The proliferation of such particles led to attempts to extend the concept of isospin, which had been successful in pion physics, to strange particles in order to yield new predictions without detailed knowledge of the underlying dynamics of the strong interaction. Large mass differences between particles made it unclear which should be grouped together into the extended isospin multiplets, members of which should have equal mass in the case of unbroken isospin symmetry. Early attempts due to Ikeda, Ohnuki and Ogawa included grouping the proton, neutron and $\Lambda$ into a single representation $\mathbf{3}$ of $SU(3)$, their antiparticles into the $\bar{\mathbf{3}}$ representation and regarding all other hadrons as composites of these particles[16]. This allowed the known pseudoscalar mesons to be grouped into an octet via the decomposition

$$\mathbf{3} \otimes \bar{\mathbf{3}} = \mathbf{8} \oplus \mathbf{1} \tag{1.2}$$

Though incorrect, this led Gell-Mann, Ne'eman, Speiser and Tarski to independently posit the correct resolution that the nucleon $N$ and the $\Lambda, \Sigma$ and $\Xi$ form an $\mathbf{8}$ of $SU(3)$[17–19]. The $\rho, \omega$ and $K(892)$ could be grouped similarly. From such considerations there followed useful relations such as the Gell-Mann-Okubo mass formula for $SU(3)$ multiplets of baryons, leading to the prediction[20]

$$M_N + M_\Xi = \frac{1}{2}(M_\Lambda + 3M_\Sigma) \tag{1.3}$$

which was found to match data well.

The grouping of hadrons into the adjoint representation $\mathbf{8}$ of $SU(3)$ was suggestive that there might be a use for the fundamental representation, and indeed in 1964 Gell-Mann and Zweig independently proposed[21,22] that baryons and mesons were made up of particles of non-integral electric charges which could be grouped into a **3**, the $u$,$d$ and $s$[3]. Mesons could then be interpreted as bound $q\bar{q}$ states and baryons $qqq$ states, with mass differences accomodated by breaking of the $SU(3)$ through an $s$, $(u, d)$ mass splitting. The $SU(3)$ could even be further extended into an $SU(6)$ by incorporating the spin, $(u\uparrow, u\downarrow, d\uparrow, d\downarrow, s\uparrow, s\downarrow)$ which contains both the $SU(3)$ flavour group but also the $SU(2)$ spin for fixed flavour.

Concerns about the quark picture grew from an unease about fractionally charged states but also from a disturbing implication of the $SU(6)$ grouping. It was necessary to neglect orbital angular momentum when grouping quarks in order to produce the best results, but this resulted in the three-quark wavefunction being completely symmetric under interchange of the particles. This observation was in contradiction with Fermi-Dirac statistics, which dictate that such a wavefunction be completely antisymmetric. Several resolutions were put forward, including the suggestion that a kind of 'parastatistics' goverened quark dynamics, but the successful answer was to introduce a new degree of freedom to the quarks, taking three values, which would render the overall wavefunction antisymmetric. This idea from Han and Nambu[23] led them to associate a new $SU(3)$ symmetry group with the quarks which began to be known as colour—each of the $u, d, s$ was conceived as a triplet in the colour space while the bound states of hadrons were singlets and thus colourless. In 1973 Fritzsch, Gell-Mann and Leutwyler[24] combined these ideas with the long-standing Yang-Mills concept of a non-Abelian gauge theory[25] and regarded the colour degree of freedom as the charge of a field. They thus created the theory we now know as quantum chromodynamics (QCD), which has been extremely well verified.

Despite this, the growth of the strong coupling at low energies means that no coloured particle has ever been directly observed—quarks exist as bound states of hadrons, confined by the strong force. Coloured particles produced in proton-proton collisions at the LHC are immediately grouped into colourless objects created from the vacuum by their surrounding colour field, and 'jets' of mesons and baryons are detected by the experiments. Our evidence for the veracity of QCD as a theory of the strong interaction comes largely from the study of its behaviour in the perturbative regime, at energies where the coupling is sufficiently small that it is possible to perform calculations of observables. At present very little is known about the behaviour of the theory in regions where it is non-perturbative, i.e. at scales $Q \sim \Lambda_{\text{QCD}} \sim 200\,\text{MeV}$. Numerical approaches to tackle this regime exist and are based on discretisation of spacetime onto which gauge fields are placed—this approach

[3]The name quark is due to Gell-Mann, and originates in James Joyce's book Finnegans Wake:

> *Three quarks for Muster Mark!/Sure he hasn't got much of a bark/And sure any he has it's all beside the mark.*

is known as lattice QCD. Studies of this kind require extensive computing resources and are growing in capability.

## 1.3. Thesis content and structure

In this thesis, we present analytic calculations of higher order QCD corrections to processes of phenomenological relevance at hadron colliders at the LHC. Knowledge of such corrections is crucial in order to match the precision of experimental measurement currently being obtained. We then consider how knowledge of higher order corrections may be applied to experimental scenarios in order to make and test predictions for physically observable quantities.

Chapter 2 provides an introduction to some fundamental concepts in quantum chromodynamics sufficient to prepare the reader for the main content of the work. In Chapter 3, we consider jet amplitudes in which all involved partons are coloured and massless and look specifically at the case of $2 \to 3$ scattering. We provide an overview of the ingredients necessary to compute loop corrections to these processes and how this is done in practice. We take the process $q\bar{q} \to Q\bar{Q}g$ and present the amplitudes at 1- and 2-loop order, illustrating the difficulties commonly encountered in the computation of such amplitudes and how they may be overcome. In Chapter 4, we present higher order corrections to the heavy-light quark vertex and apply these to top quark production and decay processes. We provide full analytic results for the amplitudes for these processes.

We then take a more phenomenological tack and focus on the properties of heavy quarks. In Chapter 5, we use knowledge of the NNLO corrections to top quark pair production together with data obtained at the ATLAS and CMS experiments to perform a simultaneous extraction of two important parameters of the Standard Model, the strong coupling $\alpha_s$ and the top quark mass $m_t$. In particular, we consider measurements which are fully differential in some kinematic variables and minimise a goodness-of-fit parameter in order to obtain best-fit values. Chapter 6 deals with flavour scheme choice in the context of heavy-quark initiated processes and discusses when a heavy quark (such as the bottom or the top) can be considered a constituent of the proton. The specific class of processes dealt with are those in which a boson (Higgs or $Z$) is produced in association with two heavy quarks. We assess the reliability of the schemes and analyse theoretical predictions in the context of LHC experiments and of future colliders. Finally, we present some general comments and conclusions and paths towards future work.

# Chapter 2.

# QCD in a nutshell

The purpose of this section is not to provide a comprehensive overview of the subject nor to cover every aspect of the theoretical background, but rather to provide an introduction to some broad and general concepts that will be used in later chapters.

## 2.1. The QCD Lagrangian

QCD is a quantum Yang-Mills theory including fermions and built on the gauge group $SU(3)$. We begin, as is customary, with the Lagrangian

$$\mathcal{L}_{\text{QCD}} = -\frac{1}{4}F^{a\mu\nu}F^{a}_{\mu\nu} + \sum_f (\bar{q}_f)_i(i\not{D} - m_f)_{ij}(q_f)_j - \frac{1}{2\xi}(\partial^\mu A^a_\mu)^2 + \partial^\mu\bar{\eta}^a D^{bc}_\mu \eta^c \tag{2.1}$$

which is composed of terms for the gauge fields, fermionic fields, a gauge fixing term and a ghost term. Some explanation is due. We adopt the Feynman slash notation so that $\not{p} \equiv p_\mu\gamma^\mu$ and the Dirac gamma matrices satisfy the Clifford algebra

$$\{\gamma^\mu, \gamma^\nu\} = 2g^{\mu\nu}. \tag{2.2}$$

We identify the fermionic fields as being spin-$\frac{1}{2}$ quarks of flavour $f$ and mass $m_f$ which transform under the fundamental representation of the gauge group $SU(3)$—accordingly, they carry the group indices $i, j$. The kinetic term for the gluon field is given in terms of a gauge invariant object constructed from the field strength tensor

$$F^a_{\mu\nu} = \partial_\mu A^a_\nu - \partial_\nu A^a_\mu - g_s f^{abc} A^b_\mu A^c_\nu \tag{2.3}$$

where the spin-1 gluon fields $A^a_\mu$ live in the adjoint representation. The final term, not present in QED, indicates the non-Abelian nature of the gauge group and gives rise to self-interactions of the gluons. We note the appearance of the QCD coupling $g_s$ and the structure constants of the group $f^{abc}$.

Interactions between the quarks and gluons are encoded in the covariant derivative

$$(D_\mu)_{ij} = \partial_\mu \delta_{ij} + i g_s T^a_{ij} A^a_\mu \tag{2.4}$$

where the $T^a$ are traceless, Hermitian matrices in the fundamental representation and furnish the Lie algebra

$$[T^a, T^b] = i f^{abc} T^c. \tag{2.5}$$

### 2.1.1. Gauge invariance

In constructing our field theory, we demanded local invariance of the quark sector of the Lagrangian under $SU(3)$ transformations

$$q(x) \to q'(x) = \exp(igT \cdot \theta(x)) q(x) \equiv V(x) q(x), \tag{2.6}$$

'local' indicating that they differ at every point in spacetime. This was made possible by the introduction of the covariant derivative, transforming in the same way as the field

$$D_\mu q(x) \to D'_\mu q'(x) = V(x) D_\mu q(x) \tag{2.7}$$

where we have dropped colour labels. The implication is then that the gluon field transforms as

$$T \cdot A'_\mu = V(x)(T \cdot A_\mu) V^{-1}(x) + \frac{i}{g_s}(\partial_\mu V(x)) V^{-1}(x) \tag{2.8}$$

and so

$$A'^a_\mu = A^a_\mu - \partial_\mu \theta^a(x) - g_s f^{bca} \theta^b(x) A^c_\mu \tag{2.9}$$

in order to render the entire Lagrangian invariant.

A gauge theory with the first two terms in 2.1 alone corresponding to matter and gauge fields encounters a problem. We still have a gauge freedom under

$$A'^a_\mu = A^a_\mu - D^{ab}_\mu \theta^b(x) \tag{2.10}$$

which is just a rewriting of 2.9 using the covariant derivative in the adjoint representation

$$(D_\mu)^{ab} = \partial_\mu \delta^{ab} + g_s f^{bca} A^c_\mu. \tag{2.11}$$

The interpretation is that our definition of the field $A^a_\mu$ is not unique and we have a redundancy in our description (ultimately related to our insistence on embedding the two physical degrees of freedom, the transverse gluon polarisations, in a four component vector field). We can remedy this by imposing a condition on $A^a_\mu$, for example $\partial^\mu A^a_\mu = 0$, a choice called gauge fixing. This is easiest to do by introducing an auxiliary field to the Lagrangian which acts as a Lagrange multiplier,

$$\mathcal{L}_{\text{gauge-fixing}} = -\frac{1}{2\xi}(\partial^\mu A^a_\mu)^2 \tag{2.12}$$

to enforce the gauge choice. This particular form is known as covariant or $R_\xi$ gauge and has the advantage of being Lorentz invariant. There remains a choice of $\xi$ to be made—common choices are Feynman gauge ($\xi = 1$), Lorenz gauge ($\xi = 0$) and unitary gauge ($\xi \to \infty$). Each has their own advantage depending on the situation, and all physical quantities calculated are naturally independent of the gauge choice.

In an Abelian theory such as QED, this is sufficient to fix the dynamics of the fields. In QCD, however, it is necessary to add a term

$$\mathcal{L}_{\text{ghost}} = \partial^\mu \bar{\eta}^a D^{bc}_\mu \eta^c \tag{2.13}$$

where the $\eta^a$ are anti-commuting scalar fields known as Fadeev-Popov ghosts. Without entering into the details of the path integral derivation of gauge invariance, it suffices to say that these unphysical particles are necessary in order to cancel the unphysical timelike and longitudinal polarisations of gluons which may propagate within loops.

The origin of the ghost term in the Lagrangian results from our insistence that the gluon propagator resulting from the Lagrangian be Lorentz covariant. An alternative is to make a gauge choice in which the ghosts decouple from the physical states—in the spirit of 'conservation of difficulty', the unfortunate consequence is that such gauge choices violate Lorentz invariance. An example of a non-covariant gauge is known as the axial gauge and results in a term in the Lagrangian

$$\mathcal{L}_{\text{gauge-fixing}} + \mathcal{L}_{\text{ghost}} = -\frac{1}{2\xi}(n^\mu A^a_\mu)^2 + \bar{\eta}^a n^\mu (\delta^{ac}\partial_\mu + g_s f^{abc} A^b_\mu)\eta^c \tag{2.14}$$

where we have introduced a reference vector $n^\mu$ as well as the parameter $\xi$ which plays a similar rôle to that in the covariant gauge. Comparing the forms of the gluon propagator, in covariant gauge we have

$$i\Pi^{\mu\nu ab}_{\text{covariant}} = \frac{-i\delta^{ab}}{p^2 + i\mathbf{0}} \left[ g_{\mu\nu} - (1-\xi)\frac{p_\mu p_\nu}{p^2} \right] \tag{2.15}$$

while in axial gauge

$$i\Pi^{\mu\nu ab}_{\text{axial}} = \frac{-i\delta^{ab}}{p^2 + i\mathbf{0}} \left[ g_{\mu\nu} - \frac{n^\mu p^\nu + n^\nu p^\mu}{n \cdot p} + \frac{(n^2 + \xi p^2) p^\mu p^\nu}{(n \cdot p)^2} \right]. \tag{2.16}$$

For $\xi = 0$, we see that the latter satisfies $n_\mu \Pi^{\mu\nu ab}_{\text{axial}} = 0$ and since the ghost vertex is proportional to $n^\mu$, the ghosts decouple completely from the theory. A particular case in which $n^2 = 0$ is known as lightcone gauge, in which only polarisations transverse to the $n$-$p$ plane propagate—such a gauge can be useful in processes involving multiple gluons.

### 2.1.2. Ward identities

A massless, spin-1 gauge boson state is specified by both its momentum and its polarisation $|p, \varepsilon^j\rangle$ so that we have

$$\langle 0| A_\mu(x) |p, \varepsilon^j\rangle = \varepsilon^j_\mu e^{-ip\cdot x}. \tag{2.17}$$

In the massless case, the index $j = 1, 2$—the polarisation state of the boson can be expressed as a linear combination of two physical basis states which are transverse to the momentum. Under Lorentz transformations however, these two polarisation states are in general mixed with the momentum so that

$$\varepsilon^i_\mu(p) \to c^{ij}(\Lambda)\varepsilon^j_\mu + c^{i3}(\Lambda) p_\mu \tag{2.18}$$

where $\Lambda$ is a general Lorentz transformation. The transformed polarisation vector lies outside of the Hilbert space spanned by $\varepsilon^1$,$\varepsilon^2$. To see what consequence this has, consider an amplitude involving an external gluon which we can write as

$$\mathcal{M} = \mathcal{M}_\mu \varepsilon^\mu. \tag{2.19}$$

Under a Lorentz transformation we find

$$\mathcal{M} \to \varepsilon'^\mu \mathcal{M}'_\mu + c(\Lambda) p^\mu \mathcal{M}'_\mu \tag{2.20}$$

where $\mathcal{M}'_\mu = \Lambda_{\mu\nu}\mathcal{M}_\nu$ and $\varepsilon'^\mu$ is a linear combination of the $\varepsilon_1, \varepsilon_2$. At present, it appears that there exists no physical polarisation for which the transformed matrix element is the same as in the original frame. If we insist on Lorentz invariance, we must have that

$$p^\mu \mathcal{M}_\mu = 0, \tag{2.21}$$

which is known as the Ward identity. Put differently, replacing the polarisation vectors in an amplitude with an external gluon by the gluon momentum recovers zero in a Lorentz invariant field theory.

In processes with several external gluons, the situation is more complicated than that in QED and actually the amplitude need not vanish but is rather related to processes involving external ghosts[26]. We will not discuss this further.

### 2.1.3. Colour algebra

In Equation 2.1, we saw that the quark fields carried an index $i$ in the fundamental representation of the gauge group which runs from 1 to the dimension $N_c$ (3 in the case of QCD). Said index is usually referred to as the 'colour' and called red, green or blue (hence 'chromodynamics'). Similarly, the gluons carry an index $a$ in the adjoint representation of the group, which runs from 1 to $N_c^2 - 1$ (8, in QCD).

The traceless, Hermitian generators of $SU(3)$ in the fundamental representation $T^a$ obey certain properties that can be deduced from the algebra and will prove useful when evaluating Feynman diagrams. In the following and in order to elucidate the group structure, we work in a generic $SU(N)$ group (the case of QCD where $N = 3$ follows trivially). We have

$$\text{Lie algebra} : [T^a, T^b] = if^{abc}T^c \tag{2.22}$$

$$\text{Tracelessness} : \text{Tr}(T^a) = 0 \tag{2.23}$$

$$\text{Normalisation} : \text{Tr}(T^a T^b) = \frac{1}{2}\delta^{ab} \tag{2.24}$$

$$\text{Anticommutator} : \{T^a, T^b\} = \frac{1}{N}\delta^{ab} + d^{abc}T^c \tag{2.25}$$

$$\text{Fierz identity}: T^a_{ij}T^a_{kl} = \frac{1}{2}\left(\delta_{il}\delta_{jk} - \frac{1}{N}\delta_{ij}\delta_{kl}\right) \tag{2.26}$$

where $f^{abc}$ and $d^{abc}$ are the antisymmetric and symmetric structure constants respectively.

A Casimir operator is one which commutes with all generators of $SU(N)$. The object $T^aT^a$ is such an operator which we can evaluate using a special case of the Fierz identity:

$$T^a_{ij}T^a_{jk} = C_F\delta_{ik} \tag{2.27}$$

where $C_F \equiv \frac{N^2-1}{2N}$ is the Casimir invariant in the fundamental representation. Similarly, in the adjoint we have

$$f^{abc}f^{abd} = C_A\delta^{cd} \tag{2.28}$$

and for the symmetric structure constants

$$d^{abc}d^{abd} = \frac{N^2-4}{N}\delta^{cd} \tag{2.29}$$

where $C_A \equiv N$ is the Casimir invariant in the adjoint representation. Note that in all cases the operator is proportional to the identity matrix of dimension $N$ or $N^2 - 1$ and so commutes with all generators in the corresponding representation.

Further identities involving the colour matrices can be found in Appendix A, as well as a proof of the Fierz identity.

### 2.1.4. Feynman rules

We are now in a position to write down the Feynman rules following from the Lagrangian. These will allow us to calculate the amplitude for any process in QCD for specified initial and final state particles simply by drawing all permitted diagrams (to a given order in the coupling $g_s$) and assigning them factors for each vertex, propagator and external leg. Conventionally, fermions are denoted with solid lines, gluons with 'springs' and ghosts with dotted lines.

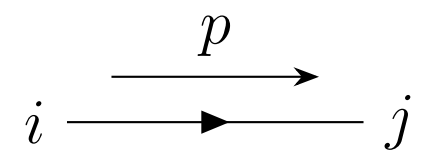

$$i\delta_{ij}\frac{(\not{p}+m)}{p^2-m^2+i\mathbf{0}}$$

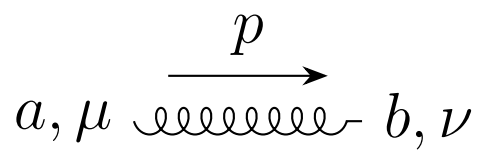

$$\frac{-i\delta^{ab}}{p^2+i\mathbf{0}}\left[g_{\mu\nu}-(1-\xi)\frac{p_\mu p_\nu}{p^2}\right]$$

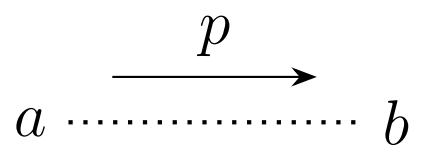

$$\frac{i\delta^{ab}}{p^2+i\mathbf{0}}$$

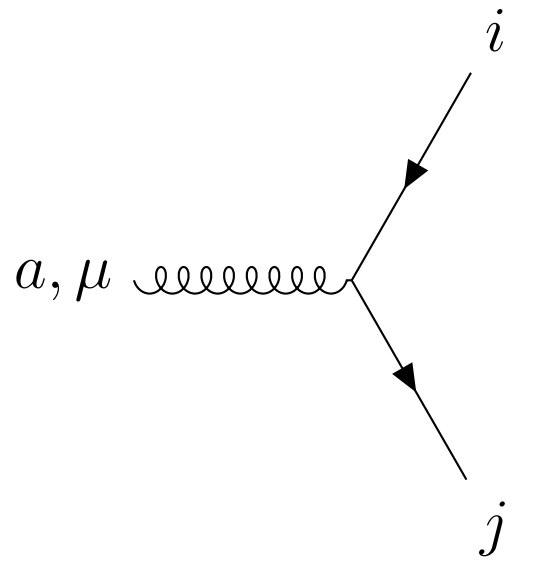

$$ig_s\gamma_\mu T^a_{ji}$$

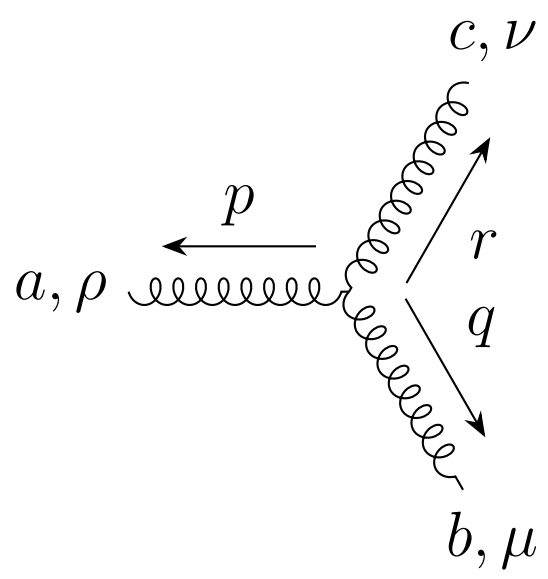

$$-\,g_s f^{abc}[(p-q)_\nu g_{\rho\mu}+(q-r)_\rho g_{\mu\nu}+(r-p)_\mu g_{\nu\rho}]$$

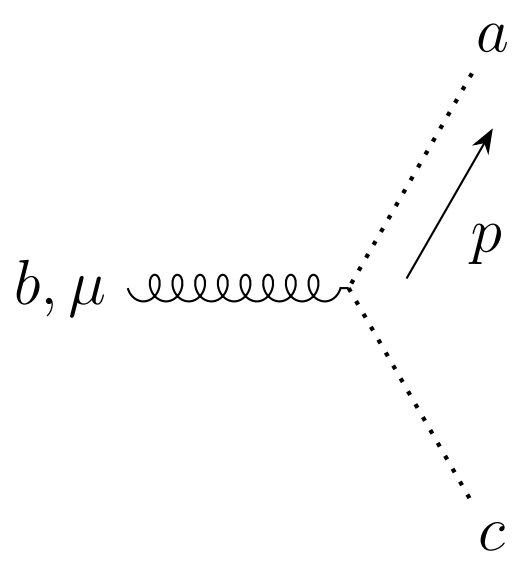

$$g_s f^{abc} p_\mu$$

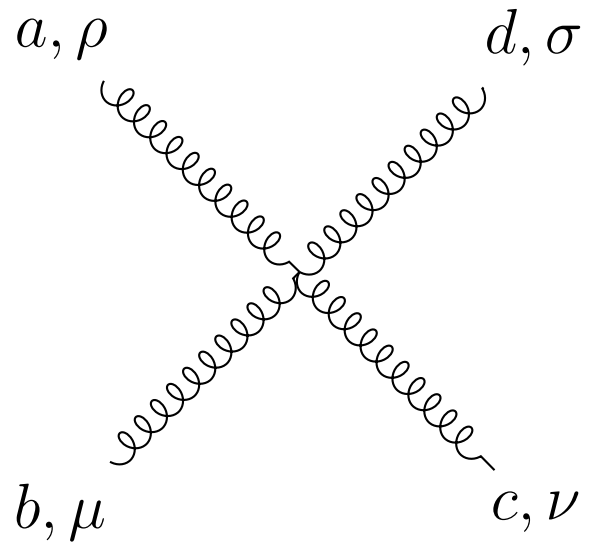

$$
\begin{aligned}
&- ig_s^2 f^{abe} f^{cde} (g_{\rho\nu} g_{\mu\sigma} - g_{\rho\sigma} g_{\mu\nu}) \\
&- ig_s^2 f^{ace} f^{bde} (g_{\rho\mu} g_{\nu\sigma} - g_{\rho\sigma} g_{\mu\nu}) \\
&- ig_s^2 f^{ade} f^{cbe} (g_{\rho\nu} g_{\mu\sigma} - g_{\rho\mu} g_{\sigma\nu})
\end{aligned}
$$

## 2.2. UV Renormalisation and IR Regularisation

A source of bafflement to our forerunners was the appearance of a multitude of divergences in quantum field theory when calculating in perturbation theory beyond the leading order. These divergences occurred when calculating loop diagrams which require an integral over an unobserved momentum—for example, consider the correction to the gluon propagator

$i\mathcal{M}^{\mu\nu} =$

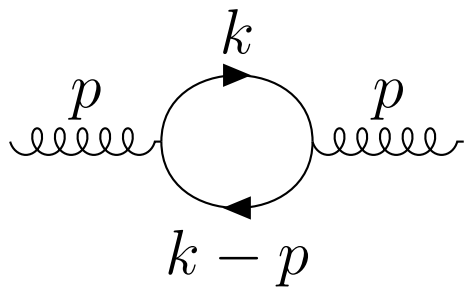

$$
= \quad g_s^2 \mathrm{Tr}(T^a T^b) \int \frac{d^4k}{(2\pi)^4} \frac{i}{k^2 - m^2 + i\mathbf{0}} \frac{i}{(k-p)^2 + m^2 + i\mathbf{0}} \mathrm{Tr}[\gamma^\mu (\not{k} - \not{p} + m) \gamma^\nu (\not{k} + m)].
$$

The loop integral we have to evaluate here is, for large $k$,

$$
\int \frac{d^4k}{(2\pi)^4} \frac{k^2}{k^4} \sim \int k\, dk \to \infty \tag{2.30}
$$

and hence is ultraviolet (UV) divergent. Another similar kind of divergence can occur in a loop integral, called infrared (IR) because it occurs at small values of the loop momentum—an example would be the integral

$$
\int \frac{d^4k}{(2\pi)^4} \frac{1}{(k-p)^2 k^2} \sim \int \frac{dk}{k} \tag{2.31}
$$

which is divergent at small $k$.

It was eventually realised that the UV divergences could be systematically tamed using a process called renormalisation in order to obtain a finite prediction for any physical quantity. The IR singularities could similarly be regularised such that the regulator dependence cancelled when a fully inclusive physical observable was calculated. There are several approaches to this problem. One, called the Pauli-Villars regularisation, relies on the introduction of new ghost particles of mass $\Lambda$ which cancel the divergent contributions to loop integrals from physical particles—the mass $\Lambda$ then acts as the UV regulator. It is also possible to introduce a hard cut-off to the loop momentum, though this violates Lorentz invariance. In practice, the most useful and ubiquitous method is known as dimensional regularisation[27] and involves continuing the number of spacetime dimensions from 4 to $d = 4 - 2\epsilon$. The loop integrals in this dimension are now convergent[1]. This has the advantage of regulating both UV and IR divergences and being gauge invariant. Divergences are manifested as poles in the regulator $\epsilon$, which is sent to zero once the poles have cancelled at the end of the calculation. We will treat each kind of divergence separately and make some brief comments on each.

### 2.2.1. UV divergences

The mistake at leading order in perturbation theory was our assumption that quantities in the Lagrangian, such as the strong coupling, were physically observable parameters. In fact, these quantities are formally infinite and can be renormalised to give finite values at a given order in perturbation theory. This is achieved by introducing similarly infinite renormalisation factors to absorb the divergences. We are only entitled to introduce one factor for each parameter in the theory—thus in QED, for example, we can introduce mass, electric charge, vacuum energy density, and electron and photon field factors.

To this end, we define field renormalisations

$$q^b = \sqrt{Z_2}\, q^R, \quad A^b_\mu = \sqrt{Z_3}\, A^R_\mu, \quad \eta^b = \sqrt{Z_{3\eta}}\, \eta^R \tag{2.32}$$

where the superscripts $b$ and $R$ denote bare and renormalised quantities respectively and the $Z_2, Z_3$ are field renormalisation factors. We have dropped colour labels for clarity. Similarly for the mass, we have

$$m^b = Z_m m^R \tag{2.33}$$

[1] While the UV divergent integrals may be made convergent for $d < 4$ and so $\epsilon_{\rm UV} > 0$, the IR divergent integrals are convergent in $d > 4$ and hence $\epsilon_{\rm IR} < 0$. Nevertheless, since we will find separate cancellations of UV and IR divergences we may just work with $\epsilon_{\rm UV} = \epsilon_{\rm IR} = \epsilon$.

and the strong coupling

$$g_s^b = Z_g g_s^R \tag{2.34}$$

By convention we also define $Z_1 \equiv Z_g Z_2 \sqrt{Z_3}$.

At this point, the renormalised quantities are finite at each order in perturbation theory. We are now able to compute physical observables which are free of UV divergences at a given order. We work in dimensional regularisation and analytically continue the number of spacetime dimensions to $d = 4 - 2\epsilon$. We expand the $Z_i$ around their leading order values so that we have

$$Z_i \equiv 1 + \delta_i \tag{2.35}$$

for the counterterms $\delta_i$ starting at order $g_s^2$. We may then introduce additional Feynman rules in our now renormalised perturbation theory which correspond to counterterm insertions on the field lines. It is also necessary to rescale our coupling to ensure it remains dimensionless, which involves introducing a mass scale $\mu$ via

$$g_s \to \mu^{\frac{4-d}{2}} g_s. \tag{2.36}$$

The scale $\mu$ is often called the subtraction point. The counterterms themselves are derived by considering loop corrections to the propagators and vertices and isolating the divergent parts. There are a great number of different conventions adopted by different authors differing by numerical factors and functions of $\epsilon$. We will refrain from giving a complete set of counterterms here, but instead introduce and define them as needed in later chapters. Note that there is a certain ambiguity in determining the counterterms—adding any finite piece will still produce a finite result. The choice of the finite part is known as the subtraction scheme, a choice which any calculated observable must be independent of. It is often convenient to use the modified minimal substraction ($\overline{\text{MS}}$) scheme, in which all finite parts are subtracted along with factors of $\log 4\pi$ and $\gamma_E$ which arise in dimensional regularisation. There exist alternative choices—an important case is known as the on-shell scheme, in which each renormalised mass in the theory $m^R$ is identified with the location of the pole in the propagator. This defines the pole mass and results in counterterms with a non-trivial finite part, as opposed to the $\overline{\text{MS}}$ case. We will use combinations of the schemes in later chapters.

### 2.2.2. IR divergences

Unlike the ultraviolet case, in which divergences were associated with large values of the internal momenta and hence short distance physics, infrared divergences are concerned with long distance behaviour. To demonstrate, we give a specific example of more general behaviour and consider the emission of a gluon from a quark leg:

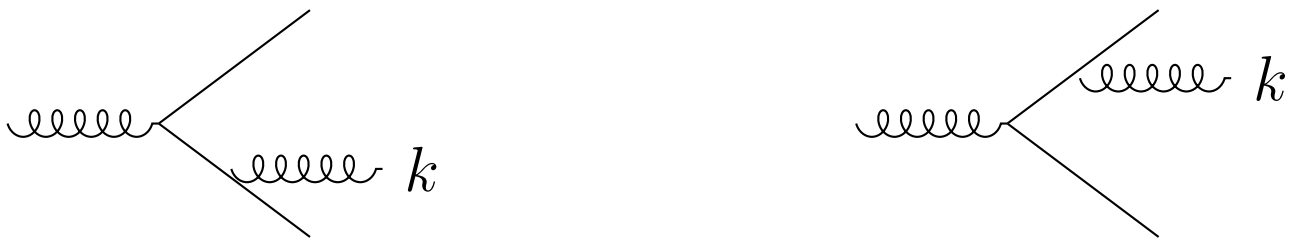

The amplitude for such a process can be obtained from the Feynman rules and is given by

$$\mathcal{M}^{\mu}_{q\bar{q}g} = (ig_s)^2 T^a T^b \left( \bar{u}(p_1)\not{\varepsilon}\frac{i}{\not{p}_1 + \not{k}}\gamma^{\mu} v(p_2) - \bar{u}(p_1)\gamma^{\mu}\frac{i}{\not{p}_2 + \not{k}}\not{\varepsilon} v(p_2) \right). \tag{2.37}$$

Taking the eikonal approximation in which $k \ll p_1, p_2$ we have

$$\mathcal{M}^{\mu}_{q\bar{q}g} \approx i g_s^2 T^a T^b \bar{u}(p_1)\gamma^{\mu} v(p_2) \left( \frac{p_1 \cdot \varepsilon}{p_1 \cdot k} - \frac{p_2 \cdot \varepsilon}{p_2 \cdot k} \right) \tag{2.38}$$

where we have made use of the equations of motion for a massless (anti)spinor $\not{p}u(p) = 0$ and the Dirac algebra. The squared amplitude is then given by

$$|\mathcal{M}_{q\bar{q}g}|^2 \approx |\mathcal{M}_{q\bar{q}}|^2 g_s^2 C_F \frac{2p_1 \cdot p_2}{(p_1 \cdot k)(p_2 \cdot k)} \tag{2.39}$$

where $|\mathcal{M}_{q\bar{q}}|^2$ is the corresponding amplitude without the emitted gluon. We see the gluon emission factorises out of the squared amplitude. Rewriting in terms of observables, the 'eikonal' factor is

$$\frac{2p_1 \cdot p_2}{(p_1 \cdot k)(p_2 \cdot k)} \propto \frac{1}{E^2(1 - \cos^2\theta)} \tag{2.40}$$

where $E$ and $\theta$ are the energy and angle of the emitted gluon respectively. Including the phase space, the expression for emission in the eikonal approximation is then

$$\frac{2\alpha_s C_F}{\pi}\frac{dE}{E}\frac{d\theta}{\sin\theta}\frac{d\phi}{2\pi} \tag{2.41}$$

which is clearly divergent in two limits—either as $E \to 0$ or as $\theta \to 0$. The former is known as a soft divergence and the latter a collinear divergence.

While these divergences may initially seem problematic, once again they can be remedied by full consideration of the definition of an observable. Since soft and collinear gluons are not resolvable, any detector will not 'see' them and the observed process will be the same as that without the emission. We should therefore treat these emissions as $\mathcal{O}(\alpha_s)$ corrections to the 'bare' process and so on the same footing as loop corrections. In fact, a theorem due to Kinoshita, Lee and Nauenberg[28,29] states that infrared divergences will cancel in calculating an observable when all possible initial and final states are summed over. In the case of massless QCD, this means that beyond leading order in perturbation theory we must consider $l$-loop corrections with $n$ particles in the final state as well as $(l-1)$-loop corrections with $n+1$ particles, $(l-2)$-loop corrections with $n+2$ particles *&c.*

It should be noted that unlike in the UV case, IR singularities only cancel once the full cross section is computed, the integral over the phase space has been performed and initial state collinear singularities are factored out. It is also the case that in situations where multiple gluon emissions must be considered, the pleasing factorisation property of the soft emission that we observed in the single gluon case is no longer manifest. The amplitude must then be reorganised into parts which are finite and those which are divergent in only one limit.

### 2.2.3. Varieties of dimensional regularisation

Thus far we have been rather imprecise in our definition of dimensional regularisation. The choice of how to continue the number of dimensions is not unique and three different schemes are commonly used. In Conventional Dimensional Regularisation (CDR), all quantities are treated as $d$-dimensional, including internal and external momenta and all polarisation vectors. In the 't Hooft-Veltman scheme (tHV), while momenta and helicities of the internal particles are $d$-dimensional, those of the external particles are 4-dimensional. Finally, in Dimensional Reduction (DR) it is only the momenta of the internal particles that are $d$-dimensional, and all other momenta and helicities are 4-dimensional. Properties of the schemes are summarised in Table 2.1.

There are certain subtleties associated with the treatment of the matrix $\gamma^5 = i\gamma^0\gamma^1\gamma^2\gamma^3$ in $d$ dimensions[30]. An example of a difficulty that may arise is the incompatibility of the properties

$$\{\gamma^\mu, \gamma^5\} = 0 \tag{2.42}$$

and

$$\mathrm{Tr}\left(\gamma^\mu\gamma^\nu\gamma^\rho\gamma^\sigma\gamma^5\right) \neq 0 \tag{2.43}$$

| | CDR | tHV | DR |
|---|---|---|---|
| | $\gamma^\mu = \hat{\gamma}^\mu + \tilde{\gamma}^\mu$ | $\gamma^\mu = \hat{\gamma}^\mu + \tilde{\gamma}^\mu$ | $\gamma^\mu = \hat{\gamma}^\mu, \quad \tilde{\gamma}^\mu = 0$ |
| $\{\gamma^5, \gamma^\mu\}$ | 0 | Eq. 2.45 | 0 |
| Internal mom. | $k = \hat{k} + \tilde{k}$ | $k = \hat{k} + \tilde{k}$ | $k = \hat{k} + \tilde{k}$ |
| External mom. | $p_i = \hat{p}_i + \tilde{p}_i$ | $p_i = \hat{p}_i, \quad \tilde{p}_i = 0$ | $p_i = \hat{p}_i, \quad \tilde{p}_i = 0$ |
| Int. gluon pol. | $d-2$ | $d-2$ | 2 |
| Ext. gluon pol. | $d-2$ | 2 | 2 |

**Table 2.1.:** Varieties of dimensional regularisation.

in $d$ dimensions. In 4 dimensions the former would hold and the latter trace would give

$$\mathrm{Tr}\left(\gamma^\mu \gamma^\nu \gamma^\rho \gamma^\sigma \gamma^5\right) = 4i\varepsilon^{\mu\nu\rho\sigma}. \tag{2.44}$$

We might therefore expect a smooth continuation to the $d$-dimensional case, but apparently this can only be true if the anticommutator is non-vanishing. The consequence is the breaking of gauge invariance, clearly undesirable. A pragmatic approach is to split the gamma matrices into 4 and $2\epsilon$ dimensional parts $\gamma^\mu = \hat{\gamma}^\mu + \tilde{\gamma}^\mu$ where $\hat{\gamma}^\mu$ is 4-dimensional and $\tilde{\gamma}^\mu$ is $2\epsilon$-dimensional. We may then leave $\gamma^5$ in 4 dimensions and

$$\{\gamma^\mu, \gamma^5\} = \begin{cases} 0 & \mu \in \{0,1,2,3\} \\ 2\tilde{\gamma}^\mu \gamma^5 & \text{otherwise.} \end{cases} \tag{2.45}$$

The interpretation is that $\gamma^5$ acts normally in the physical dimensions and trivially otherwise, $[\gamma^5, \tilde{\gamma}^\mu] = 0$. When also applied to the metric and momenta, this dimension splitting separates the $d$-dimensional space into orthogonal subspaces.

### 2.2.4. The running coupling

The mass scale introduced in Equation 2.36 in order to keep the coupling dimensionless is clearly an unphysical object. It does not appear in the bare QCD Lagrangian and is therefore an artefact of our regularisation scheme. Ergo, any observable $O$ should be independent of the choice of $\mu$. Consider a dimensionless observable which depends on a large scale $Q \gg m_q$. If it were true that $Q$ were the only scale of the problem, the dependence of $O$ should be completely flat—the introduction of the scale $\mu$, however, means in general $O$ is a function of the ratio $\frac{Q^2}{\mu^2}$. Since $\alpha_s$ is also renormalised, it too should gain some $\mu$ dependence. Insisting on independence of the observable on $\mu$ implies

that

$$\mu^2 \frac{d}{d\mu^2} O(Q^2/\mu^2, \alpha_s) = \left( \mu^2 \frac{\partial}{\partial \mu^2} + \mu^2 \frac{\partial \alpha_s}{\partial \mu^2} \frac{\partial}{\partial \alpha_s} \right) O = 0. \tag{2.46}$$

Introducing a compact notation

$$t = \log \left( \frac{Q^2}{\mu^2} \right), \qquad \beta(\alpha_s) = \mu^2 \frac{\partial \alpha_s}{\partial \mu^2}, \tag{2.47}$$

we may rewrite Equation 2.46 as

$$\left( -\frac{\partial}{\partial t} + \beta(\alpha_s) \frac{\partial}{\partial \alpha_s} \right) O(e^t, \alpha_s) = 0. \tag{2.48}$$

We may solve this equation by defining the running coupling $\alpha_s(Q^2)$ as

$$t = \int_{\alpha_s}^{\alpha_s(Q^2)} \frac{dx}{\beta(x)}, \qquad \alpha_s(\mu^2) \equiv \alpha_s \tag{2.49}$$

and differentiating we find

$$\frac{\partial \alpha_s(Q^2)}{\partial t} = \beta(\alpha_s(Q^2)), \qquad \frac{\partial \alpha_s(Q^2)}{\partial \alpha_s} = \frac{\beta(\alpha_s(Q^2))}{\beta(\alpha_s)}. \tag{2.50}$$

We see that all the scale dependence of $O$ is absorbed by the running of the coupling since $O(1, \alpha_s(Q^2))$ solves this renormalisation group equation. We can therefore predict the value of $O$ at any scale $Q$ if we know the behaviour of $\alpha_s(Q^2)$.

We consider the calculation of the $\beta$ function. Expressing the bare coupling in terms of the renormalised coupling

$$\alpha_s^b = \alpha_s \frac{Z_1}{Z_2 \sqrt{Z_3}} \mu^{2\epsilon} \tag{2.51}$$

where it is now understood that $\alpha_s$ is renormalised, we note the $\mu$ independence of the bare coupling and write

$$\mu^2 \frac{d}{d\mu^2} \alpha_s^b = \mu^2 \frac{d}{d\mu^2} \left( \alpha_s \frac{Z_1}{Z_2 \sqrt{Z_3}} \mu^{2\epsilon} \right) = 0. \tag{2.52}$$

We now may expand the $Z_i$ to arbitrary order and, using the definition of the $\beta$ function and the expressions for the counterterms, we find that

$$\beta(\alpha_s) = -\alpha_s \left[ \beta_0 \left( \frac{\alpha_s}{4\pi} \right) + \beta_1 \left( \frac{\alpha_s}{4\pi} \right)^2 + \dots \right] \tag{2.53}$$

where

$$\beta_0 = \frac{11}{3}C_A - \frac{2}{3}n_f \tag{2.54}$$

and

$$\beta_1 = \frac{34}{3}C_A^2 - \frac{10}{3}C_A n_f - 2C_F n_f. \tag{2.55}$$

Truncating the series at first order and solving, we have

$$\alpha_s(Q^2) = \frac{\alpha_s(\mu^2)}{1 + \alpha_s(\mu^2)\frac{\beta_0 t}{4\pi}}. \tag{2.56}$$

We see that as $t$ increases $\alpha_s(Q^2)$ decreases as long as the number of light flavours $n_f < 17$, a property known as asymptotic freedom. It is this property that means we can resort to perturbative calculations at high energies (where $\alpha_s$ is small) and that results in confinement at low energies (where $\alpha_s$ is large). This behaviour follows from the sign of $\beta_0$ and comes ultimately from the non-Abelian nature of QCD.

Note that it is common in the $\overline{\text{MS}}$ scheme to take

$$\mu^2 = \frac{\mu_R^2 e^{\gamma_E}}{4\pi} \tag{2.57}$$

which defines the $\overline{\text{MS}}$ renormalisation scale $\mu_R$. This has the advantage of removing factors of $\gamma_E - \log(4\pi)$ which accompany the poles arising in loop integrals.

### 2.2.5. Decoupling of heavy particles

In cases where a single mass scale $M$ in a problem is much greater than other typical scales, it can be useful to consider an effective theory for scales $\Lambda \ll M$ where the corrections due to the large mass entering in loops are of order $1/M$ and all other interactions are local. An example would be the top quark in QCD, which with a mass $m_t \sim 173.3$ GeV lies far above the other quarks—the next heaviest is the bottom quark, with a mass $m_b \sim 4.18$ GeV. At processes involving energy scales $\sim \mathcal{O}(1)$ GeV, the top is never produced as a final state and appears only within loops. All operators of the full theory can be expanded in $1/M$, with the relationship between the bare quantities in the two theories being given

by

$$
\begin{aligned}
q^b &= \sqrt{\zeta_q^b} q'^b, \quad A_\mu^b = \sqrt{\zeta_A^b} A_\mu'^b, \\
g_s^b &= \sqrt{\zeta_\alpha^b} g_s'^b
\end{aligned}
\tag{2.58}
$$

and for the $\overline{\text{MS}}$ renormalised quantities

$$
\begin{aligned}
q(\mu) &= \sqrt{\zeta_q(\mu)} q'(\mu), \quad A(\mu) = \sqrt{\zeta_A(\mu)} A'(\mu), \\
\alpha_s(\mu) &= \zeta_\alpha(\mu) \alpha_s'(\mu)
\end{aligned}
\tag{2.59}
$$

where

$$
\begin{aligned}
\zeta_q(\mu) &= \frac{Z_q'(\alpha_s'(\mu))}{Z_q(\alpha_s(\mu))} \zeta_q^b, \quad \zeta_A(\mu) = \frac{Z_A'(\alpha_s'(\mu))}{Z_A(\alpha_s(\mu))} \zeta_A^b, \\
\zeta_\alpha(\mu) &= \frac{Z_\alpha'(\alpha_s'(\mu))}{Z_\alpha(\alpha_s(\mu))} \zeta_\alpha^b
\end{aligned}
\tag{2.60}
$$

In order to determine the matching between the theories, we consider the renormalised propagators, vertices *&c.* near the mass shell where $p^2 = 0$—in this way, corrections of $\mathcal{O}(1/M^2)$ play no rôle. We can determine the $\zeta_i$ to arbitrary order[31]; we will later be interested in the effect of heavy quarks on the running of the coupling and so discuss this here. For QCD with $n_l$ light quarks and a top quark of mass $m_t$, the couplings in the effective and full theories are related by

$$
\alpha_s^{(n_l+1)}(\mu) = \zeta_\alpha(\mu) \alpha_s^{(n_l)}(\mu). \tag{2.61}
$$

To calculate $\zeta_\alpha$, we require the loop corrections to the gluon and ghost propagators and the gluon-ghost vertex. We find that[32]

$$
\zeta_\alpha(\mu) = 1 + \frac{\alpha_s^{(n_l)}(\mu)}{4\pi} \left[ -\frac{1}{6} \log\left(\frac{\mu^2}{m_t^2}\right) \right] + \mathcal{O}(\alpha_s^2). \tag{2.62}
$$

Although not apparent here, at higher orders the running coupling now has a discontinuity when crossing flavour thresholds (for example, at NNLO such a discontinuity occurs for $\mu = m_b$). It is important to use the 5 and 6 flavour couplings when appropriate—were one to use the full 6-flavour running of the coupling at lower energies, the convergence of the perturbative expansion may well be spoiled due to large logarithms.

### 2.2.6. Scale variation and uncertainty

Calculations truncated at fixed order will display some degree of scale dependence, both explicitly and also through the running coupling. The arbitrariness associated with the choice of the unphysical scale $\mu_R$ is problematic as it leads to an arbitrariness in a prediction for a physical observable. The scale is normally chosen to be of the order of other hard scales in the process such as the mass of the final state particles (this prevents the possibility of logarithms $\log(\mu^2/Q^2)$ becoming dangerously large), but there is no particular reason *a priori* that one should not choose that scale varied by a factor of 2, say. In practice what is done is to examine the behaviour of the theoretical prediction as a function of the scale over some interval, perhaps $Q/2 < \mu_R < 2Q$. This provides an estimate of the effect of missing higher order corrections (since at all orders, the scale should disappear completely from physical predictions). As the order of the perturbative calculation is increased, the dependence should become increasingly flat if the perturbative series is well-behaved. We consider the variation in the prediction to be a source of theoretical error (the scale uncertainty), though of a somewhat unusual kind.

There exist claims that the choice of scale can be made unambiguously based on considerations of the Principle of Maximum Conformality, which aims to remove the $n_f$ dependence from the running of the coupling at each order[33,34]. Recent work has shown, however, that this does not result in a substantial reduction of the ambiguity when truncating the series at finite order.

## 2.3. QCD at colliders

### 2.3.1. Cross sections and decay rates

We turn to a more phenomenologically focussed discussion. The observable quantities in an experiment are not the amplitudes obtained from the Feynman rules themselves, but rather interaction rates within a finite spacetime interval. We must also include the number density of final states and normalise to a given flux of incident particles when constructing a measurable quantity. Each incident beam in a collider experiment presents an effective area related to the probability of any two constituents of the beam interacting, known as a cross section, which accounts for the aforementioned considerations:

$$d\sigma = \frac{|\mathcal{M}|^2}{4\sqrt{(p_1 \cdot p_2)^2 - m_1^2 m_2^2}} \left( \prod_f \int \frac{d^3 p_f}{(2\pi)^3 2E_f} \right) (2\pi)^4 \delta^{(4)} \left(p_1 + p_2 - \Sigma p_f\right) \tag{2.63}$$

where $p_1$ and $p_2$ are the four momenta of the incoming particles and $f$ runs over all final state configurations. In the case that there is only one unstable particle in the initial state, what we measure instead is a decay rate

$$d\Gamma = \frac{1}{2m}|\mathcal{M}|^2 \left( \prod_f \int \frac{d^3 p_f}{(2\pi)^3 2E_f} \right) (2\pi)^4 \delta^{(4)} \left(p - \Sigma p_f\right). \tag{2.64}$$

Upon integration over the entire phase space, we may relate this to the average lifetime of the particle:

$$\tau = \frac{1}{\Gamma}. \tag{2.65}$$

### 2.3.2. The parton model and factorisation

In collider experiments, the objects that are dealt with are not the fundamental quarks and gluons which appear in the Lagrangian but rather composite hadronic states (at the LHC, the proton). The proton structure was initially deduced from deep inelastic scattering experiments, in which electron beams were fired at a stationary proton target. At high energies it was found that the scattering cross section was largely scale invariant, suggestive of the existence of point-like objects within the proton from which electrons were elastically scattered. The most successful description, known as the parton model, supposes that it is composed of quasi-free constituents known as partons which, for most intents and purposes, can be identified with the quarks and gluons of QCD. There exist three valence quarks in each proton, which give rise to its quantum numbers, and a host of other sea partons which include the gluons in the colour field binding the quarks and quark-antiquark pairs created from splittings of said gluons. The model is formulated in the so-called 'infinite momentum frame' where the proton is assumed to have very high energy and hence a negligible mass. The partons themselves move collinearly with the proton and each carry a fraction $x$ of its longitudinal momentum. The distributions of the momenta of the partons of different kinds are given by their Parton Distribution Functions (PDFs) $f_i(x)$, non-perturbative quantities which must be extracted from experiment. Various extractions exist and differ in their methodology, either utilising a parametrisation of the functions or resorting to more data-driven techniques[35–37].

In the 'naïve' parton model, we mentioned that there was an assumption that the transverse momenta of the partons were small. Due to the emission of gluons from the initial parton, however, this need not necessarily be the case and we should in general include these corrections. This involves an integral over the transverse momentum of the

gluon $k_T$:

$$\int_{\mu_0^2}^{Q^2} \frac{dk_T^2}{k_T^2} = \log\left(\frac{Q^2}{\mu_0^2}\right) \tag{2.66}$$

where the upper limit is set by some physical scale. The lower limit should properly be zero, but to avoid divergent behaviour it is necessary to introduce a cut-off at an arbitrary scale $\mu_0$. It can be shown[38] that, in order to render these corrections to the proton structure functions finite in the collinear limit, one can introduce a running parton density

$$f(x,\mu_F^2) = f(x,\mu_0^2) + \int_x^1 \frac{dy}{y} f(y,\mu_F^2)\frac{\alpha_s}{2\pi}\left[P\left(\frac{x}{y}\right)\log\left(\frac{\mu_F^2}{\mu_0^2}\right)\right] + \mathcal{O}(\alpha_s^2) \tag{2.67}$$

We see that the logarithmic behaviour arises from the transverse momentum integration and that there is a universal function $P$ (known as a splitting function) determined by the form of the quark-gluon vertex. The collinear singularities associated with the gluon splitting have been absorbed (or resummed) into the running of the PDF at a scale $\mu_F$, much in the same way as the virtual corrections to the gluon propagator were absorbed into the running coupling $\alpha_s$ at a scale $\mu_R$. Following by analogy, we may insist on independence of the scale $\mu_0$ to obtain the (Dokshitzer)-Gribov-Lipatov-Altarelli-Parisi (GLAP) equation[39–42]

$$t\frac{\partial f(x,t)}{\partial t} = \frac{\alpha_s(t)}{2\pi}\int_x^1 \frac{dy}{y} P\left(\frac{x}{y}\right) f(y,t) \tag{2.68}$$

where here $t = \mu^2$. The parton distributions themselves cannot be calculated ab initio, being objects defined at a non-perturbative scale[2]. However, once extracted at a given energy scale it is possible to deduce their scale dependence using the GLAP equations.

In general, the different species of parton present in the proton each have their own PDF $f_i(x,\mu_F^2)$ and these mix under scale evolution. We can calculate the splitting functions perturbatively for each possible splitting ($q \to qg$,$g \to gg$ *&c.*) and write the GLAP equations as

$$t\frac{\partial}{\partial t}\begin{pmatrix} f_i(x,\mu_F) \\ f_g(x,\mu_F)\end{pmatrix} = \sum_j \frac{\alpha_s(t)}{2\pi}\int_x^1 \frac{dy}{y}\begin{pmatrix} P_{q_i q_j}\left(\frac{x}{y}\right) & P_{q_i g}\left(\frac{x}{y}\right) \\ P_{g q_j}\left(\frac{x}{y}\right) & P_{gg}\left(\frac{x}{y}\right)\end{pmatrix}\begin{pmatrix} f_j(y,\mu_F) \\ f_g(y,\mu_F)\end{pmatrix}. \tag{2.69}$$

[2]Recent work in lattice QCD has made progress towards this through numerical simulation[43], but it is reasonable to say that the dynamics of a strongly coupled field theory are not well understood.

for quark flavours $i$. At leading order, the $P_{ij}(x)$ have a pleasing physical meaning—they may be interpreted as the probabilities of finding a parton of type $i$ in a parton of type $j$ with a fraction $x$ of the parent parton momentum. At this order, they are given by

$$\begin{aligned}
P_{qq}^{(0)}(x) &= C_F \left[ \frac{1+x^2}{[1-x]_+} + \frac{3}{2}\delta(1-x) \right], \\
P_{qg}^{(0)}(x) &= \frac{1}{2}[x^2 + (1-x)^2], \\
P_{gq}^{(0)}(x) &= C_F \left[ \frac{1+(1-x)^2}{x} \right], \\
P_{gg}^{(0)}(x) &= 2C_A \left[ \frac{x}{[1-x]_+} + \frac{(1-x)}{x} + x(1-x) \right] + \frac{\beta_0}{2}\delta(1-x)
\end{aligned} \tag{2.70}$$

where the 'plus prescription' is defined so that

$$\int_0^1 dx \frac{f(x)}{[1-x]_+} \equiv \int_0^1 dx \frac{f(x) - f(1)}{1-x} \tag{2.71}$$

to exclude the divergence at $x = 1$.

Bearing these considerations in mind, we are led to write down the form of the interaction cross section between two hadrons as

$$\sigma = \sum_{i,j} \int_{x_{1,\mathrm{min}}}^1 \int_{x_{2,\mathrm{min}}}^1 dx_1 dx_2 f_i(x_1, \mu_F^2) f_j(x_2, \mu_F^2) \hat{\sigma}_{ij}(x_1 p_1, x_2 p_2, Q, \ldots; \mu_F^2) \tag{2.72}$$

where we distinguish between the partonic cross section $\hat{\sigma}$, defined in the centre of mass frame of the colliding partons, and the hadronic cross section $\sigma$ which is the measured quantity, a convolution of the partonic case with the PDFs. Typically $x_{\mathrm{min}} \gtrsim Q^2/\hat{s}$ for $\hat{s} = (p_1 + p_2)^2$. This is an example of a factorisation theorem. We have been able to separate the short and long distance physics at a factorisation scale $\mu_F$[3]—on the one hand, we have a set of universal PDFs which characterise the proton and are process independent, and on the other we have a hard function which is process dependent and is characterised by the dynamics of the theory. Although factorisation is assumed to hold for all processes (and in fact all calculations at hadron colliders depend on it holding), it has only been proven to hold in a limited number of cases, including inclusive deep inelastic scattering and in Drell-Yan processes ($pp \to l^+ l^-$)[44].

[3]While the factorisation and renormalisation scales $\mu_F$ and $\mu_R$ are formally independent objects, in practice they are often taken to be the same.

### 2.3.3. Collider observables

Since in a given proton-proton collision we do not know what fractions of the proton momenta the interacting partons carry, it follows that the initial state partonic momenta in the centre of mass frame are not known. In measurement, it is therefore useful to restrict ourselves to variables which are defined with respect to boosts along the beam axis or practical to use even when Lorentz transformed. We will assume a cylindrical geometry orientated along the $z$-axis and begin by defining

$$\begin{aligned} p^\mu &= (E, p_x, p_y, p_z) \\ p_T^2 &= p_x^2 + p_y^2 \\ m_T &= \sqrt{p_T^2 + m^2} \end{aligned} \tag{2.73}$$

where $p_T$ is the momentum of the particle transverse to the beam axis and $m_T$ a quantity known as the transverse mass. These latter two quantities are invariant with respect to boosts along $z$, a useful property since in general the centre of mass frame of the partons is moving along the beam. We may also define the rapidity

$$y = \frac{1}{2}\log\left(\frac{E+p_z}{E-p_z}\right) = \text{artanh}\left(\frac{p_z}{E}\right), \tag{2.74}$$

a variable related to the angle of emission in the $x$-$z$ plane. We see that it varies from 0 (where the particle is emitted transverse to the beam axis) to $\pm\infty$ (where the particle is (anti)parallel to the beam). Under a Lorentz transformation, it behaves as

$$y \to y - \text{artanh}\beta, \tag{2.75}$$

also a useful property since this implies the difference between two rapidities is Lorentz invariant.

In practice, the rapidity is difficult to measure—at high values, the beam pipe can prevent a precise measurement of $p_z$. We therefore define the pseudorapidity

$$\eta = -\log\left(\tan\frac{\theta}{2}\right) \tag{2.76}$$

where $\theta$ is the angle to the $z$ axis. This requires measurement of only one angle. For highly relativistic particles, the pseudorapidity approximates the rapidity (and in fact $\eta|_{m=0} = y$).

# Chapter 3.

# Five parton multiloop amplitudes

In this chapter we describe the requisite ingredients for a computation in QCD beyond leading order and detail the general process for computing a loop amplitude. We present an application—the computation of a 1-loop, five-point jet amplitude in massless QCD which is of phenomenological relevance at hadron colliders such as the LHC. We also discuss the computation of five-point amplitudes at 2-loop order and present new results for the planar amplitude.

## 3.1. Beyond leading order in QCD

### 3.1.1. The perturbative expansion of amplitudes

The fact that QCD is both non-linear and strongly coupled means that, at present, no complete solution of the theory is known. It is only through the property of asymptotic freedom that we are able to calculate relevant quantities—at high energies $Q \gg \Lambda_{\text{QCD}}$, the smallness of the coupling $\alpha_s(Q) = g_s^2/(4\pi)$ allows us to perform a perturbative expansion which we can truncate at a fixed order to obtain an approximate result. We are generally interested in physical observables and so we consider objects such as scattering amplitudes, elements of the $S$-matrix which give the dynamical contribution to cross sections. We write

$$|\mathcal{M}\rangle = 4\pi\alpha_s^n \left( \left|\mathcal{M}^{(0)}\right\rangle + \left(\frac{\alpha_s}{4\pi}\right) \left|\mathcal{M}^{(1)}\right\rangle + \left(\frac{\alpha_s}{4\pi}\right)^2 \left|\mathcal{M}^{(2)}\right\rangle + \dots \right) \tag{3.1}$$

where the amplitude is a vector in the colour and spin space of the external particles and $n$ is the power of $\alpha_s$ present at lowest order for the process under consideration. The expansion may also be represented pictorially using Feynman diagrams. One sees that the lowest order 'Born' term $\left|\mathcal{M}^{(0)}\right\rangle$ is usually represented by tree-level diagrams while higher order terms are represented by diagrams containing an increasing number of loops of internal particles[1].

The object that enters into the calculation of cross sections is not the amplitude itself, but rather its square $|\mathcal{M}|^2 = \langle\mathcal{M}\,|\,\mathcal{M}\rangle$. Remembering that the cancellation of infrared singularities for an $N$ parton amplitude at order $n$ requires inclusion of information about the $N+1$ parton amplitude at order $n-1$, we see that at leading order we have

$$\sigma^{\mathrm{LO}} = \int d\Phi_N \left\langle \mathcal{M}_N^{(0)} \middle| \mathcal{M}_N^{(0)} \right\rangle \tag{3.2}$$

while at next-to-leading order

$$\sigma^{\mathrm{NLO}} = \int d\Phi_N \, 2\mathrm{Re}\left\{\left\langle \mathcal{M}_N^{(0)} \middle| \mathcal{M}_N^{(1)} \right\rangle\right\} + \int d\Phi_{N+1} \left\langle \mathcal{M}_{N+1}^{(0)} \middle| \mathcal{M}_{N+1}^{(0)} \right\rangle \tag{3.3}$$

and at next-to-next-to-leading order

$$\begin{aligned} \sigma^{\mathrm{NNLO}} = &\int d\Phi_N \, 2\mathrm{Re}\left\{\left\langle \mathcal{M}_N^{(0)} \middle| \mathcal{M}_N^{(2)} \right\rangle\right\} + \int d\Phi_N \, 2\mathrm{Re}\left\{\left\langle \mathcal{M}_N^{(1)} \middle| \mathcal{M}_N^{(1)} \right\rangle\right\} \\ &+ \int d\Phi_{N+1} \, 2\mathrm{Re}\left\{\left\langle \mathcal{M}_{N+1}^{(0)} \middle| \mathcal{M}_{N+1}^{(1)} \right\rangle\right\} + \int d\Phi_{N+2} \left\langle \mathcal{M}_{N+2}^{(0)} \middle| \mathcal{M}_{N+2}^{(0)} \right\rangle \end{aligned} \tag{3.4}$$

where we have collected all phase space and numerical factors in $d\Phi$. We see that the NLO corrections are of two kinds—we have the virtual contribution from loop diagrams $\left|\mathcal{M}_N^{(1)}\right\rangle$ and the real emission contribution $\left|\mathcal{M}_{N+1}^{(0)}\right\rangle$. Similarly, at NNLO we have double virtual, double real and real-virtual contributions.

### 3.1.2. Ultraviolet renormalisation of the coupling

In the $\overline{\mathrm{MS}}$ scheme, the bare coupling $\alpha_s^b$ is related to the running coupling $\alpha_s \equiv \alpha_s(\mu_R^2)$ at renormalisation scale $\mu_R$ by

$$\alpha_s^b = \mu_R^{2\epsilon} S_\epsilon^{-1} Z_{\alpha_s} \zeta_{\alpha_s} \alpha_s \tag{3.5}$$

where $S_\epsilon = (4\pi)^\epsilon e^{-\epsilon\gamma_E}$, $Z_{\alpha_s}$ is the $\overline{\mathrm{MS}}$ renormalisation constant and $\zeta_{\alpha_s}$ is the heavy-quark decoupling constant (which for the purposes of this chapter, where all particles are massless, we neglect). To perform the UV renormalisation we make the replacement 3.5 in Equation 3.1 and expand $Z_{\alpha_s}$, which is given to 2-loop order by

$$Z_{\alpha_s} = 1 - \frac{\beta_0}{\epsilon}\left(\frac{\alpha_s}{4\pi}\right) + \left(\frac{\beta_0^2}{\epsilon^2} - \frac{\beta_1}{2\epsilon}\right)\left(\frac{\alpha_s}{4\pi}\right)^2 + \dots \tag{3.6}$$

[1]There exist in fact cases where the lowest order term for a process is also represented by diagrams containing loops since simpler diagrams cannot be drawn, for example the pair production of Higgs bosons by gluons.

### 3.1.3. Infrared divergences of virtual amplitudes

Having performed the ultraviolet renormalisation of an amplitude by introducing appropriate counterterms, the $\left|\mathcal{M}^{(n)}\right\rangle$ still possess infrared singularities manifest as poles in the dimensional regulator $\epsilon$. The order of the pole increases with $n$ so that

$$\left|\mathcal{M}^{(n)}\right\rangle \sim \left(\frac{1}{\epsilon}\right)^{2n} + \dots \tag{3.7}$$

where the ellipsis represents less divergent pieces. We may factorise these singularities like so

$$|\mathcal{M}\rangle = \mathcal{Z}(\epsilon, \{p_i\}, \{m_i\}, \mu_R)\,|\mathcal{F}\rangle \tag{3.8}$$

where we have introduced an infrared renormalisation constant $\mathcal{Z}$ which is a function of the masses and momenta of the external particles and acts as an operator in the colour space. This factor absorbs all the infrared divergent behaviour of the amplitude $|\mathcal{M}\rangle$ such that the remainder $|\mathcal{F}\rangle$ is finite in the limit $\epsilon \to 0$[45,46]. It too has a perturbative expansion

$$\mathcal{Z} = 1 + \left(\frac{\alpha_s}{4\pi}\right)\mathcal{Z}^{(1)} + \left(\frac{\alpha_s}{4\pi}\right)^2\mathcal{Z}^{(2)} + \dots \tag{3.9}$$

so that, expanding 3.8, we find

$$\left|\mathcal{M}^{(0)}\right\rangle = \left|\mathcal{F}^{(0)}\right\rangle \tag{3.10}$$

$$\left|\mathcal{M}^{(1)}\right\rangle = \mathcal{Z}^{(1)}\left|\mathcal{M}^{(0)}\right\rangle + \left|\mathcal{F}^{(1)}\right\rangle \tag{3.11}$$

$$\left|\mathcal{M}^{(2)}\right\rangle = \left(\mathcal{Z}^{(2)} - \mathcal{Z}^{(1)}\mathcal{Z}^{(1)}\right)\left|\mathcal{M}^{(0)}\right\rangle + \mathcal{Z}^{(1)}\left|\mathcal{M}^{(1)}\right\rangle + \left|\mathcal{F}^{(2)}\right\rangle. \tag{3.12}$$

The $\mathcal{Z}$-operator satisfies a renormalisation group equation

$$\frac{d}{d\log\mu_R}\mathcal{Z}(\epsilon, \{p_i\}, \{m_i\}, \mu_R) = -\Gamma(\{p_i\}, \{m_i\}, \mu_R)\mathcal{Z}(\epsilon, \{p_i\}, \{m_i\}, \mu_R) \tag{3.13}$$

where $\Gamma$ is the anomalous dimension operator whose expression is well-known at 2-loop order[47–52]. In the case that the partons are all massless, it is given by

$$\Gamma(\{p_i\},\{m_i\},\mu_R) = \sum_{i,j} \frac{T_i \cdot T_j}{2} \gamma_{\text{cusp}}(\alpha_s) \log\left(\frac{\mu_R^2}{-s_{ij}}\right) + \sum_i \gamma^i(\alpha_s) + \mathcal{O}(\alpha_s^3) \tag{3.14}$$

where $s_{ij} = (p_i + p_j)^2$ and the sums run over all partons in the process. We note the presence of the colour matrices, where here $T_i$ refers to a matrix insertion on leg $i$ rather than an element of the matrix. This reflects the fact that $\mathcal{Z}$ is an operator in the colour space. To define the action of the $T_i$ on $|\mathcal{M}\rangle$ precisely, we note that for an $m$-parton amplitude the structure in colour space can be made explicit by defining

$$\mathcal{M}_m^{c_1,\dots,c_m}(p_1,\dots,p_m) \equiv \langle c_1,\dots,c_m \,|\, \mathcal{M}_m(p_1,\dots,p_m)\rangle \tag{3.15}$$

where $c_1,\dots,c_m$ are the colours of the $m$ partons and the $\{|c_1,\dots,c_m\rangle\}$ form an orthogonal basis in the $m$-parton colour space. We then see that the action of the $T_i$ is a projection

$$\langle c_1,\dots,c_i,\dots,c_m \,|\, T_i^a \,|\, b_1,\dots,b_i,\dots,b_m\rangle = \delta_{c_1 b_1} \dots T^a_{c_i b_i} \dots \delta_{c_m b_m}. \tag{3.16}$$

We understand that where $T_i \cdot T_j$ terms are not present explicitly, an identity matrix in the colour space is implied.

The solution to 3.13 is then given by[53]

$$\mathcal{Z} = 1 + \left(\frac{\alpha_s}{4\pi}\right)\left(\frac{\Gamma_0'}{4\epsilon^2} + \frac{\Gamma_0}{2\epsilon}\right) + \left(\frac{\alpha_s}{4\pi}\right)^2 \left[\frac{(\Gamma_0')^2}{32\epsilon^4} + \frac{\Gamma_0'}{8\epsilon^3}\left(\Gamma_0 - \frac{3}{2}\beta_0\right) + \frac{\Gamma_0}{8\epsilon^2}(\Gamma_0 - 2\beta_0) + \frac{\Gamma_1'}{16\epsilon^2} + \frac{\Gamma_1}{4\epsilon}\right] + \dots \tag{3.17}$$

where the primes denote derivatives $d(\log \mu_R)$, the massless cusp anomalous dimensions controlling the collinear singularities associated with emission from each leg are given by[54]

$$\gamma_0^{\text{cusp}} = 4, \tag{3.18}$$

$$\gamma_1^{\text{cusp}} = \left(\frac{268}{9} - \frac{4\pi^2}{3}\right) C_A - \frac{40}{9} n_f \tag{3.19}$$

and the massless anomalous dimensions for the quarks and gluons are given by

$$\gamma_0^q = -3C_F, \tag{3.20}$$

$$\gamma_1^q = C_F^2\left(-\frac{3}{2}+2\pi^2-24\zeta_3\right)+C_FC_A\left(-\frac{961}{54}-\frac{11\pi^2}{6}+26\zeta_3\right)+\frac{C_Fn_f}{2}\left(\frac{130}{27}+\frac{2\pi}{3}\right), \tag{3.21}$$

$$\gamma_0^g = -\beta_0, \tag{3.22}$$

$$\gamma_1^g = C_A^2\left(-\frac{692}{27}+\frac{11\pi^2}{18}+2\zeta_3\right)+\frac{C_An_f}{2}\left(\frac{256}{27}-\frac{2\pi^2}{9}\right)+2C_Fn_f. \tag{3.23}$$

This in hand, we are now able to extract the finite remainders from our virtual amplitudes.

### 3.1.4. Loop diagrams and loop integrals

The expression for an $l$-loop Feynman diagram with $T$ internal lines and $N$ independent external momenta, obtained from the Feynman rules of the theory, will be a function of propagators of the $l$ internal loop momenta and of dot products of the external momenta. Parton species aside, in this section we define a propagator as an object of the form

$$P = \frac{1}{q^2-a} = \frac{1}{D} \tag{3.24}$$

where the momentum $q$ is a linear combination of the loop momenta $k_i$ and the external momenta $p_i$, and $a$ is a constant which will normally be a function of the masses of the internal lines. We are able to form $N(N+1)/2$ scalar products involving external momenta only (kinematic invariants) and $n = l(l+1)/2 + Nl$ scalar products with loop momenta which will be integration variables, either of the form $k_i \cdot k_j$ or of the form $k_i \cdot p_j$.

We call the set of $T$ propagators which appear in a diagram a topology. In addition to these propagators, we also introduce additional irreducible numerators so that we can express any of the integration variables in terms of a set $\{P_1, ..., P_n\}$. In contrast with the reducible numerators, the irreducible numerators will never cancel with any denominator since they cannot be written as a linear combination of propagators appearing in the topology. This complete set of propagators we call an integral family. Any subset of $t$ propagators of the family we call a sector of weight $t$ and multiplicity $\binom{n}{t}$ [2]. There are in total $2^n$ sectors in a family. A sector whose propagators form a subset of the propagators of another sector is called a subsector of that sector. A scalar integral appearing in a

[2]It is clear that, defined this way, a topology is a special kind of sector which corresponds directly to a diagram and has exactly $T$ propagators.

$t$-propagator sector is generically of the form

$$I = \int \left( \prod_{i=1}^{l} \frac{d^d k_i}{(2\pi)^d} \right) P_{j_1}^{r_1} \dots P_{j_t}^{r_t} P_{j_{t+1}}^{-s_1} \dots P_{j_n}^{-s_{n-t}} \tag{3.25}$$

for integer $r_i \geq 1$ and $s_i \geq 0$. The corner integral of the sector is that for which $r \equiv \sum_{i=1}^{t} r_i = t$ and $s \equiv \sum_{i=1}^{n-t} s_i = 0$.

An integral belonging to a family $F$ can be uniquely identified by the powers $\{\nu_i\}$ of its propagators, positive powers indicating the presence of denominators and negative powers indicating the presence of irreducible numerators.

It is often the case that two or more sectors will be related by a shift of loop momentum $k_i \to \sum_{j=1}^{l} A_{ij} k_j + \sum_{j=1}^{N} B_{ij} p_j$, in which case all integrals belonging in one sector can be expressed in terms of integrals belonging to the other. This is known as a sector relation. It is also possible that all integrals belonging to a sector are zero which can occur if the corner integral is zero (scaleless integrals, for example, are zero in dimensional regularisation). In this case, the sector is known as a zero sector.

### 3.1.5. Integration by parts identities

The evaluation of loop diagrams will in general result in an expression which is a function of many scalar loop integrals multiplied by rational coefficients which are functions of the kinematic invariants. It is common, however, that many of the loop integrals which appear are not truly independent of one another but are in fact related by a set of equations known as integration by parts (IBP) identities. Application of these identities allows us to express the many integrals which appear in an amplitude in terms of a smaller number of master integrals which are the objects we evaluate, either analytically or numerically.

The IBP relations[55] follow from the observation that in dimensional regularisation the integral of a total derivative is zero,

$$\int \prod_{i}^{l} \frac{d^d k_i}{(2\pi)^d} \frac{\partial}{\partial k_j^\mu} \left( \frac{q^\mu}{\prod_\kappa D_\kappa} \right) = 0 \tag{3.26}$$

where $q^\mu$ can be a linear combination of the $k_i$ and $p_i$. Evaluating the derivative leads to a sum of terms and hence a relation between integrals. The only effect of the derivative will be to shift powers of the denominators of the integrand on the left hand side, and so the IBP equation will relate integrals belonging to the same integral family—it essentially acts as a ladder operator on the powers of the propagators.

To find a general recurrence relation using said ladder operators for general values of the powers of the propagators $\nu_i$ is feasible for only the simplest cases. In the Laporta approach[56], the process is automated by making specific choices for the vector $\nu$ (seed integrals) and constructing a system of equations. Since the number of relations grows faster than the number of integrals to be solved for, for a sufficiently large system a Gaussian elimination can be performed in order to express any scalar integral within the range of seeds in terms of a finite basis of master integrals which cannot be solved for further[57]. This procedure has been implemented in many publicly available programs.[58–61]

A related set of identities utilise the fact that all scalar integrals $I(\{p_i\})$ are Lorentz invariant under an infinitesimal transformation of the external momenta $p^\mu \to p^\mu + \delta\omega^\mu_\nu p^\nu$:

$$I(\{p_i\}) = I(\{p_i + \delta p_i\}). \tag{3.27}$$

Expanding the right hand side, we find that

$$I(\{p_i + \delta p_i\}) = I(\{p_i\}) + \delta p_i \sum_{n=1}^{N} \frac{\partial I(\{p_i\})}{\partial p_n^\mu} = I(\{p_i\}) + \delta\omega^\mu_\nu \sum_{n=1}^{N} p_n^\nu \frac{\partial I(\{p_i\})}{\partial p_n^\mu} \tag{3.28}$$

and utilising the antisymmetry of $\delta\omega^\mu_\nu$

$$\sum_{n=1}^{N} \left( p_n^\nu \frac{\partial}{\partial p_{n\mu}} - p_n^\mu \frac{\partial}{\partial p_{n\nu}} \right) I(\{p_i\}) = 0 \tag{3.29}$$

which can then be contracted with all manner of $p_i p_j$ to obtain new relations. It has been shown that these are not, in fact, independent of the IBP identities but are of interest in that a Lorentz Invariance (LI) identity generated from a given seed alone is independent of the IBP identities generated from the same seed alone[62].

The master integrals form a basis of a vector space $V$ of dimension $M$ which is spanned by the infinite number of Feynman integrals. It can be proved that a finite basis can always be found—the basis is not, however, unique and moreover it may happen that two or more master integrals are linearly related to each other when evaluated directly. It is, however, important to distinguish between distinct elements of the basis even when the integrals are equivalent—this distinction will become crucial later in this chapter.

Denoting a generic Feynman integral by its indices,

$$I(\nu_1, \ldots, \nu_n) \equiv \int \prod_i^l \frac{d^d k_i}{(2\pi)^d} \frac{1}{\prod_\kappa^n D_\kappa^{\nu_\kappa}} \tag{3.30}$$

the solutions to the IBP equations take the form

$$I(\nu_1, \ldots, \nu_n) = \sum_{m=1}^{M} c_m(\nu_1, \ldots, \nu_n) \hat{I}_m \tag{3.31}$$

where the $\hat{I}_m$ are the master integrals. Finding the solution of the IBP equations is synonymous with deriving the set of coefficients $c_{i,m}$ for $m = 1, \ldots, M$ for any required integral $I_i$ (we will switch between using the full set of indices explicitly and the compound index $i$ for convenience).

Solving the IBP equations becomes increasingly computationally expensive as the number of scales in the problem increases which may correspond to either an increase in multiplicity of the partons or in the number of massive lines to be considered. Such problems typically involve polynomials with a large number of variables (the mass scales and the dimension $d$) of high degree. Complexity clearly increases too with the number of loops. In cases where one requires solutions for integrals with high powers of the propagators[3] it may be that the entire approach is unfeasible for modern computers due to the size of the system of equations to be solved—either the memory of the machine is exhausted or the running time is prohibitively long.

To remedy this difficulty, several variants of the Laporta algorithm have been proposed that exploit the redundancy of the IBP equations to reduce the number to be solved[62] or solve only the equations necessary for a set of requested integrals[59]. Alternatively, approaches have been developed which partially circumvent the problem through use of numerical unitarity[63]. At the present time, however, there exists no single methodology which represents a significant improvement in increasing the range of problems which can be solved.

The IBP approach can also be used to determine the analytic expressions for the master integrals themselves using what is called the 'method of differential equations'. Taking the derivative of a master integral with respect to a loop momentum, one can use the IBP solutions to write the right hand side in terms of other master integrals. By repeating for each master, one obtains a system of differential equations for the masters which can be solved. We will not discuss this further here, nor will we comment on the means of evaluating Feynman integrals directly—it suffices to say that there exist authoritative texts on this issue, see for example [64].

[3]The definition of 'high' is clearly process dependent but for 2-loop problems could typically be considered $r + s > 3$.

## 3.2. The $q\bar{q} \to Q\bar{Q}g$ amplitude at 1-loop

In this section we discuss perhaps the simplest 3-jet amplitude, that for the process $q\bar{q} \to Q\bar{Q}g$, where $q$ and $Q$ represent distinct flavours of light quark. Knowledge of this amplitude is necessary not only as an NLO correction to the 3-jet process but also for the NNLO corrections to the $2 \to 2$ process $q\bar{q} \to Q\bar{Q}$ (a real-virtual contribution). The amplitude itself has previously been computed[65], as have the other $2 \to 3$ QCD amplitudes[66,67], in the spinor-helicity formalism[4] and using various techniques including some derived from string theory. The squared amplitude, however, is not itself publicly available. There are certain advantages to having $|\mathcal{M}|^2$ rather than $|\mathcal{M}\rangle$—although the size of the expression is generally larger, speed of numerical evaluation can be improved since only one number need be calculated for each phase space point (one would have to square the complex number returned from numerical evaluation of the amplitude alone).

### 3.2.1. Kinematics of $2 \to 3$ processes

We begin by considering a $2 \to N$ process

$$p_a + p_b \to p_1 + \cdots + p_N \tag{3.32}$$

where the four-vector $p_i = (E_i, \vec{p}_i)$. In order to satisfy energy-momentum conservation, we have

$$\begin{aligned} E_a + E_b &= \sum_{i=1}^{N} E_i \\ \vec{p}_a + \vec{p}_b &= \sum_{i=1}^{N} \vec{p}_i \end{aligned} \tag{3.33}$$

with particles on-shell so that $E_i^2 = \vec{p}_i^{\,2} + m_i^2$. The final state 3-momentum vectors $\vec{p}_i$ inhabit a $3N$ dimensional space, which energy-momentum conservation restricts to a $3N-4$ dimensional phase space. In the case of $2 \to 3$ scattering, we therefore have available 5 independent kinematic invariants with which to parametrise our phase space.

We change notation a little and assign momenta such that

$$q(p_1) + \bar{q}(p_2) \to Q(p_3) + \bar{Q}(p_4) + g(p_5) \tag{3.34}$$

[4]This will be discussed in the following chapter.

with the momenta $p_1, p_2$ considered incoming and the rest outgoing. All particles are considered massless. We choose kinematic invariants

$$\begin{aligned} s_{12} &= (p_1 + p_2)^2 \\ s_{23} &= (p_2 + p_3)^2 \\ s_{34} &= (p_3 + p_4)^2 \\ s_{45} &= (p_4 + p_5)^2 \\ s_{15} &= (p_1 + p_5)^2. \end{aligned} \tag{3.35}$$

Clearly these choices are not unique and in fact we are able to relate any other invariants or dot products of momenta (e.g. $s_{35}$, $p_1 \cdot p_4$) to this complete set.

### 3.2.2. Computation of the amplitude

We make extensive use of the program `Reduze 2`[60], including its features for performing the IBP reduction and computation of amplitudes. We first generate diagrams for the process using the program `QGRAF`[68] and, including 5 light flavours within the loops, find 126 1-loop diagrams and 5 tree level diagrams.

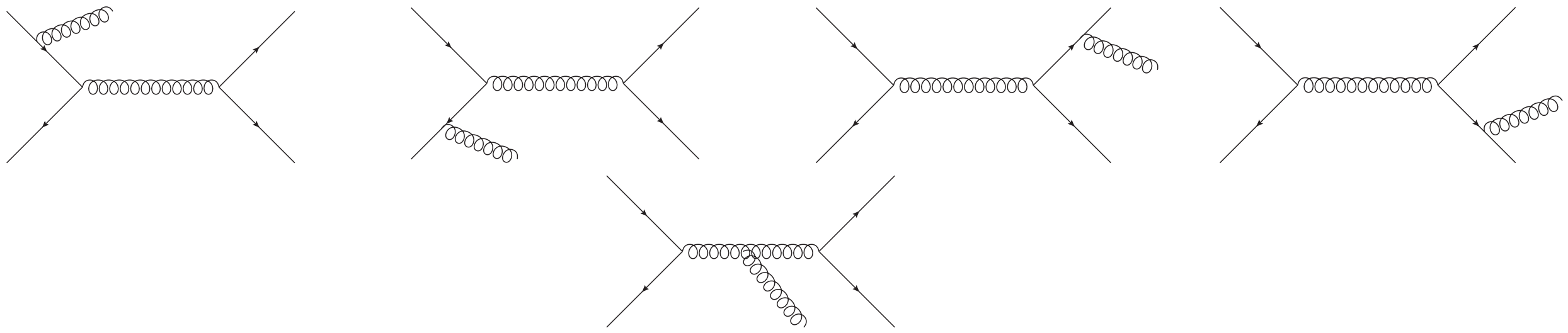

**Figure 3.1.:** The tree level diagrams for the process $q\bar{q} \to Q\bar{Q}g$.

We next define integral families to which the loop diagrams can be mapped. At 1-loop, the number of propagators that need to be defined happens to be equal to the maximum number of propagators present in a diagram, and so no additional numerators need be added—our families will coincide with the highest weight topologies. It happens that in our case only a single family is needed. We define the family $A$ in Table 3.1.

Once the kinematics and integral families have been defined in the input files, `Reduze` analyses the provided families and identifies sector relations, symmetries and zero sectors. Relationships between sectors belonging to crossed families (those related to the defined families by cyclic permutations of the external legs) are also found. Provided with the diagrams generated by `QGRAF`, `Reduze` then maps these to sectors of the original and crossed integral families in such a way as to minimise the number of sectors that need to be reduced. Using user-specified Feynman rules, amplitudes are generated from the mapped diagrams. Although the program has built-in functionality that allows it to square the diagrams using the library `GiNaC`[69] as a backend, in order to improve speed of evaluation we have interfaced a custom script to perform this task. This utilises the `Mathematica` package `ColorMath`[70] to evaluate the traces over the colour matrices and feeds the output to the computer algebra program `FORM`[71] which evaluates the Dirac traces. We work in Feynman gauge and use CDR throughout. The evaluated diagrams are passed back to `Reduze`, which for each replaces the dot products with kinematic invariants and propagators of the mapped sector.

| Family $A$ |
|---|
| $k$ |
| $k + p_1$ |
| $k + p_1 + p_2$ |
| $k + p_1 + p_2 - p_3$ |
| $k + p_1 + p_2 - p_3 - p_4$ |

**Table 3.1.:** An integral family to capture all 1-loop diagrams in the process $q\bar{q} \to Q\bar{Q}g$.

At this stage the squared diagrams are expressed in terms of many scalar Feynman integrals multiplied by coefficients which are rational functions of the kinematic invariants and the dimension of spacetime $d$. The IBP reduction is then run, with ranges of $r$ and $s$ set to values based on the highest powers of denominators and numerators which appear in the unreduced integrals of the amplitude. Once complete, the results of the reduction are inserted into the squared amplitude to produce a result which is a sum of coefficients multiplying master integrals. Knowledge of the analytic form of the master integrals is needed to complete the computation.

### 3.2.3. 1-loop master integrals

After the reduction, 83 master integrals are present in the amplitude though only 3 are distinct, the rest being related by crossings of the external legs. We make use of known analytic results for the 1-loop bubble and box integrals available in the library `QCDLoop`[72] as well as those for the pentagon integrals[73][5]. We verified that mappings to our particular case have been performed correctly by numerical evaluation of the integrals using the program `SecDec`[74].

[5]It is in fact possible to write the massless pentagon integral as a sum of box integrals, since it can be shown that the tadpole, bubble, triangle and box integrals form a complete basis at 1-loop up to $\mathcal{O}(\epsilon^0)$[73]. Nevertheless, since the result for the pentagon is relatively compact and readily available, we choose to leave the expression in this form.

From henceforth we shall drop everywhere the Feynman prescription $+i\mathbf{0}$, understanding it to be implicit. We define

$$A(\nu_1,\nu_2,\nu_3,\nu_4,\nu_5) = \frac{\mu^{2\epsilon}}{i\pi^{\frac{d}{2}} r_\Gamma} \int d^dk \frac{1}{(k^2)^{\nu_1}((k+p_1)^2)^{\nu_2}((k+p_{12})^2)^{\nu_3}((k+p_{123})^2)^{\nu_4}((k+p_{1234})^2)^{\nu_5}} \tag{3.36}$$

where $p_{ij} \equiv \sum_{k=i}^{j} \sigma_k p_k$ ($\sigma_k = +1$ if the momentum is incoming and $-1$ if outgoing) and the overall constant appearing from the integration over a sphere in $d$ dimensions is given by

$$r_\Gamma = \frac{\Gamma^2(1-\epsilon)\Gamma(1+\epsilon)}{\Gamma(1-2\epsilon)}. \tag{3.37}$$

For convenience, we introduce the dilogarithm (also known as Spence's function) here, which is defined as the integral

$$\mathrm{Li}_2(x) = -\int_0^x \frac{dt}{t} \log{(1-t)} \tag{3.38}$$

and possesses a branch cut along the real axis from $x = 1$ to $\infty$. Values often encountered in practical applications include

$$\mathrm{Li}_2(1) = \frac{\pi^2}{6},\ \mathrm{Li}_2(-1) = -\frac{\pi^2}{12}. \tag{3.39}$$

The dilogarithm occurs frequently in 1-loop integrals, and in fact at higher loops its generalised form, the polylogarithm, is ubiquitous:

$$\mathrm{Li}_{s+1}(x) = \int_0^x \frac{\mathrm{Li}_s(t)}{t} dt. \tag{3.40}$$

The uncrossed master integrals are given by

$$\begin{aligned} A(1,0,1,0,0) = {} & \frac{1}{\epsilon} + 2 + i\pi + L_{12} + \left(\frac{1}{2}\left(L_{12}+i\pi\right)^2 + 2\left(L_{12}+i\pi\right) + 4\right)\epsilon \\ & + \left(\frac{1}{6}\left(L_{12}+i\pi\right)^3 + \left(L_{12}+i\pi\right)^2 + 4\left(L_{12}+i\pi\right) + 8\right)\epsilon^2 + \mathcal{O}(\epsilon^3) \end{aligned} \tag{3.41}$$

$$A(1,1,1,1,0) = -\frac{2}{s_{12}s_{23}}\frac{1}{\epsilon^2} - \frac{2\left(L_{12}+L_{23}-L_{45}\right)}{s_{12}s_{23}}\frac{1}{\epsilon}$$
$$+\frac{3L_{45}^2 - 6i\pi L_{45} + 6i\left(\pi + iL_{12}\right)L_{23} + 6\mathrm{Li}_2\left(1-\frac{s_{45}}{s_{12}}\right) + 6\mathrm{Li}_2\left(1+\frac{s_{45}}{s_{23}}\right) - 2\pi^2}{3s_{12}s_{23}} + \mathcal{O}\left(\epsilon\right) \tag{3.42}$$

$$A(1,1,1,1,1) = \frac{-s_{12}s_{15} - s_{12}s_{23} - s_{23}s_{34} + s_{34}s_{45} - s_{45}s_{15}}{s_{12}s_{15}s_{23}s_{34}s_{45}}\frac{1}{\epsilon^2}$$
$$+\frac{1}{s_{12}s_{15}s_{23}s_{34}s_{45}}\frac{1}{\epsilon}\Big[\left(L_{12}+L_{15}-L_{23}-L_{34}-L_{45}+i\pi\right)s_{12}s_{15}$$
$$+\left(L_{12}-L_{15}+L_{23}-L_{34}-L_{45}+i\pi\right)s_{12}s_{23} + \left(-L_{12}-L_{15}+L_{23}+L_{34}-L_{45}+i\pi\right)s_{23}s_{34}$$
$$+\left(-L_{12}+L_{15}-L_{23}-L_{34}+L_{45}+i\pi\right)s_{45}s_{15} + \left(L_{12}+L_{15}+L_{23}-L_{34}-L_{45}+i\pi\right)s_{34}s_{45}\Big]$$
$$+\frac{1}{6s_{12}s_{15}s_{23}s_{34}s_{45}}\Bigg\{\Bigg[-3L_{12}^2 + 6\left(-L_{15}+L_{23}+L_{34}+L_{45}-i\pi\right)L_{12}$$
$$-3L_{15}^2 - 3L_{23}^2 - 3L_{34}^2 - 3L_{45}^2 - 12\mathrm{Li}_2\left(1-\frac{s_{23}}{s_{15}}\right) - 12\mathrm{Li}_2\left(1-\frac{s_{45}}{s_{12}}\right)$$
$$+6i\pi L_{23} - 6L_{23}L_{34} + 6i\pi L_{34} - 6L_{23}L_{45} - 6L_{34}L_{45} + 6i\pi L_{45} + 6L_{15}\left(L_{23}+L_{34}+L_{45}-i\pi\right) + 4\pi^2\Bigg]s_{12}s_{15}$$
$$+\Bigg[-3L_{12}^2 + 6\left(L_{15}-L_{23}+L_{34}+L_{45}-i\pi\right)L_{12}$$
$$-3L_{15}^2 - 3L_{23}^2 - 3L_{34}^2 - 3L_{45}^2 - 12\mathrm{Li}_2\left(1-\frac{s_{15}}{s_{23}}\right) - 12\mathrm{Li}_2\left(1-\frac{s_{34}}{s_{12}}\right)$$
$$-6i\pi L_{23} + 6L_{23}L_{34} + 6i\pi L_{34} + 6L_{15}\left(L_{23}-L_{34}-L_{45}+i\pi\right) + 6L_{23}L_{45} - 6L_{34}L_{45} + 6i\pi L_{45} + 4\pi^2\Bigg]s_{12}s_{23}$$
$$+\Bigg[-3L_{12}^2 + 6\left(-L_{15}+L_{23}+L_{34}-L_{45}+i\pi\right)L_{12}$$
$$-3L_{15}^2 - 3L_{23}^2 - 3L_{34}^2 - 3L_{45}^2 - 12\mathrm{Li}_2\left(1-\frac{s_{12}}{s_{34}}\right) - 12\mathrm{Li}_2\left(1+\frac{s_{45}}{s_{23}}\right)$$
$$-6i\pi L_{23} - 6L_{23}L_{34} - 6i\pi L_{34} + 6L_{15}\left(L_{23}+L_{34}-L_{45}+i\pi\right) + 6L_{23}L_{45} + 6L_{34}L_{45} + 6i\pi L_{45} + 4\pi^2\Bigg]s_{23}s_{34}$$
$$+\Bigg[-3L_{12}^2 + 6\left(L_{15}-L_{23}-L_{34}+L_{45}+i\pi\right)L_{12}$$
$$-3L_{15}^2 - 3L_{23}^2 - 3L_{34}^2 - 3L_{45}^2 - 12\mathrm{Li}_2\left(1+\frac{s_{34}}{s_{15}}\right) - 12\mathrm{Li}_2\left(1-\frac{s_{12}}{s_{45}}\right)$$
$$+6i\pi L_{23} - 6L_{23}L_{34} + 6i\pi L_{34} + 6L_{15}\left(L_{23}+L_{34}-L_{45}-i\pi\right) + 6L_{23}L_{45} + 6L_{34}L_{45} - 6i\pi L_{45} + 4\pi^2\Bigg]s_{45}s_{15}$$

$$+\left[3L_{12}^2+6\left(L_{15}+L_{23}-L_{34}-L_{45}+i\pi\right)L_{12}\right.$$
$$+3L_{15}^2+3L_{23}^2+3L_{34}^2+3L_{45}^2+12\mathrm{Li}_2\left(1+\frac{s_{15}}{s_{34}}\right)+12\mathrm{Li}_2\left(1+\frac{s_{23}}{s_{45}}\right)$$
$$\left.+6i\pi L_{23}-6L_{23}L_{34}-6i\pi L_{34}+6L_{15}\left(L_{23}-L_{34}-L_{45}+i\pi\right)-6L_{23}L_{45}+6L_{34}L_{45}-6i\pi L_{45}-4\pi^2\right]s_{45}s_{34}\Bigg\}$$
$$+\mathcal{O}\left(\epsilon\right) \quad (3.43)$$

where we have expanded in $\epsilon$ and made explicitly the analytic continuation

$$(-s_{ij})^{-\epsilon}\rightarrow|s_{ij}|e^{-i\pi\Theta(s_{ij})} \quad (3.44)$$

$$\log\left(-s_{ij}\right)\rightarrow\log|s_{ij}|+i\pi\Theta(s_{ij}) \quad (3.45)$$

(where $\Theta(x)$ is the Heaviside step function) so that all invariants are in the physical region. We have also defined $L_{ij}\equiv\log\frac{\mu^2}{s_{ij}}$.

### 3.2.4. IR subtraction

The form of the $\mathcal{Z}$ operator for a given process is dictated by the number and type of external partons. For the four quark, one gluon case and to $\mathcal{O}(\alpha_s)$ it is given by

$$\mathcal{Z}^{(1)}=1+\frac{\alpha_s}{4\pi}\left\{\frac{1}{\epsilon}\left[-\frac{3}{2}C_F+\sum_{\substack{i=1\\j>i}}^{5}T_i\cdot T_j\log\left(\frac{\sigma_{ij}\mu^2}{s_{ij}}\right)\right]+\frac{1}{\epsilon^2}\sum_{\substack{i=1\\j>i}}^{5}T_i\cdot T_j\right\} \quad (3.46)$$

where

$$\sigma_{ij}=\begin{cases}+1 & \text{if } i,j \text{ opposite sign}\\ -1 & \text{if } i,j \text{ same sign}\end{cases}. \quad (3.47)$$

Equation 3.46 implies that when evaluating $\left\langle\mathcal{M}^{(0)}\,\middle|\,\mathcal{Z}\,\middle|\,\mathcal{M}^{(0)}\right\rangle$, we will need matrix elements of the form

$$\left\langle\mathcal{M}^{(0)}\,\middle|\,T_i\cdot T_j\,\middle|\,\mathcal{M}^{(0)}\right\rangle \quad (3.48)$$

which are known as colour-correlated amplitudes. We can make use of the property of colour conservation in order to reduce the number of amplitudes that need to be calculated—specifically, we have that

$$\sum_{i=1} T_i \left|\mathcal{M}\right\rangle = 0 \tag{3.49}$$

where the sum is over all partons. We also note that, from the colour algebra,

$$T_i \cdot T_j = T_j \cdot T_i \quad \text{if } i \neq j, \quad T_i^2 = C_i \tag{3.50}$$

where the $C_i$ are the appropriate Casimir operators. The consequence is that not only are not all the correlators independent, those which are diagonal in the colour space are simply given by the Born amplitude multiplied by the Casimir factor ($C_F$ or $C_A$ depending on the leg). The remaining independent correlators were calculated using a modified version of `Reduze` to generate the amplitudes with the colour matrix insertions and the aforementioned `FORM` script for evaluation.

### 3.2.5. Results

We are now in a position to construct the 1-loop finite remainder. It is given by

$$\left\langle \mathcal{M}^{(0)} \,\middle|\, \mathcal{F}^{(1)} \right\rangle = S_\epsilon^{-1} \left\langle \mathcal{M}^{(0)} \middle| \mathcal{M}_b^{(1)} \right\rangle - \left\langle \mathcal{M}^{(0)} \,\middle|\, \mathcal{Z}^{(1)} \,\middle|\, \mathcal{M}^{(0)} \right\rangle - \frac{\beta_0}{\epsilon} \left\langle \mathcal{M}^{(0)} \,\middle|\, \mathcal{M}^{(0)} \right\rangle \tag{3.51}$$

where the subscript $b$ indicates the bare amplitude and the last term on the right hand side provides the UV renormalisation of the coupling. We reiterate that the left hand side should be entirely free of poles and finite in the limit $\epsilon \to 0$. This requirement provides a preliminary check on the 1-loop amplitude.

A full expression for the Born amplitude is given in the ancillary files. Note that although the Born amplitude is finite in the limit $\epsilon \to 0$, in CDR it is necessary to retain the subleading terms in $\epsilon$ as these will provide contributions of $\mathcal{O}(\epsilon^0)$ during the UV renormalisation and IR subtraction.

The 1-loop finite remainder $\left|\mathcal{F}^{(1)}\right\rangle$ is also included in the ancillary files accompanying this thesis. The result has been checked numerically against the output of the program `NJet`[75] which calculates multi-jet amplitudes at NLO and agreement found to within the Monte Carlo error of that program.

## 3.3. The $q\bar{q} \to Q\bar{Q}g$ amplitude at 2-loops

Despite being a frequently measured event at hadron colliders such as the LHC, at the time of writing there is very limited knowledge of the amplitudes for 3-jet production at NNLO. In fact, no complete 2-loop amplitude is known for any $2 \to 3$ process. This is due solely to the lack of availability of a solution to the IBP equations for the relevant topologies. Partial results have become available in the past few years for the 2-loop five gluon amplitude, initially only for a particular helicity configuration of the external legs[76–80] but more recently for the complete planar amplitude[81,82]. Some results for the non-planar topologies are also known[83]. In much of this work, the numerical unitarity approach has proved extremely useful[63,84,85].

In this section we present results for the 2-loop, planar $q\bar{q} \to Q\bar{Q}g$ amplitude. The IBP reduction bottleneck that had prevented previous calculation of this process has been overcome by use of a new strategy for implementing the Laporta algorithm which is based on the independent calculation of different projections of the IBP equations onto the vector space of master integrals.

### 3.3.1. Solving IBP equations by projection

We detail here our strategy. We begin with the assumption that the IBP system has a solution, i.e. that every loop integral can be expressed through a set of basis master integrals and that such a basis is known. The determination of a finite basis of masters is a well-studied problem, but here we take a pragmatic viewpoint which is informed by the observation that all problems known to us do possess such a basis. There are several ways to construct the basis, such as solving the IBP system over a restricted set of integrals and/or using numerical values for the kinematic invariants.

The novelty in our approach lies in the way in which we determine the coefficients $c_{i,m}$ appearing in Equation 3.31. In existing approaches, the full set of coefficients for a given $i$ is derived simultaneously to obtain a final expression for $I_i$. Here, we pursue a different strategy in which the projection of $I_i$ onto each master is derived independently. Put differently, we split the problem of solving the system of IBP equations into $M$ independent problems, one for each of the $M$ projections. In order to implement this, we apply the usual set of IBP identities to a modification of the space $V$ such that $M-1$ of its elements (corresponding to all but one of the masters) are set to zero beforehand. For example, in order to derive the projection onto master $\hat{I}_1$ of any integral $I_i$ one first sets $\hat{I}_2 = \hat{I}_3 = \cdots = \hat{I}_M = 0$ and then solves the IBP equations. In this way, once the IBP system has been fully solved, one obtains a result that is of the following form

$$I(\nu_1, \ldots, \nu_n) = c_1(\nu_1, \ldots, \nu_n)\hat{I}_1 \tag{3.52}$$

i.e. one will have derived the coefficients $c_1(\nu_1, \ldots, \nu_n)$ which are the projection of the full solution onto the master $\hat{I}_1$. Repeating the same approach but setting $\hat{I}_1 = \hat{I}_3 = \cdots = \hat{I}_M = 0$ one derives the coefficients $c_2(\nu_1, \ldots, \nu_n)$ and so on. To obtain the complete solution of the IBP system one simply needs to add all $M$ independently derived projections.

The validity of this method of solution follows from the fact that each integral $I_i$ has an expansion in the set of masters $\hat{I}_m$, i.e. at each step the IBP equations can be rewritten as a homogeneous linear combination of all master integrals. Since the IBP equations are themselves linear and homogeneous in terms of the integrals $I_i$, one can see that in essence the IBP equations never mix projections belonging to different master integrals. Our proposal simply states that each of these projections can be computed in isolation from the others.

The IBP solving strategy described here is independent of the approach used for solving the equations. In practice we have implemented the standard Laporta algorithm but this is not a necessity. The strategy can lead to a more efficient solving of the system for several reasons. First, once $M - 1$ masters are set to zero, many sectors become zero sectors and thus do not need to be computed. This is a major simplification in practice. Second, setting masters to zero at the outset of the calculation simplifies the intermediate steps. The reason is that the IBP equations which are solved first are generated from seeds that are in some sense close to the master integrals. In this way the information about vanishing masters is incorporated into the resulting IBP equations at an early stage in the solving process. In large systems with many masters our strategy could lead to a significant reduction in the size of the intermediate expressions which in turn reduces the computer memory requirement that is the limiting factor in large problems. Third, by solving for one master at a time it is possible to parallelise the problem by simultaneous computation of several projections. The amount of parallelisation achieved is only restricted by the available memory and CPU.

### 3.3.2. Diagrams, families and topologies

Two 2-loop integral families are sufficient to map all the diagrams in the amplitude—they are listed in Table 3.2.

| Family $B$ | Family $C$ |
|---|---|
| $k_1$ | $k_1$ |
| $k_2$ | $k_2$ |
| $k_1 + p_1$ | $k_1 + p_1 + p_2$ |
| $k_1 + p_1 + p_2$ | $k_1 - k_2$ |
| $k_2 - p_3$ | $k_2 + p_1$ |
| $k_2 - k_1 - p_3$ | $k_2 + p_1 + p_2$ |
| $k_2 - k_1 - p_1 - p_2 + p_4$ | $k_2 - p_3$ |
| $k_2 + p_4$ | $k_1 + p_1 + p_2 - p_3$ |
| $k_2 + p_1 + p_2$ | $k_1 + p_1 + p_2 - p_3 - p_4$ |
| $k_2 + p_1$ | $k_2 - p_3 - p_4$ |
| $k_1 + p_3$ | $k_1 + p_1$ |

**Table 3.2.:** Integral families required to capture all 2-loop diagrams for $q\bar{q} \to Q\bar{Q}g$

Within the families $B$ and $C$, diagrams are mapped to 7 distinct topologies of 8 propagators or fewer; we specify these with a vector whose entries are 1 if the propagator appears and 0 if it does not according to the ordering in Table 3.2. The topologies are

$$\begin{aligned} B_1 &= (1,1,1,1,1,1,1,1,0,0,0) \\ B_2 &= (1,1,1,1,0,1,1,1,0,0,1) \\ B_3 &= (0,1,0,1,0,1,1,1,1,1,0) \\ B_4 &= (1,1,1,1,1,1,0,0,0,0,1) \end{aligned} \qquad \begin{aligned} C_1 &= (1,1,1,1,1,1,0,1,1,0,0) \\ C_2 &= (1,1,1,1,1,0,0,1,1,0,1) \\ C_3 &= (1,0,1,1,1,1,0,1,0,0,1) \end{aligned} \tag{3.53}$$

The $C$ topologies are planar while the $B$ topologies are non-planar—we therefore prioritise $C$ above $B$ in the sense that diagrams which could be mapped to subtopologies of either the $C_i$ or $B_i$ are preferentially mapped to the $C_i$. Diagrammatic representations of the highest weight topologies are shown in Figure 3.2.

We determine the basis of master integrals for each topology by running the IBP reduction using `Reduze` with numerical values for the kinematic invariants. We find 113 masters in $B_1$, 75 in $B_2$, 62 in $C_1$, 28 in $C_2$ and 10 in $C_3$. The master integrals are listed in Appendix B. Using the strategy outlined above, we are able to compute all coefficients $c_{i,m}$ belonging to the planar topologies $C_1, C_2$ and $C_3$ needed for the evaluation of the $q\bar{q} \to Q\bar{Q}g$ amplitude. This includes the results for all required integrals with irreducible numerator powers as high as -5 and/or squared denominators. The master integrals themselves are known analytically[86] for the $C_1$ topology (sometimes called the pentabox in the literature). It happens that the masters of the topologies $C_2$ and $C_3$ can be expressed as linear combinations and/or crossings of the masters of $C_1$—all required integrals are therefore known.

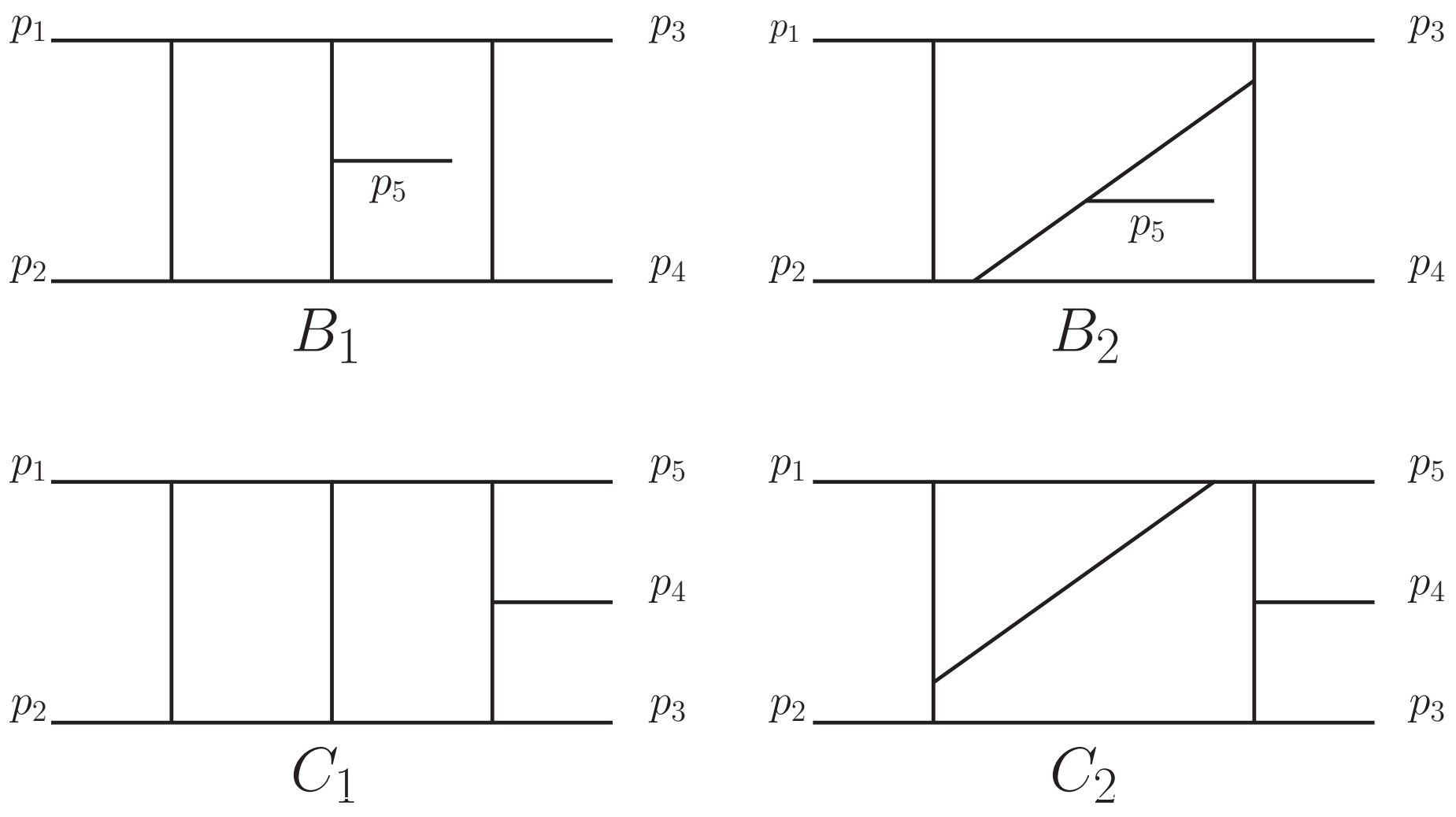

**Figure 3.2.:** The highest weight 2-loop topologies needed to fully map the $q\bar{q} \to Q\bar{Q}g$ amplitude.

### 3.3.3. Targeted IBP equations

Our implementation of the Laporta algorithm is able to solve a range of $r, s$ values $r : [t, t+1]$ and $s : [0, 5]$ which is sufficient to reduce all of the integrals present in the amplitude. However, it happens that of the $\mathcal{O}(10^4)$ integrals that appear in the amplitude, there are only 5 distinct integrals (i.e. those not related by crossings) which have powers of the irreducible numerators totalling -5. Since increasing the range of desired solutions results in a factorial-type growth in the number of equations to be solved, it would be convenient if we could suffice with solving the equations in a range $s : [0, 4]$ and then find some other way to simplify the small number of remaining integrals. In fact, such a strategy exists as we now explain.

For a generic choice of $q^\mu$, the differentiation in the IBP identity Equation 3.26 will result in a linear combination of integrals with doubled propagators in the denominator. This occurs when the derivative operator acts on the denominator factors of the integrals. Since this type of integral occurs in amplitudes only in very specific cases (2-loop self-energy type corrections), this intermediate result is somewhat unfortunate—no useful expression is obtained and further IBPs must be solved in order to express the integral in terms of masters. However, for specific choices of the $q^\mu$ an IBP identity may be derived which features integrals with no higher denominator powers, thus obviating the need for reduction results for these integrals. This could be attained by choosing the $q^\mu$ such that upon differentiation of a propagator factor $1/D_i$, the resultant numerator vanishes upon taking the dot product with $q$. For example, differentiating the simple case $D_i = (k-p)^2$ one would obtain

$$\frac{\partial}{\partial k^\mu}\frac{1}{(k-p)^2} = -2(k^\mu - p^\mu)\frac{1}{(k-p)^4} \tag{3.54}$$

which features the unpleasant doubled propagator, but then might choose a $q^\mu$ such that

$$q \cdot (k-p) = 0 \tag{3.55}$$

so that the right hand side vanishes completely. A sufficient condition for the vanishing of the doubled propagators is that

$$q \cdot (k-p) \propto (k-p)^2 \tag{3.56}$$

or equivalently for multiple loops

$$\sum_{i=1}^{l} q_i^\mu \frac{\partial}{\partial k_i^\mu} D_i \propto D_i \tag{3.57}$$

which would result in a cancellation between numerator and denominator, thus also removing the doubling. The constraint must be enforced for every denominator factor so that, for an integral with $n_d$ denominators of the form

$$\prod_{i=1}^{n_d} D_i = \prod_{i=1}^{n_d}(c_{i1}k_1 + c_{i2}k_2 - v_i)^2 \tag{3.58}$$

where $v_i$ is some linear combination of the external momenta, we may write

$$\mathrm{Rem}\frac{\left[q_1\cdot\frac{\partial}{\partial k_1} + q_2\cdot\frac{\partial}{\partial k_2}\right](c_{i1}k_1 + c_{i2}k_2 - v_i)^2}{(c_{i1}k_1 + c_{i2}k_2 - v_i)^2} = 0 \tag{3.59}$$

where Rem denotes the remainder upon synthetic division. This is an equation for the $q_i^\mu$ which can now be solved. We desire the subset of solutions such that the resulting IBP equation features integrals with irreducible numerators, the cases with reducible numerators being trivial. Such solutions have been found for a number of different topologies[87], including our case of interest, the pentabox. Following notation in [88], we denote summation over the vectors $q_i$ for each loop with capital Latin letters

$$q_A\frac{\partial}{\partial k_A} \equiv \sum_{i=1}^{l} q_{i\mu}\frac{\partial}{\partial k_{i\mu}}. \tag{3.60}$$

and so, given a pair of the $q_i^\mu$ and an integral with numerator $\mathcal{N}$ and denominator $\mathcal{D} = \prod_i D_i$, write the IBP as

$$\int \prod_{i=1}^{l} \frac{d^d k_i}{(2\pi)^d} \frac{\partial}{\partial k_A} \frac{q_A \mathcal{N}}{\mathcal{D}} = 0. \tag{3.61}$$

Expanding and multiplying, we obtain

$$\mathcal{D} \sum_{r=1}^{n_q} \partial_A \frac{q_{rA} \mathcal{N}_r}{\mathcal{D}} = \sum_{r=1}^{n_q} \mathcal{N}_r \mathcal{D} \partial_A \frac{q_{rA}}{\mathcal{D}} + \sum_{r=1}^{n_q} q_{rA} \partial_A \mathcal{N}_r \tag{3.62}$$

where $r$ labels each solution pair $\{q_1^\mu, q_2^\mu\}$ from 1 to the number of solutions of Equation 3.59—in the case of the pentabox, this happens to be $n_q = 9$. We may choose general forms for the numerator $\mathcal{N}$ of appropriate mass dimension in order to derive different IBPs which may be of use.

The 5 integrals with powers of irreducible numerators totalling 5 are

$$\begin{aligned}
I_1^{(-5)} &\equiv C(1,1,1,1,1,1,-1,1,1,-1,-3) \\
I_2^{(-5)} &\equiv C(1,1,1,1,1,1,-2,1,1,0,-3) \\
I_3^{(-5)} &\equiv C(1,1,1,1,1,1,-2,1,1,-1,-2) \\
I_4^{(-5)} &\equiv C(1,1,1,1,1,1,-1,1,1,-2,-2) \\
I_5^{(-5)} &\equiv C(1,1,1,1,1,1,0,1,1,-2,-3).
\end{aligned} \tag{3.63}$$

In order to completely reduce these 5 integrals, it is sufficient to express them in terms of integrals with $s = 4, r = t$ and then use our existing reduction results to write them in terms of masters. To this end, we apply the technique described above and make use of the results provided in [87] for the $q_i^\mu$ in the pentabox topology. We work with the integral family defined in that paper which we call $C'$ before mapping the final results to our own family $C$. The family $C'$ is defined

$$\begin{aligned}
C' = \{&k_1, (k_1 - p_1), (k_1 - p_1 - p_2), (k_1 - p_1 - p_2 - p_3), (k_1 + p_5), k_2, (k_2 - p_1), \\
&(k_2 - p_1 - p_2), (k_2 - p_4 - p_5), (k_2 - p_5), (k_1 + k_2)\}
\end{aligned} \tag{3.64}$$

The integrals in 3.63 which belong to $C$ can be expressed as a linear combination of integrals belonging to $C'$ which have the same numerator powers or lower. Using `Reduze`, we perform a basis change from the crossed topology $C$x14253 to $C'$ and obtain said expressions [6]. The $C'$ integrals which have the highest numerator power are

$$\begin{aligned} J_1^{(-5)} &\equiv C'(1,1,1,1,-1,1,1,-1,-3,1,1) \\ J_2^{(-5)} &\equiv C'(1,1,1,1,-2,1,1,0,-3,1,1) \\ J_3^{(-5)} &\equiv C'(1,1,1,1,-2,1,1,-2,-1,1,1) \\ J_4^{(-5)} &\equiv C'(1,1,1,1,-2,1,1,-1,-2,1,1) \\ J_5^{(-5)} &\equiv C'(1,1,1,1,-3,1,1,0,-2,1,1). \end{aligned} \tag{3.65}$$

Adopting similar notation to that in [88], we define

$$\begin{aligned} r_{ij} &= k_i \cdot k_j & u_{25} &= k_2 \cdot p_5 \\ u_{11} &= k_1 \cdot p_1 & t_{15} &= k_1 \cdot p_5 \\ u_{12} &= k_1 \cdot p_2 & t_{21} &= k_2 \cdot p_1 \\ u_{13} &= k_1 \cdot p_3 & t_{22} &= k_2 \cdot p_2 \\ u_{24} &= k_2 \cdot p_4 & \chi_{ij} &= \frac{s_{ij}}{s_{12}} \end{aligned} \tag{3.66}$$

where the $r_{ij}, u_{ij}$ are reducible invariants (in the sense that when appearing in the numerator of an integral they can be rewritten in terms of propagators which cancel denominator factors) and the $t_{ij}$ are irreducible numerator factors. Armed with our 9 sets of $q^\mu_{rA}$, it is now left to choose appropriate $\mathcal{N}_r$ to generate the desired IBP equations. As the final 3 sets of vectors are redundant[87], we discard these solutions and choose $\mathcal{N}_{7,8,9} = 0$. We are interested in solutions featuring powers of the $t_{ij}$ which sum to 5; accordingly we choose the following (where the superscript labels the integral in 3.65):

$$\begin{aligned} \mathcal{N}_r^1 &= (0,0,\xi_1 t_{15}^2 t_{22}, \xi_2 t_{15}^2 t_{22}, \xi_3 t_{15}^2 t_{22}, 0) \\ \mathcal{N}_r^2 &= (0,\xi_4 t_{15}^2 t_{21}, 0, 0, \xi_5 t_{15}^2 t_{21}, 0) \\ \mathcal{N}_r^3 &= (0,\xi_6 t_{15} t_{21}^2 + \xi_7 t_{15} t_{21} t_{22}, \xi_8 t_{15} t_{21}^2, 0, \xi_9 t_{15} t_{21} t_{22}, 0) \\ \mathcal{N}_r^4 &= (0,\xi_{10} t_{15} t_{21}^2 + \xi_{11} t_{15} t_{21} t_{22}, \xi_{12} t_{15} t_{21}^2 + \xi_{13} t_{15} t_{21} t_{22}, \xi_{14} t_{15} t_{21} t_{22}, 0, 0) \\ \mathcal{N}_r^5 &= (0,\xi_{15} t_{15}^2 t_{22}, \xi_{16} t_{15}^2 t_{22}, 0, \xi_{17} t_{15}^2 t_{22}, 0) \end{aligned} \tag{3.67}$$

[6]The crossed topology name $C\mathrm{x}ij\ldots$ refers to a cyclic permutation of the external momenta

where, to isolate the terms corresponding to the desired integrals, we take

$$\begin{aligned}
\xi_1 &= 1 \\
\xi_2 &= \frac{\chi_{45}\left(2\chi_{15}-\chi_{23}+\chi_{45}\right)}{\chi_{23}\left(\chi_{23}-\chi_{45}\right)} \\
\xi_3 &= \frac{2\left(\chi_{15}-\chi_{34}+1\right)\chi_{45}}{\chi_{23}\left(\chi_{23}-\chi_{45}\right)} \\
\xi_4 &= -\frac{2\chi_{23}\chi_{34}}{\chi_{15}-\chi_{34}+1} \\
\xi_5 &= 1 \\
\xi_6 &= \frac{2\chi_{34}\chi_{45}}{\chi_{15}} \\
\xi_7 &= -\frac{\chi_{34}\chi_{45}\left(2\chi_{34}+\chi_{45}-1\right)}{\chi_{15}\left(\chi_{15}-\chi_{34}+1\right)} \\
\xi_8 &= 1 \\
\xi_9 &= \frac{\chi_{45}\left(2\chi_{34}+\chi_{45}-1\right)}{2\chi_{15}\chi_{23}} \\
\xi_{10} &= \frac{2\chi_{34}\chi_{45}}{\chi_{15}} \\
\xi_{11} &= -\frac{2\chi_{34}\chi_{45}\left(-2\chi_{15}+\chi_{23}+4\chi_{34}+\chi_{45}-2\right)}{\chi_{15}\left(2\chi_{15}-\chi_{23}-2\chi_{34}+1\right)} \\
\xi_{12} &= 1 \\
\xi_{13} &= \frac{4\chi_{15}^2-2\left(\chi_{23}+4\chi_{34}+\chi_{45}-2\right)\chi_{15}+\left(2\chi_{34}+\chi_{45}-1\right)^2}{2\chi_{15}\left(2\chi_{15}-\chi_{23}-2\chi_{34}+1\right)} \\
\xi_{14} &= -\frac{\chi_{45}\left(2\chi_{34}+\chi_{45}-1\right)^2}{2\chi_{15}\chi_{23}\left(2\chi_{15}-\chi_{23}-2\chi_{34}+1\right)} \\
\xi_{15} &= \frac{\left(\chi_{23}+1\right)\chi_{34}\chi_{45}}{\chi_{15}\left(\chi_{15}-\chi_{34}+1\right)} \\
\xi_{16} &= 1 \\
\xi_{17} &= \frac{\left(2\chi_{15}-\chi_{23}-2\chi_{34}+1\right)\chi_{45}}{2\chi_{15}\chi_{23}}.
\end{aligned} \tag{3.68}$$

We thus obtain direct IBP equations for the integrals 3.65 in terms of integrals with $s \leq 4$, which we can transform back to equations for the integrals 3.63.

Although we have applied to it to a particular case and not used it to perform a complete reduction to master integrals, the method outlined in this section actually provides a path towards a completely analytic solution of the IBP equations. Using the results in [88], it is possible (for certain topologies) to express an integral with a single, generic irreducible numerator power $\rho$ in terms of masters and coefficients which are functions of $\rho$. This is, however, contingent on being able to determine the special set of vectors satisfying Equation 3.62 and does not admit the possibility of multiple generic powers. Progress in this direction could eventually lead to a completely analytic solution for any IBP system.

### 3.3.4. Results

Our strategy has been implemented in `C++` code which uses the program `Fermat`[89] to manipulate the large rational expressions which occur during solving. We have made available the complete solutions to the IBP equations for topologies $C_1, C_2$ and $C_3$ online. The full results are available at

`http://www.precision.hep.phy.cam.ac.uk/results/amplitudes/`.

The implementation of the method described in Subsection 3.3.1 has been checked against `Reduze` for the four-point, 2-loop amplitude and complete agreement found. A number of non-trivial 2-loop checks have also been performed. We have verified that our calculation for the topology $B_2$ agrees with the results in [83] by comparing all integrals with numerator powers of -4 (which is the highest numerator power computed in that paper). We have also applied the method described in Subsection 3.3.3 to obtain expressions for the integrals 3.63 in terms of integrals with lower numerator powers. Using our calculation for those several dozen integrals with lower powers, we find agreement with our direct calculation of the projections of the integrals 3.63 onto all masters.

We have also inserted the results of the integral reduction into an analytic expression for the planar part of the $q\bar{q} \to Q\bar{Q}g$ amplitude, obtained using the method detailed in Section 3.2. Unlike the 1-loop case, we are unable to verify pole subtraction using the $\mathcal{Z}$ operator as we have only the planar part of the amplitude. The result is split into separate files, one for the projection of the amplitude onto each set of masters (crossed and uncrossed) defined by the set of indices $\nu_i$.

The gigabyte-size of the resulting expressions makes their numerical evaluation non-trivial. In addition, the analytic expressions for the relevant master integrals involve the class of functions known as multiple polylogarithms. The computation of these functions requires the evaluation of iterated integrals, itself a time-consuming procedure. A dedicated effort will be required if one is to use the provided amplitudes for collider phenomenology.

# Chapter 4.

# Real-virtual corrections to top quark production and decay

In this chapter we apply some of the techniques described in Chapter 3 to calculate the real-virtual corrections to both the decay of the top quark and also single top production. We also introduce a more modern amplitude technology known as the spinor-helicity formalism, which removes some of the redundancy in the computation of amplitudes with particles of definite helicity. We provide analytic expressions for the helicity amplitudes in a structure function decomposition.

## 4.1. The top quark

The existence of a third generation of quarks was first posited by Kobayashi and Masakawa, who invoked the GIM mechanism[90] in order to explain the observation of CP violation in kaon decay.[91] The discovery of the bottom quark in 1977 by the E288 experiment[92] implied the existence of its weak isospin doublet partner the top, but it was not until 1995 that discovery was confirmed at the Tevatron experiments[93,94]. Its mass was measured to be $174.3 \pm 5.1$ GeV[95], considerably higher than any other Standard Model particle known at the time.

Because of this large mass, the top quark is the only known coloured particle which decays before it is able to hadronise—it does so with a rate given at leading order by

$$\Gamma(t \to bW^+) = \frac{G_F m_t^3}{8\pi\sqrt{2}} |V_{tb}|^2 \left(1 - \frac{M_W^2}{m_t^2}\right)^2 \left(1 + 2\frac{M_W^2}{m_t^2}\right) \quad \sim \, 1.5\,\text{GeV} \tag{4.1}$$

which leads to a lifetime an order of magnitude shorter than the hadronisation time set by $\Lambda_{\text{QCD}}$. The decay follows almost exclusively a single channel due to the CKM suppression of cross-generational transitions between quark flavours, $|V_{tb}|\sim 1$[1]. The $W^+$ may decay either leptonically or hadronically, with the hadronic channel being the dominant one. Experimentally however, it is the leptonic channel which is easiest to identify amongst a large background of coloured particles and therefore the final state $l^+\nu_l b$ is most often used to reconstruct a top. The measured branching fractions are[96]

$$\begin{aligned} BR(t \to e\nu_e b) &= (13.3 \pm 0.6)\% \\ BR(t \to \mu\nu_\mu b) &= (13.4 \pm 0.6)\% \\ BR(t \to q\bar{q}b) &= (66.5 \pm 1.4)\% \end{aligned}$$

The similarity of the electron and muon channel rates reflects lepton universality, the 'flavour blindness' of the weak interaction to leptons.

This unique property of the top allows us to probe the properties of a 'bare' quark that has not been confined by the strong force—the spin, for example, is preserved in the decay and accessible by measurement of the products. Top quarks are therefore of great interest to phenomenologists. Processes involving tops also provide a large background to signals of New Physics at the LHC and accordingly a good understanding of their behaviour is necessary in order to disentangle the known from the exotic. Running at energies of 7,8 and 13 TeV, the LHC pair-produces tops copiously—the cross section for this process is known at NNLO[97–100] and has been implemented in a publicly available program[101]. Single top production and decay are also known at NNLO[102,103], but at the time of writing the combined pair production and decay is not available. Here, we provide the real-virtual vertex corrections which will contribute to the full calculation of this process.

## 4.2. The spinor-helicity formalism

Before we enter into the details of the calculation proper, it will be instructive to introduce the spinor-helicity formalism which we will later use to write the amplitudes. The ideas therein have made simple computations which would be highly impractical if one resorted to traditional Feynman diagram approaches, such as those we took in the previous chapter. The approach is based on considering spinors for states of definite helicity and representing four momenta as outer products of these spinors. We consider first massless particles, for which the helicity is a well-defined, Lorentz invariant quantity, and then see how properties of massive particles may be incorporated into the formalism. We forgo a discussion of the important topic of colour ordering as in our applications we will work with a combination of coloured and colourless particles—there are, however, several authoritative reviews on the subject[104,105].

[1]The Particle Data Group[96] report a value of $|V_{tb}|= 1.009 \pm 0.031$. In light of this and to prevent undue proliferation of notation, in the remainder of this chapter we will take $|V_{tb}|= 1$.

### 4.2.1. Massless spin-1/2

The Dirac equation for a massless fermion of momentum $p$ is given by

$$i\partial\!\!\!/\psi(p) = 0 \tag{4.2}$$

or, in momentum space,

$$p\!\!\!/U(p) = 0. \tag{4.3}$$

We notice that in the massless limit no distinction is made between particles and antiparticles in the equations of motion. The object $U(p)$ is a four-component Dirac spinor which can in fact be decomposed into two irreducible representations of the Lorentz group, forming two two-component Weyl spinors: we write

$$U(p) = \begin{pmatrix} u_L(p) \\ u_R(p) \end{pmatrix} \tag{4.4}$$

and take the $\gamma$ matrices in the Weyl representation,

$$\gamma^\mu = \begin{pmatrix} 0 & \sigma^\mu \\ \bar{\sigma}^\mu & 0 \end{pmatrix} \tag{4.5}$$

where $\sigma^\mu = (1, \vec{\sigma})$, $\bar{\sigma}^\mu = (1, -\vec{\sigma})$ and the Pauli matrices are given by

$$\sigma_1 = \begin{pmatrix} 0 & 1 \\ 1 & 0 \end{pmatrix}, \quad \sigma_2 = \begin{pmatrix} 0 & -i \\ i & 0 \end{pmatrix}, \quad \sigma_3 = \begin{pmatrix} 1 & 0 \\ 0 & -1 \end{pmatrix}. \tag{4.6}$$

The left- and right-handed Weyl spinors $u_L$ and $u_R$ can then be projected out using the operators

$$P_L = \frac{1-\gamma_5}{2}, \quad P_R = \frac{1+\gamma_5}{2} \tag{4.7}$$

where

$$\gamma_5 = \begin{pmatrix} -1 & 0 \\ 0 & 1 \end{pmatrix}. \tag{4.8}$$

We may then obtain the equations of motion for the Weyl spinors

$$p_\mu \sigma^\mu u_R(p) = 0, \quad p_\mu \bar{\sigma}^\mu u_L(p) = 0. \tag{4.9}$$

The two solutions are related by the charge conjugation operation—to obtain a right-handed solution from a left-handed, we write

$$u_R(p) = i\sigma_2 u_L^*(p). \tag{4.10}$$

We may also identify left- and right-handed antiparticle solutions with right- and left-handed particle solutions, viz. $v_R = u_L$, $v_L = u_R$. The right- and left-handed spinor solutions describe incoming particles with spin parallel and antiparallel to the direction of motion respectively.

Constructing the four-component spinors of definite helicity from the two-component,

$$U_L(p) = \sqrt{2p_0} \begin{pmatrix} u_L(p) \\ 0 \end{pmatrix}, \quad U_R(p) = \sqrt{2p_0} \begin{pmatrix} 0 \\ u_R(p) \end{pmatrix} \tag{4.11}$$

and their conjugates

$$\overline{U}_L(p) = \sqrt{2p_0} \begin{pmatrix} 0 \\ u_L^\dagger(p) \end{pmatrix}, \quad \overline{U}_R(p) = \sqrt{2p_0} \begin{pmatrix} -u_R^\dagger(p) \\ 0 \end{pmatrix} \tag{4.12}$$

we introduce the notation

$$U_L(p) = p], \, U_R(p) = p\rangle, \, \overline{U}_L(p) = \langle p, \, \overline{U}_R(p) = [p. \tag{4.13}$$

The spinors are related to their four-momenta as

$$p\rangle[p = U_R(p)\overline{U}_R(p) = \not{p}\frac{1}{2}\left(1 - \gamma^5\right), \quad p]\langle p = U_L(p)\overline{U}_L(p) = \not{p}\frac{1}{2}\left(1 + \gamma^5\right). \tag{4.14}$$

We may then write the Lorentz-invariant spinor products compactly as

$$\overline{U}_L(p)U_R(q) = \langle pq\rangle, \quad \overline{U}_R(p)U_L(q) = [pq]. \tag{4.15}$$

We observe that the angle and square brackets exhibit some useful properties, for example that

$$\overline{U}_L(p)U_L(q) = \langle pq] = 0, \quad \overline{U}_R(p)U_R(q) = [pq\rangle = 0, \tag{4.16}$$

that

$$\langle pq\rangle = [pq]^*. \tag{4.17}$$

and also that

$$\langle pq\rangle[qp] = \mathrm{Tr}\left\{\not{q}\not{p}\frac{1}{2}\left(1+\gamma^5\right)\right\} = 2p\cdot q. \tag{4.18}$$

The last identity implies that the angle and square brackets are square roots of the kinematic invariants up to an undefined phase,

$$\langle pq\rangle = \sqrt{|2p\cdot q|}e^{i\phi_{pq}}, \quad [qp] = \sqrt{|2p\cdot q|}e^{-i\phi_{pq}}. \tag{4.19}$$

They are also antisymmetric in their arguments, to wit

$$\langle pq\rangle = -\langle qp\rangle, \quad [pq] = -[qp]. \tag{4.20}$$

Amplitudes obtained from Feynman diagrams contain products of spinors with gamma matrices sandwiched in between. There exist some useful relations between objects of this type[105], for example that

$$[p\gamma^\mu q\rangle = \langle q\gamma^\mu p] \tag{4.21}$$

and, using the Fierz identity for the $\sigma$ matrices, that

$$\langle p\gamma^\mu q]\langle k\gamma_\mu l] = 2\langle pk\rangle[lq], \quad \langle p\gamma^\mu q][k\gamma_\mu l\rangle = 2\langle pl\rangle[kq]. \tag{4.22}$$

Finally, we have the Schouten identities

$$\begin{aligned}&\langle ij\rangle\langle kl\rangle + \langle ik\rangle\langle lj\rangle + \langle il\rangle\langle jk\rangle = 0,\\ &[ij][kl] + [ik][lj] + [il][jk] = 0\end{aligned} \tag{4.23}$$

which follow from the antisymmetry of the spinor products. Provided with all identities described above, we are in a position to reduce an amplitude to a product of angle and square brackets which we can then numerically evaluate and square.

### 4.2.2. Massless spin-1

A massless spin-1 particle is described by its four momentum $p^\mu$ and polarisation vector $\varepsilon^\mu$. Working in lightcone gauge with a reference vector $r^\mu$, we have that

$$\varepsilon^\mu p_\mu = 0,\ \varepsilon^\mu r_\mu = 0,\ \varepsilon^*_\mu \varepsilon^\mu = -1 \tag{4.24}$$

where the first condition enforces the polarisation to be transverse to the four momentum, the second enforces the gauge choice $r_\mu A^\mu = 0$ and the last condition is a choice of normalisation. We can represent the outgoing polarisation vectors using our bracket notation as

$$\varepsilon^{*\mu}_R(p) = \frac{\langle r\gamma^\mu p]}{\sqrt{2}\langle rp\rangle}, \quad \varepsilon^{*\mu}_L(p) = -\frac{[r\gamma^\mu p\rangle}{\sqrt{2}[rp]} \tag{4.25}$$

where $r$ cannot be collinear with $p$. It is clear that these representations satisfy the properties in Equation 4.24. Considering the transformation upon changing the reference vector from $r$ to $s$, say, we find that

$$\varepsilon^\mu_R(p,r) - \varepsilon^\mu_R(p,s) = -\sqrt{2}p^\mu \frac{\langle rs\rangle}{\langle rp\rangle\langle sp\rangle} \tag{4.26}$$

from which we arrive at the Ward identity: that, for $\mathcal{M} = \mathcal{M}^\mu \varepsilon_\mu$, we have that

$$\mathcal{M}^\mu p_\mu = 0. \tag{4.27}$$

This allows us to choose a different reference vector for each external spin-1 particle without changing the result for the amplitude, a property that can enable great simplification in the final expression.

### 4.2.3. Massive spin-1/2

For a massive particle, helicity is no longer a good quantum number—the mass term in the Dirac equation connects left- and right-handed terms and helicity is no longer necessarily preserved in an interaction. However, we may still accommodate this type of particle in our formalism. For a massive quark of momentum $p_t$ and mass $m_t$ such that $p_t^2 = m_t^2$, we may decompose the massive momentum into two massless momenta, $p_t = p_u + \eta$ with $p_u^2 = \eta^2 = 0$. The massive spinor obeys the Dirac equation $(\not{p}_t - m_t)u_t(p_t)$ and can be

decomposed as

$$u_+(p_t) = |p_u\rangle + \frac{m_t|\eta]}{[p_u\eta]}, \quad u_-(p_t) = |p_u] + \frac{m_t|\eta\rangle}{\langle p_u\eta\rangle}. \tag{4.28}$$

The reference vector $\eta$ is again arbitrary and may be set to one of the massless momenta in the problem—when the amplitude is squared, all $\eta$ dependence should vanish.

## 4.3. NNLO corrections to the heavy-light quark vertex

We turn to a discussion of the higher order corrections to the heavy-light quark vertex. The 2-loop corrections to this vertex are well known and have been calculated in the context of decays of B mesons[106–108], but the real-virtual correction which we consider here is not publicly available in a convenient form (though it may be obtained via a crossing of the results presented in [109]). Our discussion is not limited to a particular process, as calculating the quark current with an off-shell $W$ boson will in fact be of use in several different cases. For example, with the top in the initial state and the $W$ boson in the s-channel we may couple the free Lorentz index to a leptonic line and obtain the amplitude for the top decay, but we could also consider the case with the top in the final state and the W in the t-channel. This latter case, when coupled to a light quark line, would give the amplitude for single top production—the two are related by a simple crossing of the external legs. For concreteness, however, in this section we will detail the calculation for the decay case, before providing results for both the top decay and the single top production amplitudes.

### 4.3.1. Set-up of the calculation

We will compute the helicity amplitudes rather than the squared amplitude summed over spins—this will allow us to isolate the contributions from the different spin states of the top. We work in Feynman gauge and use the t'Hooft-Veltman regularisation scheme, in which external states are four-dimensional and loop momenta are $d$-dimensional. We consider the process

$$t(p_1) \to b(p_2) \,+\, W^*(p_3) \,+\, g(p_4) \tag{4.29}$$

where the top has a mass $m_t$, the $W$ has a four-momentum $Q$ and we neglect the mass of the $b$ quark. We define kinematic invariants

$$\begin{aligned} s_1 &= (p_1 - p_2)^2 \\ s_2 &= (p_2 + p_4)^2. \end{aligned} \tag{4.30}$$

The most general tensor for the heavy-light quark current can be written down by considering the independent Dirac structures left after imposing the equations of motion and the commutation relations for the gamma matrices. It is given by

$$S^\mu(t, b, W, g) = \sum_{i=1}^{26} C_i S_i^\mu \tag{4.31}$$

where the structure functions $C_i$ are functions of the kinematic invariants and the 26 Dirac structures are

$$\begin{aligned}
S_1^\mu &= \bar{u}(p_2)\not{\varepsilon}_4 \not{p}_4 \gamma^\mu u(p_1) & S_{14}^\mu &= \bar{u}(p_2)\not{\varepsilon}_4 p_4^\mu u(p_1) \\
S_2^\mu &= \bar{u}(p_2)\gamma^\mu \not{p}_4 \not{\varepsilon}_4 u(p_1) & S_{15}^\mu &= \bar{u}(p_2)\not{p}_4 (p_1 \cdot \varepsilon_4) p_1^\mu u(p_1) \\
S_3^\mu &= \bar{u}(p_2)\not{p}_4 \gamma^\mu (p_1 \cdot \varepsilon_4) m_t u(p_1) & S_{16}^\mu &= \bar{u}(p_2)\not{p}_4 (p_1 \cdot \varepsilon_4) p_2^\mu u(p_1) \\
S_4^\mu &= \bar{u}(p_2)\not{p}_4 \gamma^\mu (p_2 \cdot \varepsilon_4) m_t u(p_1) & S_{17}^\mu &= \bar{u}(p_2)\not{p}_4 (p_1 \cdot \varepsilon_4) p_4^\mu u(p_1) \\
S_5^\mu &= \bar{u}(p_2)\gamma^\mu \not{\varepsilon}_4 m_t u(p_1) & S_{18}^\mu &= \bar{u}(p_2)\not{p}_4 (p_2 \cdot \varepsilon_4) p_1^\mu u(p_1) \\
S_6^\mu &= \bar{u}(p_2)\not{\varepsilon}_4 \gamma^\mu m_t u(p_1) & S_{19}^\mu &= \bar{u}(p_2)\not{p}_4 (p_2 \cdot \varepsilon_4) p_2^\mu u(p_1) \\
S_7^\mu &= \bar{u}(p_2)\not{\varepsilon}_4 \not{p}_4 p_1^\mu m_t u(p_1) & S_{20}^\mu &= \bar{u}(p_2)\not{p}_4 (p_2 \cdot \varepsilon_4) p_4^\mu u(p_1) \\
S_8^\mu &= \bar{u}(p_2)\not{\varepsilon}_4 \not{p}_4 p_2^\mu m_t u(p_1) & S_{21}^\mu &= \bar{u}(p_2)(\varepsilon_4 \cdot p_1) p_1^\mu m_t u(p_1) \\
S_9^\mu &= \bar{u}(p_2)\not{\varepsilon}_4 \not{p}_4 p_4^\mu m_t u(p_1) & S_{22}^\mu &= \bar{u}(p_2)(\varepsilon_4 \cdot p_1) p_2^\mu m_t u(p_1) \\
S_{10}^\mu &= \bar{u}(p_2)\gamma^\mu (p_1 \cdot \varepsilon_4) u(p_1) & S_{23}^\mu &= \bar{u}(p_2)(\varepsilon_4 \cdot p_1) p_4^\mu m_t u(p_1) \\
S_{11}^\mu &= \bar{u}(p_2)\gamma^\mu (p_2 \cdot \varepsilon_4) u(p_1) & S_{24}^\mu &= \bar{u}(p_2)(\varepsilon_4 \cdot p_2) p_1^\mu m_t u(p_1) \\
S_{12}^\mu &= \bar{u}(p_2)\not{\varepsilon}_4 p_1^\mu u(p_1) & S_{25}^\mu &= \bar{u}(p_2)(\varepsilon_4 \cdot p_2) p_2^\mu m_t u(p_1) \\
S_{13}^\mu &= \bar{u}(p_2)\not{\varepsilon}_4 p_2^\mu u(p_1) & S_{26}^\mu &= \bar{u}(p_2)(\varepsilon_4 \cdot p_2) p_4^\mu m_t u(p_1).
\end{aligned} \tag{4.32}$$

We have imposed the tranversality constraint $\varepsilon_4 \cdot p_4 = 0$ here. We may further reduce the number of structures to a truly independent set by imposing the QCD Ward identity,

$$S^\mu(t, b, W, g)(\varepsilon_4 \to p_4) = 0, \tag{4.33}$$

which yields the relations

$$
\begin{aligned}
C_5 - C_6 &= C_3(p_1 \cdot p_4) + C_4(p_2 \cdot p_4) \\
C_{11} &= -C_{10}\frac{p_1 \cdot p_4}{p_2 \cdot p_4} \\
C_{12} &= -C_{15}(p_1 \cdot p_4) - C_{18}(p_2 \cdot p_4) \\
C_{13} &= -C_{16}(p_1 \cdot p_4) - C_{19}(p_2 \cdot p_4) \\
C_{14} &= -C_{17}(p_1 \cdot p_4) - C_{20}(p_2 \cdot p_4) \\
C_{24} &= -C_{21}\frac{p_1 \cdot p_4}{p_2 \cdot p_4} \\
C_{25} &= -C_{22}\frac{p_1 \cdot p_4}{p_2 \cdot p_4} \\
C_{26} &= -C_{23}\frac{p_1 \cdot p_4}{p_2 \cdot p_4} - 2C_5\frac{1}{p_2 \cdot p_4}.
\end{aligned}
\tag{4.34}
$$

The electroweak vertex that couples the $b$ and the $t$ has a factor of $\frac{1}{2}(1-\gamma^5)$, which, as we discussed in Chapter 2, may cause some difficulty when working in $d$-dimensions. To obviate this inconvenience, we enforce the spinor for the $b$ to be left-handed and consequently remove the axial coupling from the vertex. This is possible because the chiral and helicity states coincide for the massless $b$. The helicity amplitudes we present later will only involve left-handed $b$ quarks, with vanishing amplitudes for the right-handed cases.

## 4.3.2. UV renormalisation

The structure functions may be perturbatively expanded as

$$C_i = \frac{\sqrt{4\pi\alpha_s}g_W}{\sqrt{2}}T^a_{ij}\left[C_i^{(0)} + \left(\frac{\alpha_s}{4\pi}\right)C_i^{(1)} + \left(\frac{\alpha_s}{4\pi}\right)^2 C_i^{(2)} + \mathcal{O}(\alpha_s^3)\right] \tag{4.35}$$

In addition to the UV renormalisation of the coupling we performed in Chapter 3 and the wavefunction renormalisations, the introduction of the top quark means we also have a mass to renormalise. Once again we renormalise $\alpha_s$ in the $\overline{\text{MS}}$ scheme but now renormalise the mass and heavy quark wavefunction in the on-shell (OS) scheme. We treat each quantity to be renormalised in turn. We consider the wavefunction renormalisation and, setting aside the mass and coupling renormalisation temporarily, the renormalised structure functions are given by

$$C_i = \sqrt{Z_{2,b}^{\overline{\text{MS}}}Z_{2,t}^{OS}}C_i^b(\alpha_s^b). \tag{4.36}$$

where the renormalisation constants are given at 1-loop by

$$\begin{aligned} Z_{2,t}^{OS} &= 1 + \frac{\alpha_s}{4\pi} C_F \left[ -\frac{3}{\epsilon} - 4 - 3\log\left(\frac{\mu^2}{m_t^2}\right)\right] + \mathcal{O}(\alpha_s^2) \\ Z_{2,b}^{\overline{\text{MS}}} &= 1 + \mathcal{O}(\alpha_s^2) \end{aligned} \tag{4.37}$$

With the inclusion of the top quark, we are now working in the full theory of QCD with 6 quark flavours. However, given that we consider processes with scales $\mathcal{O}(m_t)$, it would be inappropriate to include contributions from the top itself to the running of the coupling which is of course a loop effect. With this in mind, we renormalise $\alpha_s$ in the $\overline{\text{MS}}$ scheme with 6 active flavours and then decouple the top from the running using the relationship in Equation 2.62.

In all, we have before mass renormalisation at 1-loop

$$C_i^{(1)} = C_i^{(1),b} + \left( -\frac{\beta_0}{\epsilon} - \frac{1}{6}\log\left(\frac{\mu^2}{m_t^2}\right) + \frac{1}{2}\delta_{2,t}^{(1)} \right) C_i^{(0)}. \tag{4.38}$$

We recall that the Feynman rules for a renormalised field theory require us to include diagrams with counterterm insertions. Anticipating the diagram generation, we require the counterterm Feynman rule for a heavy quark line

$$—\!\!\otimes\!\!— = (\not{p}\delta_{2,t} - (\delta_m^{OS} + \delta_{2,t}) m_t) \tag{4.39}$$

and find that, as chance would have it, at 1-loop (and only at 1-loop)

$$Z_m^{OS} = Z_{2,t} \tag{4.40}$$

### 4.3.3. Extraction of the structure functions

The past few decades have seen much progress in the development of technology to compute amplitudes, some based on rather sophisticated mathematics—they may be derived as a limiting case of some string scattering processes, through the amplituhedron, *&c*. We will not detail these cutting-edge approaches here, which often work best in problems with a high degree of symmetry, though good reviews exist[110]. Instead, we will discuss two approaches to computing the structure functions through traditional Feynman diagram based techniques. The first was implemented in the calculation of 2-loop corrections to quark-quark, quark-gluon, 3-jet scattering and other processes by Glover *& al.*[111–113] and is known as a projection method. In this approach, projection operators are constructed to isolate the coefficients of the Dirac structures. We write the expansion of a generic amplitude as

$$\mathcal{M} = \sum_{i=1}^{n} A_i\left(\{s_k\}\right) S_i \tag{4.41}$$

where the $A_i$ are the structure functions and the $S_i$ are the Dirac structures, and expand also the $A_i$ in a basis of colour structures $\mathcal{C}_i$

$$A_i = \sum_{I=1}^{N} A_i^{[I]}\left(\{s_k\}\right) \mathcal{C}_i \tag{4.42}$$

where the capital Latin indices now label the colour basis. We now form projection operators

$$\begin{aligned} \mathcal{P}_j \mathcal{M} &= S_j^\dagger \mathcal{M} \\ &= \sum_{i=1}^{n} A_i\left(\{s_k\}\right) \left(S_j^\dagger S_i\right) \end{aligned} \tag{4.43}$$

and, defining the matrix $O_{ij} \equiv S_j^\dagger S_i$, we may isolate the coefficients as

$$A_i\left(\{s_k\}\right) = \sum_{j=1}^{n} O_{ij}^{-1} S_j^\dagger \mathcal{M}. \tag{4.44}$$

This approach is not limited by the number of loops and involves a straightforward Feynman-diagrammatic calculation with evaluation of traces over Dirac and colour matrices as usual. However, as the number of independent Dirac structures becomes large the inversion of the matrix $O_{ij}$ becomes computationally difficult and the method is therefore impractical to apply to the case we consider here. The second tactic, which we adopt, involves writing down the amplitude from the Feynman rules and reducing the tensor integrals which appear to scalar ones before applying the IBP identities. The method of reduction of 1-loop tensor integrals is originally due to Passarino and Veltman[114], and we detail the procedure below. We should mention that, because tensor reductions for 2-loop integrals are known only up to 3-point functions[115], this approach is often only useful at 1 loop.

The Passarino-Veltman reduction is based on considerations of Lorentz covariance. Defining a general 1-loop tensor integral as

$$T^N_{\mu_1 \ldots \mu_P}\left(p_1, \ldots, p_{N-1}, m_0, \ldots, m_{N-1}\right) = \frac{\mu^{2\epsilon}}{i\pi^{\frac{d}{2}} r_\Gamma} \int d^d k \frac{k_{\mu_1} \ldots k_{\mu_P}}{D_0 D_1 \ldots D_{N-1}} \tag{4.45}$$

with denominators

$$D_0 = k^2 - m_0^2,\ D_i = (k + p_i)^2 - m_i^2,\ i = 1, \ldots, N-1 \tag{4.46}$$

we may decompose the tensor into a linear combination of tensors constructed from the external momenta and the metric tensor. With $p_{ij} = p_i + p_j$ and $p_{i0} = p_i$, we introduce an artificial momentum $p_0$ to write[116]

$$T^N_{\mu_1 \ldots \mu_P}(p_1, \ldots, p_{N-1}, m_0, \ldots, m_{N-1}) = \sum_{i_1, \ldots, i_P}^{N-1} T^N_{i_1, \ldots, i_P} p_{i_1 \mu_1} \ldots p_{i_P \mu_P} \tag{4.47}$$

where the $g_{\mu\nu}$ terms are obtained by omitting all terms containing an odd factor of $p_0$ and even factors of $p_0$ are replaced by the completely symmetric tensor constructed from the $g_{\mu\nu}$, viz.

$$\begin{aligned} p_{0\mu_1} p_{0\mu_2} &\to g_{\mu_1\mu_2}, \\ p_{0\mu_1} p_{0\mu_2} p_{0\mu_3} p_{0\mu_4} &\to g_{\mu_1\mu_2} g_{\mu_3\mu_4} + g_{\mu_1\mu_3} g_{\mu_2\mu_4} + g_{\mu_1\mu_4} g_{\mu_2\mu_3}. \end{aligned} \tag{4.48}$$

The coefficients $T^N_{i_1, \ldots, i_P}$ are then obtained by projection. Dotting with all external momenta and the metric, we rewrite scalar products of the loop momentum with external momenta as denominator factors which cancel below. To demonstrate, we consider the rank-1 massless 3-point integral

$$\begin{aligned} T^3_\mu(p_1, p_2, 0, 0) &= \int_{-\infty}^{\infty} d^d \bar{k} \frac{k_\mu}{k^2 (k + p_1)^2 (k + p_{12})^2} \\ &= T^3_1 l_{1\mu} + T^3_2 l_{2\mu} \end{aligned} \tag{4.49}$$

where $l_1 = p_1, l_2 = p_{12}$ and we have absorbed factors in $d^d\bar{k}$. Dotting each term with the $l_{i\mu}$ and using the identity

$$k \cdot l_i = \frac{1}{2}\left[(k + l_i)^2 - k^2 - l_i^2\right] \tag{4.50}$$

we have the matrix equation

$$\begin{aligned} &\begin{pmatrix} 2l_1 \cdot l_1 & 2l_1 \cdot l_2 \\ 2l_2 \cdot l_1 & 2l_2 \cdot l_2 \end{pmatrix} \begin{pmatrix} T^3_1 \\ T^3_2 \end{pmatrix} = \begin{pmatrix} \lambda_1 \\ \lambda_2 \end{pmatrix} \\ &\lambda_1 = T^2_0(l_2) - T^2_0(l_2 - l_1) - l_1^2 T^3_0(l_1, l_2) \\ &\lambda_2 = T^2_0(l_2) - T^2_0(l_2 - l_1) - l_2^2 T^3_0(l_1, l_2) \end{aligned} \tag{4.51}$$

which we can solve by inverting the Gram matrix [2]. The tensor integral is now expressed in terms of a linear combination of scalar integrals $T_0^N$.

The calculation proceeds as follows. We generate diagrams using `QGRAF` and feed the output through a modified version of `Reduze` to map the diagrams to topologies and produce analytic expressions for the amplitudes following from the Feynman diagrams. We pass the amplitudes to `FORM`, which decomposes the amplitude into tensor structures and then inserts expressions for the tensor integrals obtained in `Mathematica` via a Passarino-Veltman reduction. `FORM` then performs the algebraic manipulations necessary to rewrite the resulting tensor structures in terms of those defined in Equation 4.32. This last step involves commuting gamma matrices past each other and implementing the equations of motion for the spinors. The output, a set of scalar integrals multiplying Dirac structures, is passed back to `Reduze` and the results of the IBP reduction for the topologies are inserted. Finally, we obtain a full expression for $S^\mu(t, W, b, g)$ where the structure functions are linear combinations of master integrals.

### 4.3.4. Diagrams, topologies and master integrals

We find two diagrams at leading order (see Figure 4.1) and 11 at 1-loop order (see Figure 4.2), as well as a single non-vanishing counterterm diagram. The loop diagrams can be mapped to three topologies, which we list in Table 4.1.

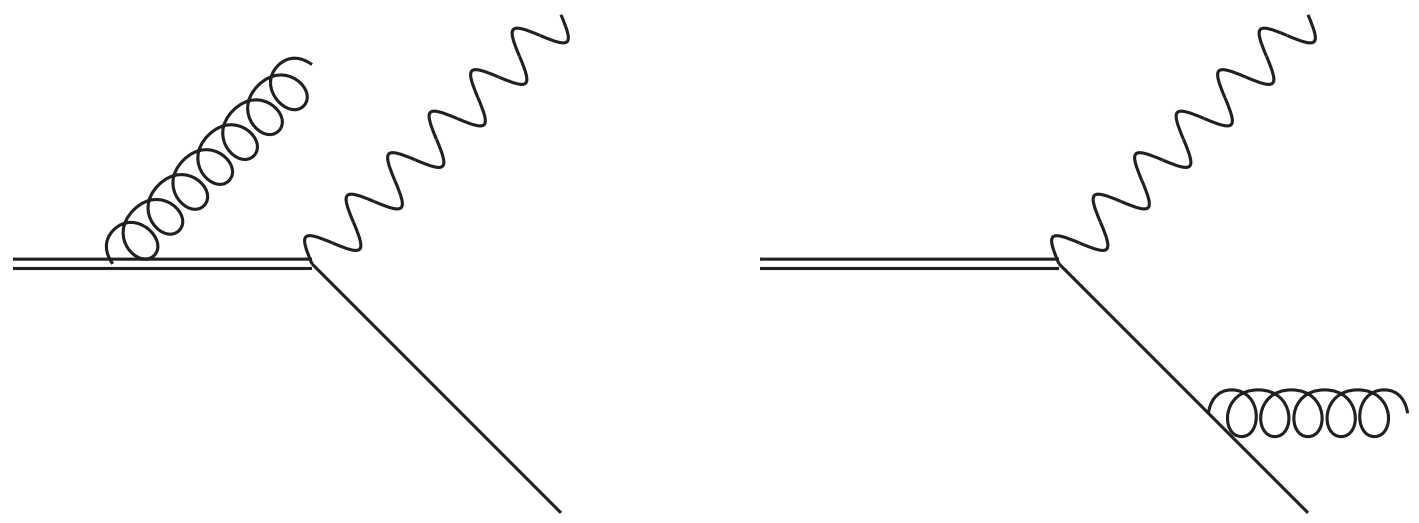

**Figure 4.1.:** The leading order $tWbg$ vertex diagrams.

We require a Passarino-Veltman reduction of integrals up to rank 3 for 2-, 3- and 4-point functions. After the tensor and IBP reductions, we find 11 master integrals in total: 6 belonging to family D, 2 to E and 3 to F. The integrals were obtained from `QCDLoop`, with the exception of $D(1,1,1,0)$ for which no result was available—we took the result for the 3-point function with massive lines presented in [116]. All results were again checked numerically against the output of the program `SecDec`.

[2]This can become problematic when the matrix becomes singular—if this is due to some of the momenta being linearly dependent, we may simply omit these momenta in the decomposition. If the momenta are not linearly dependent but the determinant is zero, an alternative reduction strategy must be found[117].

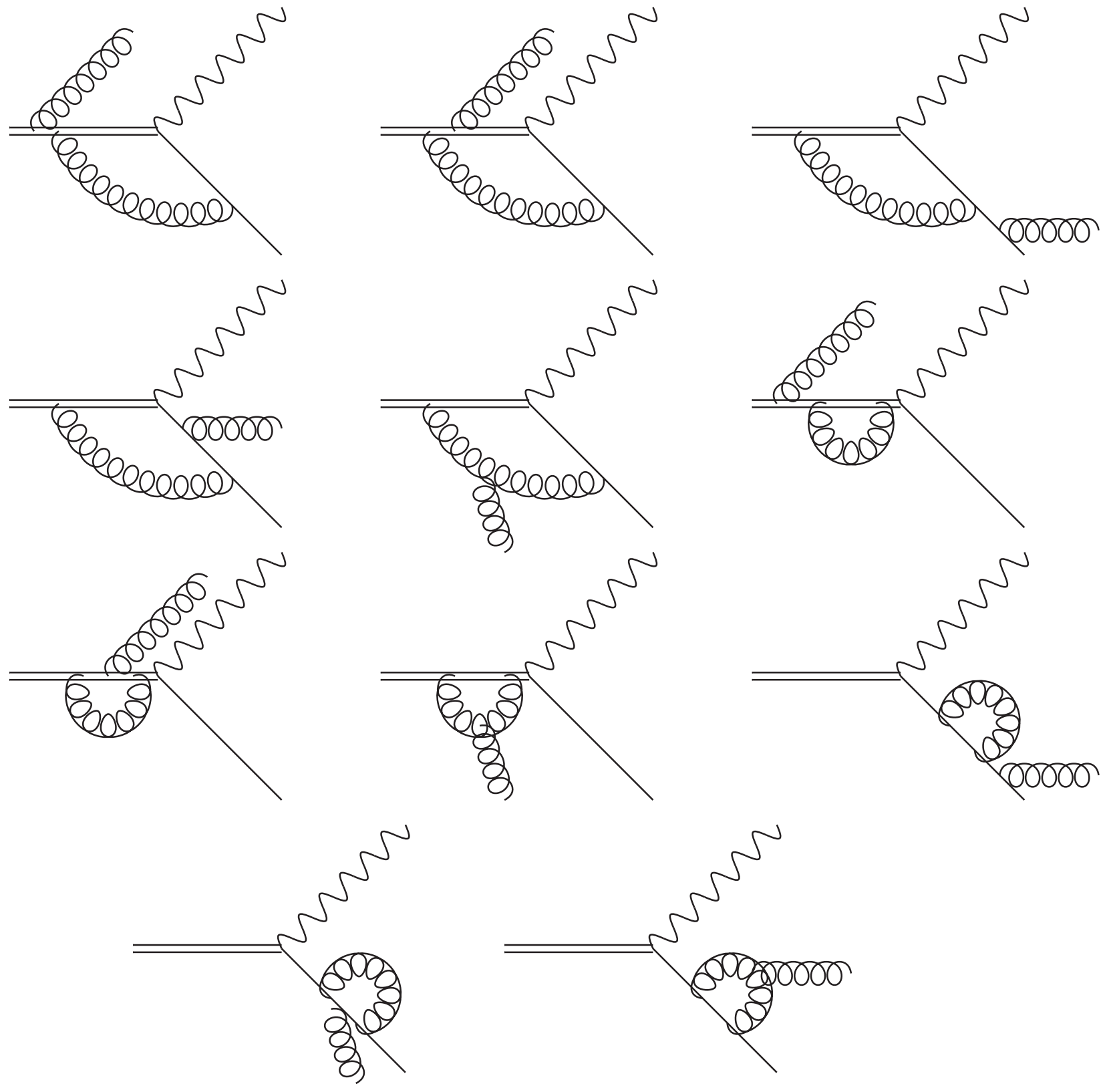

**Figure 4.2.:** The 1-loop corrections to the $tWbg$ vertex.

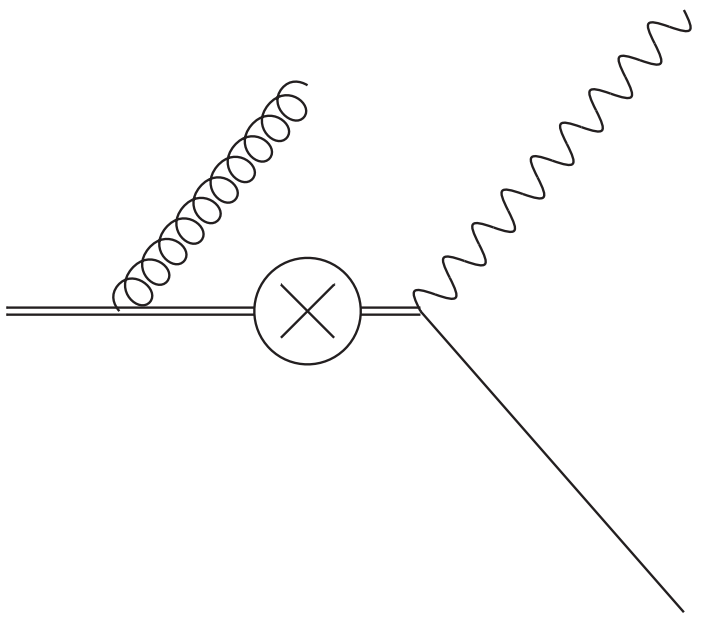

**Figure 4.3.:** The counterterm diagram required for the UV renormalisation of the amplitude.

| Family D | Family E | Family F |
|---|---|---|
| $(k, 0)$ | $(k, 0)$ | $(k, 0)$ |
| $(k - p_1, m_t)$ | $(k - p_1, m_t)$ | $(k - p_1, m_t)$ |
| $(k - p_2 - p_4, 0)$ | $(k - p_2 - p_4, 0)$ | $(k - p_1 + p_4, m_t)$ |
| $(k - p_4, 0)$ | $(k - p_2, 0)$ | $(k - p_2, 0)$ |

**Table 4.1.:** Integral families required to capture all 1-loop diagrams for the vertex $tWbg$.

Using the notation

$$L_1 = \log\left(\frac{\mu^2}{m_t^2 - s_1}\right),\ L_2 = \log\left(\frac{\mu^2}{s_2}\right),\ L_{12} = \log\left(\frac{\mu^2}{s_1 + s_2 - M_W^2}\right),\ L_t = \log\left(\frac{\mu^2}{m_t^2}\right),$$
$$L_{tw} = \log\left(\frac{\mu^2}{m_t^2 - M_W^2}\right),\ \kappa = \sqrt{m_t^4 + (M_W^2 - s_2)^2 - 2m_t^2(M_W^2 + s_2)} \quad (4.52)$$

the master integrals are given by

$$D(0,1,0,0) = \frac{m_t^2}{\epsilon} + (L_t + 1)\, m_t^2 + \mathcal{O}(\epsilon) \quad (4.53)$$

$$D(1,0,1,0) = \frac{1}{\epsilon} + (L_2 + i\pi + 2) + \mathcal{O}(\epsilon) \quad (4.54)$$

$$D(0,1,1,0) = \frac{1}{\epsilon} + \left[L_{\text{tw}}\left(1 - \frac{m_t^2}{M_W^2}\right) + \frac{L_t m_t^2}{M_W^2} + 2\right] + \mathcal{O}(\epsilon) \quad (4.55)$$

$$D(0,1,0,1) = \frac{1}{\epsilon} + \frac{(L_t + 2)\, m_t^2 + (L_{12} + 2)\,(M_W^2 - s_1 - s_2)}{m_t^2 + M_W^2 - s_1 - s_2} + \mathcal{O}(\epsilon) \quad (4.56)$$

$$\begin{aligned}
D(1,1,1,0) = \frac{1}{2\kappa}\Bigg\{ & 4\text{Li}_2\left(\frac{m_t^2 - M_W^2 - \kappa + s_2}{m_t^2 - M_W^2 + \kappa + s_2}\right) \\
& + 2\text{Li}_2\left(\frac{m_t^4 - (2M_W^2 + \kappa + s_2)\, m_t^2 + M_W^2\,(M_W^2 - \kappa - s_2)}{m_t^4 + (-2M_W^2 + \kappa - s_2)\, m_t^2 + M_W^2\,(M_W^2 - \kappa - s_2)}\right) \\
& + 2\text{Li}_2\left(\frac{m_t^4 - (2M_W^2 + \kappa + s_2)\, m_t^2 + M_W^2\,(M_W^2 - \kappa - s_2)}{m_t^4 - (2M_W^2 + \kappa + s_2)\, m_t^2 + M_W^2\,(M_W^2 + \kappa - s_2)}\right) \\
& - 2\text{Li}_2\left(\frac{m_t^4 - (2M_W^2 + \kappa + s_2)\, m_t^2 + M_W^2\,(M_W^2 + \kappa - s_2)}{m_t^4 + (-2M_W^2 + \kappa - s_2)\, m_t^2 + M_W^2\,(M_W^2 - \kappa - s_2)}\right) \\
& + \log^2\left(\frac{\kappa + m_t^2 - M_W^2 + s_2}{-\kappa + m_t^2 - M_W^2 + s_2}\right) - 2\pi i \log\left(\frac{\kappa + m_t^2 - M_W^2 + s_2}{-\kappa + m_t^2 - M_W^2 + s_2}\right) \\
& - 4\text{Li}_2\left(1 - \frac{\kappa}{s_2}\right) - \frac{\pi^2}{3}\Bigg\} + \mathcal{O}(\epsilon)
\end{aligned} \quad (4.57)$$

$$
\begin{aligned}
D(1,1,1,1) = & \frac{3}{2s_2\left(M_W^2-s_1-s_2\right)}\frac{1}{\epsilon^2} \\
& + \frac{L_t+2L_2-4\log\left(\frac{-M_W^2+s_1+s_2}{m_t^2}\right)+2\log\left(1-\frac{M_W^2}{m_t^2}\right)+2\pi i}{2s_2\left(M_W^2-s_1-s_2\right)}\frac{1}{\epsilon} \\
& - \frac{1}{12s_2\left(M_W^2-s_1-s_2\right)}\Bigg\{-12L_2\left[L_t-2\log\left(\frac{-M_W^2+s_1+s_2}{m_t^2}\right)\right] \\
& - 12L_t\left[\log\left(1-\frac{M_W^2}{m_t^2}\right)+i\pi\right]+3L_t^2+24\mathrm{Li}_2\left(\frac{-m_t^2+s_1+s_2}{-M_W^2+s_1+s_2}\right) \\
& + 24i\pi\log\left(\frac{-M_W^2+s_1+s_2}{m_t^2}\right)+12\log^2\left(1-\frac{M_W^2}{m_t^2}\right)+5\pi^2\Bigg\}+\mathcal{O}\left(\epsilon\right)
\end{aligned} \tag{4.58}
$$

$$
E(0,1,0,1) = \frac{1}{\epsilon}+\left[\frac{L_t m_t^2}{s_1}+L_1\left(1-\frac{m_t^2}{s_1}\right)+2\right]+\mathcal{O}\left(\epsilon\right) \tag{4.59}
$$

$$
\begin{aligned}
E(1,1,1,1) = & \frac{3}{2s_2\left(s_1-m_t^2\right)\epsilon^2}\frac{1}{\epsilon^2}+\frac{\frac{1}{2}L_t+L_2+\log\left(1-\frac{M_W^2}{m_t^2}\right)-2\log\left(1-\frac{s_1}{m_t^2}\right)+i\pi}{s_2\left(s_1-m_t^2\right)}\frac{1}{\epsilon} \\
& + \frac{1}{12s_2\left(m_t^2-s_1\right)}\Bigg\{-12\pi i L_t-12L_t\log\left(1-\frac{M_W^2}{m_t^2}\right) \\
& + 12L_2\left[2\log\left(1-\frac{s_1}{m_t^2}\right)-L_t\right]+3L_t^2+24\mathrm{Li}_2\left(\frac{M_W^2-s_1}{m_t^2-s_1}\right) \\
& + 12\log^2\left(1-\frac{M_W^2}{m_t^2}\right)+24\pi i\log\left(1-\frac{s_1}{m_t^2}\right)+5\pi^2\Bigg\}+\mathcal{O}\left(\epsilon\right)
\end{aligned} \tag{4.60}
$$

$$
F(1,1,1,0) = \frac{\pi^2-6\mathrm{Li}_2\left(\frac{m_t^2+M_W^2-s_1-s_2}{m_t^2}\right)}{6\left(M_W^2-s_1-s_2\right)}+\mathcal{O}\left(\epsilon\right) \tag{4.61}
$$

$$
F(0,1,1,1) = \frac{\mathrm{Li}_2\left(\frac{s_1}{m_t^2}\right)-\mathrm{Li}_2\left(\frac{M_W^2}{m_t^2}\right)}{M_W^2-s_1}+\mathcal{O}\left(\epsilon\right) \tag{4.62}
$$

$$
\begin{aligned}
F(1,1,1,1) = {} & \frac{1}{2\epsilon^2\left(m_t^2-s_1\right)\left(-M_W^2+s_1+s_2\right)} \\
& + \frac{L_t - 2\log\left(\frac{-M_W^2+s_1+s_2}{m_t^2}\right) + 2\log\left(1-\frac{M_W^2}{m_t^2}\right) - 2\log\left(1-\frac{s_1}{m_t^2}\right)}{2\epsilon\left(m_t^2-s_1\right)\left(-M_W^2+s_1+s_2\right)} \\
& - \frac{1}{12\left(m_t^2-s_1\right)\left(-M_W^2+s_1+s_2\right)}\Bigg\{ -12L_t\Bigg[-\log\left(\frac{-M_W^2+s_1+s_2}{m_t^2}\right) \\
& + \log\left(1-\frac{M_W^2}{m_t^2}\right) - \log\left(1-\frac{s_1}{m_t^2}\right)\Bigg] - 3L_t^2 + 24\mathrm{Li}_2\left(\frac{M_W^2-s_1}{m_t^2-s_1}\right) \\
& + 24\mathrm{Li}_2\left(-\frac{-m_t^2+s_1+s_2}{M_W^2-s_1-s_2}\right) - 24\log\left(1-\frac{s_1}{m_t^2}\right)\log\left(\frac{-M_W^2+s_1+s_2}{m_t^2}\right) \\
& + 12\log^2\left(1-\frac{M_W^2}{m_t^2}\right) + \pi^2\Bigg\} + \mathcal{O}\left(\epsilon\right).
\end{aligned}
\tag{4.63}
$$

An explicit analytic continuation has been performed to the physical region for the process in which the top is in the initial state. While crossing the legs of the vertex in order to obtain contributions to, say, single top production does not change the amplitude directly, the different kinematics will result in a different physical region for the kinematic invariants and so the master integrals need to be analytically continued accordingly.

### 4.3.5. Tree-level results

We present the tree-level results for the structure functions; the 1-loop results are included in the ancillary material that accompanies this thesis. At this order there are only four non-zero structure functions:

$$
\begin{aligned}
C_1^{(0)} &= \frac{1}{s_2} \\
C_2^{(0)} &= \frac{1}{s_1+s_2-M_W^2} \\
C_{10}^{(0)} &= -\frac{2}{s_1+s_2-M_W^2} \\
C_{11}^{(0)} &= \frac{2}{s_2}.
\end{aligned}
\tag{4.64}
$$

## 4.4. Top quark decay

Given the vertex correction derived above, we are now in a position to calculate the real-virtual correction to the top quark decay. We consider only the leptonic decay of the $W$ (since this is primarily how it is identified experimentally) and accordingly assign momenta:

$$t(p_1) \to b(p_2) \,+\, g(p_4) \,+\, \nu_l(p_5) \,+\, l^+(p_6). \tag{4.65}$$

We now assign definite helicity states to the external particles to obtain the helicity amplitudes. For the $b$ and the leptons, the helicity structure is fixed by the electroweak vertex—the $b$ and the $\nu_l$ must be in negative helicity states while the $l^+$ must be in a positive helicity state. The top and the gluon, however, can be in either state and we label them $\lambda_1$, $\lambda_2$. The helicity amplitudes are obtained by contracting the leptonic tensor $L_\mu$ with the quark line

$$|\mathcal{M}_{\lambda_1\lambda_2}\rangle = L_\mu(\nu_l, l^+) S^\mu_{\lambda_1\lambda_2}(t, W, b, g) P_W \tag{4.66}$$

where the leptonic tensor is given in spinor-helicity notation by

$$L_\mu = \langle p_5 \gamma_\mu p_6] \tag{4.67}$$

and the $W$ propagator factor $P_W = 1/(Q^2 - M_W^2)$.

We assign the auxiliary momenta $q_1$,$r_1$ and $r_2$ to the spinor-helicity representations of the massive spinor and the gluon polarisation vector, defining

$$\begin{aligned} p_1^\mu &= q_1^\mu + \frac{m_t^2}{2p_1 \cdot r_2} r_2^\mu \\ \varepsilon_R^{*\mu}(p_4) &= \frac{1}{\sqrt{2}} \frac{\langle r_1 \gamma^\mu p_4]}{\langle r_1 p_4\rangle} \\ \varepsilon_L^{*\mu}(p_4) &= -\frac{1}{\sqrt{2}} \frac{[r_1 \gamma^\mu p_4\rangle}{[r_1 p_4]}. \end{aligned} \tag{4.68}$$

The `FORM` library `Spinney`[118] was used to compute the helicity amplitudes for each helicity combination at tree level and at 1-loop and to perform simplification of the spinor bracket structures where possible. Expressions for the UV renormalised form factors were then inserted to obtain the amplitude. The IR subtraction was performed using the

operator

$$\mathcal{Z} = 1 + \frac{\alpha_s}{4\pi}\left[-\frac{13}{3\epsilon^2} + \frac{-6\log\left(-\frac{\mu m_t}{M_W^2 - s_1 - s_2}\right) + \frac{2}{3}\log\left(-\frac{\mu m_t}{s_1 - m_t^2}\right) - 6\log\left(-\frac{\mu^2}{s_2}\right) - \frac{43}{3}}{2\epsilon}\right] + \mathcal{O}(\alpha_s^2). \tag{4.69}$$

It should be noted, however, that since we are working in the t'Hooft-Veltman scheme rather than CDR as in Chapter 3, all external states are considered four-dimensional and so the Born term has no subleading $\epsilon$ dependence whatsoever. The IR subtraction therefore serves only to remove the poles and the finite remainder of the amplitude is the same as the finite piece of the amplitude. The results were verified numerically against the program `GoSam`[119] and full agreement found.

The Born amplitudes are

$$\begin{aligned}\left|\mathcal{M}^{(0)}_{++}\right\rangle = &\frac{m_t^3 C_{10}^{(0)} \langle p_5 p_2\rangle [r_2 p_4] [r_2 p_6] \langle r_2 r_1\rangle}{(p_1\cdot r_2)[r_2 q_1]\langle r_1 p_4\rangle} + \frac{2m_t C_{10}^{(0)} \langle p_5 p_2\rangle [q_1 p_4] [r_2 p_6] \langle q_1 r_1\rangle}{[r_2 q_1]\langle r_1 p_4\rangle} \\ &- \frac{2m_t C_{11}^{(0)} [p_2 p_4] \langle p_5 p_2\rangle [r_2 p_6] \langle r_1 p_2\rangle}{[r_2 q_1]\langle r_1 p_4\rangle} + \frac{4m_t C_2^{(0)} [p_6 p_4] \langle p_5 p_2\rangle [r_2 p_4]}{[r_2 q_1]}\end{aligned} \tag{4.70}$$

$$\begin{aligned}\left|\mathcal{M}^{(0)}_{+-}\right\rangle = &-\frac{m_t^3 C_{10}^{(0)} [r_2 r_1] \langle p_5 p_2\rangle [r_2 p_6] \langle r_2 p_4\rangle}{(p_1\cdot r_2)\,[r_1 p_4]\,[r_2 q_1]} + \frac{4m_t C_1^{(0)} \langle p_2 p_4\rangle \langle p_5 p_4\rangle [r_2 p_6]}{[r_2 q_1]} \\ &- \frac{2m_t C_{10}^{(0)} \langle p_5 p_2\rangle [r_2 p_6] [q_1 r_1] \langle q_1 p_4\rangle}{[r_1 p_4]\,[r_2 q_1]} + \frac{2m_t C_{11}^{(0)} \langle p_2 p_4\rangle \langle p_5 p_2\rangle [r_1 p_2] [r_2 p_6]}{[r_1 p_4]\,[r_2 q_1]}\end{aligned} \tag{4.71}$$

$$\begin{aligned}\left|\mathcal{M}^{(0)}_{-+}\right\rangle = &-\frac{m_t^2 C_{10}^{(0)} \langle p_5 p_2\rangle \langle r_2 r_1\rangle [q_1 p_6] [r_2 p_4]}{(p_1\cdot r_2)\,\langle r_1 p_4\rangle} - \frac{2C_{10}^{(0)} \langle p_5 p_2\rangle [q_1 p_4] [q_1 p_6] \langle q_1 r_1\rangle}{\langle r_1 p_4\rangle} \\ &+ \frac{2C_{11}^{(0)} [p_2 p_4] \langle p_5 p_2\rangle [q_1 p_6] \langle r_1 p_2\rangle}{\langle r_1 p_4\rangle} - 4C_2^{(0)} [p_6 p_4] \langle p_5 p_2\rangle [q_1 p_4]\end{aligned} \tag{4.72}$$

$$\begin{aligned}\left|\mathcal{M}^{(0)}_{--}\right\rangle = &\frac{m_t^2 C_{10}^{(0)} [r_2 r_1] \langle p_5 p_2\rangle [q_1 p_6] \langle r_2 p_4\rangle}{(p_1\cdot r_2)\,[r_1 p_4]} + \frac{2C_{10}^{(0)} \langle p_5 p_2\rangle [q_1 p_6] [q_1 r_1] \langle q_1 p_4\rangle}{[r_1 p_4]} \\ &- \frac{2C_{11}^{(0)} \langle p_2 p_4\rangle \langle p_5 p_2\rangle [q_1 p_6] [r_1 p_2]}{[r_1 p_4]} - 4C_1^{(0)} \langle p_2 p_4\rangle \langle p_5 p_4\rangle [q_1 p_6]\end{aligned} \tag{4.73}$$

Writing

$$\left|\mathcal{M}^{(1)}_{\lambda_1\lambda_2}\right\rangle = \sum_{i=1}^{26} \Omega_i^{\lambda_1\lambda_2} C_i^{(1)}, \tag{4.74}$$

the 1-loop amplitudes read

$$\Omega_1^{++} = 0$$

$$\Omega_2^{++} = \frac{4m_t\,[p_4p_6]\,\langle p_2p_5\rangle\,[r_2p_4]}{[r_2q_1]}$$

$$\Omega_3^{++} = -\frac{m_t^3\,[p_4p_6]\,\langle p_2p_4\rangle\,\langle r_1r_2\rangle\,[r_2p_4]\,\langle q_1p_5\rangle}{p_1\cdot r_2\,\langle r_1p_4\rangle} - \frac{2m_t\,[p_4p_6]\,\langle p_2p_4\rangle\,[q_1p_4]\,\langle q_1p_5\rangle\,\langle r_1q_1\rangle}{\langle r_1p_4\rangle}$$

$$\Omega_4^{++} = -\frac{2m_t\,[p_2p_4]\,[p_4p_6]\,\langle p_2p_4\rangle\,\langle q_1p_5\rangle\,\langle r_1p_2\rangle}{\langle r_1p_4\rangle}$$

$$\Omega_5^{++} = -\frac{4m_t\,[p_4p_6]\,\langle p_2p_5\rangle\,\langle r_1q_1\rangle}{\langle r_1p_4\rangle}$$

$$\Omega_6^{++} = \frac{4m_t\,[p_4p_6]\,\langle q_1p_5\rangle\,\langle r_1p_2\rangle}{\langle r_1p_4\rangle}$$

$$\Omega_7^{++} = 0$$

$$\Omega_8^{++} = 0$$

$$\Omega_9^{++} = 0$$

$$\Omega_{10}^{++} = \frac{m_t^3\,\langle p_2p_5\rangle\,\langle r_1r_2\rangle\,[r_2p_4]\,[r_2p_6]}{p_1\cdot r_2\,[r_2q_1]\,\langle r_1p_4\rangle} + \frac{2m_t\,\langle p_2p_5\rangle\,[q_1p_4]\,[r_2p_6]\,\langle r_1q_1\rangle}{[r_2q_1]\,\langle r_1p_4\rangle}$$

$$\Omega_{11}^{++} = \frac{2m_t\,[p_2p_4]\,\langle p_2p_5\rangle\,[r_2p_6]\,\langle r_1p_2\rangle}{[r_2q_1]\,\langle r_1p_4\rangle}$$

$$\Omega_{12}^{++} = \frac{m_t^3\,[r_2p_4]\,[r_2p_6]\,\langle r_1p_2\rangle\,\langle r_2p_5\rangle}{p_1\cdot r_2\,[r_2q_1]\,\langle r_1p_4\rangle} + \frac{2m_t\,[q_1p_6]\,[r_2p_4]\,\langle q_1p_5\rangle\,\langle r_1p_2\rangle}{[r_2q_1]\,\langle r_1p_4\rangle}$$

$$\Omega_{13}^{++} = \frac{2m_t\,[p_2p_6]\,\langle p_2p_5\rangle\,[r_2p_4]\,\langle r_1p_2\rangle}{[r_2q_1]\,\langle r_1p_4\rangle}$$

$$\Omega_{14}^{++} = \frac{2m_t\,[p_4p_6]\,\langle p_4p_5\rangle\,[r_2p_4]\,\langle r_1p_2\rangle}{[r_2q_1]\,\langle r_1p_4\rangle}$$

$$\begin{aligned}\Omega_{15}^{++} = &-\frac{m_t^5\,\langle p_2p_4\rangle\,\langle r_1r_2\rangle\,[r_2p_4]^2\,[r_2p_6]\,\langle r_2p_5\rangle}{4\,(p_1\cdot r_2)^2\,[r_2q_1]\,\langle r_1p_4\rangle} - \frac{m_t^3\,\langle p_2p_4\rangle\,\langle r_1r_2\rangle\,[q_1p_6]\,[r_2p_4]^2\,\langle q_1p_5\rangle}{2p_1\cdot r_2\,[r_2q_1]\,\langle r_1p_4\rangle}\\ &-\frac{m_t^3\,\langle p_2p_4\rangle\,[q_1p_4]\,[r_2p_4]\,[r_2p_6]\,\langle r_2p_5\rangle\,\langle r_1q_1\rangle}{2p_1\cdot r_2\,[r_2q_1]\,\langle r_1p_4\rangle}\\ &-\frac{m_t\,\langle p_2p_4\rangle\,[q_1p_4]\,[q_1p_6]\,[r_2p_4]\,\langle q_1p_5\rangle\,\langle r_1q_1\rangle}{[r_2q_1]\,\langle r_1p_4\rangle}\end{aligned}$$

$$\Omega_{16}^{++} = -\frac{m_t^3\,[p_2p_6]\,\langle p_2p_4\rangle\,\langle p_2p_5\rangle\,\langle r_1r_2\rangle\,[r_2p_4]^2}{2p_1\cdot r_2\,[r_2q_1]\,\langle r_1p_4\rangle} - \frac{m_t\,[p_2p_6]\,\langle p_2p_4\rangle\,\langle p_2p_5\rangle\,[q_1p_4]\,[r_2p_4]\,\langle r_1q_1\rangle}{[r_2q_1]\,\langle r_1p_4\rangle}$$

$$\Omega_{17}^{++} = -\frac{m_t^3\,[p_4p_6]\,\langle p_2p_4\rangle\,\langle p_4p_5\rangle\,\langle r_1r_2\rangle\,[r_2p_4]^2}{2p_1\cdot r_2\,[r_2q_1]\,\langle r_1p_4\rangle} - \frac{m_t\,[p_4p_6]\,\langle p_2p_4\rangle\,\langle p_4p_5\rangle\,[q_1p_4]\,[r_2p_4]\,\langle r_1q_1\rangle}{[r_2q_1]\,\langle r_1p_4\rangle}$$

$$\Omega_{18}^{++}$$
$$= -\frac{m_t^3\,[p_2p_4]\,\langle p_2p_4\rangle\,[r_2p_4]\,[r_2p_6]\,\langle r_1p_2\rangle\,\langle r_2p_5\rangle}{2p_1\cdot r_2\,[r_2q_1]\,\langle r_1p_4\rangle} - \frac{m_t\,[p_2p_4]\,\langle p_2p_4\rangle\,[q_1p_6]\,[r_2p_4]\,\langle q_1p_5\rangle\,\langle r_1p_2\rangle}{[r_2q_1]\,\langle r_1p_4\rangle}$$

$$\Omega_{19}^{++} = -\frac{m_t\,[p_2p_4]\,[p_2p_6]\,\langle p_2p_4\rangle\,\langle p_2p_5\rangle\,[r_2p_4]\,\langle r_1p_2\rangle}{[r_2q_1]\,\langle r_1p_4\rangle}$$

$$\Omega_{20}^{++} = -\frac{m_t\,[p_2p_4]\,[p_4p_6]\,\langle p_2p_4\rangle\,\langle p_4p_5\rangle\,[r_2p_4]\,\langle r_1p_2\rangle}{[r_2q_1]\,\langle r_1p_4\rangle}$$

$$\Omega_{21}^{++} = \frac{m_t^5\,\langle r_1r_2\rangle\,[r_2p_4]\,[r_2p_6]\,\langle q_1p_2\rangle\,\langle r_2p_5\rangle}{4\,(p_1\cdot r_2)^2\,\langle r_1p_4\rangle} + \frac{m_t^3\,\langle r_1r_2\rangle\,[q_1p_6]\,[r_2p_4]\,\langle q_1p_2\rangle\,\langle q_1p_5\rangle}{2p_1\cdot r_2\,\langle r_1p_4\rangle}$$
$$+ \frac{m_t^3\,[q_1p_4]\,[r_2p_6]\,\langle q_1p_2\rangle\,\langle r_2p_5\rangle\,\langle r_1q_1\rangle}{2p_1\cdot r_2\,\langle r_1p_4\rangle} + \frac{m_t\,[q_1p_4]\,[q_1p_6]\,\langle q_1p_2\rangle\,\langle q_1p_5\rangle\,\langle r_1q_1\rangle}{\langle r_1p_4\rangle}$$

$$\Omega_{22}^{++} = \frac{m_t^3\,[p_2p_6]\,\langle p_2p_5\rangle\,\langle r_1r_2\rangle\,[r_2p_4]\,\langle q_1p_2\rangle}{2p_1\cdot r_2\,\langle r_1p_4\rangle} + \frac{m_t\,[p_2p_6]\,\langle p_2p_5\rangle\,[q_1p_4]\,\langle q_1p_2\rangle\,\langle r_1q_1\rangle}{\langle r_1p_4\rangle}$$

$$\Omega_{23}^{++} = \frac{m_t^3\,[p_4p_6]\,\langle p_4p_5\rangle\,\langle r_1r_2\rangle\,[r_2p_4]\,\langle q_1p_2\rangle}{2p_1\cdot r_2\,\langle r_1p_4\rangle} + \frac{m_t\,[p_4p_6]\,\langle p_4p_5\rangle\,[q_1p_4]\,\langle q_1p_2\rangle\,\langle r_1q_1\rangle}{\langle r_1p_4\rangle}$$

$$\Omega_{24}^{++} = \frac{m_t^3\,[p_2p_4]\,[r_2p_6]\,\langle q_1p_2\rangle\,\langle r_1p_2\rangle\,\langle r_2p_5\rangle}{2p_1\cdot r_2\,\langle r_1p_4\rangle} + \frac{m_t\,[p_2p_4]\,[q_1p_6]\,\langle q_1p_2\rangle\,\langle q_1p_5\rangle\,\langle r_1p_2\rangle}{\langle r_1p_4\rangle}$$

$$\Omega_{25}^{++} = \frac{m_t\,[p_2p_4]\,[p_2p_6]\,\langle p_2p_5\rangle\,\langle q_1p_2\rangle\,\langle r_1p_2\rangle}{\langle r_1p_4\rangle}$$

$$\Omega_{26}^{++} = \frac{m_t\,[p_2p_4]\,[p_4p_6]\,\langle p_4p_5\rangle\,\langle q_1p_2\rangle\,\langle r_1p_2\rangle}{\langle r_1p_4\rangle} \tag{4.75}$$

$$\Omega_{1}^{+-} = -\frac{4m_t\,\langle p_2p_4\rangle\,\langle p_4p_5\rangle\,[r_2p_6]}{[r_2q_1]}$$

$$\Omega_{2}^{+-} = 0$$

$$\Omega_{3}^{+-} = \frac{m_t^3\,[p_4p_6]\,[r_1r_2]\,\langle p_2p_4\rangle\,\langle q_1p_5\rangle\,\langle r_2p_4\rangle}{p_1\cdot r_2\,[r_1p_4]} + \frac{2m_t\,[p_4p_6]\,\langle p_2p_4\rangle\,[r_1q_1]\,\langle q_1p_4\rangle\,\langle q_1p_5\rangle}{[r_1p_4]}$$

$$\Omega_{4}^{+-} = \frac{2m_t\,[p_4p_6]\,\langle p_2p_4\rangle^2\,[r_1p_2]\,\langle q_1p_5\rangle}{[r_1p_4]}$$

$$\Omega_{5}^{+-} = -\frac{4m_t\,\langle p_2p_5\rangle\,[r_1p_6]\,\langle q_1p_4\rangle}{[r_1p_4]}$$

$$\Omega_{6}^{+-} = \frac{4m_t\,\langle p_2p_4\rangle\,[r_1p_6]\,\langle q_1p_5\rangle}{[r_1p_4]}$$

$$\Omega_7^{+-} = -\frac{m_t^3 \langle p_2 p_4\rangle [r_2 p_6] \langle q_1 p_4\rangle \langle r_2 p_5\rangle}{p_1 \cdot r_2} - 2m_t \langle p_2 p_4\rangle [q_1 p_6] \langle q_1 p_4\rangle \langle q_1 p_5\rangle$$

$$\Omega_8^{+-} = -2m_t [p_2 p_6] \langle p_2 p_4\rangle \langle p_2 p_5\rangle \langle q_1 p_4\rangle$$

$$\Omega_9^{+-} = -2m_t [p_4 p_6] \langle p_2 p_4\rangle \langle p_4 p_5\rangle \langle q_1 p_4\rangle$$

$$\Omega_{10}^{+-} = -\frac{m_t^3 [r_1 r_2] \langle p_2 p_5\rangle [r_2 p_6] \langle r_2 p_4\rangle}{p_1 \cdot r_2 [r_1 p_4] [r_2 q_1]} - \frac{2m_t \langle p_2 p_5\rangle [r_2 p_6] [r_1 q_1] \langle q_1 p_4\rangle}{[r_1 p_4] [r_2 q_1]}$$

$$\Omega_{11}^{+-} = -\frac{2m_t \langle p_2 p_4\rangle \langle p_2 p_5\rangle [r_1 p_2] [r_2 p_6]}{[r_1 p_4] [r_2 q_1]}$$

$$\Omega_{12}^{+-} = -\frac{m_t^3 [r_1 r_2] \langle p_2 p_4\rangle [r_2 p_6] \langle r_2 p_5\rangle}{p_1 \cdot r_2 [r_1 p_4] [r_2 q_1]} - \frac{2m_t [r_1 r_2] \langle p_2 p_4\rangle [q_1 p_6] \langle q_1 p_5\rangle}{[r_1 p_4] [r_2 q_1]}$$

$$\Omega_{13}^{+-} = -\frac{2m_t [p_2 p_6] [r_1 r_2] \langle p_2 p_4\rangle \langle p_2 p_5\rangle}{[r_1 p_4] [r_2 q_1]}$$

$$\Omega_{14}^{+-} = -\frac{2m_t [p_4 p_6] [r_1 r_2] \langle p_2 p_4\rangle \langle p_4 p_5\rangle}{[r_1 p_4] [r_2 q_1]}$$

$$\begin{aligned}\Omega_{15}^{+-} = &\frac{m_t^5 [r_1 r_2] \langle p_2 p_4\rangle [r_2 p_4] [r_2 p_6] \langle r_2 p_4\rangle \langle r_2 p_5\rangle}{4 (p_1 \cdot r_2)^2 [r_1 p_4] [r_2 q_1]} + \frac{m_t^3 [r_1 r_2] \langle p_2 p_4\rangle [q_1 p_6] [r_2 p_4] \langle q_1 p_5\rangle \langle r_2 p_4\rangle}{2p_1 \cdot r_2 [r_1 p_4] [r_2 q_1]} \\ &+ \frac{m_t^3 \langle p_2 p_4\rangle [r_2 p_4] [r_2 p_6] [r_1 q_1] \langle q_1 p_4\rangle \langle r_2 p_5\rangle}{2p_1 \cdot r_2 [r_1 p_4] [r_2 q_1]} \\ &+ \frac{m_t \langle p_2 p_4\rangle [q_1 p_6] [r_2 p_4] [r_1 q_1] \langle q_1 p_4\rangle \langle q_1 p_5\rangle}{[r_1 p_4] [r_2 q_1]}\end{aligned}$$

$$\Omega_{16}^{+-} = \frac{m_t^3 [p_2 p_6] [r_1 r_2] \langle p_2 p_4\rangle \langle p_2 p_5\rangle [r_2 p_4] \langle r_2 p_4\rangle}{2p_1 \cdot r_2 [r_1 p_4] [r_2 q_1]} + \frac{m_t [p_2 p_6] \langle p_2 p_4\rangle \langle p_2 p_5\rangle [r_2 p_4] [r_1 q_1] \langle q_1 p_4\rangle}{[r_1 p_4] [r_2 q_1]}$$

$$\Omega_{17}^{+-} = \frac{m_t^3 [p_4 p_6] [r_1 r_2] \langle p_2 p_4\rangle \langle p_4 p_5\rangle [r_2 p_4] \langle r_2 p_4\rangle}{2p_1 \cdot r_2 [r_1 p_4] [r_2 q_1]} + \frac{m_t [p_4 p_6] \langle p_2 p_4\rangle \langle p_4 p_5\rangle [r_2 p_4] [r_1 q_1] \langle q_1 p_4\rangle}{[r_1 p_4] [r_2 q_1]}$$

$$\Omega_{18}^{+-} = \frac{m_t^3 \langle p_2 p_4\rangle^2 [r_1 p_2] [r_2 p_4] [r_2 p_6] \langle r_2 p_5\rangle}{2p_1 \cdot r_2 [r_1 p_4] [r_2 q_1]} + \frac{m_t \langle p_2 p_4\rangle^2 [q_1 p_6] [r_1 p_2] [r_2 p_4] \langle q_1 p_5\rangle}{[r_1 p_4] [r_2 q_1]}$$

$$\Omega_{19}^{+-} = \frac{m_t [p_2 p_6] \langle p_2 p_4\rangle^2 \langle p_2 p_5\rangle [r_1 p_2] [r_2 p_4]}{[r_1 p_4] [r_2 q_1]}$$

$$\Omega_{20}^{+-} = \frac{m_t [p_4 p_6] \langle p_2 p_4\rangle^2 \langle p_4 p_5\rangle [r_1 p_2] [r_2 p_4]}{[r_1 p_4] [r_2 q_1]}$$

$$\begin{aligned}\Omega_{21}^{+-} = &-\frac{m_t^5 [r_1 r_2] [r_2 p_6] \langle q_1 p_2\rangle \langle r_2 p_4\rangle \langle r_2 p_5\rangle}{4 (p_1 \cdot r_2)^2 [r_1 p_4]} - \frac{m_t^3 [r_1 r_2] [q_1 p_6] \langle q_1 p_2\rangle \langle q_1 p_5\rangle \langle r_2 p_4\rangle}{2p_1 \cdot r_2 [r_1 p_4]} \\ &- \frac{m_t^3 [r_2 p_6] [r_1 q_1] \langle q_1 p_2\rangle \langle q_1 p_4\rangle \langle r_2 p_5\rangle}{2p_1 \cdot r_2 [r_1 p_4]} - \frac{m_t [q_1 p_6] [r_1 q_1] \langle q_1 p_2\rangle \langle q_1 p_4\rangle \langle q_1 p_5\rangle}{[r_1 p_4]}\end{aligned}$$

$$\Omega_{22}^{+-} = -\frac{m_t^3\,[p_2p_6]\,[r_1r_2]\,\langle p_2p_5\rangle\,\langle q_1p_2\rangle\,\langle r_2p_4\rangle}{2p_1\cdot r_2\,[r_1p_4]} - \frac{m_t\,[p_2p_6]\,\langle p_2p_5\rangle\,[r_1q_1]\,\langle q_1p_2\rangle\,\langle q_1p_4\rangle}{[r_1p_4]}$$

$$\Omega_{23}^{+-} = -\frac{m_t^3\,[p_4p_6]\,[r_1r_2]\,\langle p_4p_5\rangle\,\langle q_1p_2\rangle\,\langle r_2p_4\rangle}{2p_1\cdot r_2\,[r_1p_4]} - \frac{m_t\,[p_4p_6]\,\langle p_4p_5\rangle\,[r_1q_1]\,\langle q_1p_2\rangle\,\langle q_1p_4\rangle}{[r_1p_4]}$$

$$\Omega_{24}^{+-} = -\frac{m_t^3\,\langle p_2p_4\rangle\,[r_1p_2]\,[r_2p_6]\,\langle q_1p_2\rangle\,\langle r_2p_5\rangle}{2p_1\cdot r_2\,[r_1p_4]} - \frac{m_t\,\langle p_2p_4\rangle\,[q_1p_6]\,[r_1p_2]\,\langle q_1p_2\rangle\,\langle q_1p_5\rangle}{[r_1p_4]}$$

$$\Omega_{25}^{+-} = -\frac{m_t\,[p_2p_6]\,\langle p_2p_4\rangle\,\langle p_2p_5\rangle\,[r_1p_2]\,\langle q_1p_2\rangle}{[r_1p_4]}$$

$$\Omega_{26}^{+-} = -\frac{m_t\,[p_4p_6]\,\langle p_2p_4\rangle\,\langle p_4p_5\rangle\,[r_1p_2]\,\langle q_1p_2\rangle}{[r_1p_4]} \tag{4.76}$$

$$\Omega_{1}^{-+} = 0$$

$$\Omega_{2}^{-+} = -4\,[p_4p_6]\,\langle p_2p_5\rangle\,[q_1p_4]$$

$$\Omega_{3}^{-+} = \frac{m_t^4\,[p_4p_6]\,\langle p_2p_4\rangle\,\langle r_1r_2\rangle\,[r_2p_4]\,\langle r_2p_5\rangle}{p_1\cdot r_2\,\langle r_1p_4\rangle\,\langle r_2q_1\rangle} + \frac{2m_t^2\,[p_4p_6]\,\langle p_2p_4\rangle\,[q_1p_4]\,\langle r_2p_5\rangle\,\langle r_1q_1\rangle}{\langle r_1p_4\rangle\,\langle r_2q_1\rangle}$$

$$\Omega_{4}^{-+} = \frac{2m_t^2\,[p_2p_4]\,[p_4p_6]\,\langle p_2p_4\rangle\,\langle r_1p_2\rangle\,\langle r_2p_5\rangle}{\langle r_1p_4\rangle\,\langle r_2q_1\rangle}$$

$$\Omega_{5}^{-+} = \frac{4m_t^2\,[p_4p_6]\,\langle p_2p_5\rangle\,\langle r_1r_2\rangle}{\langle r_1p_4\rangle\,\langle r_2q_1\rangle}$$

$$\Omega_{6}^{-+} = -\frac{4m_t^2\,[p_4p_6]\,\langle r_1p_2\rangle\,\langle r_2p_5\rangle}{\langle r_1p_4\rangle\,\langle r_2q_1\rangle}$$

$$\Omega_{7}^{-+} = 0$$

$$\Omega_{8}^{-+} = 0$$

$$\Omega_{9}^{-+} = 0$$

$$\Omega_{10}^{-+} = -\frac{m_t^2\,\langle p_2p_5\rangle\,\langle r_1r_2\rangle\,[q_1p_6]\,[r_2p_4]}{p_1\cdot r_2\,\langle r_1p_4\rangle} - \frac{2\,\langle p_2p_5\rangle\,[q_1p_4]\,[q_1p_6]\,\langle r_1q_1\rangle}{\langle r_1p_4\rangle}$$

$$\Omega_{11}^{-+} = -\frac{2\,[p_2p_4]\,\langle p_2p_5\rangle\,[q_1p_6]\,\langle r_1p_2\rangle}{\langle r_1p_4\rangle}$$

$$\Omega_{12}^{-+} = -\frac{m_t^2\,[q_1p_4]\,[r_2p_6]\,\langle r_1p_2\rangle\,\langle r_2p_5\rangle}{p_1\cdot r_2\,\langle r_1p_4\rangle} - \frac{2\,[q_1p_4]\,[q_1p_6]\,\langle q_1p_5\rangle\,\langle r_1p_2\rangle}{\langle r_1p_4\rangle}$$

$$\Omega_{13}^{-+} = -\frac{2\,[p_2p_6]\,\langle p_2p_5\rangle\,[q_1p_4]\,\langle r_1p_2\rangle}{\langle r_1p_4\rangle}$$

$$\Omega_{14}^{-+} = -\frac{2\,[p_4p_6]\,\langle p_4p_5\rangle\,[q_1p_4]\,\langle r_1p_2\rangle}{\langle r_1p_4\rangle}$$

$$\Omega_{15}^{-+} = \frac{m_t^4\,\langle p_2p_4\rangle\,\langle r_1r_2\rangle\,[q_1p_4]\,[r_2p_4]\,[r_2p_6]\,\langle r_2p_5\rangle}{4\,(p_1\cdot r_2)^2\,\langle r_1p_4\rangle} + \frac{m_t^2\,\langle p_2p_4\rangle\,\langle r_1r_2\rangle\,[q_1p_4]\,[q_1p_6]\,[r_2p_4]\,\langle q_1p_5\rangle}{2p_1\cdot r_2\,\langle r_1p_4\rangle}$$
$$+ \frac{m_t^2\,\langle p_2p_4\rangle\,[q_1p_4]^2\,[r_2p_6]\,\langle r_2p_5\rangle\,\langle r_1q_1\rangle}{2p_1\cdot r_2\,\langle r_1p_4\rangle} + \frac{\langle p_2p_4\rangle\,[q_1p_4]^2\,[q_1p_6]\,\langle q_1p_5\rangle\,\langle r_1q_1\rangle}{\langle r_1p_4\rangle}$$

$$\Omega_{16}^{-+} = \frac{m_t^2\,[p_2p_6]\,\langle p_2p_4\rangle\,\langle p_2p_5\rangle\,\langle r_1r_2\rangle\,[q_1p_4]\,[r_2p_4]}{2p_1\cdot r_2\,\langle r_1p_4\rangle} + \frac{[p_2p_6]\,\langle p_2p_4\rangle\,\langle p_2p_5\rangle\,[q_1p_4]^2\,\langle r_1q_1\rangle}{\langle r_1p_4\rangle}$$

$$\Omega_{17}^{-+} = \frac{m_t^2\,[p_4p_6]\,\langle p_2p_4\rangle\,\langle p_4p_5\rangle\,\langle r_1r_2\rangle\,[q_1p_4]\,[r_2p_4]}{2p_1\cdot r_2\,\langle r_1p_4\rangle} + \frac{[p_4p_6]\,\langle p_2p_4\rangle\,\langle p_4p_5\rangle\,[q_1p_4]^2\,\langle r_1q_1\rangle}{\langle r_1p_4\rangle}$$

$$\Omega_{18}^{-+} = \frac{m_t^2\,[p_2p_4]\,\langle p_2p_4\rangle\,[q_1p_4]\,[r_2p_6]\,\langle r_1p_2\rangle\,\langle r_2p_5\rangle}{2p_1\cdot r_2\,\langle r_1p_4\rangle} + \frac{[p_2p_4]\,\langle p_2p_4\rangle\,[q_1p_4]\,[q_1p_6]\,\langle q_1p_5\rangle\,\langle r_1p_2\rangle}{\langle r_1p_4\rangle}$$

$$\Omega_{19}^{-+} = \frac{[p_2p_4]\,[p_2p_6]\,\langle p_2p_4\rangle\,\langle p_2p_5\rangle\,[q_1p_4]\,\langle r_1p_2\rangle}{\langle r_1p_4\rangle}$$

$$\Omega_{20}^{-+} = \frac{[p_2p_4]\,[p_4p_6]\,\langle p_2p_4\rangle\,\langle p_4p_5\rangle\,[q_1p_4]\,\langle r_1p_2\rangle}{\langle r_1p_4\rangle}$$

$$\Omega_{21}^{-+} = -\frac{m_t^6\,\langle r_1r_2\rangle\,[r_2p_4]\,[r_2p_6]\,\langle r_2p_2\rangle\,\langle r_2p_5\rangle}{4\,(p_1\cdot r_2)^2\,\langle r_1p_4\rangle\,\langle r_2q_1\rangle} - \frac{m_t^4\,\langle r_1r_2\rangle\,[q_1p_6]\,[r_2p_4]\,\langle q_1p_5\rangle\,\langle r_2p_2\rangle}{2p_1\cdot r_2\,\langle r_1p_4\rangle\,\langle r_2q_1\rangle}$$
$$- \frac{m_t^4\,[q_1p_4]\,[r_2p_6]\,\langle r_2p_2\rangle\,\langle r_2p_5\rangle\,\langle r_1q_1\rangle}{2p_1\cdot r_2\,\langle r_1p_4\rangle\,\langle r_2q_1\rangle} - \frac{m_t^2\,[q_1p_4]\,[q_1p_6]\,\langle q_1p_5\rangle\,\langle r_2p_2\rangle\,\langle r_1q_1\rangle}{\langle r_1p_4\rangle\,\langle r_2q_1\rangle}$$

$$\Omega_{22}^{-+} = -\frac{m_t^4\,[p_2p_6]\,\langle p_2p_5\rangle\,\langle r_1r_2\rangle\,[r_2p_4]\,\langle r_2p_2\rangle}{2p_1\cdot r_2\,\langle r_1p_4\rangle\,\langle r_2q_1\rangle} - \frac{m_t^2\,[p_2p_6]\,\langle p_2p_5\rangle\,[q_1p_4]\,\langle r_2p_2\rangle\,\langle r_1q_1\rangle}{\langle r_1p_4\rangle\,\langle r_2q_1\rangle}$$

$$\Omega_{23}^{-+} = -\frac{m_t^4\,[p_4p_6]\,\langle p_4p_5\rangle\,\langle r_1r_2\rangle\,[r_2p_4]\,\langle r_2p_2\rangle}{2p_1\cdot r_2\,\langle r_1p_4\rangle\,\langle r_2q_1\rangle} - \frac{m_t^2\,[p_4p_6]\,\langle p_4p_5\rangle\,[q_1p_4]\,\langle r_2p_2\rangle\,\langle r_1q_1\rangle}{\langle r_1p_4\rangle\,\langle r_2q_1\rangle}$$

$$\Omega_{24}^{-+} = -\frac{m_t^4\,[p_2p_4]\,[r_2p_6]\,\langle r_1p_2\rangle\,\langle r_2p_2\rangle\,\langle r_2p_5\rangle}{2p_1\cdot r_2\,\langle r_1p_4\rangle\,\langle r_2q_1\rangle} - \frac{m_t^2\,[p_2p_4]\,[q_1p_6]\,\langle q_1p_5\rangle\,\langle r_1p_2\rangle\,\langle r_2p_2\rangle}{\langle r_1p_4\rangle\,\langle r_2q_1\rangle}$$

$$\Omega_{25}^{-+} = -\frac{m_t^2\,[p_2p_4]\,[p_2p_6]\,\langle p_2p_5\rangle\,\langle r_1p_2\rangle\,\langle r_2p_2\rangle}{\langle r_1p_4\rangle\,\langle r_2q_1\rangle}$$

$$\Omega_{26}^{-+} = -\frac{m_t^2\,[p_2p_4]\,[p_4p_6]\,\langle p_4p_5\rangle\,\langle r_1p_2\rangle\,\langle r_2p_2\rangle}{\langle r_1p_4\rangle\,\langle r_2q_1\rangle} \tag{4.77}$$

$$\Omega_1^{--} = 4\,\langle p_2p_4\rangle\,\langle p_4p_5\rangle\,[q_1p_6]$$

$$\Omega_2^{--} = 0$$

$$\Omega_3^{--} = -\frac{m_t^4\,[p_4p_6]\,[r_1r_2]\,\langle p_2p_4\rangle\,\langle r_2p_4\rangle\,\langle r_2p_5\rangle}{p_1\cdot r_2\,[r_1p_4]\,\langle r_2q_1\rangle} - \frac{2m_t^2\,[p_4p_6]\,\langle p_2p_4\rangle\,[r_1q_1]\,\langle q_1p_4\rangle\,\langle r_2p_5\rangle}{[r_1p_4]\,\langle r_2q_1\rangle}$$

$$\Omega_4^{--} = -\frac{2m_t^2\,[p_4p_6]\,\langle p_2p_4\rangle^2\,[r_1p_2]\,\langle r_2p_5\rangle}{[r_1p_4]\,\langle r_2q_1\rangle}$$

$$\Omega_5^{--} = \frac{4m_t^2\,\langle p_2p_5\rangle\,[r_1p_6]\,\langle r_2p_4\rangle}{[r_1p_4]\,\langle r_2q_1\rangle}$$

$$\Omega_6^{--} = -\frac{4m_t^2\,\langle p_2p_4\rangle\,[r_1p_6]\,\langle r_2p_5\rangle}{[r_1p_4]\,\langle r_2q_1\rangle}$$

$$\Omega_7^{--} = \frac{m_t^4\,\langle p_2p_4\rangle\,[r_2p_6]\,\langle r_2p_4\rangle\,\langle r_2p_5\rangle}{p_1\cdot r_2\,\langle r_2q_1\rangle} + \frac{2m_t^2\,\langle p_2p_4\rangle\,[q_1p_6]\,\langle q_1p_5\rangle\,\langle r_2p_4\rangle}{\langle r_2q_1\rangle}$$

$$\Omega_8^{--} = \frac{2m_t^2\,[p_2p_6]\,\langle p_2p_4\rangle\,\langle p_2p_5\rangle\,\langle r_2p_4\rangle}{\langle r_2q_1\rangle}$$

$$\Omega_9^{--} = \frac{2m_t^2\,[p_4p_6]\,\langle p_2p_4\rangle\,\langle p_4p_5\rangle\,\langle r_2p_4\rangle}{\langle r_2q_1\rangle}$$

$$\Omega_{10}^{--} = \frac{m_t^2\,[r_1r_2]\,\langle p_2p_5\rangle\,[q_1p_6]\,\langle r_2p_4\rangle}{p_1\cdot r_2\,[r_1p_4]} + \frac{2\,\langle p_2p_5\rangle\,[q_1p_6]\,[r_1q_1]\,\langle q_1p_4\rangle}{[r_1p_4]}$$

$$\Omega_{11}^{--} = \frac{2\,\langle p_2p_4\rangle\,\langle p_2p_5\rangle\,[q_1p_6]\,[r_1p_2]}{[r_1p_4]}$$

$$\Omega_{12}^{--} = \frac{m_t^2\,\langle p_2p_4\rangle\,[r_2p_6]\,[r_1q_1]\,\langle r_2p_5\rangle}{p_1\cdot r_2\,[r_1p_4]} + \frac{2\,\langle p_2p_4\rangle\,[q_1p_6]\,[r_1q_1]\,\langle q_1p_5\rangle}{[r_1p_4]}$$

$$\Omega_{13}^{--} = \frac{2\,[p_2p_6]\,\langle p_2p_4\rangle\,\langle p_2p_5\rangle\,[r_1q_1]}{[r_1p_4]}$$

$$\Omega_{14}^{--} = \frac{2\,[p_4p_6]\,\langle p_2p_4\rangle\,\langle p_4p_5\rangle\,[r_1q_1]}{[r_1p_4]}$$

$$\begin{aligned}\Omega_{15}^{--} = &-\frac{m_t^4\,[r_1r_2]\,\langle p_2p_4\rangle\,[q_1p_4]\,[r_2p_6]\,\langle r_2p_4\rangle\,\langle r_2p_5\rangle}{4\,(p_1\cdot r_2)^2\,[r_1p_4]} \\ &- \frac{m_t^2\,[r_1r_2]\,\langle p_2p_4\rangle\,[q_1p_4]\,[q_1p_6]\,\langle q_1p_5\rangle\,\langle r_2p_4\rangle}{2p_1\cdot r_2\,[r_1p_4]} \\ &- \frac{m_t^2\,\langle p_2p_4\rangle\,[q_1p_4]\,[r_2p_6]\,[r_1q_1]\,\langle q_1p_4\rangle\,\langle r_2p_5\rangle}{2p_1\cdot r_2\,[r_1p_4]} - \frac{\langle p_2p_4\rangle\,[q_1p_4]\,[q_1p_6]\,[r_1q_1]\,\langle q_1p_4\rangle\,\langle q_1p_5\rangle}{[r_1p_4]}\end{aligned}$$

$$\Omega_{16}^{--} = -\frac{m_t^2\,[p_2p_6]\,[r_1r_2]\,\langle p_2p_4\rangle\,\langle p_2p_5\rangle\,[q_1p_4]\,\langle r_2p_4\rangle}{2p_1\cdot r_2\,[r_1p_4]} - \frac{[p_2p_6]\,\langle p_2p_4\rangle\,\langle p_2p_5\rangle\,[q_1p_4]\,[r_1q_1]\,\langle q_1p_4\rangle}{[r_1p_4]}$$

$$\Omega_{17}^{--} = -\frac{m_t^2\,[p_4p_6]\,[r_1r_2]\,\langle p_2p_4\rangle\,\langle p_4p_5\rangle\,[q_1p_4]\,\langle r_2p_4\rangle}{2p_1\cdot r_2\,[r_1p_4]} - \frac{[p_4p_6]\,\langle p_2p_4\rangle\,\langle p_4p_5\rangle\,[q_1p_4]\,[r_1q_1]\,\langle q_1p_4\rangle}{[r_1p_4]}$$

$$\Omega_{18}^{--} = -\frac{m_t^2 \langle p_2p_4\rangle^2 [q_1p_4] [r_1p_2] [r_2p_6] \langle r_2p_5\rangle}{2p_1\cdot r_2 [r_1p_4]} - \frac{\langle p_2p_4\rangle^2 [q_1p_4] [q_1p_6] [r_1p_2] \langle q_1p_5\rangle}{[r_1p_4]}$$

$$\Omega_{19}^{--} = -\frac{[p_2p_6] \langle p_2p_4\rangle^2 \langle p_2p_5\rangle [q_1p_4] [r_1p_2]}{[r_1p_4]}$$

$$\Omega_{20}^{--} = -\frac{[p_4p_6] \langle p_2p_4\rangle^2 \langle p_4p_5\rangle [q_1p_4] [r_1p_2]}{[r_1p_4]}$$

$$\Omega_{21}^{--} = \frac{m_t^6 [r_1r_2] [r_2p_6] \langle r_2p_2\rangle \langle r_2p_4\rangle \langle r_2p_5\rangle}{4(p_1\cdot r_2)^2 [r_1p_4] \langle r_2q_1\rangle} + \frac{m_t^4 [r_1r_2] [q_1p_6] \langle q_1p_5\rangle \langle r_2p_2\rangle \langle r_2p_4\rangle}{2p_1\cdot r_2 [r_1p_4] \langle r_2q_1\rangle}$$
$$+ \frac{m_t^4 [r_2p_6] [r_1q_1] \langle q_1p_4\rangle \langle r_2p_2\rangle \langle r_2p_5\rangle}{2p_1\cdot r_2 [r_1p_4] \langle r_2q_1\rangle} + \frac{m_t^2 [q_1p_6] [r_1q_1] \langle q_1p_4\rangle \langle q_1p_5\rangle \langle r_2p_2\rangle}{[r_1p_4] \langle r_2q_1\rangle}$$

$$\Omega_{22}^{--} = \frac{m_t^4 [p_2p_6] [r_1r_2] \langle p_2p_5\rangle \langle r_2p_2\rangle \langle r_2p_4\rangle}{2p_1\cdot r_2 [r_1p_4] \langle r_2q_1\rangle} + \frac{m_t^2 [p_2p_6] \langle p_2p_5\rangle [r_1q_1] \langle q_1p_4\rangle \langle r_2p_2\rangle}{[r_1p_4] \langle r_2q_1\rangle}$$

$$\Omega_{23}^{--} = \frac{m_t^4 [p_4p_6] [r_1r_2] \langle p_4p_5\rangle \langle r_2p_2\rangle \langle r_2p_4\rangle}{2p_1\cdot r_2 [r_1p_4] \langle r_2q_1\rangle} + \frac{m_t^2 [p_4p_6] \langle p_4p_5\rangle [r_1q_1] \langle q_1p_4\rangle \langle r_2p_2\rangle}{[r_1p_4] \langle r_2q_1\rangle}$$

$$\Omega_{24}^{--} = \frac{m_t^4 \langle p_2p_4\rangle [r_1p_2] [r_2p_6] \langle r_2p_2\rangle \langle r_2p_5\rangle}{2p_1\cdot r_2 [r_1p_4] \langle r_2q_1\rangle} + \frac{m_t^2 \langle p_2p_4\rangle [q_1p_6] [r_1p_2] \langle q_1p_5\rangle \langle r_2p_2\rangle}{[r_1p_4] \langle r_2q_1\rangle}$$

$$\Omega_{25}^{--} = \frac{m_t^2 [p_2p_6] \langle p_2p_4\rangle \langle p_2p_5\rangle [r_1p_2] \langle r_2p_2\rangle}{[r_1p_4] \langle r_2q_1\rangle}$$

$$\Omega_{26}^{--} = \frac{m_t^2 [p_4p_6] \langle p_2p_4\rangle \langle p_4p_5\rangle [r_1p_2] \langle r_2p_2\rangle}{[r_1p_4] \langle r_2q_1\rangle} \tag{4.78}$$

## 4.5. Single top production

We turn to the case of the production of a single top quark and consider the process

$$b(p_1) + q(p_6) \to t(p_2) + g(p_4) + q'(p_5) \tag{4.79}$$

where $q, q'$ are light quarks. At NNLO, the process receives contributions from 2-loop corrections to the heavy-light vertex, 2-loop corrections to the light-light vertex and real-virtual corrections to each. Non-factorisable diagrams in which gluon loops join the two quark lines also contribute, but are colour-suppressed by factors of order $1/N_C$ and so are generally considered sub-leading. The amplitudes for the real-virtual correction to the heavy-light line case can be obtained from the decay amplitudes provided in the previous section by crossing external legs and including the appropriate CKM matrix factor $V_{qq'}$ to account for the difference in coupling between leptons and quarks. Care must be taken, however, to accommodate the effect of the changed kinematics on the analytic continuation of the loop integrals—the change in sign of some invariants leads to different imaginary parts associated with the logarithms and a direct effect on the real parts of the amplitude via $\log^2$ terms.

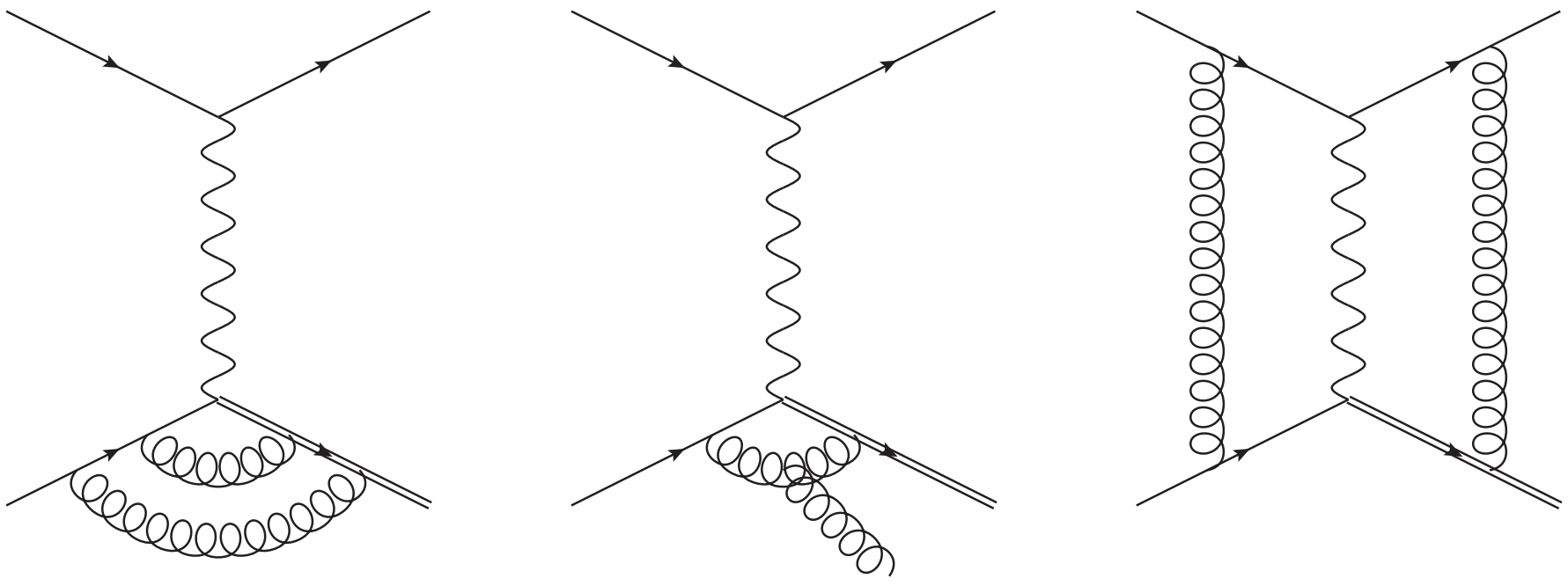

**Figure 4.4.:** Classes of NNLO correction to single top production. We consider the second kind here—contributions from the third are considered colour-suppressed. Corrections to the light quark line are not shown.

The light quark helicities are again fixed by the electroweak vertex and we obtain the full matrix element by contracting the heavy-light line with a light quark tensor

$$|\mathcal{M}_{\lambda_1\lambda_2}\rangle = H_\mu(q,q')S^\mu_{\lambda_1\lambda_2}(t,W,b,g)P_W. \tag{4.80}$$

where the light quark tensor $H_\mu$ is given by

$$H_\mu = [p_5\gamma_\mu p_6\rangle. \tag{4.81}$$

The auxiliary momenta are assigned as follows:

$$\begin{aligned} p_2^\mu &= q_2^\mu + \frac{m_t^2}{2p_2\cdot r_2} r_2^\mu \\ \varepsilon_R^{*\mu}(p_4) &= \frac{1}{\sqrt{2}}\frac{\langle r_1\gamma^\mu p_4]}{\langle r_1 p_4\rangle} \\ \varepsilon_L^{*\mu}(p_4) &= -\frac{1}{\sqrt{2}}\frac{[r_1\gamma^\mu p_4\rangle}{[r_1 p_4]}. \end{aligned} \tag{4.82}$$

We relegate the results to an ancillary file. Numerical checks have been performed against the output of `GoSam` and full agreement found.

# Chapter 5.

# Extraction of $\alpha_s$ and $m_t$ from differential top distributions

In this chapter we consider the phenomenological application of precision corrections in processes involving heavy quarks. We compare NNLO predictions for differential $t\bar{t}$ distributions with 8 TeV ATLAS and CMS data in order to extract values for two important Standard Model parameters, the top quark mass $m_t$ and the strong coupling constant $\alpha_s$.

## 5.1. Introduction

Together, the strong coupling constant $\alpha_s$ and the top quark mass $m_t$ comprise the most important parameters of QCD. Indeed, in processes at high energy scales $\gg \Lambda_{\text{QCD}}$ where the masses of the lighter quarks can be neglected they are essentially the only free parameters. The former is present in every perturbative calculation as the expansion parameter while the latter plays an important rôle in governing the stability of the electroweak vacuum as well as entering into theory predictions of important Standard Model backgrounds to LHC processes such as single and pair top production. Despite its importance, the value of $\alpha_s$ is one of the least well known parameters in the Standard Model and a dominant uncertainty in many precision calculations. The measured value of the top mass is also subject to assumptions made by event generators and it has been suggested that the pole mass could be of order 1 GeV higher compared to current implementations.

While the pole mass is defined to be independent of $\mu_R$, in general $\alpha_s$ will show $\mu_R$ dependence and is therefore not itself a physical observable. It is conventional to take the value evaluated at $\mu_R = M_Z$ where $M_Z$ is the Z boson mass. The current world average is $\alpha_s(M_Z) = 0.1181 \pm 0.0011$ which has been obtained from many different data sets and via several methods of extraction (for a review, see [96]). A high degree of precision in calculation is needed to obtain these values, stressing again the importance of NNLO and higher order corrections. Recently, the dependence of the total top pair production cross section has been used to extract a value[120] and following this a determination was made by the NNPDF collaboration using a PDF analysis[121]. In addition, lattice QCD has been used to obtain values from the non-perturbative regime of QCD[122–124].

Measurements of the top mass may proceed either by direct reconstruction of the decay products or by examination of the dependence of the total cross section $\sigma_{t\bar{t}}$ on the mass. Numerous analyses by the D0 and CDF experiments at the Tevatron and the ATLAS and CMS experiments at the LHC have published measurements, both direct and indirect. The current world average is $173.0 \pm 0.4$ GeV from direct measurements and $173.1 \pm 0.9$ GeV from indirect measurements[96].

In this work we take an alternative approach and exploit the joint dependence of differential distributions in top pair production on $m_t$ and $\alpha_s$ to extract a value for these parameters. Information about the total $t\bar{t}$ cross section is in itself insufficient to constrain both parameters, as the dependence is correlated. Shape information from the kinematic distributions, however, allows a simultaneous extraction to be performed. For the first time, fully differential NNLO calculations for several values of $m_t$ are used and compared with data from the ATLAS and CMS experiments collected during run 1 at $\sqrt{s} = 8$ TeV. Specifically, we look at four differential distributions in top pair production events: the transverse momentum of the top, $p_T^t$, the invariant mass of the top pair $M_{t\bar{t}}$, the rapidity of the top $y_t$ and the rapidity of the pair $y_{t\bar{t}}$.

## 5.2. Differential $t\bar{t}$ at the 8 TeV LHC

### 5.2.1. Experimental measurements

We consider differential cross section measurements of top-quark pair production at $\sqrt{s} = 8$ TeV by the ATLAS[125] and CMS[126] experiments. The final state considered is lepton+jets, in which the $t\bar{t}$ pair is reconstructed from its semileptonic decay products ($t\bar{t} \to W^+ b W^- \bar{b} \to l^\pm \nu_l b q q' \bar{b}$). We use results in the full phase space in terms of top and top-pair kinematic variables due to the lack of availability of NNLO predictions for leptonic and jet observables, though measurements in this fiducial phase space are available. Current work on a full treatment of top pair production at NNLO including full corrections to the decay will remedy this. While ATLAS provide both absolute and normalised distributions, only the normalised are available from CMS requiring reconstruction of the absolute distribution from the total cross section. Of the distributions available, we choose those for which NNLO calculations are available, namely the transverse momentum of the top, $p_T^t$, the invariant mass of the top pair $M_{t\bar{t}}$, the rapidity of the top $y_t$ and the rapidity of the pair $y_{t\bar{t}}$. Binnings are common to the experiments and are listed in Table 5.1.

| Observable | Bin edges |
|---|---|
| $p_T^t$ | {0, 60, 100, 150, 200, 260, 320, 400, 500} GeV |
| $M_{t\bar{t}}$ | {345, 400, 470, 550, 650, 800, 1100, 1600} GeV |
| $y_t$ | {-2.5, -1.6, -1.2, -0.8, -0.4, 0.0, 0.4, 0.8, 1.2, 1.6, 2.5} |
| $y_{t\bar{t}}$ | {-2.5, -1.3, -0.9, -0.6, -0.3, 0.0, 0.3, 0.6, 0.9, 1.3, 2.5} |

**Table 5.1.:** ATLAS and CMS common bin edges for measurements of $p_T^t$, $M_{t\bar{t}}$, $y_t$ and $y_{t\bar{t}}$ at 8 TeV.

The correlations between bins of a distribution are encoded in an experimental covariance matrix, constructed using knowledge of the systematic and statistical uncertainties. We make use of the results presented in [127] for both ATLAS and CMS.

We use measurements of the total cross section from ATLAS[128] and CMS[129], also at $\sqrt{s} = 8$ TeV, in the dileptonic $e\mu$ decay channel. These measurements correspond to an ATLAS dataset with integrated luminosity of 20.3fb$^{-1}$ and a CMS dataset with 19.7fb$^{-1}$.

## 5.2.2. Theoretical predictions

To quantitatively compare against experimental measurements and make extractions of parameters, theoretical predictions of these measured quantities, at the highest possible precision, are required. For the total $t\bar{t}$ cross section we use $\sigma^{\text{theory}}(\alpha_s, m_t) = \sigma_{t\bar{t}}^{NNLO+NNLL}(\alpha_s, m_t)$, i.e. the NNLO-QCD fixed-order prediction supplemented with soft-gluon resummation[97–100,130] . The fast computation of this is provided through `top++`[101] . For the total cross section we set the renormalisation and factorisation scales to the common scale $\mu = \mu_R = \mu_F = m_t$. For the differential cross section we use NNLO-QCD predictions $d\sigma^{\text{theory}}(\alpha_s, m_t) = d\sigma_{t\bar{t}}^{NNLO}(\alpha_s, m_t)$[131–133] computed using the `Stripper` framework[54,134] . The renormalisation and factorisation scales are chosen to be the optimal ones found in [133], namely

$$\mu_R = \mu_F = H_T/4, \quad \text{for } M_{t\bar{t}},\ y_t,\ y_{t\bar{t}}\,, \tag{5.1}$$

$$\mu_R = \mu_F = M_T/2, \quad \text{for } p_T^t \tag{5.2}$$

where

$$H_T = \sqrt{m_t^2 + (p_T^t)^2} + \sqrt{m_t^2 + (p_T^{\bar{t}})^2}\,. \tag{5.3}$$

For clarity, we note that normalised distributions are obtained from the corresponding absolute distributions by dividing the weight of each bin by the sum of weights in all bins.

Finding the best fit values for $\alpha_s$ and $m_t$ requires theoretical predictions made with many input values of these fundamental parameters. Furthermore, since the value of $\alpha_s$ is typically taken from the associated input PDF set used to evaluate cross sections, these predictions must be made with multiple PDF sets. To allow for the fast computation of the relevant differential quantities, we have made use of the `fastNLO` interface[135–137] to `Stripper`, producing tables that allow, for a fixed set of observables and binnings and a fixed input top-quark mass, $\mathcal{O}$(1 second) evaluations of the NNLO differential cross section with any desired PDF set. We have produced `fastNLO` tables for the distributions and binnings of Table 5.1 and for the following values of the top-quark mass,

$$m_t = \{169.0, 171.0, 172.5, 173.3, 175.0\}\ \text{GeV}. \tag{5.4}$$

Predictions for different input $\alpha_s$ values are obtained by convoluting the `fastNLO` tables with PDF sets of the same family which have been fit using different values of $\alpha_s$.

### Dependence on $\alpha_s$ and $m_t$ of differential observables

In Figures 5.1 and 5.2 we show the sensitivity on $\alpha_s$ and $m_t$ of the absolute and normalised distributions respectively, at NNLO-QCD accuracy. It is the non-trivial shape of these dependences that will allow us to perform simultaneous extractions of $\alpha_s$ and $m_t$. The plots have been produced using the CT14 PDF set—however, very similar patterns (in particular the size of the effects of varying $\alpha_s$ and $m_t$) are observed when using other PDF sets[1].

For the absolute distributions, we observe that the sensitivity on $\alpha_s$ closely follows the $\alpha_s$-dependence of the total cross section, namely $\sim \mathcal{O}(\alpha_s^2)$. More precisely, changing the values of $\alpha_s$ by $x\%$ results in roughly a $x^2\%$ change in the weights of each bin. For all absolute distributions, the sensitivity increases in the tails of distributions, where corrections beyond the Born level in the cross section will have larger effects. A similar pattern can be observed in the $m_t$ sensitivity of the rapidity distributions, $y_t$ and $y_{t\bar{t}}$. Namely, the behaviour of the weights of each bin roughly follow the $\sim \mathcal{O}(m_t^{-4})$ dependence of the total cross section, and once again with the tails seeing slightly increased sensitivity. On the other hand, the $p_T^t$ and $M_{t\bar{t}}$ distributions display very different sensitivities. The bins most sensitive to changes in $m_t$ are the lowest bins in the ranges where the bulk of the cross section lies. In the tails however, the sensitivity is dramatically reduced — this can be understood by the fact that in the tails the top-quark is effectively massless (the finite-mass effects being suppressed by powers of large-$p_T^t$ or large-$M_{t\bar{t}}$).

[1]The effects of varying the value of the top quark mass have been also discussed, at NLO in [127, 138].

In the case of normalised distributions, the sensitivity plots shown in Figure 5.2 show a far more involved dependence on $\alpha_s$ and $m_t$. These patterns arise because the bin-weights of the absolute distributions are divided by the cross-section over all bins. In general we observe that, except in the tails of distributions, the effect of varying $\alpha_s$ by $\pm 0.003$ around the baseline value of $\alpha_s = 0.118$ is around the 1%-level or less. On the other hand, variations of $m_t$ by $\pm 2 - 3$ GeV around the value $m_t = 173.3$ GeV can result in much larger effects: for $p_T^t$ the variations in $m_t$ lead to $\sim 2\%$ effects and for $M_{t\bar{t}}$ the effects are $\gtrsim 0 - 5\%$ in the bins with largest weight. For $y_t$ and $y_{t\bar{t}}$ the effects of varying $m_t$ are much smaller, below 0.5% in the bins with largest weight. If experimental uncertainties were negligible, then we would expect that these patterns were reflected in the uncertainties on the extracted values of $\alpha_s$ and $m_t$. That is, we could expect that given that the magnitude of $\alpha_s$ dependence is roughly the same across all distributions, this should translate to roughly similar uncertainties on the values of $\alpha_s$ extracted from each distribution. In contrast, we would expect that the uncertainties on the extracted values of $m_t$ would be largest when using the normalised $y_t$ and $y_{t\bar{t}}$ data (and significantly larger than the uncertainties obtained when using the normalised $p_T^t$ and particularly $M_{t\bar{t}}$ data), due to the relative insensitivity to the mass of these distributions. We indeed find evidence of such behaviour in Section 5.4.

### Choice of input PDFs

In order to make an extraction of $\alpha_s$ we require a (publicly available) PDF family to have PDF sets that have been fit with different values of $\alpha_s$. To be able to construct a robust fit of the theoretical predictions, (see Section 5.2.2) it is preferable that the PDF family contains $\geq 3$ values of $\alpha_s$. The PDFs considered in this study (all NNLO fits) are the CT14,[35] NNPDF3.0[139] and NNPDF3.1[37] sets, which come with 13, 5 and 11 values of $\alpha_s$ respectively. We have also considered the MMHT14[36] PDF family. However, since the set does not pass our internal consistency checks, we do not present the corresponding extractions. Further details are set out later in this section.

Of these families, CT14 alone was fit with no top quark data. This is desirable in our extraction in order to avoid potential bias. NNPDF3.0 has used measurements of the top pair cross section at the LHC, whilst NNPDF3.1 additionally includes the ATLAS measurements of $y_t$ and the CMS measurements of $y_{t\bar{t}}$ at 8 TeV. Given the inclusion of the differential data in this latter case, we consider results from the NNPDF3.1 set with some caution and include them mainly to compare PDF consistency.

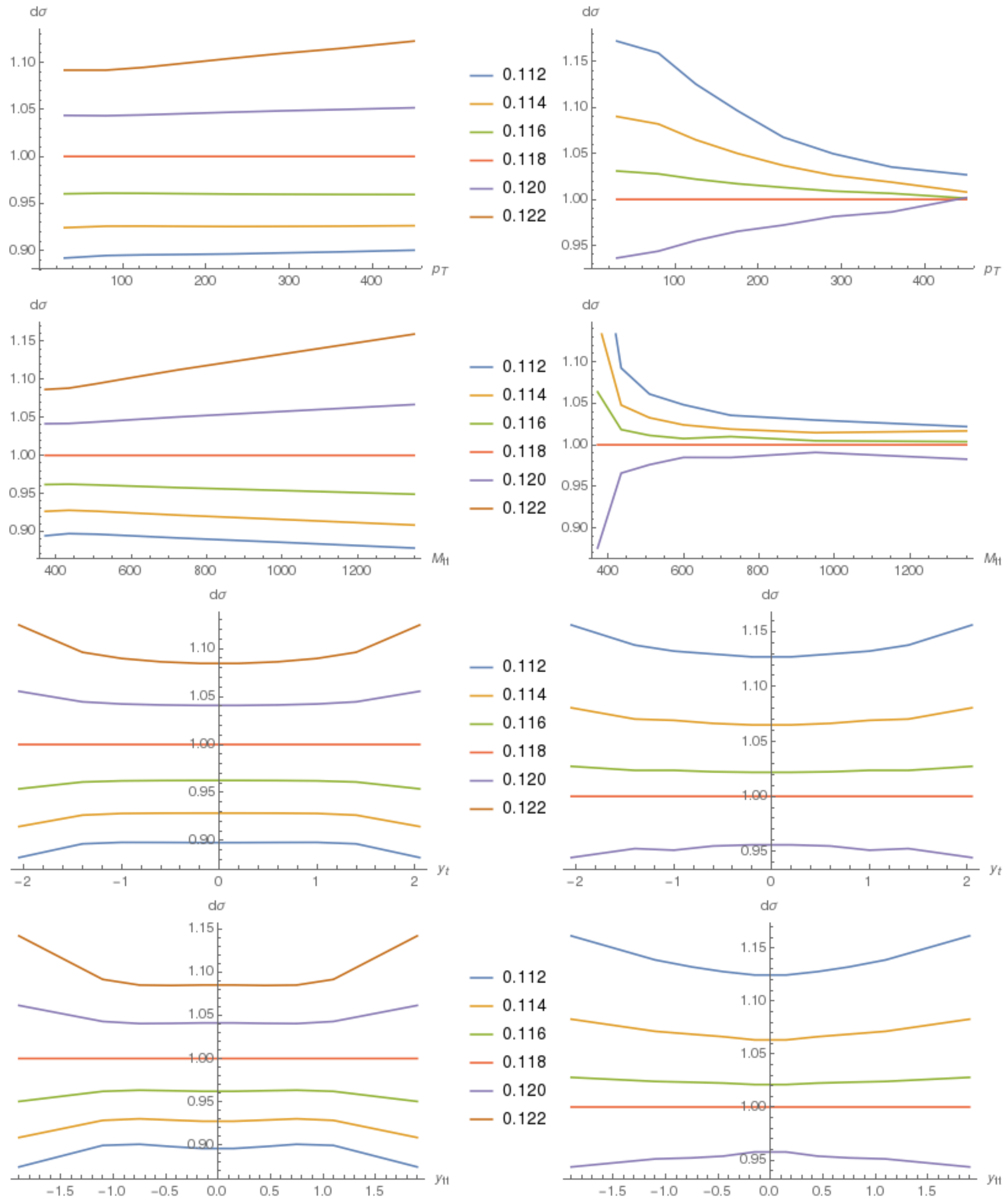

**Figure 5.1.:** The sensitivity on $\alpha_s$ (left) and $m_t$ (right) of the $p_T^t$ (row 1), $M_{t\bar{t}}$ (row 2), $y_t$ (row 3) and $y_{t\bar{t}}$ (row 4) absolute distributions at NNLO-QCD. The plots illustrate the ratios of the normalised distributions for different values of $\alpha_s$ or $m_t$ to the absolute distributions computed with the world average values of $\alpha_s = 0.118$ and $m_t = 173.3$ GeV. The CT14 PDF set has been used to produce these curves, however, the patterns of the dependence are similar for other PDF sets.

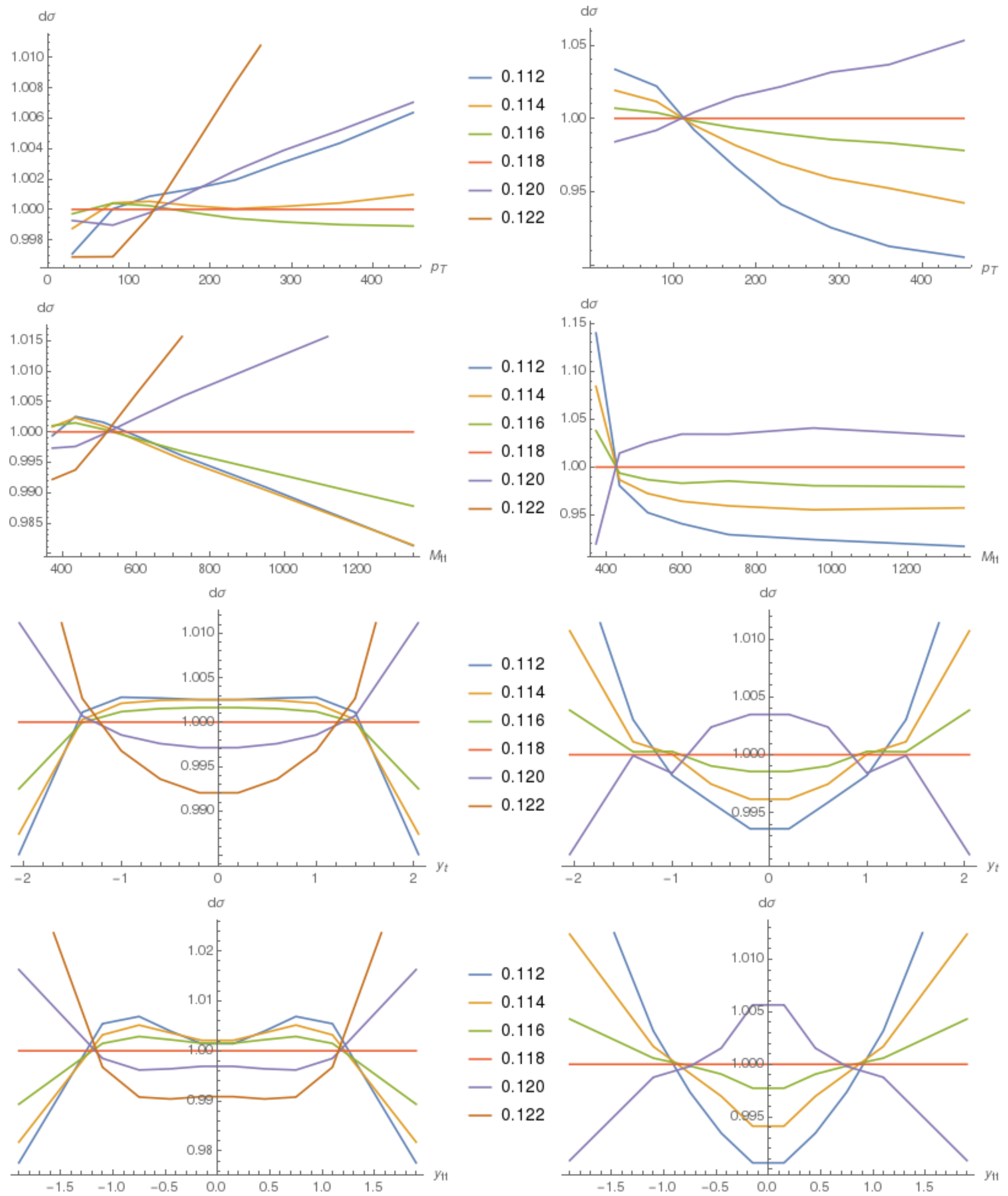

**Figure 5.2.:** The sensitivity on $\alpha_s$ (left) and $m_t$ (right) of the $p_T^t$ (row 1), $M_{t\bar{t}}$ (row 2), $y_t$ (row 3) and $y_{t\bar{t}}$ (row 4) normalised distributions at NNLO-QCD. The plots illustrate the ratios of the normalised distributions for different values of $\alpha_s$ or $m_t$ to the normalised distributions computed with the world average values of $\alpha_s = 0.118$ and $m_t = 173.3$ GeV. The CT14 PDF set has been used to produce these curves, however, the patterns of the dependence are similar for other PDF sets.

### Two-dimensional interpolations in $\alpha_s$ and $m_t$

To aid the parameter extraction via minimisation of a chi-squared objective function, for each PDF family (CT14, NNPDF3.0, NNPDF3.1) it is convenient to move from a discrete grid of predictions in $\alpha_s$ and $m_t$ to a smooth two-dimensional surface. This is achieved by fitting functions to the sets of $(\alpha_s, m_t)$ points we have available, both for the total cross section as well as for the weight of each bin of every distribution. For the one-dimensional fits we fit functions in $\alpha_s$ ($g(\alpha_s)$) or $m_t$ ($f(m_t)$), while in the two-dimensional case the fit takes a factorised form as the product of a function of $\alpha_s$ with a function of $m_t$. The forms of these functions are detailed next.

The total cross section is fit as a function of $\alpha_s$ using polynomials of order 1, 2 and 3 for the NNPDF3.0, CT14 and NNPDF3.1 sets respectively, while for the mass dependence we use the functional form recommended in [100]

$$\sigma_i^{\text{theory}}(m) = \sigma_i^{\text{theory}}(m_{\text{ref}}) \left(\frac{m_{\text{ref}}}{m}\right)^4 \left(1 + a_1 \frac{(m - m_{\text{ref}})}{m_{\text{ref}}} + a_2 \frac{(m - m_{\text{ref}})^2}{m_{\text{ref}}^2}\right), \tag{5.5}$$

where we take $m_{\text{ref}} = 172.5$ GeV as a reference value for the top-quark mass.

The functional forms of the fits for the differential cross sections are chosen based on the number of available points in $m_t$ and $\alpha_s$ (the latter depending on the PDF set used) and based on the quality and stability of the interpolation and extrapolation. To parametrise the $\alpha_s$ dependence of each bin, we choose a polynomial of order 4 for the CT14 and NNPDF3.1 sets and a linear fit for the NNPDF3.0 set[2]. For the mass dependences of the distributions, we use the forms

$$w^i_{M_{t\bar{t}}}(m) = w^i_{M_{t\bar{t}}}(m_{\text{ref}}) \left(\frac{m_{\text{ref}}}{m}\right)^4 \left(1 + a_1 \frac{(m - m_{\text{ref}})}{m_{\text{ref}}} + a_2 \frac{(m - m_{\text{ref}})^2}{m_{\text{ref}}^2} + a_3 \left(1 - \frac{1}{\cosh^6(m - m_{\text{ref}})}\right)\right) \tag{5.6}$$

$$w^i_{p_T^t}(m) = w^i_{p_T^t}(m_{\text{ref}}) \left(\frac{m_{\text{ref}}}{m}\right)^4 \left(1 + a_1 \frac{(m - m_{\text{ref}})}{m_{\text{ref}}} + \frac{a_2}{m}\right) \tag{5.7}$$

$$w^i_y(m) = w^i_y(m_{\text{ref}}) \left(1 + a_1 e^{-a_2(m - m_{\text{ref}})}\right) \tag{5.8}$$

where $w^i(m)$ is the weight of the $i^{\text{th}}$ bin and we take a reference mass $m_{\text{ref}} = 172.5$ GeV. The same functional form is used for both rapidity distributions $y_t$ and $y_{t\bar{t}}$. The coefficients $a_j$ are fit independently for the results of each distribution and PDF set.

[2]Since for NNPDF3.0 only five values of $\alpha_s$ are available, we impose a linear dependence to avoid overfitting—in this way we can fit the predictions to within $3\sigma$ for the rapidity distributions and to within $2\sigma$ for the $p_T^t$ and $M_{t\bar{t}}$ distributions.

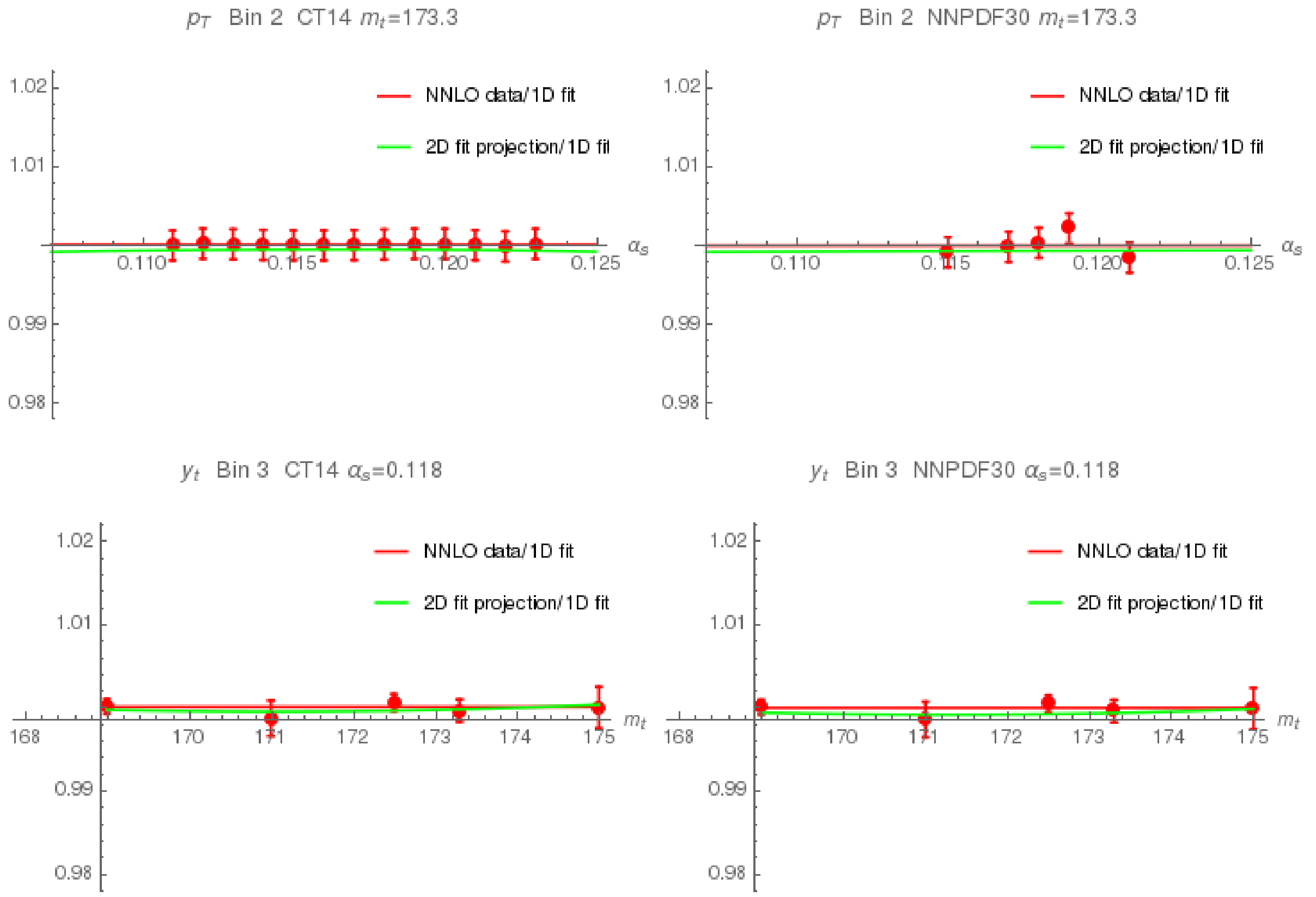

**Figure 5.3.:** Example plots showing the interpolations in $\alpha_s$ and $m_t$. The upper (lower) plots show the interpolations of theoretical predictions in $\alpha_s$ ($m_t$) for the cross sections in the second bin of $p_T^t$ (third bin of $y_t$), using the CT14 (left) and NNPDF30 (right) PDF sets. The red points display the ratio of the computed NNLO predictions to the one-dimensional interpolation. The green line illustrates the projection of the two-dimensional fit in $\alpha_s$ and $m_t$ onto one-dimension.

We note that for our extractions we always use the two-dimensional fits, projecting out one dimension as required for the one-dimensional extractions of $\alpha_s$ or $m_t$. We perform, however, both one- and two-dimensional fits independently—this allows us to check the consistency of the fits, namely, that the projections of the two-dimensional fits onto one dimension (by fixing either $m_t$ or $\alpha_s$ to some fixed value) are, as expected, equivalent to fitting in one dimension to within the Monte Carlo error of the theory predictions. Some examples of the quality of the fits and of the consistency between one- and two-dimensional fits can be seen in Figure 5.3. Requiring good quality fits and extrapolations is one of the aspects determining which PDF set we have chosen to use — our results for the CT14, NNPDF3.0 and NNPDF3.1 sets all show very good agreement, as a function of $\alpha_s$ and $m_t$, between the bin weight from the fit and the discrete set of the calculated points to within Monte Carlo error. This is not the case for the MMHT PDF set, where we have observed a discontinuity in the slope of the fit as a function of $\alpha_s$ at $\alpha_s = 0.118$ and hence omit results from this PDF family.

## 5.3. Extraction Methodology

In order to quantitatively compare experimental measurements to theoretical predictions we use a least-squares method. We consider two sets of measurements: (a) measurements of normalised distributions supplemented with measurements of the total cross section, or (b) measurements of absolute distributions. In both of these cases we first include data only from a single experiment in the $\chi^2$ objective (we comment on using data from both ATLAS and CMS a little further on). The corresponding definitions of the $\chi^2$ variables we use are

$$\chi^2_{\text{norm}} = \frac{1}{(N_{\text{data}}-1)} \sum_{i,j=1}^{N_{\text{data}}-1} \zeta_i C^{-1}_{ij} \zeta_j \quad + \frac{(\sigma_{\text{theory}} - \sigma_{\text{data}})^2}{\delta\sigma^2_{\text{data}}} \,, \tag{5.9}$$

$$\chi^2_{\text{abs}} = \frac{1}{N_{\text{data}}} \sum_{i,j=1}^{N_{\text{data}}} \zeta_i C^{-1}_{ij} \zeta_j \,, \tag{5.10}$$

where $\zeta_i = \zeta_i^{\text{data}} - \zeta_i^{\text{theory}}$ is the difference between the measured and predicted weights in bin $i$ and $C_{ij}$ a covariance matrix encoding experimental sources of error and correlations between bins $i$ and $j$ of the distributions. We use the so-called experimental definition of the covariance matrix, mirroring the analysis performed in Section 3.3 of [127]. In the cases where normalised shape distributions have been used the final bin of each distribution has been removed in order to render the covariance matrix invertible[3]. A possible advantage of using normalised distributions rather than absolute distributions is the improved control of systematic errors on the experimental measurements of the total cross section. There is also a partial cancellation of uncertainties such as the luminosity which takes place in normalising the distributions.

The sources of uncertainty included in the covariance matrices, and hence in our our extractions, are at present purely experimental. These experimental uncertainties are statistical and systematic, also including uncertainties due to luminosity and beam energy and the bin-by-bin correlations of these is fully retained. In a more complete treatment of uncertainties, sources of theoretical uncertainty such as that due to scale variation and PDF uncertainties should also be included. At the present time, however, there is no consensus on how scale uncertainties should be included in a correlated fashion across bins. Since from studies of the total cross section we expect the PDF errors to be an order of magnitude smaller than those due to the scale variation, we neglect both contributions in this work.

[3]The choice of bin to be removed does not affect the value of the $\chi^2$.

We consider individual extractions of the parameters in which $m_t$ is fixed at the world average $m_t = 173.3$ GeV and we extract $\alpha_s$ or, conversely, fits where we fix $\alpha_s$ to $\alpha_s = 0.118$ and extract $m_t$. In both of these cases, the resulting $\chi^2$ is one-dimensional and the best fit value of $m_t$ or $\alpha_s$ is that which minimises the $\chi^2$ objective. The uncertainties on the extracted values of the parameters are estimated through the standard $\Delta\chi^2 = \pm 1$ variations, which naturally correspond to $1\sigma$ deviations from best fit value.

While it is standard practice to fix $\alpha_s$ and extract $m_t$ (or vice-versa), this could potentially lead to a bias in the extracted value of $m_t$ (or $\alpha_s$) or not actually yield extracted parameters that minimise the $\chi^2$ objective in the full $(m_t, \alpha_s)$ plane. In order to process beyond the one-dimensional studies, we therefore study also the case where both parameters, $m_t$ *and* $\alpha_s$, are allowed to vary independently resulting in a two-dimensional $\chi^2$. Once again the best fit value is taken to be that which minimises this two-dimensional $\chi^2$. Estimating the uncertainties on the extracted values is slightly more involved in this case, since in the $|\Delta\chi^2| = 1$ condition traces out a contour in the $(m_t, \alpha_s)$ plane (in the ideal case, an ellipse). The estimate of the uncertainties on the extracted values of $\alpha_s$ and $m_t$ are derived from these contours by setting $m_t$ and $\alpha_s$ respectively to their best fit value deducing the range corresponding to $\Delta\chi^2 = \pm 1$ once this choice has been made.

Hitherto we have only discussed extractions from a single distribution measured by a single experiment. Given that the shape of the dependence on $\alpha_s$ or $m_t$ is distribution-dependent, combining information from different distributions in an extraction could provide complementary constraints. In doing so, it may be possible to obtain more reliable best fit parameters as well as smaller overall uncertainties. For the study presented here, it is not possible to consistently perform an extraction based on measurements of multiple distributions from a single experiment since the experimental correlated uncertainties between distributions are not publicly available at the present time for either experiment. Neglecting these correlations would undoubtedly bias the results and underestimate the extraction uncertainties. However, since it is realistic to assume that, to a large extent, measurements performed by *different* experiments are largely independent, we do consider the combination of a single distribution from ATLAS with one from CMS[4].

[4] The assumption of negligible correlated uncertainties is not completely above question since uncertainties due to luminosity may to some extent be correlated between experiments. In order to make some modicum of progress, however, we argue that these correlations are likely to be negligible compared to other sources of uncertainty present.

To perform such an extraction from combined distributions we define straightforward extensions of our $\chi^2$ definitions in Equations (5.9, 5.10),

$$\chi^2_{\text{norm}} = \frac{1}{(N_{\text{ATLAS}} + N_{\text{CMS}} - 2)} \left( \sum_{i,j=1}^{N_{\text{ATLAS}}-1} \zeta_{i,\text{ATLAS}} C^{-1}_{ij,\text{ATLAS}} \zeta_{j,\text{ATLAS}} + \sum_{i,j=1}^{N_{\text{CMS}}-1} \zeta_{i,\text{CMS}} C^{-1}_{ij,\text{CMS}} \zeta_{j,\text{CMS}} \right)$$
$$+ \frac{(\sigma_{\text{NNLO}} - \sigma_{\text{ATLAS}})^2}{\delta\sigma^2_{\text{ATLAS}}} + \frac{(\sigma_{\text{NNLO}} - \sigma_{\text{CMS}})^2}{\delta\sigma^2_{\text{CMS}}}, \tag{5.11}$$

$$\chi^2_{\text{abs}} = \frac{1}{(N_{\text{ATLAS}} + N_{\text{CMS}})} \left( \sum_{i,j=1}^{N_{\text{ATLAS}}} \zeta_{i,\text{ATLAS}} C^{-1}_{ij,\text{ATLAS}} \zeta_{j,\text{ATLAS}} + \sum_{i,j=1}^{N_{\text{CMS}}} \zeta_{i,\text{CMS}} C^{-1}_{ij,\text{CMS}} \zeta_{j,\text{CMS}} \right). \tag{5.12}$$

## 5.4. Results and discussion

In this section we present and discuss the results of our extractions of $\alpha_s$ and $m_t$, both separately from one-dimensional $\chi^2$ functions, as well as simultaneously letting both parameters vary. We focus on the extractions using measurements of normalised distributions and the cross sections.

### 5.4.1. Criteria for extractions of $\alpha_s$ and $m_t$

Upon performing the extractions using the definitions of $\chi^2$ introduced in the previous section, it quickly becomes apparent that some extractions can be of low quality (indicated by a high value of $\chi^2$) and/or can be non-sensible (the extracted parameters are very different from current world average values). Although we present results for all extractions, we introduce a set of constraints on the extractions in order to guide the discussion. We find it realistic to restrict ourselves to extractions that yield best fit values of $\alpha_s$ and $m_t$ that lie roughly within $\pm 3\ \sigma$ of the current world-average values, i.e.

$$0.115 \le \alpha_s \le 0.120 \quad \text{and} \quad 170.0 \text{ GeV} \le m_t \le 175.0 \text{ GeV}. \tag{5.13}$$

Furthermore, given our definitions of $\chi^2$, reasonable agreement between measurement and theory would be represented by the restricted set of values in $\chi^2$,

$$\chi^2 < 1.8. \tag{5.14}$$

In the next section, we present tables of minimum $\chi^2$ values from each extraction, together with the corresponding best fit values of $\alpha_s$ and $m_t$. To aid the reader, in these tables we will highlight in grey the extractions that satisfy the criteria we have set out above.

## 5.4.2. Separate extraction of $\alpha_s$ and $m_t$ using single distributions

The first set of extractions we present are the separate extractions of $\alpha_s$ and $m_t$ using the one-dimensional form of our $\chi^2$ objective in Equation 5.9. The results of the $\alpha_s$ extraction, where the value of top mass has been fixed to $m_t = 173.3$ GeV, are shown in Table 5.2. The results of the $m_t$ extraction, where the value of the strong coupling constant has been fixed to $\alpha_s = 0.118$, are shown in Table 5.3.

| | ATLAS | | | | | |
|---|---|---|---|---|---|---|
| | CT14 | | NNPDF30 | | NNPDF31 | |
| | $\alpha_s$ | $\chi^2_{\text{min}}$ | $\alpha_s$ | $\chi^2_{\text{min}}$ | $\alpha_s$ | $\chi^2_{\text{min}}$ |
| $p_T^t$ | $0.1154^{+0.0064}_{-0.0081}$ | 1.01 | $0.1178^{+0.0071}_{-0.0068}$ | 1.42 | $0.1170^{+0.0067}_{-0.0071}$ | 2.51 |
| $M_{t\bar{t}}$ | $0.1174^{+0.0049}_{-0.0075}$ | 1.24 | $0.1195^{+0.0063}_{-0.0061}$ | 1.67 | $0.1199^{+0.0054}_{-0.0064}$ | 3.21 |
| $y_t$ | $0.1137^{+0.0039}_{-0.0034}$ | 5.10 | $0.1131^{+0.0039}_{-0.0036}$ | 0.93 | $0.1163^{+0.0058}_{-0.0066}$ | 0.88 |
| $y_{t\bar{t}}$ | $0.1146^{+0.0033}_{-0.0031}$ | 10.97 | $0.1131^{+0.0030}_{-0.0028}$ | 1.63 | $0.1163^{+0.0054}_{-0.0069}$ | 1.50 |

| | CMS | | | | | |
|---|---|---|---|---|---|---|
| | CT14 | | NNPDF30 | | NNPDF31 | |
| | $\alpha_s$ | $\chi^2_{\text{min}}$ | $\alpha_s$ | $\chi^2_{\text{min}}$ | $\alpha_s$ | $\chi^2_{\text{min}}$ |
| $p_T^t$ | $0.1164^{+0.0036}_{-0.0042}$ | 5.16 | $0.1119^{+0.0073}_{-0.0072}$ | 3.71 | $0.1178^{+0.0056}_{-0.0054}$ | 2.63 |
| $M_{t\bar{t}}$ | $0.1146^{+0.0035}_{-0.0036}$ | 9.84 | $0.1100^{+0.0052}_{-0.0081}$ | 7.03 | $0.1160^{+0.0053}_{-0.0138}$ | 6.10 |
| $y_t$ | $0.1165^{+0.0054}_{-0.0077}$ | 2.43 | $0.1199^{+0.0065}_{-0.0061}$ | 3.08 | $0.1199^{+0.0077}_{-0.0068}$ | 3.92 |
| $y_{t\bar{t}}$ | $0.1154^{+0.0046}_{-0.0052}$ | 2.15 | $0.1149^{+0.0057}_{-0.0052}$ | 0.89 | $0.1175^{+0.0064}_{-0.0066}$ | 0.83 |

**Table 5.2.:** Tabulated values of best fit $\alpha_s$ (with uncertainties) and associated $\chi^2_{\text{min}}$ from extractions of $\alpha_s$ using ATLAS (upper table) and CMS (lower table) measurements of normalised distributions and the total cross section. Results are shown for three different PDF sets and $m_t$ has been set to the world average value of 173.3 GeV. The cells highlighted in grey correspond to extractions that satisfy the conditions of Equations (5.13, 5.14).

Focussing first on Table 5.2, we discuss the results using the measurements of each experiment separately. For ATLAS, we find that

- Using CT14, $p_T^t$ and $M_{t\bar{t}}$ are described well, whilst $y_t$ and $y_{t\bar{t}}$ are poorly described. The extractions using measurements of $p_T^t$ and $M_{t\bar{t}}$ satisfy the criteria of Equations (5.13, 5.14), and agree within their uncertainties, however the best fit values of $\alpha_s$ from these do differ by 0.0020.

- Using NNPDF30, all distributions are described reasonably well by theory, however, the best fit values obtained from $y_t$ and $y_{t\bar{t}}$ are significantly lower than those obtained from $p_T^t$ and $M_{t\bar{t}}$. The extractions using measurements of $p_T^t$ and $M_{t\bar{t}}$ satisfy the criteria of Equations (5.13, 5.14), and agree within their uncertainties, however the best fit values of $\alpha_s$ from these differ by 0.0017.
- Using NNPDF31, the opposite pattern is observed, where $y_t$ and $y_{t\bar{t}}$ are well-described by theory, whereas $p_T^t$ and $M_{t\bar{t}}$ are not. The extractions using measurements of $y_t$ and $y_{t\bar{t}}$ satisfy the criteria of Equations (5.13, 5.14), however the best fit values of $\alpha_s$ from these are consistent with each other.
- In general, for the extractions satisfying Equations (5.13, 5.14), we find a spread in the best fit values of $\alpha_s \in \{0.1154, 0.1195\}$.

For CMS, we find that

- Using CT14, no distributions are described well by the theory.
- Using NNPDF30, only $y_{t\bar{t}}$ is described well by theory, however the best fit value of $\alpha_s$ extracted is just outside the lower bound on $\alpha_s$ we have decided is acceptable in Equation 5.13.
- Using NNPDF31, only $y_{t\bar{t}}$ is described well by theory, with the $\chi^2$ value and best fit value of $\alpha_s$ satisfying the criteria of Equations (5.13, 5.14).

We notice that for a given experiment and distribution, there can be non-trivial differences between extracted values of $\alpha_s$. For example for $p_T^t$ from ATLAS, there is a 0.0023 difference in the best fit values when using CT14 and NNPDF30, although these values are consistent within uncertainties. Comparing the results obtained when using measurements from the two experiments, it is clear that there are significant differences, particularly in the values of $\chi^2_{\text{min}}$ for $p_T^t$ and $M_{t\bar{t}}$ using CT14 or NNPDF30. As detailed in [127] this indicates a certain tension between the two sets of measurements.

Turning our attention to the top-mass extractions shown in Table 5.3, we again first discuss the results using the measurements of each experiment separately, finding similar patterns to the $\alpha_s$ extraction described above. For ATLAS, we find that

- Using CT14, $p_T^t$ and $M_{t\bar{t}}$ are described well, whilst $y_t$ and $y_{t\bar{t}}$ are poorly described. The extractions using measurements of $p_T^t$ and $M_{t\bar{t}}$ satisfy the criteria of Equations (5.13, 5.14), and are consistently within uncertainties, although the best fit values of $m_t$ from these do differ by $> 1$ GeV. The best fit values of $m_t$ from the $p_T^t$ distribution come uncertainties that are roughly a factor $\sim 3$ larger than those when using the $M_{t\bar{t}}$ distribution.

| | ATLAS | | | | | |
|---|---|---|---|---|---|---|
| | CT14 | | NNPDF30 | | NNPDF31 | |
| | $m_t$ | $\chi^2_{\text{min}}$ | $m_t$ | $\chi^2_{\text{min}}$ | $m_t$ | $\chi^2_{\text{min}}$ |
| $p_T^t$ | $174.6^{+1.6}_{-1.6}$ | 0.51 | $174.8^{+1.6}_{-1.6}$ | 0.47 | $175.4^{+1.6}_{-1.6}$ | 0.73 |
| $M_{t\bar{t}}$ | $173.3^{+0.6}_{-0.5}$ | 1.24 | $173.3^{+0.6}_{-0.5}$ | 1.72 | $173.5^{+0.6}_{-0.5}$ | 3.21 |
| $y_t$ | $179.5^{+3.0}_{-3.5}$ | 3.35 | $177.3^{+3.3}_{-3.7}$ | 1.30 | $174.9^{+3.7}_{-3.8}$ | 0.80 |
| $y_{t\bar{t}}$ | $178.8^{+2.5}_{-2.8}$ | 8.40 | $176.2^{+2.7}_{-2.9}$ | 3.25 | $174.4^{+2.9}_{-3.0}$ | 1.46 |

| | CMS | | | | | |
|---|---|---|---|---|---|---|
| | CT14 | | NNPDF30 | | NNPDF31 | |
| | $m_t$ | $\chi^2_{\text{min}}$ | $m_t$ | $\chi^2_{\text{min}}$ | $m_t$ | $\chi^2_{\text{min}}$ |
| $p_T^t$ | $170.6^{+1.6}_{-1.5}$ | 2.34 | $170.8^{+1.6}_{-1.6}$ | 1.86 | $171.4^{+1.6}_{-1.6}$ | 1.11 |
| $M_{t\bar{t}}$ | $170.5^{+1.7}_{-1.8}$ | 7.61 | $170.6^{+1.8}_{-1.8}$ | 6.23 | $170.9^{+1.8}_{-1.8}$ | 4.04 |
| $y_t$ | $173.7^{+3.5}_{-3.3}$ | 2.47 | $173.0^{+3.5}_{-3.3}$ | 3.16 | $172.1^{+3.5}_{-3.5}$ | 3.88 |
| $y_{t\bar{t}}$ | $176.2^{+4.2}_{-4.0}$ | 1.87 | $174.8^{+4.0}_{-3.8}$ | 1.05 | $173.5^{+3.9}_{-3.6}$ | 0.83 |

**Table 5.3.:** Tabulated values of best fit $m_t$ (with uncertainties) and associated $\chi^2_{\text{min}}$ from extractions of $m_t$ using ATLAS (upper table) and CMS (lower table) measurements of normalised distributions and the total cross section. Results are shown for three different PDF sets and $\alpha_s$ has been set to the world average value of 0.118. The cells highlighted in grey correspond to extractions that satisfy the conditions of Equations (5.13, 5.14).

- Using NNPDF30, all distributions, except $y_{t\bar{t}}$, are described reasonably well by theory. The best fit values from $p_T^t$ and $M_{t\bar{t}}$ satisfy the criteria of Equations (5.13, 5.14). As observed for CT14, these values differ by $> 1$ GeV, although are consistent within uncertainties. Once again, the uncertainties when extracting $m_t$ from the $p_T^t$ distributions are roughly a factor $\sim 3$ larger than when extracting using $M_{t\bar{t}}$.

- Using NNPDF31, all distributions, except $M_{t\bar{t}}$, are described well by theory. The best fit values from $y_t$ and $y_{t\bar{t}}$ satisfy the criteria of Equations (5.13, 5.14), and differ only by 0.5 GeV. The uncertainties on the extracted values of the top mass are very large in this case, as would be expected from the small dependence on the value of $m_t$ of the $y_t$ and $y_{t\bar{t}}$ distributions, see Figure 5.2.

- In general, for the extractions satisfying Equations (5.13, 5.14), we find a spread in the best fit values of $m_t \in \{173.3, 174.9\}$ GeV.

For CMS, we find that

- Using CT14, no distributions are described well by the theory.
- Using NNPDF30, only $y_{t\bar{t}}$ is described well by theory, with the best fit value satisfying the mass criteria of Equation 5.13.
- Using NNPDF31, $p_T^t$ and $y_{t\bar{t}}$ are described well by theory and satisfy Equations (5.13, 5.14). The best fit values extracted differ from each other by 2.1 GeV, however are consistent within uncertainties.
- In general, for the extractions satisfying Equations (5.13, 5.14), we find a spread in the best fit values of $m_t \in \{171.4, 174.8\}$ GeV.

As with the separate $\alpha_s$ extraction, we find that for a given experiment and distribution there can be non-trivial differences ($\geq 1$ GeV) between extracted values of $m_t$, particularly when using the data on $y_t$ or $y_{t\bar{t}}$. Interestingly, we see that the pattern of top mass sensitivity of Figure 5.2 is to a large extent reflected in the uncertainties on the extracted values, namely, that extractions based on measurements of $M_{t\bar{t}}$ and $p_T^t$ have smaller uncertainties than extractions based on $y_t$ and $y_{t\bar{t}}$.

We note that in general, for the cases considered thus far, the quality of the extraction leaves something to be desired. Certain distributions, such as the CMS $M_{t\bar{t}}$, are particularly hard to reconcile with the theoretical predictions, returning high values of $\chi^2_{\text{min}}$. Similar discrepancies have also been observed in previous analyses (see for example [127]).

### 5.4.3. Extraction of $\alpha_s$ and $m_t$ from weights of individual bins

In order to scrutinise the patterns observed in the one-dimensional extractions we perform bin-by-bin extractions of $\alpha_s$ and $m_t$ from bins of absolute distributions and compare to the best fit values obtained from an extraction from all bins. This exercise not only serves as a check on our overall extraction methodology, but also allows us to assess the importance of the effects of correlations across different bins. In this case, the form of the $\chi^2$ objective that is minimised is the same as that of Equation 5.10, with the sum over all bins replaced by just the $\chi^2$ function for a single bin. The results for the $\alpha_s$ and $m_t$ extractions are shown in Figures 5.4 and 5.5 respectively, where the error bars indicate the best fit values using just the individual bin weight and the the blue horizontal lines indicate the overall best fit values. For comparison, the world average values of $\alpha_s$ and $m_t$ are also shown in dashed-grey. The results have been obtained using theoretical predictions computed with CT14 PDFs, but very similar results are obtained using the other PDF sets we have considered.

There are a few important features to pick out of these plots. Firstly we note that, as expected, the uncertainties on the extracted values of $\alpha_s$ and $m_t$ on the whole tend to be significantly larger for the individual-bin extractions compared to the overall extractions. The sensitivity of $\alpha_s$ is largest in the tails of distributions (see Figure 5.1), and therefore, if experimental uncertainties were negligible, we might expect that uncertainties on the extracted values of $\alpha_s$ would decrease if we use the weights of bins in the tails. The fact that this is not a generally observed pattern indicates that experimental uncertainties are not negligible, and compete directly with the $\alpha_s$-sensitivity to determine the overall size of uncertainties on the best fit values. We note that by contrast, the extractions of $m_t$ from $p_T^t$ and $M_{t\bar{t}}$ do roughly follow expectations, namely, the bin-by-bin extractions show increased uncertainties for bins with low $m_t$-sensitivities (the bins in the tails of distributions). This is due to the fact that the largest $m_t$ sensitivity for the absolute $p_T^t$ and $M_{t\bar{t}}$ distributions lies in the bins with largest cross section and therefore smallest relative experimental uncertainties.

Another interesting feature is the difference between the values of $\alpha_s$ and $m_t$ extracted from individual bin weights and the best fit values that result from an extraction using all bins. If there were no correlations between different bins in a particular distribution, then the overall extracted value would lie in the range of values from the individual-bin extractions. However, we observe that for some distributions, for example $\alpha_s$ from the ATLAS measurement of $p_T^t$, or $m_t$ from the CMS measurements of $M_{t\bar{t}}$, the overall extracted values lie outside the range of individual-bin extractions. This is a clear indicator that correlations between bins (via the off-diagonal terms in the experimental covariance matrices) can have a large effect and in some cases shift the overall extracted value significantly.

### 5.4.4. Simultaneous extraction of $\alpha_s$ and $m_t$ using single distributions

We now turn our attention to the simultaneous extraction of $\alpha_s$ and $m_t$ from differential measurements, which uses the two-dimensional form of our $\chi^2$ objective in Equation 5.9. The results of the joint $\alpha_s$ and $m_t$ extractions are shown in Table 5.4.

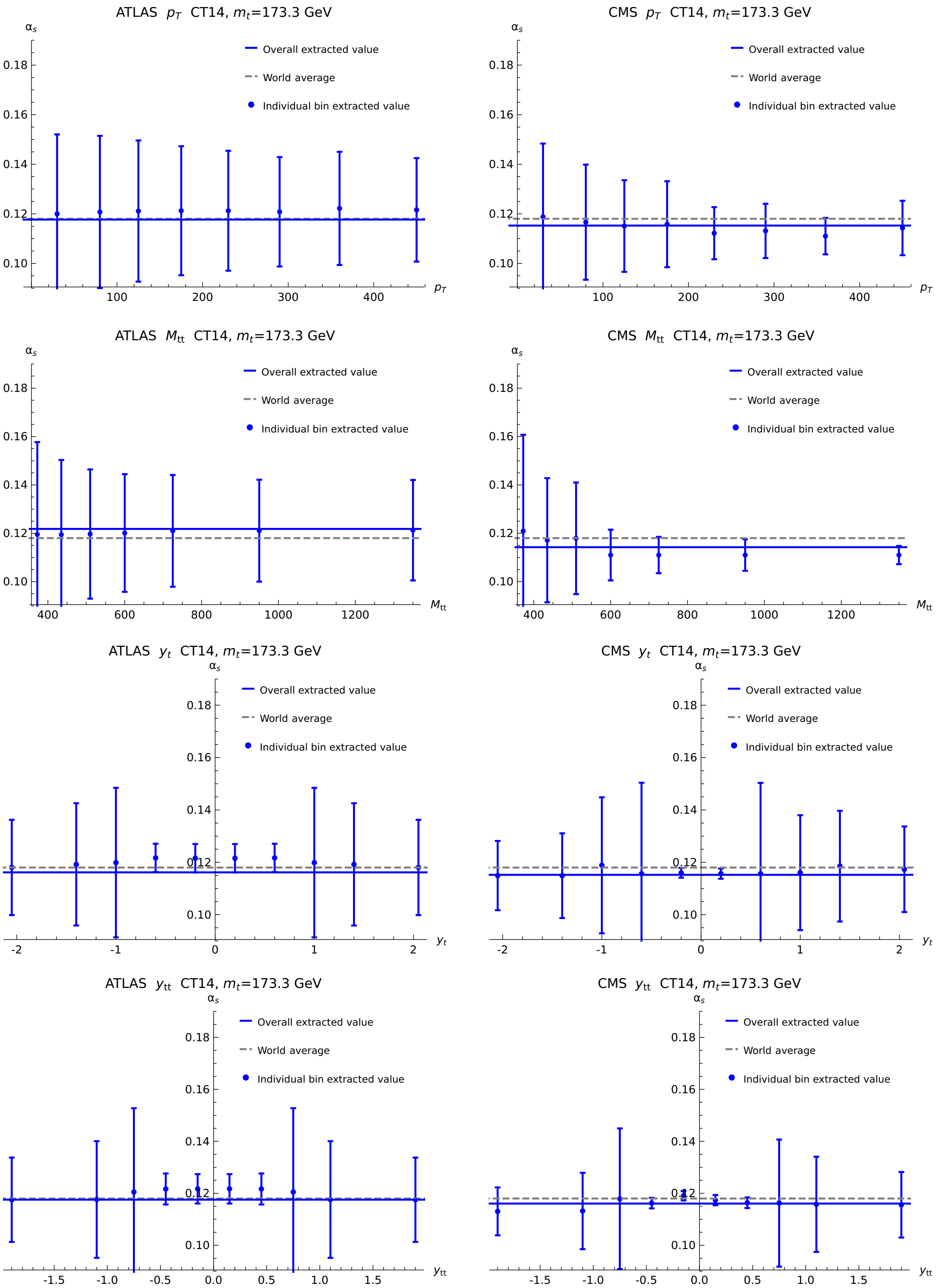

**Figure 5.4.:** Plots showing the extraction of $\alpha_s$ for $m_t = 173.3$ GeV for each individual bin of every distribution. Extractions based on ATLAS and CMS *absolute* measurements are on the left and right respectively, and extractions from measurements of $p_T^t$, $M_{t\bar{t}}$, $y_t$ and $y_{t\bar{t}}$ are shown in rows 1–4. Values extracted from all bins of the absolute distributions are shown as solid blue lines.

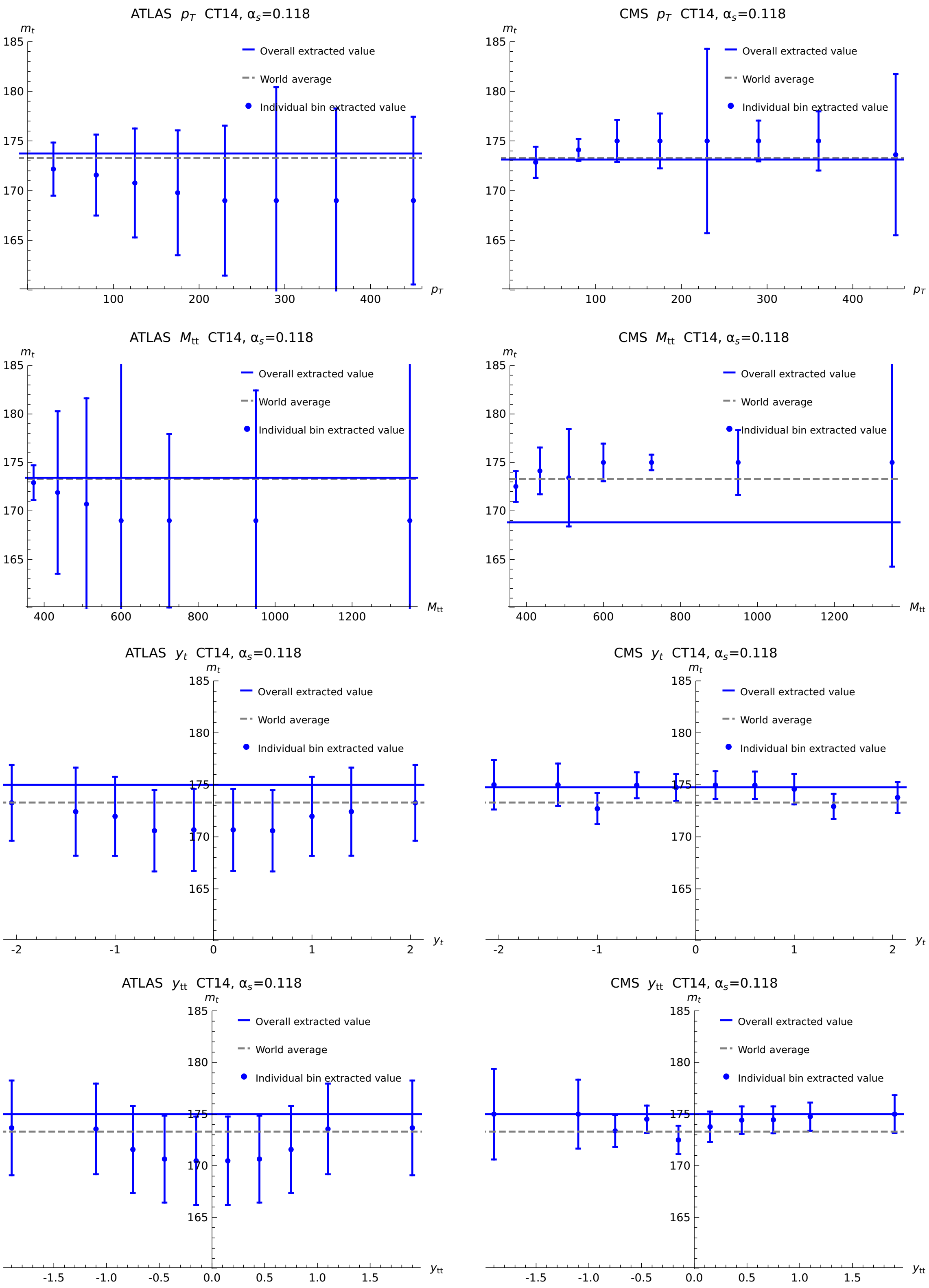

**Figure 5.5.:** Plots showing the extraction of $m_t$ for $\alpha_s = 0.118$ for each individual bin of every distribution. Extractions based on ATLAS and CMS *absolute* measurements are on the left and right respectively, and extractions from measurements of $p_T^t$, $M_{t\bar{t}}$, $y_t$ and $y_{t\bar{t}}$ are shown in rows 1–4. Values extracted from all bins of the absolute distributions are shown as solid blue lines.

| | ATLAS | | | | | | | | |
|---|---|---|---|---|---|---|---|---|---|
| | CT14 | | | NNPDF30 | | | NNPDF31 | | |
| | $\alpha_s$ | $m_t$ | $\chi^2_{\min}$ | $\alpha_s$ | $m_t$ | $\chi^2_{\min}$ | $\alpha_s$ | $m_t$ | $\chi^2_{\min}$ |
| $p_T^t$ | $0.1175^{+0.0048}_{-0.0063}$ | $174.5^{+1.6}_{-1.6}$ | 0.50 | $0.1187^{+0.0076}_{-0.0073}$ | $174.9^{+1.6}_{-1.6}$ | 0.46 | $0.1211^{+0.0071}_{-0.0066}$ | $175.7^{+1.6}_{-1.6}$ | 0.53 |
| $M_{t\bar{t}}$ | $0.1174^{+0.0050}_{-0.0080}$ | $173.2^{+0.6}_{-0.5}$ | 1.23 | $0.1195^{+0.0063}_{-0.0061}$ | $173.3^{+0.6}_{-0.5}$ | 1.67 | $0.1203^{+0.0055}_{-0.0062}$ | $173.5^{+0.6}_{-0.5}$ | 3.06 |
| $y_t$ | $0.1183^{+0.0030}_{-0.0046}$ | $179.6^{+3.0}_{-3.4}$ | 3.35 | $0.1116^{+0.0037}_{-0.0034}$ | $171.4^{+4.1}_{-3.8}$ | 0.89 | $0.1230^{+0.0051}_{-0.0057}$ | $177.8^{+3.3}_{-3.4}$ | 0.59 |
| $y_{t\bar{t}}$ | $0.1172^{+0.0030}_{-0.0040}$ | $178.6^{+2.6}_{-2.8}$ | 8.35 | $0.1125^{+0.0029}_{-0.0027}$ | $172.4^{+3.3}_{-3.3}$ | 1.59 | $0.1177^{+0.0051}_{-0.0065}$ | $174.3^{+2.9}_{-3.0}$ | 1.46 |

| | CMS | | | | | | | | |
|---|---|---|---|---|---|---|---|---|---|
| | CT14 | | | NNPDF30 | | | NNPDF31 | | |
| | $\alpha_s$ | $m_t$ | $\chi^2_{\min}$ | $\alpha_s$ | $m_t$ | $\chi^2_{\min}$ | $\alpha_s$ | $m_t$ | $\chi^2_{\min}$ |
| $p_T^t$ | $0.1096^{+0.0051}_{-0.0034}$ | $168.9^{+1.6}_{-1.6}$ | 0.71 | $0.1108^{+0.0064}_{-0.0063}$ | $170.5^{+1.6}_{-1.6}$ | 0.68 | $0.1136^{+0.0055}_{-0.0064}$ | $170.7^{+1.6}_{-1.6}$ | 0.65 |
| $M_{t\bar{t}}$ | $0.1109^{+0.0036}_{-0.0029}$ | $168.7^{+2.0}_{-2.3}$ | 4.77 | $0.1054^{+0.0055}_{-0.0053}$ | $168.8^{+2.2}_{-2.5}$ | 2.15 | $0.1101^{+0.0054}_{-0.0081}$ | $169.2^{+2.0}_{-2.3}$ | 2.60 |
| $y_t$ | $0.1100^{+0.0063}_{-0.0041}$ | $169.6^{+3.8}_{-3.4}$ | 2.26 | $0.1232^{+0.0070}_{-0.0065}$ | $175.3^{+3.4}_{-3.2}$ | 2.98 | $0.1138^{+0.0059}_{-0.0067}$ | $169.9^{+3.5}_{-3.2}$ | 3.83 |
| $y_{t\bar{t}}$ | $0.1191^{+0.0036}_{-0.0053}$ | $177.0^{+4.1}_{-3.9}$ | 1.85 | $0.1132^{+0.0053}_{-0.0049}$ | $171.8^{+4.1}_{-3.7}$ | 0.85 | $0.1165^{+0.0062}_{-0.0067}$ | $172.7^{+4.0}_{-3.6}$ | 0.83 |

**Table 5.4.:** Tabulated values of best fit $\alpha_s$ and $m_t$ (with uncertainties) and associated $\chi^2_{\min}$ from simultaneous extractions of $m_t$ and $\alpha_s$ using ATLAS (upper table) and CMS (lower table) measurements of normalised distributions and the total cross section. Results are shown for three different PDF sets. The cells highlighted in grey correspond to extractions that satisfy the conditions of Equations (5.13, 5.14).

The first feature to emphasise is that, compared to the separate extractions in Tables 5.2 and 5.3, for virtually every combination of kinematic distribution and PDF set the $\chi^2_{\min}$ value is reduced, indicating improvements in the agreement between theory and data. None of the combinations show an increase in $\chi^2_{\min}$.

We next discuss the results obtained from individual experiments. For ATLAS, we find that

- Using CT14, $p_T^t$ and $M_{t\bar{t}}$ are described well, whilst $y_t$ and $y_{t\bar{t}}$ are poorly described. The extractions using measurements of $p_T^t$ and $M_{t\bar{t}}$ satisfy the criteria of Equations (5.13, 5.14). As for the separate extraction of $m_t$, the best fit values of $m_t$ here also differ by $> 1$ GeV. In contrast to the separate extraction of $\alpha_s$, the best fit values of $\alpha_s$ lie very close to each other.

- Using NNPDF30, all distributions are described well by the theoretical predictions. The best fit values from $p_T^t$ and $M_{t\bar{t}}$ satisfy the criteria of Equations (5.13, 5.14). We find a similar pattern as for CT14, namely that for the extractions based on $M_{t\bar{t}}$ and $p_T^t$, the difference in the best fit values of $m_t$ does not become smaller in the two-dimensional extraction, however, the difference in best fit values of $\alpha_s$ does decrease.

- Using NNPDF31, all distributions, except $M_{t\bar{t}}$, are described well by theory. Only the best fit values from $y_{t\bar{t}}$ satisfy the criteria of Equations (5.13, 5.14).
- In general, for the extractions satisfying Equations (5.13, 5.14), we find a spread in the best fit values of $\alpha_s \in \{0.1174, 0.1195\}$ (an improvement on the separate extraction of $\alpha_s$ with $m_t$ fixed to the world average) and $m_t \in \{173.2, 174.9\}$ GeV.

For CMS, we find that

- Using CT14, only the $p_T^t$ distribution is described well by the theory, however the best fit values of $\alpha_s$ and $m_t$ for this extraction do not pass the criteria of Equations (5.13, 5.14).
- Using NNPDF30, $p_T^t$ and $y_{t\bar{t}}$ are described well by theory, however the best fit points do not satisfy the $\alpha_s$ criterion of Equation 5.13.
- Using NNPDF31, $p_T^t$ and $y_{t\bar{t}}$ are described well by theory. Only the extraction based on $y_{t\bar{t}}$ satisfies Equations (5.13, 5.14), with the extraction based on $p_T^t$ failing the $\alpha_s$ criterion of Equation 5.13.

In Figure 5.6 we show contour plots of $\Delta\chi^2 = \chi^2 - \chi^2_{\text{min}}$ corresponding to our extractions using measurements of $p_T^t$ and $M_{t\bar{t}}$ by ATLAS and of $y_{t\bar{t}}$ by CMS. The red vertical and horizontal lines indicate the world-average values of $m_t$ and $\alpha_s$. We have indicated the regions in the $(\alpha_s, m_t)$-plane inside which we have 'exact' NNLO theory predictions (through interpolations of the discrete points computed) as the white boxes. The outer grey regions indicate the regions where extrapolation of the theory predictions is required.

The particular set of contour plots shown in Figure 5.6 correspond to extractions where the experimental data is well-described by the theoretical predictions for both CT14 and NNPDF30 PDFs and in addition, for the first two rows, satisfies the criteria of Equations (5.13, 5.14). The shapes of the $\Delta\chi^2$ contours for the extraction using the CMS $y_{t\bar{t}}$ data (3$^{\text{rd}}$ row) indicate a large amount of compensation between $\alpha_s$ and $m_t$. This is an indication that for these distributions the $\chi^2$ objective is, to a large extent, driven by the measurements of the cross section, and that the normalised distributions provide relatively weak constraining power.

### 5.4.5. Simultaneous extraction of $\alpha_s$ and $m_t$ using combination of distributions

The final extraction we discuss in this section is one that is based on using measurements of two normalised distributions, one from ATLAS and one from CMS. As mentioned in Section 5.3, we have performed these extractions assuming that the correlations between measurements of the two experiments are negligible. While this exercise may not be fully complete when it comes to including uncertainties such as luminosity systematics, it illustrates the potential of exploiting measurements from *both* experiments. For completeness we present results for all 16 possible combinations of pairs of kinematic distributions from the two experiments— however, we mainly focus our attention and draw insights from the combinations where *different* distributions are combined (avoiding potential shared systematics that may enter in measuring a given distribution). The extractions we present are two-dimensional extractions based on the definition of $\chi^2$ given in Equation 5.11. The results of these extractions are shown in Table 5.5.

| ATLAS | CMS | CT14 | | | NNPDF30 | | | NNPDF31 | | |
|---|---|---|---|---|---|---|---|---|---|---|
| | | $\alpha_s$ | $m_t$ | $\chi^2_{\min}$ | $\alpha_s$ | $m_t$ | $\chi^2_{\min}$ | $\alpha_s$ | $m_t$ | $\chi^2_{\min}$ |
| $p_T^t$ | $p_T^t$ | $0.1142^{+0.0052}_{-0.0055}$ | $172.1^{+1.6}_{-1.6}$ | 2.64 | $0.1148^{+0.0070}_{-0.0068}$ | $172.7^{+1.6}_{-1.6}$ | 2.60 | $0.1176^{+0.0060}_{-0.0062}$ | $173.3^{+1.6}_{-1.6}$ | 2.57 |
| $p_T^t$ | $M_{t\bar{t}}$ | $0.1135^{+0.0048}_{-0.0047}$ | $172.3^{+1.4}_{-1.7}$ | 4.61 | $0.1129^{+0.0067}_{-0.0064}$ | $172.6^{+1.5}_{-1.5}$ | 4.28 | $0.1158^{+0.0057}_{-0.0062}$ | $172.9^{+1.6}_{-1.4}$ | 4.14 |
| $p_T^t$ | $y_t$ | $0.1178^{+0.0050}_{-0.0068}$ | $174.4^{+2.1}_{-2.1}$ | 1.62 | $0.1209^{+0.0071}_{-0.0067}$ | $174.8^{+2.1}_{-2.1}$ | 1.93 | $0.1233^{+0.0080}_{-0.0072}$ | $175.6^{+2.1}_{-2.1}$ | 2.52 |
| $p_T^t$ | $y_{t\bar{t}}$ | $0.1172^{+0.0044}_{-0.0058}$ | $174.7^{+2.2}_{-2.2}$ | 1.33 | $0.1173^{+0.0066}_{-0.0061}$ | $174.7^{+2.2}_{-2.1}$ | 0.78 | $0.1213^{+0.0070}_{-0.0068}$ | $175.6^{+2.1}_{-2.1}$ | 0.72 |
| $M_{t\bar{t}}$ | $p_T^t$ | $0.1162^{+0.0044}_{-0.0054}$ | $172.9^{+0.7}_{-1.1}$ | 3.06 | $0.1158^{+0.0067}_{-0.0064}$ | $173.0^{+0.7}_{-1.0}$ | 2.87 | $0.1186^{+0.0058}_{-0.0059}$ | $173.2^{+0.8}_{-0.7}$ | 2.93 |
| $M_{t\bar{t}}$ | $M_{t\bar{t}}$ | $0.1148^{+0.0044}_{-0.0049}$ | $172.9^{+0.7}_{-1.0}$ | 5.25 | $0.1140^{+0.0063}_{-0.0061}$ | $172.9^{+0.7}_{-1.0}$ | 4.79 | $0.1172^{+0.0054}_{-0.0057}$ | $173.2^{+0.7}_{-0.7}$ | 4.75 |
| $M_{t\bar{t}}$ | $y_t$ | $0.1168^{+0.0053}_{-0.0080}$ | $173.2^{+0.9}_{-0.8}$ | 1.94 | $0.1198^{+0.0064}_{-0.0061}$ | $173.3^{+0.9}_{-0.8}$ | 2.50 | $0.1203^{+0.0065}_{-0.0066}$ | $173.5^{+0.9}_{-0.8}$ | 3.57 |
| $M_{t\bar{t}}$ | $y_{t\bar{t}}$ | $0.1161^{+0.0048}_{-0.0060}$ | $173.3^{+0.9}_{-0.8}$ | 1.80 | $0.1166^{+0.0060}_{-0.0056}$ | $173.3^{+0.9}_{-0.8}$ | 1.36 | $0.1190^{+0.0061}_{-0.0065}$ | $173.5^{+0.9}_{-0.8}$ | 1.79 |
| $y_t$ | $p_T^t$ | $0.1128^{+0.0039}_{-0.0032}$ | $170.8^{+2.2}_{-2.2}$ | 4.21 | $0.1109^{+0.0043}_{-0.0040}$ | $170.5^{+2.2}_{-2.2}$ | 0.80 | $0.1126^{+0.0057}_{-0.0068}$ | $170.6^{+2.2}_{-2.2}$ | 0.84 |
| $y_t$ | $M_{t\bar{t}}$ | $0.1127^{+0.0036}_{-0.0029}$ | $170.9^{+2.4}_{-2.6}$ | 6.15 | $0.1091^{+0.0040}_{-0.0037}$ | $169.3^{+2.7}_{-2.8}$ | 1.65 | $0.1087^{+0.0060}_{-0.0069}$ | $168.8^{+2.8}_{-3.1}$ | 1.61 |
| $y_t$ | $y_t$ | $0.1175^{+0.0037}_{-0.0054}$ | $176.4^{+3.6}_{-3.7}$ | 3.59 | $0.1142^{+0.0047}_{-0.0043}$ | $172.4^{+3.8}_{-3.6}$ | 2.43 | $0.1080^{+0.0060}_{-0.0076}$ | $167.5^{+3.8}_{-3.4}$ | 2.43 |
| $y_t$ | $y_{t\bar{t}}$ | $0.1186^{+0.0032}_{-0.0048}$ | $178.6^{+3.6}_{-3.8}$ | 2.78 | $0.1120^{+0.0043}_{-0.0039}$ | $171.4^{+4.1}_{-3.8}$ | 0.89 | $0.1210^{+0.0062}_{-0.0063}$ | $176.0^{+3.7}_{-3.6}$ | 0.79 |
| $y_{t\bar{t}}$ | $p_T^t$ | $0.1140^{+0.0036}_{-0.0032}$ | $171.7^{+2.1}_{-2.1}$ | 7.95 | $0.1115^{+0.0035}_{-0.0032}$ | $170.7^{+2.1}_{-2.1}$ | 1.27 | $0.1130^{+0.0058}_{-0.0063}$ | $171.0^{+2.1}_{-2.1}$ | 1.31 |
| $y_{t\bar{t}}$ | $M_{t\bar{t}}$ | $0.1139^{+0.0034}_{-0.0030}$ | $172.1^{+1.8}_{-2.5}$ | 10.07 | $0.1105^{+0.0033}_{-0.0030}$ | $169.9^{+2.5}_{-2.6}$ | 2.32 | $0.1107^{+0.0058}_{-0.0057}$ | $169.9^{+2.5}_{-2.7}$ | 2.22 |
| $y_{t\bar{t}}$ | $y_t$ | $0.1176^{+0.0033}_{-0.0047}$ | $176.7^{+3.2}_{-3.3}$ | 6.08 | $0.1134^{+0.0038}_{-0.0035}$ | $172.0^{+3.4}_{-3.3}$ | 2.80 | $0.1179^{+0.0057}_{-0.0065}$ | $173.3^{+3.2}_{-3.2}$ | 2.79 |
| $y_{t\bar{t}}$ | $y_{t\bar{t}}$ | $0.1179^{+0.0031}_{-0.0045}$ | $178.0^{+3.1}_{-3.4}$ | 5.28 | $0.1125^{+0.0036}_{-0.0033}$ | $172.0^{+3.6}_{-3.5}$ | 1.24 | $0.1180^{+0.0056}_{-0.0065}$ | $174.1^{+3.3}_{-3.3}$ | 1.16 |

**Table 5.5.:** Extracted values of $\chi^2_{\min}$ for various PDF sets at the ATLAS and CMS experiments from a simultaneous fit of $m_t$ and $\alpha_s$. Data from the two experiments has been combined assuming no correlations between distributions. Normalised distributions have been used. The cells highlighted in grey correspond to extractions that satisfy the conditions of Equations (5.13, 5.14).

As might be anticipated from the results of the simultaneous extractions of $\alpha_s$ and $m_t$ from each individual experiment, discussed in the previous section and tabulated in Table 5.4, the combination of distributions yielding the best results for the CT14 and NNPDF30 PDF sets are $p_T^t$ or $M_{t\bar{t}}$ from ATLAS, combined with $y_{t\bar{t}}$ from CMS. These two combinations give best fit values of $\alpha_s$ and $m_t$ that differ by about 1%, but are fully consistent within their uncertainties.

In Figure 5.7 we show the $\Delta\chi^2$ contours for the extractions based on the combinations of distributions $\{p_T^t, y_{t\bar{t}}\}$ and $\{M_{t\bar{t}}, y_{t\bar{t}}\}$. Compared to the extractions based on individual distributions, the contours of the extractions combining distributions appear more regular (elliptical). Compared to the contours of $y_{t\bar{t}}$ from CMS in Figure 5.6 the spread of the $\Delta\chi^2 = 1$ contours are much reduced. We also observe improved consistency between PDF sets.

### 5.4.6. Best extracted values

At this point we are in a position to present a final best estimate of the parameters $m_t$ and $\alpha_s$ from the different distributions and PDF sets. Given the reduction in errors which we observe, we use values from the combination of ATLAS and CMS distributions as described in Subsection 5.4.5. We take an unbiased approach in combining results from different PDF sets (as recommended by the PDF4LHC collaboration[140]) and average values without further weighting for both $\alpha_s$ and $m_t$. This is also the approach taken in the determination of $\alpha_s$ in [120]. Given the inclusion of the differential top data in NNPDF31, we do not consider results from this family of PDFs.

Taking the result using the ATLAS $M_{t\bar{t}}$ distribution and the CMS $y_{t\bar{t}}$ distribution, our final result is

$$\alpha_s(M_Z) = 0.1164^{+0.0054}_{-0.0058} \tag{5.15}$$

$$m_t = 173.3^{+0.9}_{-0.8} \tag{5.16}$$

It is instructive to compare these results with the only other determination of $\alpha_s$ using 8 TeV top data[120] which gives a value $\alpha_s(M_Z) = 0.1177^{+0.0034}_{-0.0036}$ and the ATLAS[128] and CMS[129] determinations of $m_t$ which use the 8 TeV $t\bar{t}$ total cross section, and return values of $m_t = 172.9^{+2.5}_{-2.6}$ GeV and $m_t = 173.8^{+1.7}_{-1.8}$ GeV respectively. We note that our best fit values are in agreement to within $1\sigma$ with values of both parameters. Our errors in the extraction of $\alpha_s$ are roughly twice as big as those in [120] and smaller in the extraction of $m_t$ that either ATLAS or CMS extractions alone, though we would expect these to grow if theoretical errors were properly taken into account. Given the variable fit quality we have observed throughout this work, we regard these results with some caution. We do not claim that this result is competitive with existing measurements, but rather aim to show that the addition of information about event shapes can lend considerable constraining power to the parameters. The potential improvements we detail in the next section could, however, mean that in the future a more precise result could be obtained.

## 5.5. Conclusions

In this chapter we have presented the first joint extraction of $\alpha_s$ and $m_t$ from measurements of both the top-pair production cross section as well as differential distributions. We have taken experimental measurements of the kinematic variables $p_T^t, M_{t\bar{t}}, y_t$ and $y_{t\bar{t}}$ in $t\bar{t}$ events provided by the ATLAS and CMS experiments at the 8 TeV LHC. These have been compared with theoretical predictions with NNLO QCD corrections for different values of the top mass and using different PDF sets. Functional forms of the bin weights of each predicted distribution have been determined by smooth interpolation. A least-squares method has been used to perform individual and simultaneous extractions of the Standard Model parameters $m_t$ and $\alpha_s$—we have also considered combining distributions from different experiments in the absence of correlations. We obtain as a final result $\alpha_s(M_Z) = 0.1164^{+0.0054}_{-0.0058}$ and $m_t = 173.3^{+0.9}_{-0.8}$.

There are a few ways in which our treatment could be further refined in a future study. Firstly, as we have highlighted in Section 5.3, the uncertainties on the extracted values of $\alpha_s$ and $m_t$ we have presented only include the effects of experimental uncertainties for single distributions. In a more complete treatment of uncertainties, theory uncertainties due to the choice of renormalisation and factorisation scales as well as PDF uncertainties should be included in the definition of the $\chi^2$ objective function. Since the scale uncertainty is greatly reduced at NNLO when considering normalised (compared to absolute) distributions, we expect that including these will not increase the uncertainties on the extracted parameters dramatically. On the other hand, in one-dimensional extractions of $\alpha_s$, the scale uncertainty on the cross section is known to contribute a significant fraction of the overall uncertainty on the best fit $\alpha_s$[120] , and therefore we would expect this to also play a rôle for our study here. Additionally, in the case of our extractions from combined measurements of ATLAS and CMS, a complete treatment would also include systematic uncertainties correlated between the two experiments, such as that of luminosity.

Going beyond that which we have presented in this study, it would be of great interest to perform extractions of $\alpha_s$ and $m_t$ by using measurements of multiple distributions from the same experiment. This would require correlations between measurements of different distributions that are not currently available, or alternatively measurements and corresponding experimental covariance matrices of two-dimensional distributions (such as those of [141]). Finally, given the constraining power that differential measurements of top-pair production have on PDFs[127,141] , a joint fit of $\alpha_s$ and $m_t$ together with PDFs using differential data would simultaneously provide useful constraints on all these three fundamental ingredients of QCD.

# Author's note

This chapter is based on a paper in preparation—the work was a result of collaboration with M. Czakon, A. Cooper-Sarkar, A. Mitov and A. Papanastasiou. The experimental data used are identical to those analysed in the work [127] which in the case of the ATLAS experiment do not match the numbers available on the HEPData repository. The forthcoming paper will, however, use the HEPData numbers.

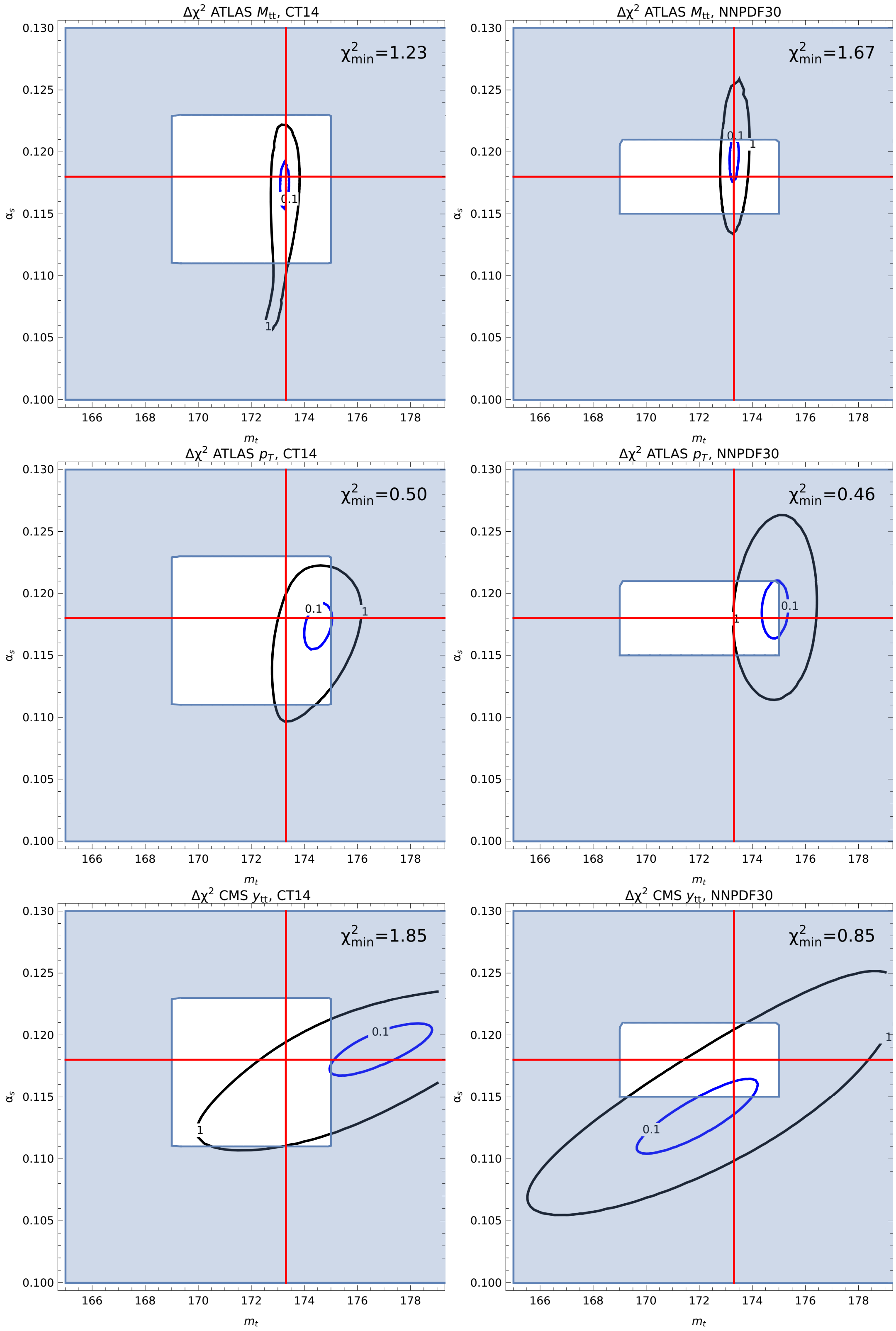

**Figure 5.6.:** $\Delta\chi^2$ contour plots from a two-dimensional extraction of $\alpha_s$ and $m_t$ using ATLAS $M_{t\bar{t}}$ (1[st] row), ATLAS $p_T^t$ (2[nd] row) and CMS $y_{t\bar{t}}$ distributions (final row). CT14 and NNPDF30 PDFs have been used in the left-hand and right-hand plots respectively. See text for further details.

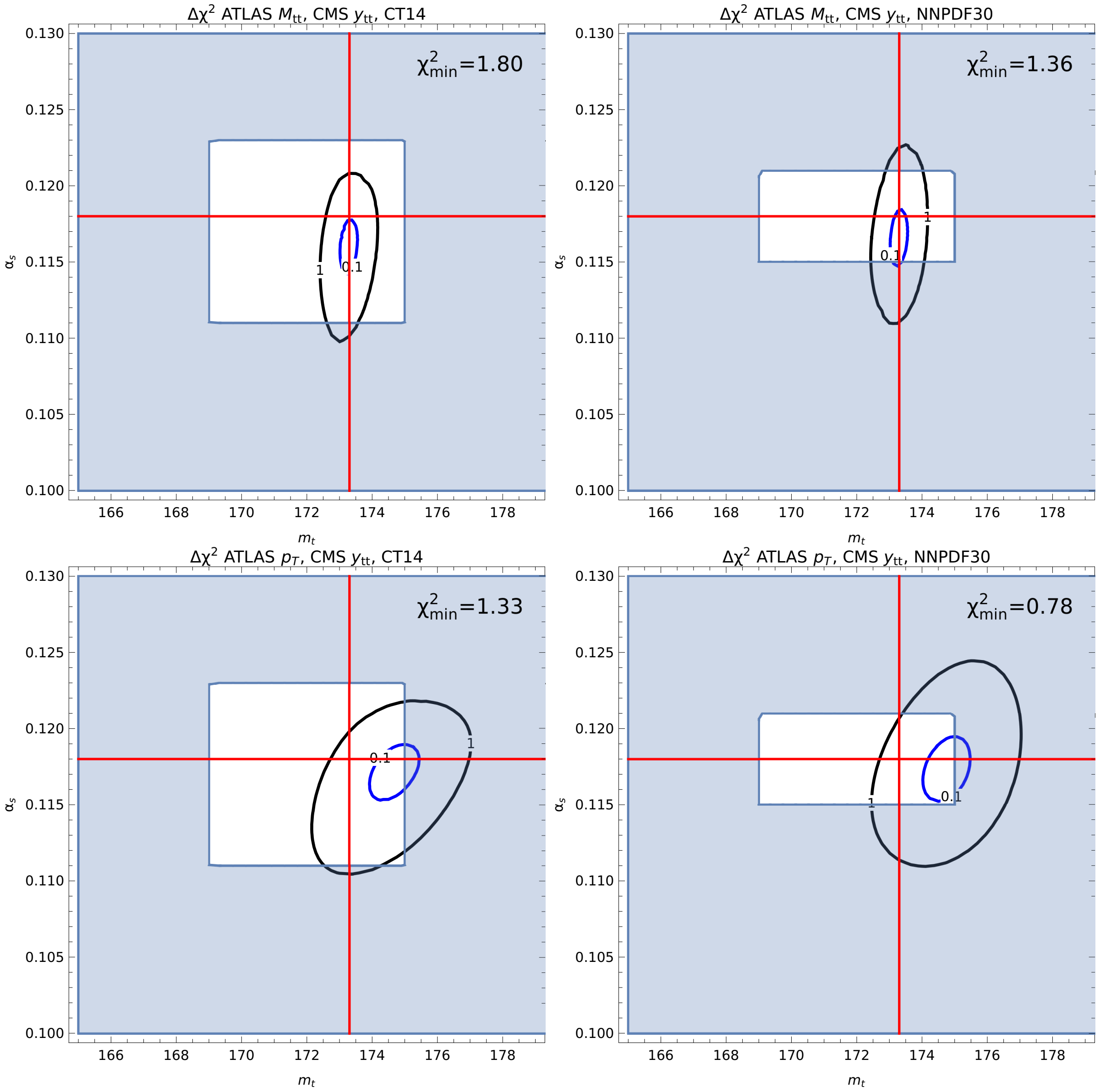

**Figure 5.7.:** $\Delta\chi^2$ contour plots from a two-dimensional extraction of $\alpha_s$ and $m_t$ using the pairs of distributions from {ATLAS, CMS}: $\{M_{t\bar{t}}, y_{t\bar{t}}\}$ (upper) and $\{p_T^t, y_{t\bar{t}}\}$. CT14 and NNPDF30 PDFs have been used in the left-hand and right-hand plots respectively.

# Chapter 6.

# The anatomy of double heavy-quark initiated processes

In this chapter we consider a different class of processes involving heavy quarks, with two in the initial rather than the final state. We study the phenomenology of processes in which heavy bosons, either a Higgs or a $Z'$, are produced and examine the physics of such processes at the LHC as well as at a future 100 TeV collider. The work in this chapter is contained in a published paper [1].

## 6.1. Introduction

The study of associated production of (possibly new) vector or scalar bosons in association with heavy quarks, such as top and bottom quarks, are among the highest priorities in LHC physics. In particular, $b$ quarks play an important role in the quest for New Physics as well as for precise SM measurements from both an experimental and a theoretical perspective. Firstly, they provide a very clean signature as they may easily be identified in a detector due to the displacement of vertices with respect to the collision point, a consequence of the $b$-quark long lifetime. Secondly, the relative strength of the Higgs Yukawa coupling (or possibly of new scalar states) to the heavy quarks is important in determining the phenomenology, both in production as well as in decay. In particular, production associated with $b$ quarks could provide the leading mode for Higgs bosons with enhanced Yukawa couplings in many scenarios beyond the Standard Model.

At hadron colliders, any process that features heavy quarks can be described according to two different and complementary approaches. In the *massive* or four-flavor (4F) scheme (in the case of $b$ quarks), the heavy quark is produced in the hard scattering and arises as a massive particle in the final state. The dependence on the heavy quark mass $m_b$ is retained in the matrix element and explicit logarithms of $Q/m_b$, $Q$ being some hard scale of the process, appear at each order in perturbation theory as a result of collinearly enhanced (yet finite) splittings $q \to qg$ or of a gluon into heavy quark pairs, $g \to q\bar{q}$. On the other hand, in the *massless* or five-flavour (5F) scheme (in the case of $b$ quarks), $Q \gg m_b$ is assumed and the heavy quark is treated on the same footing as the light quarks: it contributes to the proton wave function and enters the running of the strong coupling constant $\alpha_s$. In this scheme the heavy quark mass is neglected in the matrix element and the collinear logarithms that may spoil the convergence of the perturbative expansion of the 4F scheme cross section are resummed to all orders in the evolution of the heavy quark parton density.

In a previous work[142], processes involving a single $b$ quark in both lepton-hadron and hadron-hadron collisions were examined. It was found that, at the LHC, unless a very heavy particle is produced in the final state, the effects of initial-state collinear logarithms are always modest and such logarithms do not spoil the convergence of perturbation theory in 4F scheme calculations. This behaviour was explained by two main reasons, one of dynamical and the other of kinematical nature. The first is that the effects of the resummation of the initial-state collinear logarithms is relevant mainly at large Bjorken-$x$ and in general keeping only the explicit logs appearing at NLO is a very good approximation. The second reason is that the naïve scale $Q$ that appears in the collinear logarithms turns out to be suppressed by universal phase space factors that, at hadron colliders, reduce the size of the logarithms for processes taking place. As a result, a consistent and quantitative analysis of many processes involving one $b$ quark in the initial state was performed and a substantial agreement between total cross sections obtained at NLO (and beyond) in the two schemes found within the expected uncertainties.

In this chapter we focus on processes that can be described by two $b$ quarks in the initial state, such as $pp \to Hb\bar{b}$ or $pp \to Zb\bar{b}$. As already sketched in [142], the same arguments used for single heavy-quark initiated processes can be used to analyse the double heavy-quark case. One may naïvely expect that the resummation effects for processes with two $b$ quarks in the initial state can be simply obtained by "squaring", in some sense, those of processes with only one $b$ quark. There are, however, a number of features that are particular to the double heavy-quark processes and call for a dedicated work. One is that the lowest order contribution in the 4F scheme appears for the first time among the NNLO real corrections to the leading order 5F scheme calculation. Furthermore, due to the simplicity of the 5F description (i.e. Born amplitudes are $2 \to 1$ processes), results in the 5F scheme are now available at NNLO, while, thanks to the progress in the automation of NLO computations, 4F scheme results have become easily accessible for a wide range of final states. In fact, it is easy to understand that a meaningful comparison between the two schemes for double heavy-quark initiated processes starts to be accurate if results are taken at NNLO for the 5F and at NLO for the 4F case.

Both $pp \to Hb\bar{b}$ or $pp \to Zb\bar{b}$ have been considered in previous works. For the LHC, it was demonstrated that consistent results for both the total cross section and differential distributions for bottom-fusion initiated Higgs production can be obtained in both schemes.[143–147] Analogous studies were performed for bottom-fusion initiated $Z$ production.[144,148–151] All these studies suggested that the appropriate factorisation and renormalisation scales associated to these processes are to be chosen smaller than the mass of the final state heavy particles. In particular, scales of about $M_{H,Z}/4$ have been proposed in order to stabilise the perturbative series and make the four- and five-flavor predictions closer to each other. $(M_H + 2m_b)/4$ is the scale adopted by the LHC Higgs Cross Section Working Group (HXSWG) to match the NLO 4F and NNLO 5F scheme predictions in case of bottom-fusion initiated Higgs production via the Santander interpolation[146] and via the use of consistently matched calculations.[152–154]

While previous studies support *a posteriori* the evidence that smaller scales make the four- and five-flavor pictures more consistent, no complete analysis of the relation of the two schemes in the case of double heavy-quark initiated processes has been provided. In particular, no analytic study of the collinear enhancement of the cross section and the kinematics of this class of processes has been performed.

In this chapter, we fill this gap by extending previous work to double heavy-quark production. We first present an analytic comparison of the two schemes that allow us to unveil a clear relation between them, establish the form of the logarithmic enhancements and determine their size. We then compare the predictions for LHC phenomenology in a number of relevant cases focusing on LHC Run II. Furthermore, we expand our investigation to high energy processes involving top quarks at future colliders. At centre-of-mass energies of order 100 TeV, a new territory far beyond the reach of the LHC would be explored. At such an energy, much heavier particles could be produced at colliders and top-quark PDFs may become of relevance in processes involving top quarks in the initial state.

The structure of the chapter is as follows. In Section 6.2 we examine the kinematics of 2 to 3 body scattering and calculate the phase space factor for the particular case of $b$-initiated Higgs production—we thus derive the logarithmic contributions to the cross section which arise in a 4F scheme. We then proceed to generate kinematic distributions for the processes and use these to analyse the 4F and 5F scheme results. We conclude the section by suggesting a factorisation scale at which results from either process may be meaningfully compared. In Section 6.3 we compare the results on total cross sections obtained in both schemes for a number of phenomenologically relevant processes at the LHC and future colliders. Finally, our conclusions are presented in Section 6.4.

## 6.2. Different heavy quark schemes: analytical comparison

We start by considering Higgs boson production via $b\bar{b}$ fusion in the 4F scheme. The relevant partonic subprocess is

$$g(p_1) + g(p_2) \to b(k_1) + H(k) + \bar{b}(k_2), \tag{6.1}$$

where the $b$ quarks in the final state are treated as massive objects. Since the $b$-quark mass $m_b$ is much smaller than the Higgs boson mass $M_H$, we expect the cross section for the process (6.1) to be dominated by the configurations in which the two final-state $b$ quarks are emitted collinearly with the incident gluons. Indeed the quark-antiquark channel ($q\bar{q} \to b\bar{b}H$) that also contributes to the leading-order cross section in the 4F scheme is very much suppressed with respect to the gluon-gluon one. In order to estimate the importance of large transverse momentum $b$ quarks in the gg channel, as compared to the dominant collinear configurations, we will perform an approximate calculation of the cross section for the process (6.1) limiting ourselves to the dominant terms as $m_b \to 0$. The result will then be compared to the full leading-order 4F scheme calculation. We present here the final result; the details of the calculation can be found in Appendix D.

The differential partonic cross section can be expressed as a function of five independent invariants, which we choose to be

$$\hat{s} = (p_1 + p_2)^2; \quad t_1 = (p_1 - k_1)^2; \quad t_2 = (p_2 - k_2)^2; \quad s_1 = (k_1 + k)^2; \quad s_2 = (k_2 + k)^2. \tag{6.2}$$

Collinear singularities appear, for $m_b^2 = 0$, either when

$$t_1 \to 0; \qquad t_2 \to 0, \tag{6.3}$$

or when

$$u_1 \to 0; \qquad u_2 \to 0, \tag{6.4}$$

where

$$u_1 = (p_1 - k_2)^2; \qquad u_2 = (p_2 - k_1)^2. \tag{6.5}$$

The configuration in Equation 6.3 is achieved for

$$k_1 = (1 - z_1)p_1; \qquad k_2 = (1 - z_2)p_2; \qquad 0 \le z_i \le 1 \tag{6.6}$$

while the one in Equation 6.4 corresponds to

$$k_1 = (1 - z_1)p_2; \qquad k_2 = (1 - z_2)p_1. \tag{6.7}$$

In both cases we find

$$\hat{s} = \frac{M_H^2}{z_1 z_2}; \qquad s_1 = \frac{M_H^2}{z_1}; \qquad s_2 = \frac{M_H^2}{z_2}. \tag{6.8}$$

An explicit calculation yields

$$\hat{\sigma}^{4\mathrm{F,coll}}(\hat{\tau}) = \hat{\tau}\frac{\alpha_s^2}{4\pi^2}\frac{G_F\pi}{3\sqrt{2}}\frac{m_b^2}{M_H^2}2\int_0^1 dz_1 \int_0^1 dz_2\, P_{qg}(z_1)P_{qg}(z_2)L(z_1,\hat{\tau})L(z_2,\hat{\tau})\delta\left(z_1 z_2 - \hat{\tau}\right), \tag{6.9}$$

where

$$\hat{\tau} = \frac{M_H^2}{\hat{s}}, \tag{6.10}$$

$P_{qg}(z)$ is the leading-order quark-gluon Altarelli-Parisi splitting function

$$P_{qg}(z) = \frac{1}{2}[z^2 + (1-z)^2], \tag{6.11}$$

and

$$L(z,\hat{\tau}) = \log\left[\frac{M_H^2}{m_b^2}\frac{(1-z)^2}{\hat{\tau}}\right]. \tag{6.12}$$

The suffix "coll" reminds us that we are neglecting less singular contributions as $m_b \to 0$, i.e. either terms with only one collinear emission, which diverge as $\log m_b^2$, or terms which are regular as $m_b \to 0$.

We now observe that the leading-order partonic cross section for the process

$$b(q_1) + \bar{b}(q_2) \to H(k), \tag{6.13}$$

relevant for calculations in the 5F scheme, is given by[38]

$$\hat{\sigma}^{5\mathrm{F}}(\hat{\tau}) = \frac{G_F \pi}{3\sqrt{2}} \frac{m_b^2}{M_H^2} \delta(1-\hat{\tau}), \tag{6.14}$$

with

$$\hat{s} = (q_1 + q_2)^2. \tag{6.15}$$

Hence, the 4F scheme cross section in the collinear limit, Equation 6.9, can be rewritten as

$$\hat{\sigma}^{4\mathrm{F,coll}}(\hat{\tau}) = 2 \int_{\hat{\tau}}^{1} dz_1 \int_{\frac{\hat{\tau}}{z_1}}^{1} dz_2 \left[\frac{\alpha_s}{2\pi} P_{qg}(z_1) L(z_1, \hat{\tau})\right] \left[\frac{\alpha_s}{2\pi} P_{qg}(z_2) L(z_2, \hat{\tau})\right] \hat{\sigma}^{5\mathrm{F}} \left(\frac{\hat{\tau}}{z_1 z_2}\right). \tag{6.16}$$

The physical interpretation of the result in Equation 6.16 is straightforward: in the limit of collinear emission, the cross section for the parton process (6.1) is simply the $b\bar{b} \to H$ cross section convolved with the probability that the incident gluons split in a $b\bar{b}$ pair. This probability is logarithmically divergent as $m_b \to 0$, and this is the origin of the two factors of $L(z_i, \hat{\tau})$.

The arguments of the two collinear logarithms exhibit a dependence on the momentum fractions $z_1, z_2$, *viz.* Equation 6.12. This dependence is subleading in the collinear limit $m_b \to 0$ and indeed it could be neglected in this approximation; however, the class of subleading terms induced by the factor $(1-z_i)^2/\hat{\tau}$ in Equation 6.12 is of kinematical origin (it arises from the integration bounds on $t_1$ and $t_2$, as shown in Appendix D) and therefore universal in some sense, as illustrated in [142]. We also note that the arguments of the two collinear logs depend on both $z_1$ and $z_2$; this is to be expected, because the integration bounds on $t_1$ and $t_2$ are related to each other. However, in some cases (for example, if one wants to relate the scale choice to a change of factorisation scheme, as in [155]) a scale choice which only depends on the kinematics of each emitting line might be desirable. We have checked that the replacement

$$\log\left[\frac{M_H^2}{m_b^2} \frac{(1-z_i)^2}{z_1 z_2}\right] \to \log\left[\frac{M_H^2}{m_b^2} \frac{(1-z_i)^2}{z_i}\right] \tag{6.17}$$

has a moderate effect on physical cross sections. The replacement would make the scale at which the four- and five-flavor scheme results are comparable lower by about 20-30% but does not qualitatively modify our arguments and results below.

The corresponding 4F scheme physical cross section in hadron collisions at centre-of-mass energy $\sqrt{s}$ is given by

$$\sigma^{\rm 4F,coll}(\tau) = \int_\tau^1 dx_1 \int_{\frac{\tau}{x_1}}^1 dx_2\, g(x_1,\mu_F^2) g(x_2,\mu_F^2) \hat{\sigma}^{\rm 4F,coll}\left(\frac{\tau}{x_1 x_2}\right), \qquad (6.18)$$

where $g(x,\mu_F^2)$ is the gluon distributon function, $\mu_F$ is the factorisation scale, and

$$\tau = \frac{M_H^2}{s}. \qquad (6.19)$$

After some (standard) manipulations, we get

$$\sigma^{\rm 4F,coll}(\tau) = 2\int_\tau^1 dx_1 \int_{\frac{\tau}{x_1}}^1 dx_2\, \hat{\sigma}^{\rm 5F}\left(\frac{\tau}{x_1 x_2}\right)$$
$$\int_{x_1}^1 \frac{dz_1}{z_1}\left[\frac{\alpha_s}{2\pi} P_{qg}(z_1) L\left(z_1, z_1 z_2\right)\right] g\left(\frac{x_1}{z_1},\mu_F^2\right) \int_{x_2}^1 \frac{dz_2}{z_2}\left[\frac{\alpha_s}{2\pi} P_{qg}(z_2) L\left(z_2, z_1 z_2\right)\right] g\left(\frac{x_2}{z_2},\mu_F^2\right). \qquad (6.20)$$

We are now ready to assess the accuracy of the collinear approximation in the 4F scheme. We first consider the total cross section. In Table 6.1 we display the total 4F scheme cross section for the production of a Higgs boson at LHC 13 TeV for two values of the Higgs mass, namely $M_H = 125$ GeV and $M_H = 400$ GeV. In the first column we give the exact leading order result; the second column contains the cross section with the squared amplitude approximated by its collinear limit, but the exact expression of the phase space measure. Finally, in the third column we give the results obtained with both the amplitude and the phase-space measure in the collinear limit, which corresponds to the expression in Equation 6.20. From Table 6.1 we conclude that the production of large transverse momentum $b$ quarks, correctly taken into account in the 4F scheme, amounts to an effect of order 20% on the total cross section and tends to decrease with increasing Higgs mass.

| $M_H$ | exact | collinear ME | collinear ME and PS |
|---|---|---|---|
| 125 GeV | $4.71 \cdot 10^{-1}$ pb | $5.15 \cdot 10^{-1}$ pb | $5.82 \cdot 10^{-1}$ pb |
| 400 GeV | $5.42 \cdot 10^{-3}$ pb | $5.58 \cdot 10^{-3}$ pb | $5.91 \cdot 10^{-3}$ pb |

**Table 6.1.:** Total cross sections for Higgs boson production at the LHC 13 TeV in the 4F scheme.

We now turn to an assessment of the numerical relevance of the subleading terms included by the definition in Equation 6.12 of the collinear logarithms. To this purpose we study the distribution of $(1-z_1)^2/(z_1 z_2)$, which is the suppression factor of $M_H^2/m_b^2$ in the arguments of the logs. The results are displayed in Figure 6.1 for Higgs production at the LHC at 13 TeV and for two different values of the Higgs boson mass. The two distributions behave in a similar way: both are strongly peaked around values smaller than 1; in particular, the 68% threshold is in both cases around 0.2. This confirms that, although formally subleading with respect to $\log \frac{M_H^2}{m_b^2}$, in practice the terms proportional to $\log \frac{(1-z_i)^2}{z_1 z_2}$ give a sizeable contribution to the total cross section.

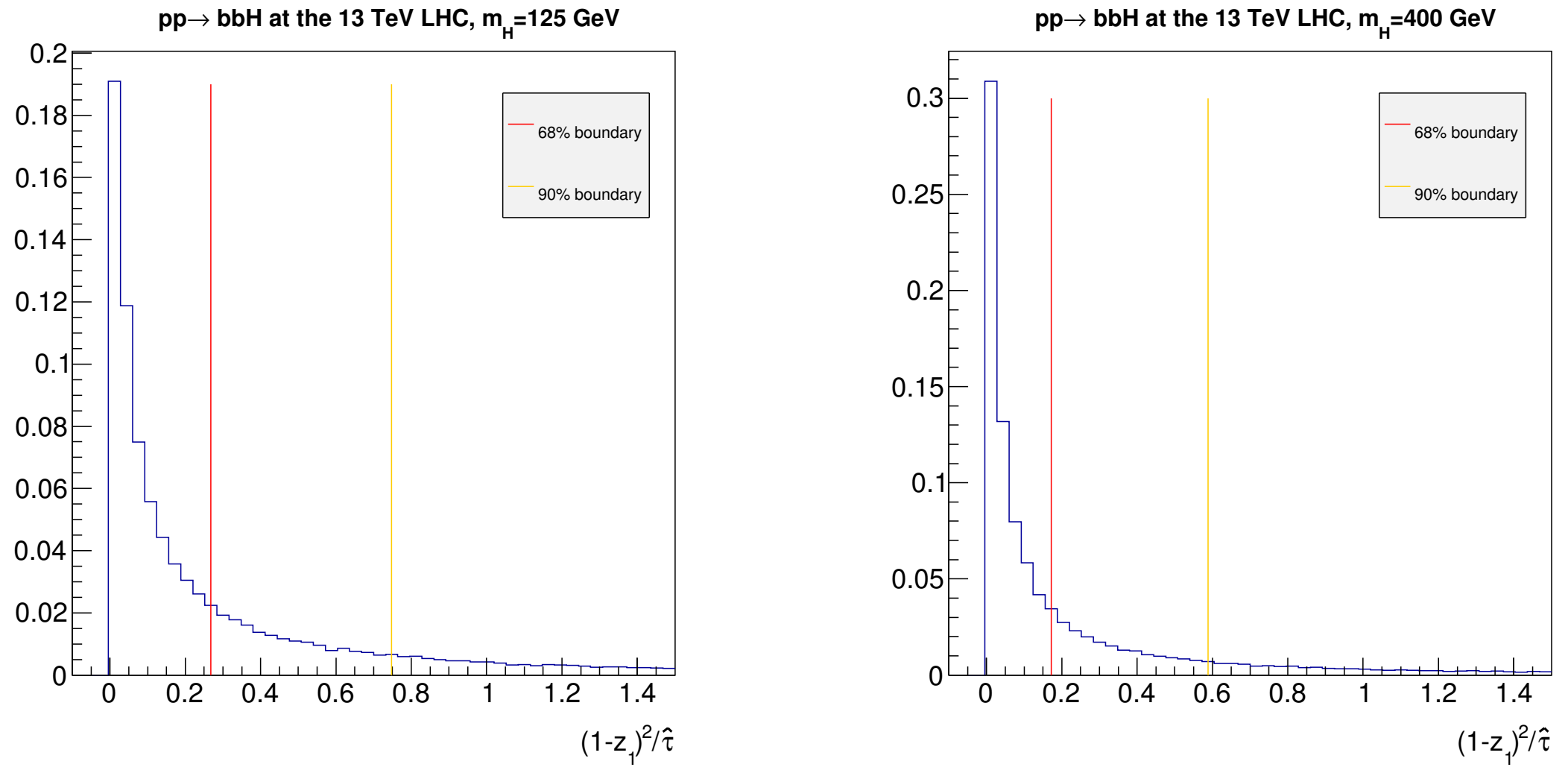

**Figure 6.1.:** Normalised distribution (events/bin) of $(1-z_1)^2/\hat{\tau}$ for $b$-initiated Higgs production in $pp$ collisions at LHC 13 TeV for $M_H = 125$ GeV (left) and $M_H = 400$ GeV (right). Both $\mu_R$ and $\mu_F$ are set to $M_H$. The vertical lines represent the values below which 68% and 90% of events lie.

A further confirmation is provided by the distributions in Figure 6.2, where the full cross sections, together with their collinear and double-collinear approximations, are plotted as functions of the partonic centre-of-mass energy. We see that the collinear cross section provides a good approximation to the full 4F scheme result. In the same picture we show the collinear cross section with the factors of $L(z_i, z_1 z_2)$ replaced by $\log \frac{M_H^2}{m_b^2}$ (solid black histogram). It is clear that in this case the collinear cross section substantially differs from the exact result.

We now consider the 5F scheme, where the $b$ quark is treated as a massless parton and collinear logarithms are resummed to all orders by the perturbative evolution of the parton distribution function. Equation 6.14 leads to a physical cross section

$$\sigma^{5\mathrm{F}}(\tau) = 2 \int_\tau^1 dx_1\, b(x_1, \mu_F^2) \int_{\frac{\tau}{x_1}}^1 dx_2\, b(x_2, \mu_F^2) \hat{\sigma}^{5\mathrm{F}} \left( \frac{\tau}{x_1 x_2} \right) . \qquad (6.21)$$

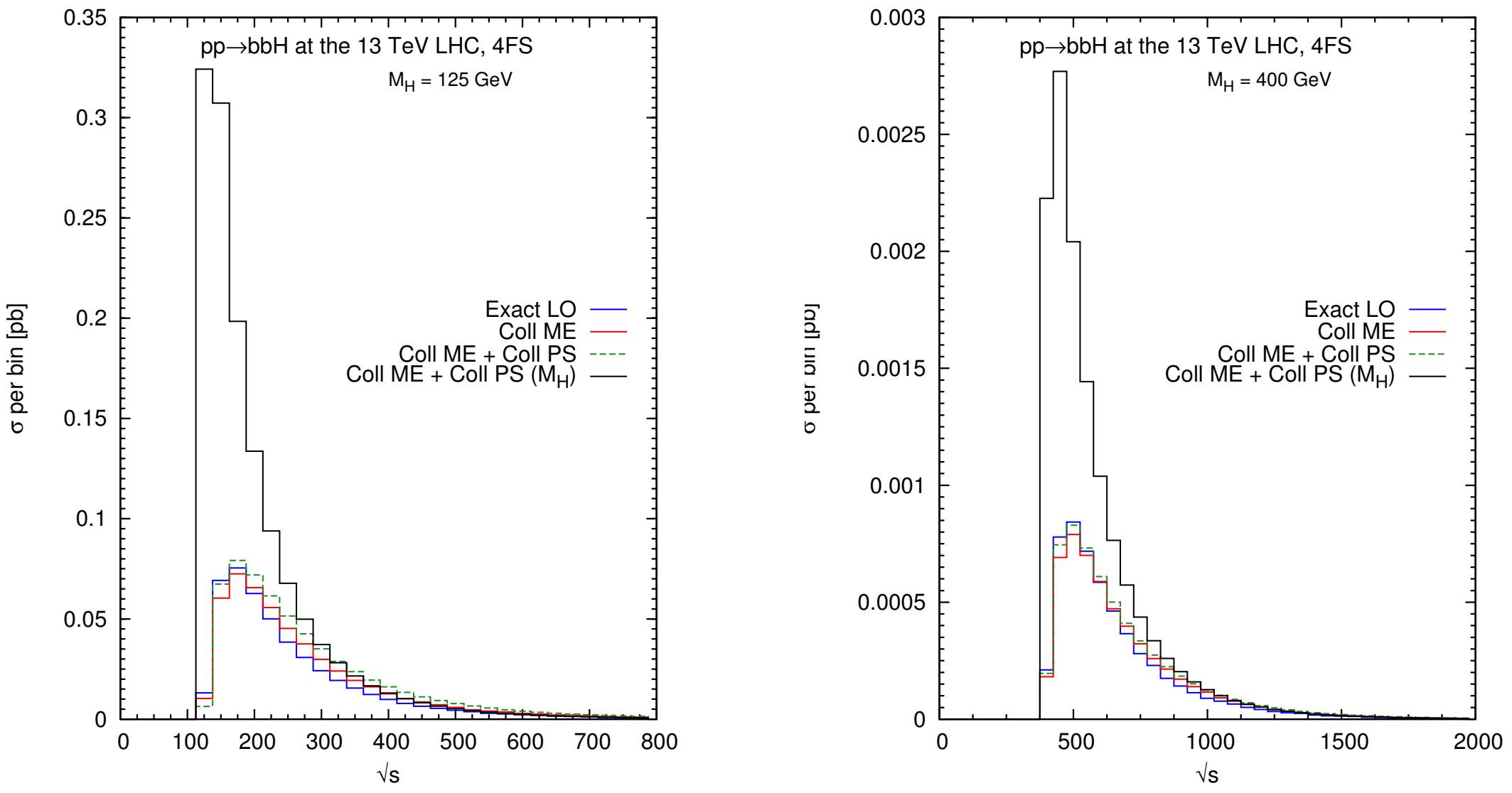

**Figure 6.2.:** Distribution of the 4F scheme cross section as a function of the partonic centre-of-mass energy $\hat{s}$ for a Higgs of mass 125 GeV (left) and of mass 400 GeV (right). The solid line represents the full cross section at leading-order, while the dashed line represents the collinear limit.

In order to make contact with the 4F scheme calculation, we observe that the $b$ quark PDF can be expanded to first order in $\alpha_s$:

$$b(x, \mu_F^2) = \frac{\alpha_s}{2\pi} L_b \int_x^1 \frac{dy}{y} P_{qg}(y) g\left(\frac{x}{y}, \mu_F^2\right) + \mathcal{O}(\alpha_s^2) = \tilde{b}^{(1)}(x, \mu_F^2) + \mathcal{O}(\alpha_s^2), \tag{6.22}$$

where

$$L_b = \log \frac{\mu_F^2}{m_b^2}. \tag{6.23}$$

Correspondingly, we may define a truncated 5F cross section $\sigma^{5\mathrm{F},(1)}(\tau)$ which contains only one power of $\log m_b^2$ for each colliding $b$ quark. This is obtained by replacing Equation 6.22 in 6.21 and performing the same manipulations that led us to 6.20: we obtain

$$\sigma^{5\mathrm{F},(1)}(\tau) = 2 \int_\tau^1 dx_1 \int_{\frac{\tau}{x_1}}^1 dx_2 \, \hat{\sigma}^{5F}\left(\frac{\tau}{x_1 x_2}\right)$$
$$\int_{x_1}^1 \frac{dy}{y} \left[\frac{\alpha_s}{2\pi} P_{qg}(y) L_b\right] g\left(\frac{x_1}{y}, \mu_F^2\right) \int_{x_2}^1 \frac{dz}{z} \left[\frac{\alpha_s}{2\pi} P_{qg}(z) L_b\right] g\left(\frac{x_2}{z}, \mu_F^2\right). \tag{6.24}$$

Equation 6.24 has exactly the same structure as the 4F scheme result in the collinear approximation Equation 6.20, except that the collinear logarithms have a constant argument. Hence, it corresponds to the solid black curve in Figure 6.2. We are therefore led to suggest that the 5F scheme results be used with a scale choice dictated by the above results, similar to what we have illustrated in [155]. Such a scale is defined so that the two schemes give the same result:

$$\sigma^{5\mathrm{F},(1)}(\tau) = \sigma^{4\mathrm{F,coll}}(\tau). \tag{6.25}$$

The explicit expression of $\tilde{\mu}_F$ is simply obtained by equating $\sigma^{5\mathrm{F},(1)}(\tau)$, Equation 6.24, which is proportional to $L_b^2 = \log^2 \frac{\mu_F^2}{m^2}$, and $\sigma^{4\mathrm{F,coll}}(\tau)$, Equation 6.20, and solving for $L_b^2$. The residual dependence on $\mu_F$ due to the gluon parton density is suppressed by an extra power of $\alpha_s$ and can therefore be neglected; we adopt the standard choice $\mu_F = M$, with $M$ either the Higgs mass or the $Z'$ mass. The size of the logarithmic terms kept explicitly in the 4F case is determined by arguments of the form $\frac{(1-z_i)^2}{\hat{\tau}}$. For $\sqrt{s} = 13$ GeV, and $m_b = 4.75$ GeV, we find the following values for $\tilde{\mu}_F$:

$$\begin{aligned} b\bar{b}H, M_H = 125\,\mathrm{GeV}: &\qquad \tilde{\mu}_F \approx 0.36\, M_H \\ b\bar{b}Z', M_{Z'} = 91.2\,\mathrm{GeV}: &\qquad \tilde{\mu}_F \approx 0.38\, M_{Z'} \\ b\bar{b}Z', M_{Z'} = 400\,\mathrm{GeV}: &\qquad \tilde{\mu}_F \approx 0.29\, M_{Z'}, \end{aligned} \tag{6.26}$$

while for $\sqrt{s} = 100$ TeV and $m_t = 173.1$ GeV, we find

$$\begin{aligned} t\bar{t}Z', M_{Z'} = 1\,\mathrm{TeV}: &\qquad \tilde{\mu}_F \approx 0.40\, M_{Z'} \\ t\bar{t}Z', M_{Z'} = 5\,\mathrm{TeV}: &\qquad \tilde{\mu}_F \approx 0.21\, M_{Z'} \\ t\bar{t}Z', M_{Z'} = 10\,\mathrm{TeV}: &\qquad \tilde{\mu}_F \approx 0.16\, M_{Z'}. \end{aligned} \tag{6.27}$$

In both cases we have used the `NNPDF30_lo_as_0130` PDF set,[139] with the appropriate number of light flavors. We have explicitly checked that the choice of $\mu_F = M_H/4$ for the gluon PDF and for the strong coupling constant does not modify in any significant way the value of $\tilde{\mu}_F$ that we obtain. This is expected given that the gluon-gluon luminosity and the dependence on $\alpha_s$ tend to compensate between numerator and denominator. We have also checked that, after the replacement in Equation 6.17, the values of $\tilde{\mu}_F$ are typically about 20-30% smaller.

We note that the scale $\tilde{\mu}_F$ is in general remarkably smaller than the mass of the produced heavy particle. As in the case of single collinear logarithm, the reduction is more pronounced for larger values of the mass of the heavy particle compared to the available hadronic centre-of-mass energy. The above results suggest that a “fair" comparison between calculations in the two schemes should be performed at factorisation/renormalisation scales smaller than the naïve choice $\mu_F = M_H$. This evidence backs up the conclusions drawn in previous studies,[144] although perhaps with a slightly larger value in the case of Higgs boson, $\tilde{\mu} \approx M_H/3$ rather than $M_H/4$.

The argument given above identifies a suitable choice for the factorisation/renormalisation scales such that, at the Born level and without resummation, the size of the logarithmic terms is correctly matched in the two schemes. At this point, further differences between the schemes can arise from the collinear resummation as achieved in the 5F scheme and from mass (power-like) terms which are present in the 4F scheme and not in the 5F one. Closely following the arguments of [144], to which we refer the interested reader for more details, we now numerically quantify the effect of the resummation. A careful study of the impact of power-like terms can be found in [152–154]. These terms have been found to have an impact no stronger than a few percent.

Starting from Equation 6.22, one can assess the accuracy of the $\mathcal{O}(\alpha_s^1)$ ($\mathcal{O}(\alpha_s^2)$) approximations compared to the full $b(x,\mu^2)$ resummed expression. The expansion truncated at order $\alpha_s^p$, often referred to as $\tilde{b}^{(p)}(x,\mu^2)$ in the literature, does not feature the full resummation of collinear logarithms, but rather it contains powers $n$ of the collinear log with $1 \leq n \leq p$.

In Figure 6.3 we display the ratio $\frac{\tilde{b}^{(p)}(x,\mu^2)}{b(x,\mu^2)}$ for $p = 1, 2$ (using the same set of PDFs adopted throughout this work) as a function of the scale $\mu^2$ for various values of the momentum fraction $x$. Deviations from one of these curves are an indication of the size of terms of order $\mathcal{O}(\alpha_s^{p+1})$ and higher, which are resummed in the QCD evolution of the bottom quark PDFs. As observed in our previous work, at LO higher-order logarithms are important and $\tilde{b}^{(1)}(x,\mu^2)$ is a poor approximation of the fully resummed distribution function. In particular, it overestimates the leading-log evolution of the $b$ PDF by 20% at very small $x$ and it underestimates it up to 30% at intermediate values of $x$. On the other hand, at NLO the explicit collinear logs present in a NLO 4F scheme calculation provide a rather accurate approximation of the whole resummed result at NLL; significant effects, of order up to 20%, appear predominantly at large values of $x$.

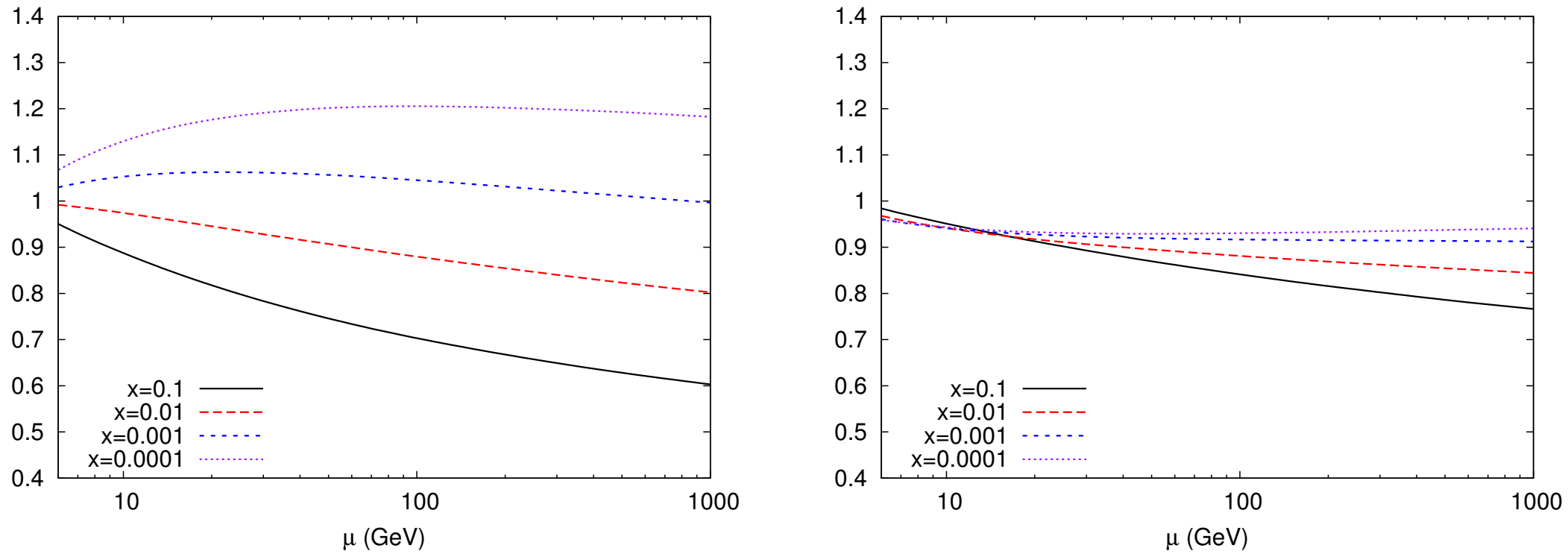

**Figure 6.3.:** The ratio $\tilde{b}^{(p)}/b$ for $p = 1$ (left) and $p = 2$ (right) as a function of the scale $\mu$ for for different values of $x$. The $n_f = 4$ and $n_f = 5$ sets of the `NNPDF3.0` family (with $\alpha_s(M_Z) = 0.118$) are associated to the $\tilde{b}$ and $b$ computations respectively.

A similar behaviour characterises the top-quark PDFs. In Figure 6.4 the ratio between the truncated top-quark PDFs $\tilde{t}$ and the evolved PDFs $t(x, \mu^2)$ is displayed for four different values of $x$ and varying the factorization scale $\mu$. We see that for the top-quark

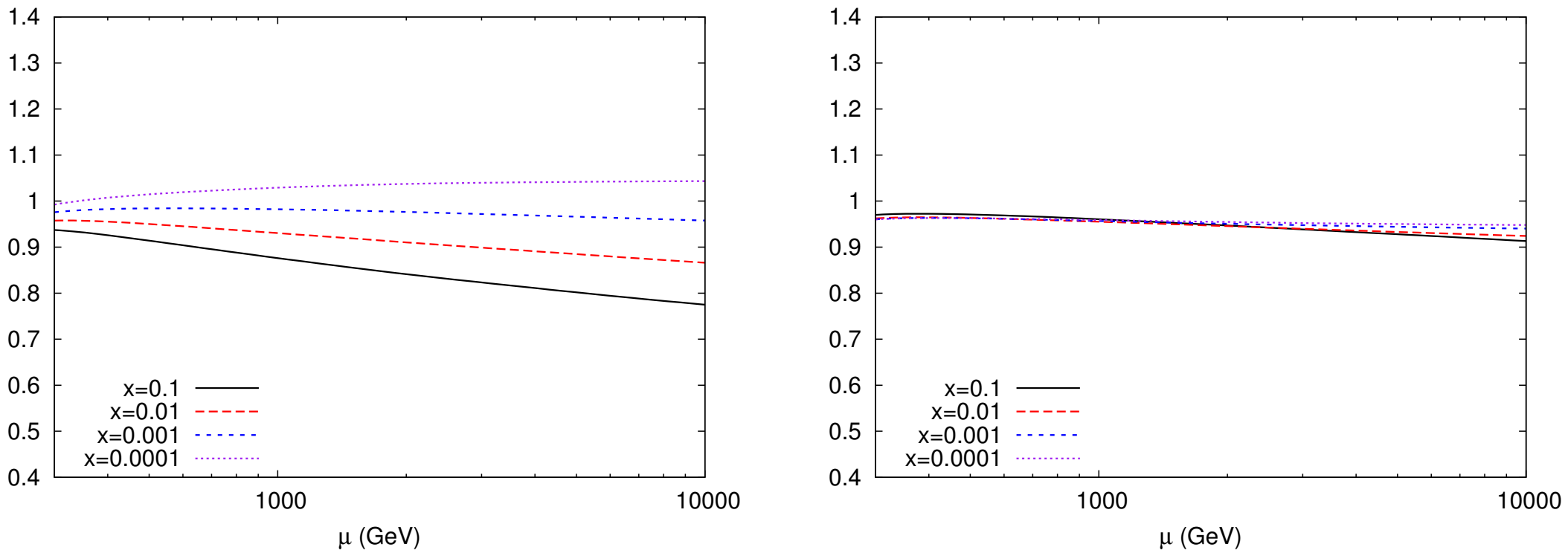

**Figure 6.4.:** Ratio $\tilde{t}/t$ at LO (left) and NLO (right) for several values of $x$ as a function of the scale $\mu$. The $n_f = 5$ and $n_f = 6$ sets of the NNPDF3.0 family (with $\alpha_s(M_Z) = 0.118$) are associated to the $\tilde{t}$ and $t$ computations respectively.

PDF at NLO, the difference between the 2-loop approximated PDF $\tilde{t}^{(2)}(x, \mu^2)$ and the fully evolved PDF $t(x, \mu^2)$ is very small (of the order of 5%) unless very high scales and large $x$ are involved. A comparable behaviour was observed in [156].

## 6.3. Different heavy quark schemes: numerical results

In this Section, we consider the production of Higgs and neutral vector bosons via $b\bar{b}$ fusion at the LHC and the production of heavy vector bosons in $t\bar{t}$ collisions at a future high energy hadron collider. We compare predictions for total rates obtained at the highest available perturbative order in the 4F and 5F schemes at the LHC and in the 5F and 6F schemes at a future 100 TeV collider.

### 6.3.1. LHC Run II

#### Bottom-fusion initiated Higgs production

Although in the SM the fully-inclusive $b\bar{b} \to H$ cross section is much smaller than the other Higgs production channels (gluon fusion, vector boson fusion, $W$ and $Z$ associated Higgs production) and its rate further decreases when acceptance cuts on the associated $b$ quarks are imposed, this production process can be important in several non-standard scenarios. For example, in supersymmetric models Higgs production in association with $b$ quarks can become a dominant production channel when couplings are enhanced with respect to the Standard Model. More specifically, in models featuring a second Higgs doublet the rate is typically increased by a factor $1/\cos^2\beta$ or $\tan^2\beta$, with $\beta = v_1/v_2$ being the ratio of two Higgs vacuum expectation values.

Calculations for $b$-initiated Higgs productions have been made available by several groups. The total cross section for this process is currently known up to next-to-next-to-leading order (NNLO) in the 5F scheme[157] and up to next-to-leading order (NLO) in the 4F scheme.[158,159] Total cross section predictions have been also obtained via matching procedures that include the resummation of the collinear logarithms on one side and the mass effects on the other, without double counting common terms. A first heuristic proposal, which has been adopted for some time by the HXSWG LHC, is based on the so-called Santander matching[146] where an interpolation between results in the 4F and in the 5F schemes is obtained by means of a weighted average of the two results. Several groups have provided properly matched calculations based on a thorough quantum field theory analysis, at NLO+NLL and beyond via the FONLL method[153] and an effective field theory approach[152,154] that yield very similar results.

Fully differential calculations in the 4F scheme up to NLO(+PS) accuracy have been recently made available[147] in MadGraph5_aMC@NLO[160] and work is in progress in the SHERPA framework.[161] These studies conclude that the 4F scheme results, thanks to the matching to parton showers, are generally more accurate than the pure 5F scheme counterparts, especially for observables which are exclusive in the $b$-quark kinematics. On the other hand, for inclusive observables the differences between 4F and 5F schemes are mild if judicious choices for scales are made. The assessment of the size of such effects and their relevance for phenomenology is the purpose of this section.

We first compare the size and the scale dependence of the 4F and 5F scheme predictions from leading-order up to the highest available perturbative order, namely NLO in the case of the 4F scheme and NNLO in the case of the 5F scheme cross sections. Results are shown in Figures 6.5 and 6.6 for the SM Higgs ($M_H = 125$ GeV) and a heavier Higgs ($M_H = 400$ GeV) respectively. The 4F scheme cross section has been generated using the public version of MadGraph5_aMC@NLO.[160] In the case of the 5F scheme calculation, the cross section has been computed with SusHi[162] and the LO and NLO results have been cross-checked against the output of MadGraph5_aMC@NLO. The input PDFs belong to the NNPDF3.0 family[139] and the $n_f = 4$ set was used in association with the 4F scheme calculation, while the $n_f = 5$ set was associated with the 5F scheme calculation, consistently with the perturbative order of the calculation, and with $\alpha_s^{5\text{F}}(M_Z) = 0.118$. Both the renormalisation and factorisation scales have been taken to be equal to $kM_H$, with $0.15 \leq k \leq 2$.

The treatment of the Higgs Yukawa coupling to $b$ quarks deserves some attention. Different settings may cause large shifts in theoretical predictions. Here we use the $\overline{\text{MS}}$ scheme; the running $b$ Yukawa $y_b(\mu)$ is computed at the scale $\mu_R$ (left plots). We have checked that computing the Yukawa at the fixed value of $M_H$ does not modify our conclusions (right plots). The numerical value of $m_b(\mu_R)$ is obtained from $m_b(m_b)$ by evolving up to $\mu_R$ at 1-loop (LO), 2-loops (NLO) or 3-loops (NNLO) with $n_f = 4$ or $n_f = 5$, depending on the scheme. The numerical value of $m_b(m_b)$ is taken to be equal to the pole mass $m_b^{\text{pole}} = 4.75$ GeV at LO (in both the 4F and 5F schemes), $m_b(m_b) = 4.16$ GeV at NLO in the 5F scheme and $m_b(m_b) = 4.34$ GeV in the 4F scheme (consistently with the settings adopted in [147]) and finally $m_b(m_b) = 4.18$ GeV at NNLO in the 5F scheme, consistently with the latest recommendation of the Higgs cross section working group[1].

The 4F and 5F scheme curves at leading order show an opposite behaviour: in the 4F scheme the scale dependence is driven by the running of $\alpha_s$ and therefore decreases with the scale, while the 5F scheme case it is determined by the scale dependence of the $b$-quark PDF which in turn leads to an increase. The inclusion of higher orders in both calculations drastically reduces the differences; nonetheless, it is clear from Figures. 6.5 and 6.6 that around the central scale $k = 1$ the best 5F scheme prediction exceeds the highest order 4F scheme prediction by a large amount, about 80%. We also observe that 4F and 5F scheme predictions are closer at lower values of the scale. The scale dependence of the 4F scheme NLO calculation is comparable in size to that of the 5F scheme NLO calculation, while it is stronger than the scale dependence of the 5F scheme NNLO calculation— this is to be expected, since in the latter the collinear logarithms are resummed.

[1]The pole mass value that we use in our calculation is slightly different from the latest recommendation $m_b^{\text{pole}} = 4.92$ GeV as well as from the value used in the PDF set adopted in our calculation $m_b^{\text{pole}} = 4.18$ GeV, however our results are not sensitive to these small variations about the current central value.

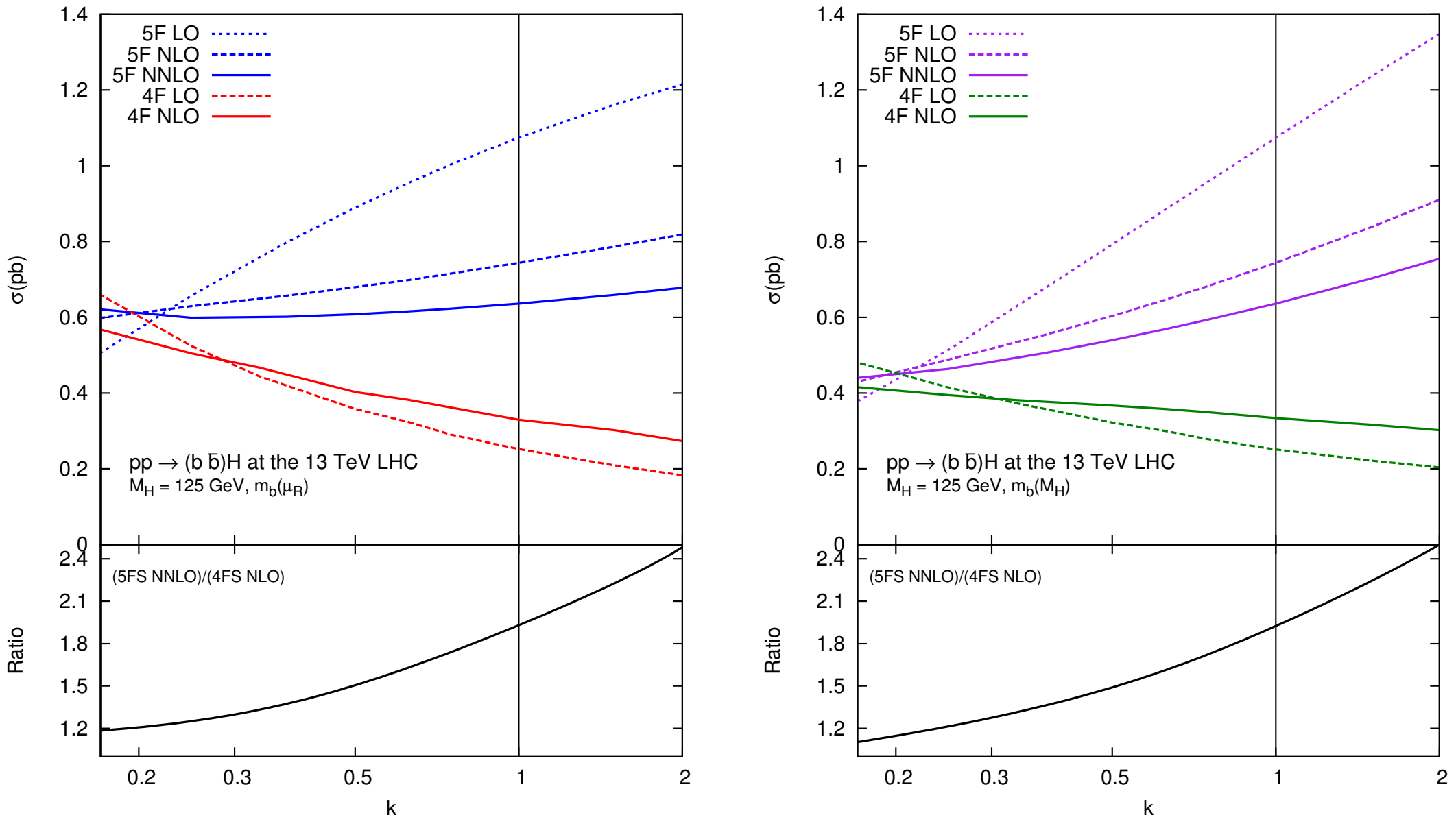

**Figure 6.5.:** Cross sections for the production of the SM Higgs boson via $b\bar{b}$ fusion ($y_b^2$ term only) in the 5F and 4F schemes for LHC 13 TeV as functions of $k = \mu/M_H$, with $\mu_F = \mu_R = \mu$. Terms proportional to $y_b y_t$ in the NLO 4F scheme have been neglected. Results with the running $b$ mass computed at a fixed scale $M_H$ are also shown (right plot). In the inset the ratio between the 5F NNLO prediction and the 4F scheme NLO prediction is displayed.

In Figure 6.6 the same curves are displayed for a heavier Higgs, $M_H = 400$ GeV. As observed in [142], for heavier final state particles differences between schemes are enhanced. In particular, at the central scale the NNLO 5F scheme prediction exceeds the 4F scheme case by a factor of two. Also in this case, at smaller values of the scale the difference is significantly reduced.

This behaviour corresponds to that expected from our analysis presented in Section 6.2. Comparing calculations at $\tilde{\mu}_F = 0.36\, M_H$ for $M_H = 125$ GeV and $\tilde{\mu}_F = 0.29\, M_H$ for $M_H = 400$ GeV, the differences between the predictions in the 4F and 5F scheme reduce to about 30-35%, a difference that can be accounted for by considering first the (positive) effects of resummation included in the 5F scheme calculation with respect to the 4F one and second the power-like quark-mass corrections that are not included the 5F calculation and estimated to be around −2-5%, see [152–154].

The effects of the resummation are easy to quantify by establishing the range of $x$ which gives the dominant contribution to Higgs production via $b\bar{b}$ collisions. To this purpose, we show in Figure 6.7 the $x$ distribution in the leading-order bottom-quark fusion Higgs production in the 5F scheme. We observe that the $x$ distribution has its maximum around $x \approx 10^{-2}$ for the Standard Model Higgs; for such values of $x$, the resummation of collinear logarithms is sizeable: the difference between the fully resummed $b$ PDF and $\tilde{b}^{(2)}$ becomes as large as 10-15% for scales between 100 and 400 GeV. Note that we expect twice the effect of a single $b$ quark in the case of processes with two $b$ quarks in the initial state, which amounts to a difference of 20-25% from resummed logarithms at $\mathcal{O}(\alpha_s^3)$ and higher between the collinear approximation of the 4F scheme calculation and the 5F scheme calculation.

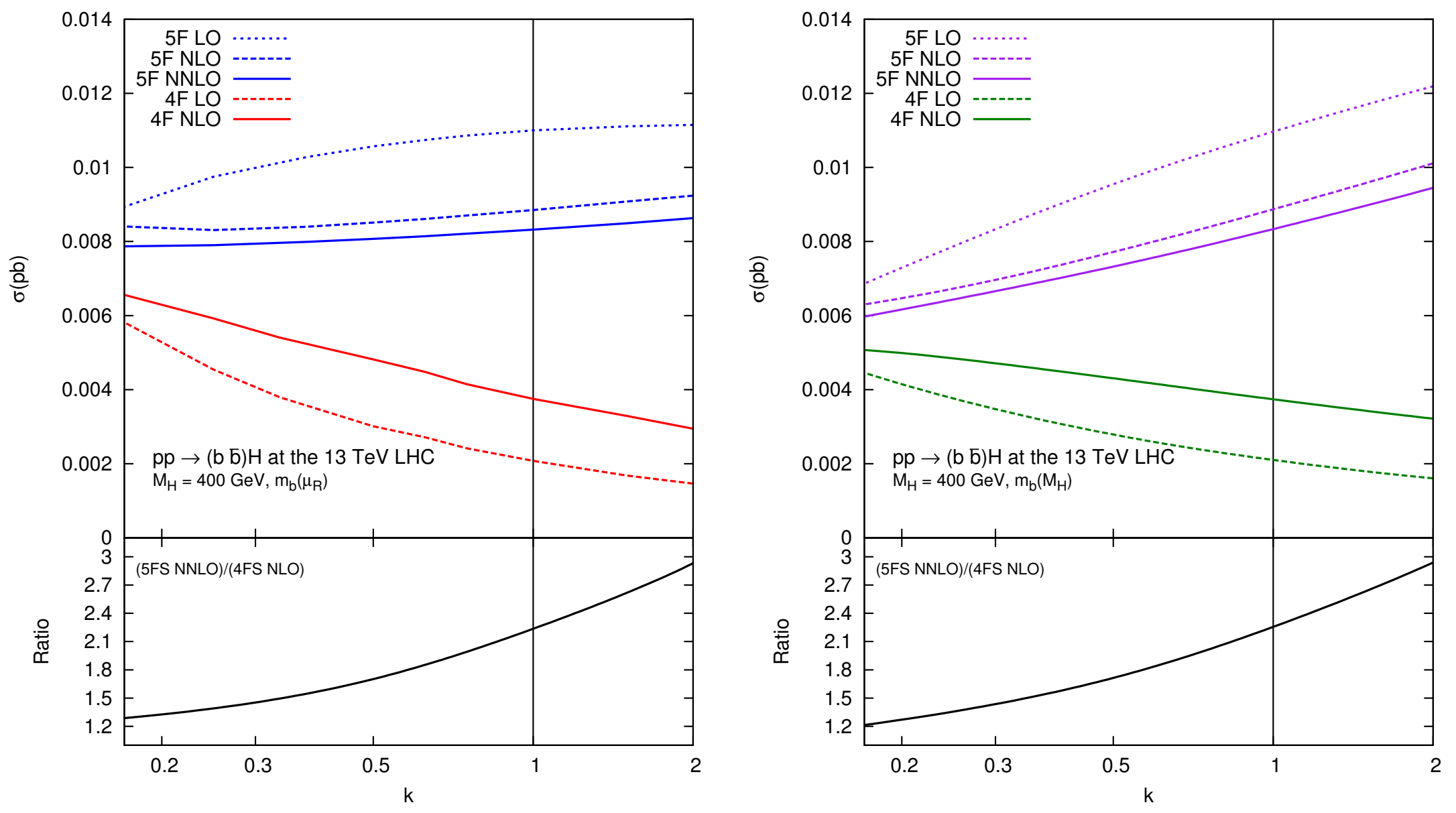

**Figure 6.6.:** Same as Figure 6.5 with $M_H = 400$ GeV

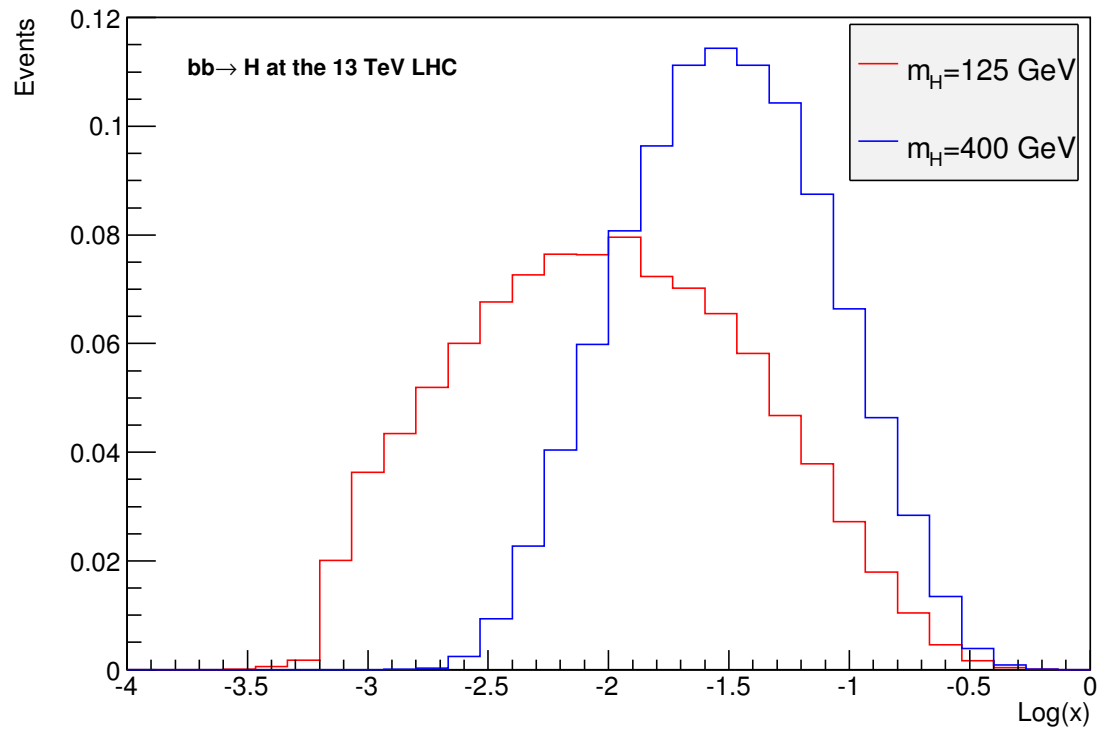

**Figure 6.7.:** Normalised distribution of the momentum fraction $x$ carried by the $b$ quark in $b\bar{b}$ initiated Higgs production, in the 5F scheme at leading order for LHC 13 TeV, for $M_H = 125$ GeV (red curve) and $M_H = 400$ GeV (blue curve).

This expectation is confirmed by the curves in Figure 6.8, where we plot the 5F scheme cross section at LO (left panel) and NLO (right panel) as a function of the Higgs mass in the range 100 GeV to 500 GeV, with $\mu_R = \mu_F = M_H/3$. The cross sections are computed with the same settings as in Figure 6.5. In the same panel we present the cross sections with the $b$ PDF replaced by the $\tilde{b}^{(p)}$ truncated PDF computed at order $p = 1$ and at order $p = 2$, together with the relevant ratios. We observe that, for a sensible value of the factorisation and renormalisation scales, as per the one suggested in this chapter $\tilde{\mu}_F \sim M_H/3$, the effect of neglecting the higher order logs resummed in the $b$ PDF evolution beyond the ones included in the second order expansion of the $b$ PDF, $\tilde{b}^{(2)}$, is smaller than 20% for the SM Higgs mass and of about 30% for a heavier Higgs. Similar conclusions are drawn if the NLO cross section is considered instead, as in the right hand-side panel. If instead we had taken as the central scale choice $\mu_R = \mu_F = M_H$ the effects of the resummation of higher order logs would appear much more significant.

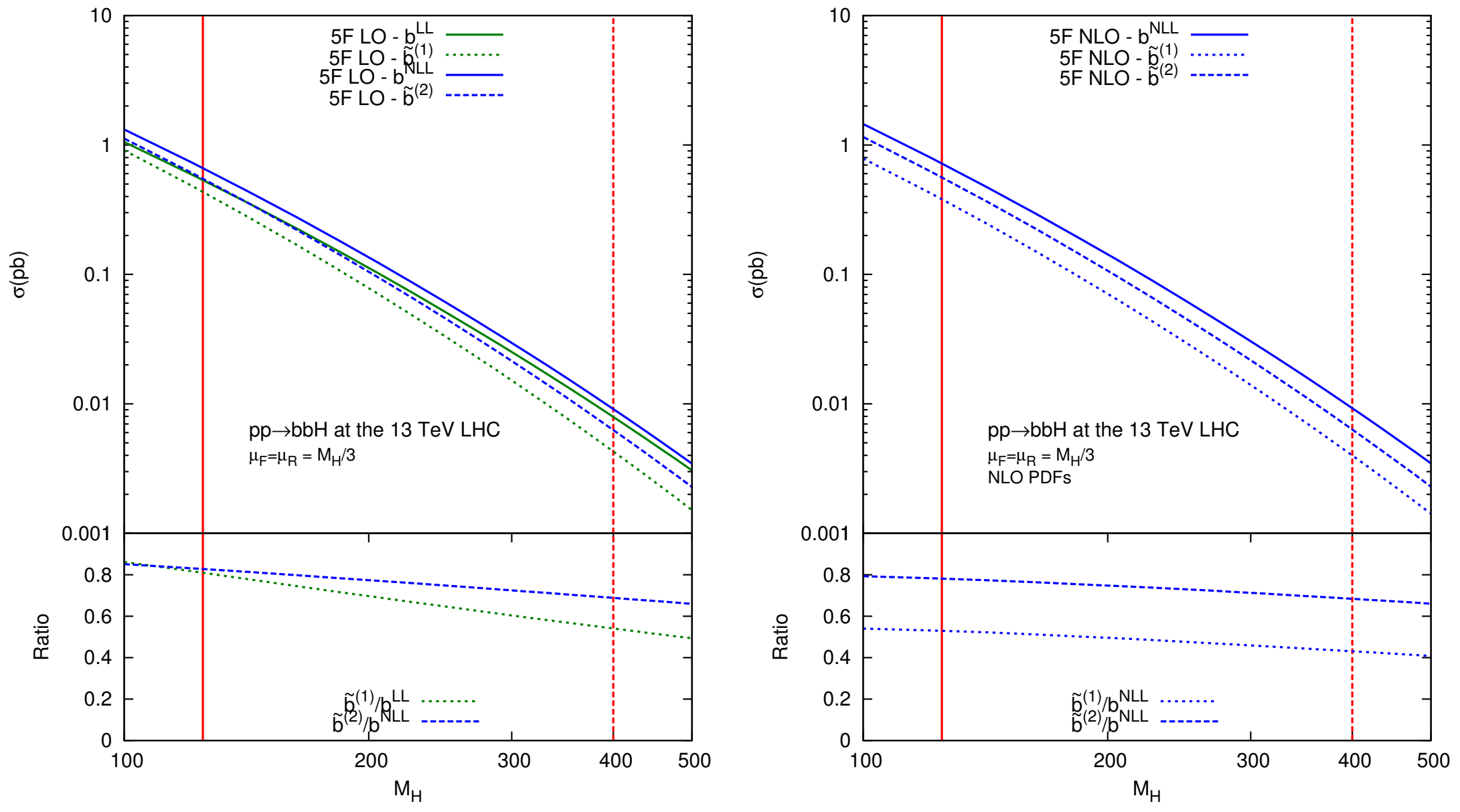

**Figure 6.8.:** Higgs production cross section via $b\bar{b}$ fusion at LO (left) and NLO (right) as a function of $M_H$, computed either with the fully resummed $b$ quark PDF at LL or NLL, or with the truncated PDF $\tilde{b}^{(p)}$ with $p = 1, 2$, with $\mu = \mu_F = \mu_R = M_H/3$.

The scale dependence of the Standard Model Higgs cross section is studied in Figure 6.9. The plots confirm the findings that the assessment of the effect of the higher-order logs resummed in a 5F scheme calculation strongly depends on the scale at which the process is computed and that at a scale close to $\tilde{\mu}_F$ the effects of higher order logs are quite moderate, while they become significant if the naïve hard scale of the process is chosen.

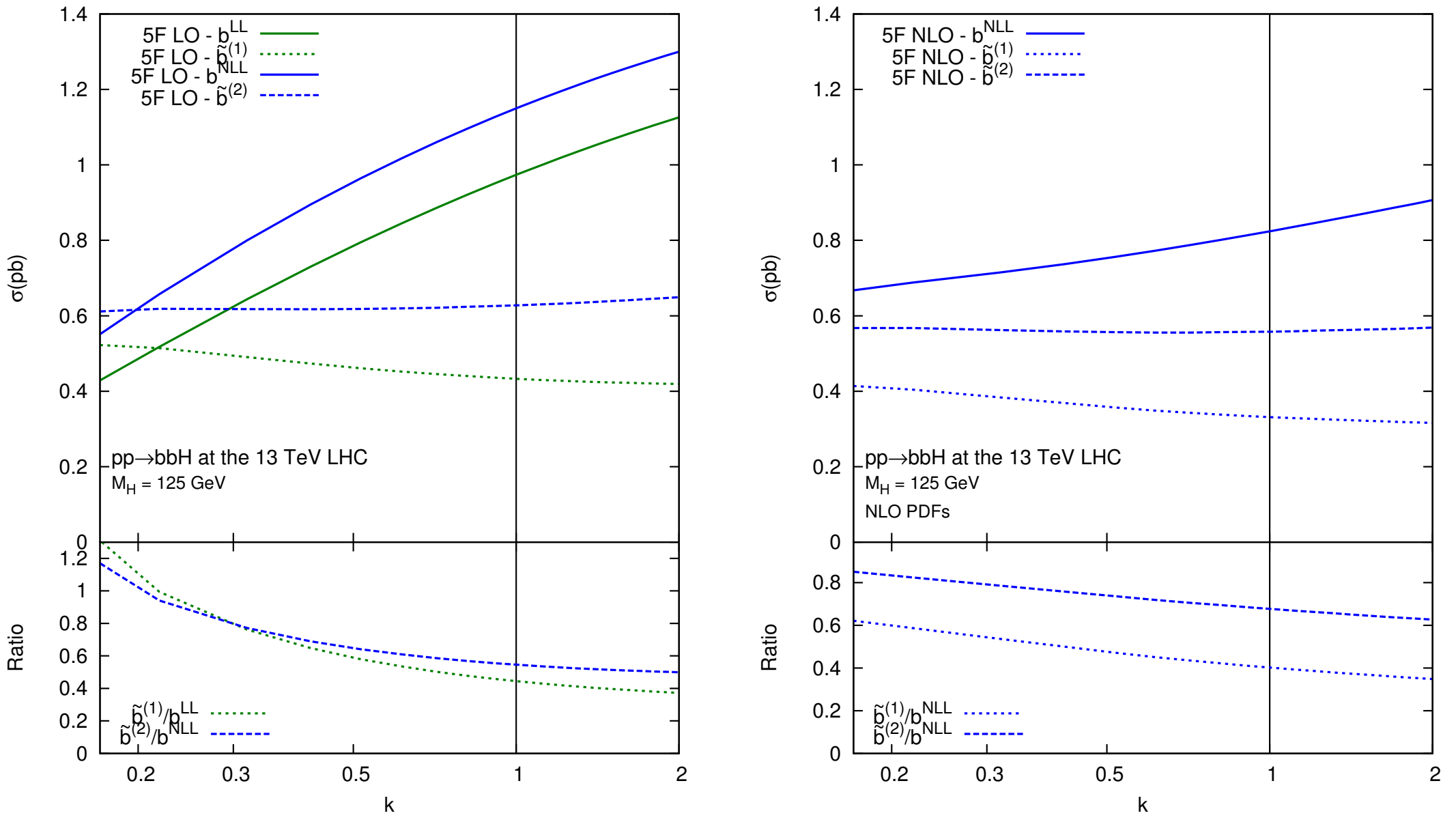

**Figure 6.9.:** Standard Model Higgs production cross section via $b\bar{b}$ fusion at LO (left) and NLO (right) as a function of $k = \mu/M_H$, with $\mu = \mu_R = \mu_F$, computed either with the fully resummed $b$ quark PDF at LL or NLL, or with the truncated PDF $\tilde{b}^{(p)}$ with $p = 1, 2$.

### Bottom-fusion initiated $Z'$ production

A similar analysis can be carried out for the case of $Z$ production. $Z$-boson production in association with one or two $b$-jets has a very rich phenomenology. It is interesting as a testbed of our understanding of QCD and it enters in precision measurements (Drell-Yan at the LHC or indirectly in the $W$ mass determination). In addition, it represents a crucial irreducible background for several Higgs production channels at the LHC. For the SM Higgs boson, $Zb\bar{b}$ production is a background to $ZH$ associated production followed by the decay of the Higgs into a bottom-quark pair. Finally, this process is a background to searches for Higgs bosons with enhanced $Hb\bar{b}$ Yukawa coupling.

Calculations for bottom-initiated $Z$ production have been made available by several groups. The $Zb\bar{b}$ production cross section was originally computed (neglecting the $b$ quark mass) in [148] for exclusive 2-jet final states. The effect of a non-zero $b$ quark mass was considered in later works[149,150] where the total cross section was also given. More recently, in [151] leptonic decays of the $Z$ boson have taken into account, together with the full correlation of the final state leptons and the parton shower and hadronisation effects. The total cross section for $Zb\bar{b}$ in the 5F scheme has been computed at NNLO accuracy for the first time in [163].

Bottom-initiated $Z$ production is in principle very different from Higgs production because the $Z$ boson has a non-negligible coupling to the light quarks. For simplicity, we will not take these couplings into account; to avoid confusion, we refer to the $Z$ boson that couples only with heavy quarks as $Z'$, even when we take its mass to be equal to 91.2 GeV as in the Standard Model.

We have calculated the 5F scheme cross sections by using a private code,[163] which has been cross-checked at LO and NLO against MadGraph5_aMC@NLO. The 4F scheme cross section has been computed with MadGraph5_aMC@NLO. Our settings are the same as in the Higgs production computation. We take the same value $\mu$ for the factorisation and renormalisation scales.

Results are presented in Figure 6.10 as functions of $k = \mu/M_{Z'}$ for $M_{Z'} = 91.2$ GeV and $M_{Z'} = 400$ GeV respectively. We observe that for $\mu = M_{Z'}$ the best 5F scheme prediction exceeds the 4F scheme prediction by almost 70%, while their difference is reduced at lower values of the scales. In this respect the behaviour of the 4F vs 5F scheme predictions reflects what we have already observed in Figure 6.5. We note, however, that the scale dependence of the 5F scheme predictions for $Zb\bar{b}$ is quite different with respect to the $Hb\bar{b}$ when $m_H = 125$ GeV. In the case of $Zb\bar{b}$ it is quite mild already at NLO and the perturbative expansion seems to converge more quickly for higher values of $\mu$ around $\mu = M_{Z'}$. The behaviour of the 5F calculations for $M_H = M_{Z'} = 400$ GeV cases, on the other hand, do not show any significant qualitative difference, apart from the fact that $Zb\bar{b}$ results have in general a milder scale dependence. The different scale sensitivity (with $\mu_R = \mu_F$) of the two processes can be traced back to the fact that while the Yukawa interaction renormalises under QCD, the EW current (and corresponding charge) is conserved, resulting in general in a milder scale dependence of the $Zb\bar{b}$ predictions.

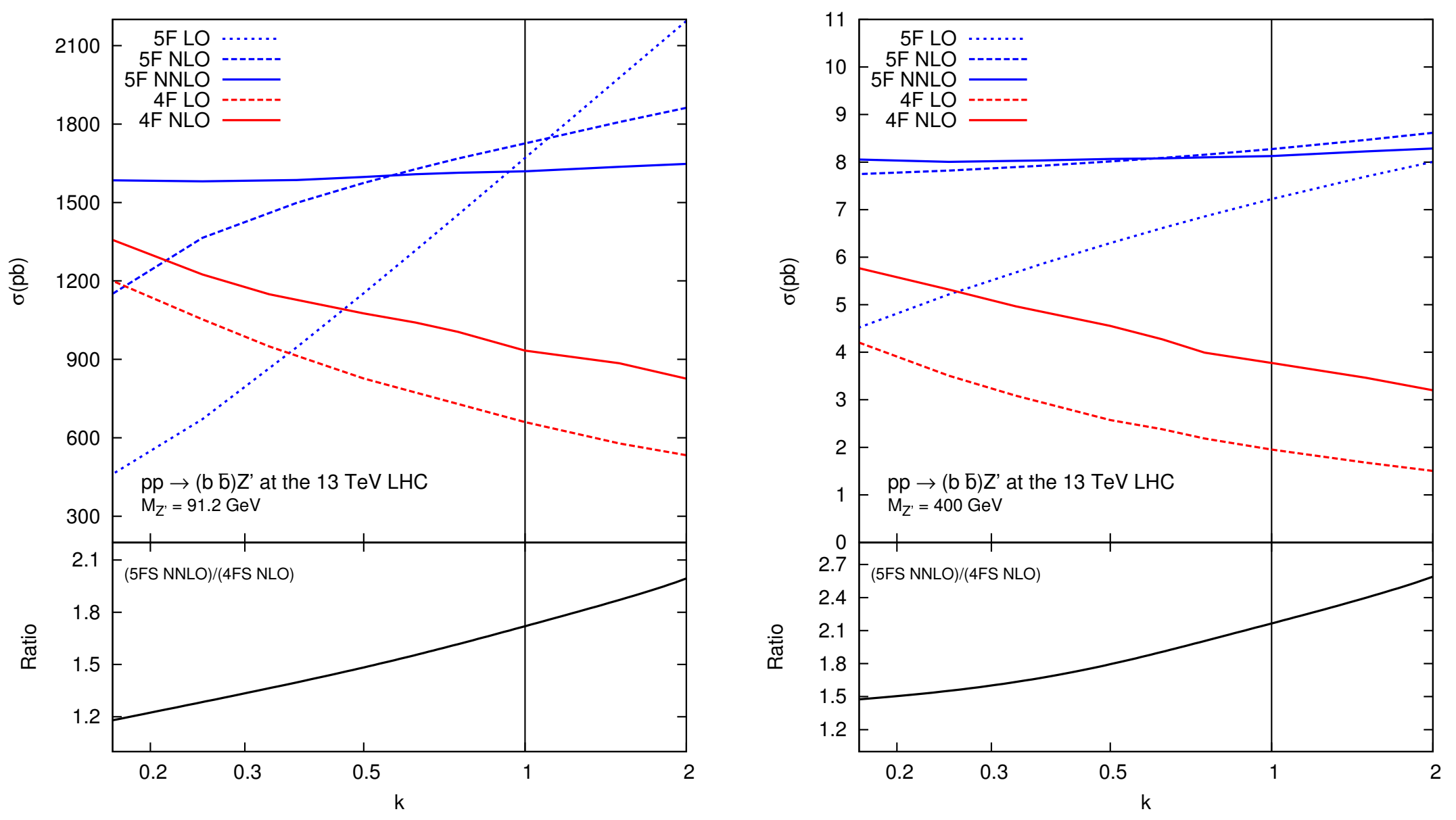

**Figure 6.10.:** Cross sections for bottom-fusion initiated $Z'$ boson production in the 5F and 4F schemes for LHC 13 TeV as functions of $k = \mu/M_{Z'}$. $M_{Z'} = 91.2$ GeV (left) and $M_{Z'} = 400$ GeV (right). Settings are specified in the text.

## 6.3.2. Future Colliders

The perspective of a proton-proton collider at a centre–of–mass energy of 100 TeV would open up a new territory beyond the reach of the LHC. New heavy particles associated with a New Physics sector may be discovered and new interactions unveiled. At such large energies, essentially all SM particles can be considered as massless, including the top quarks. We therefore expect collinear enhancements in top-quark initiated processes. In [156] the question of whether the top quark should be treated as an ordinary parton at high centre-of-mass energy, thereby defining a 6FNS, is scrutinised, and the impact of resumming collinear logs of the top quark mass is assessed. This analysis is performed in the context of charged Higgs boson production at 100 TeV. In [164], the impact of resumming initial-state collinear logarithms in the associated heavy Higgs ($M_H > 5$ TeV) and top pair production (with un-tagged top quarks) is examined and it is found to be very large at large Higgs masses.

In Figure 6.11 the total cross sections for the production of a $Z'$ boson of mass $M_{Z'} = 1$ TeV (left), $M_{Z'} = 5$ TeV (centre), $M_{Z'} = 10$ TeV (right) are plotted in the 5F and 6F schemes as a function of the renormalisation and factorisation scales, which are identified and varied between $0.2M_{Z'}$ and $2M_{Z'}$. Results are obtained by using MadGraph5_aMC@NLO for the 5F scheme and a private code for the 6F scheme. Results in the 6F scheme have been cross- checked up to NLO against MadGraph5_aMC@NLO. We have set $m_t^{\text{pole}} = 172.5$ GeV and turned off the coupling of the $Z'$ heavy boson to all lighter quarks. Firstly, we observe that the $M_{Z'} = 1$ TeV case is quite different from the $M_{Z'} = 5$ TeV and $M_{Z'} = 10$ TeV, which in turn display a very similar pattern to the $b$ initiated processes with similar $m_Q/M_{Z'}$ and $M_{Z'}/\sqrt{s}$ ratios. The behaviour of the leading-order cross section in the 6F scheme for $M_{Z'} = 1$ TeV is mitigated at higher masses and at higher orders (NLO). At NNLO the 6F-scheme cross section displays a similar scale dependence as the NLO cross section in the 5F scheme with a residual difference of about 40% between the two best predictions in the two schemes. To further investigate these differences, in Figure 6.12 we plot the distribution of the fraction of momentum carried by the top quarks for $M_{Z'} = 1$ TeV and $M_{Z'} = 5$ TeV in the 6F schemes. As expected, compared to heavier masses, the production of a $M_{Z'} = 1$ TeV is dominated by smaller values of Bjorken $x$. The ratio $M_{Z'}/m_t \simeq 6$ is not very large to start with (for comparison $M_Z/m_b \simeq 20$) and initial-state quark collinear configurations are not dominant. We conclude that in the $M_{Z'} = 1$ TeV case the differences between the two schemes are to be associated to the absence of power-like mass terms in the 6F calculation.

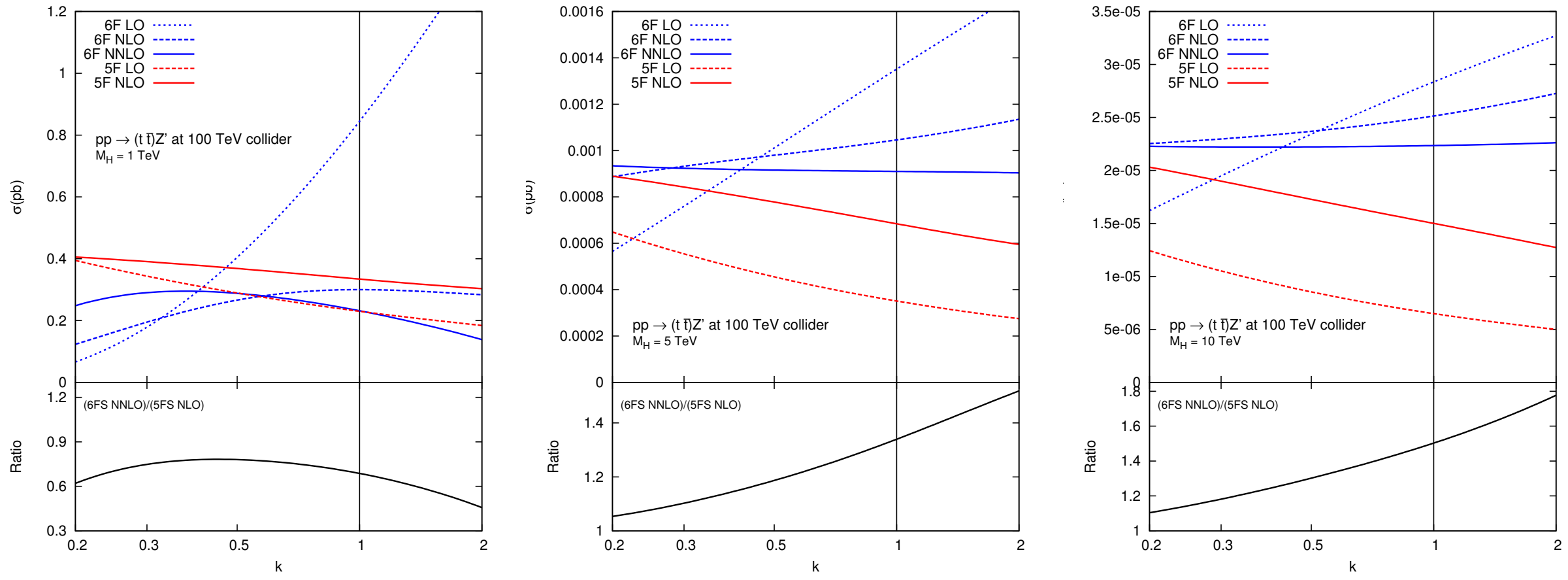

**Figure 6.11.:** Cross sections for $t\bar{t}$ initiated $Z'$ production in the 6F and 5F schemes at a 100 TeV $pp$ collider as functions of $k = \mu/M_{Z'}$. Top mass: $m_t = 173$ GeV. Mass of the heavy boson: $M_{Z'} = 1$ TeV (left), $M_{Z'} = 5$ TeV (centre), $M_{Z'} = 10$ TeV (right). The inlay below shows the ratio of the cross sections in the 6F and 5F schemes.

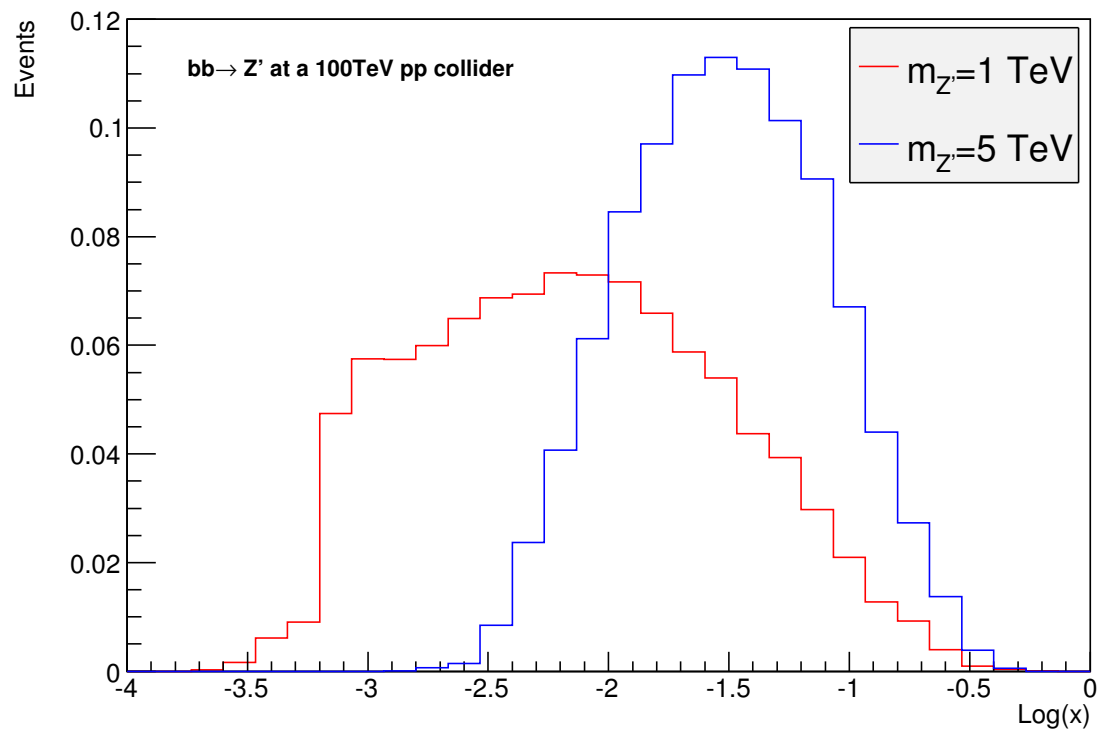

**Figure 6.12.:** Normalised distribution of momentum fraction $x$ carried by the $t\bar{t}$ initiated $Z'$ production in the 6F scheme distributions at LO in a 5F scheme for $M_{Z'} = 1$ TeV and $M_{Z'} = 5$ TeV at a 100 TeV collider. Events were generated at values of the scales $\mu_R = \mu_F = H_T/4$. Input PDF: NNPDF30 LO $n_f = 5$ ($\alpha_s(M_Z) = 0.130$).

## 6.4. Conclusions

In this work we have considered the use of four- and five-flavour schemes in precision physics at the LHC and in the context of $b$-initiated Higgs and $Z$ production. We have extended previous work done for processes involving a single $b$ quark in the initial state to cases in which two are present. We have followed a "deconstructing" methodology where the impacts of the various sources of differences between the schemes have been evaluated one by one.

Firstly, we have obtained the form of the collinear logarithms in the four-flavour scheme by performing the explicit computation of the $2 \to 3$ body scattering process and studying the collinear limit using as natural variables the $t$-channel invariants. We have then compared the resulting expression with the corresponding cross section in the 5-flavor scheme as calculated by only keeping the explicit log in the $b$-quark PDF, i.e. without resummation. This has allowed us to assess the analytic form and therefore the size of the collinear logarithms and to propose a simple procedure to identify the relevant scales in the processes where the results in the two schemes should be evaluated and compared. In so doing we have considered cases where power-like effects in the mass of the heavy quarks were assumed (and then checked a posteriori by comparing to the full result) unimportant. Secondly, we have explicitly estimated the effects of the resummation by studying fully evolved $b$ PDF with truncated expansions at finite order.

We have then applied our general approach to the case of Higgs and $Z$ boson production in association with $b$ quarks at the LHC and to heavy $Z'$ production in association with top quarks at a future 100 TeV collider. We have found that the resummation increases the cross section in most cases by about 20% (sometimes reaching as high as 70%) at the LHC and in general leads to a better precision and better convergence properties. On the other hand, for some processes the 4F scheme predictions (5F scheme in the case of associated top-quark production) at NLO also display a consistent perturbative behaviour when evaluated at suitable scales. They should therefore should be used in cases where the heavy-quark mass effects are not negligible and to predict distributions and more exclusive observables involving the heavy quarks in the final state.

# Chapter 7.

# Concluding remarks

*"But where shall wisdom be found?*
*and where is the place of understanding?"*
— Job 28:12

## 7.1. Summary

In this thesis we have discussed higher order corrections in perturbative QCD and their relevance to ongoing experiments at the LHC. In the first part of the work, we examined the computation of loop amplitudes and the integration-by-parts identities. To demonstrate the techniques used, we took the case of the 1-loop, five-point amplitude and presented fully analytic results for the finite remainder of the $q\bar{q} \to Q\bar{Q}g$ process. We then moved to the 2-loop case and commented that the limiting factor in their computation was finding the solution of the IBP equations. We presented a new approach to deriving the solution which has several advantages over traditional methods and has allowed us to reduce completely the planar 2-loop, five-point topologies to a basis of master integrals. We used these results to express the planar part of the $q\bar{q} \to Q\bar{Q}g$ 2-loop amplitude in terms of masters.

Next, we considered the important processes of top quark production and decay which are of great phenomenological relevance at the LHC. We introduced the helicity formalism for amplitudes and in said formalism derived analytic expressions for the QCD real-virtual corrections to the heavy-light quark vertex. We used this knowledge of the vertex correction to obtain the amplitudes for the top quark decay and single top production processes and extract their finite remainders. In the future, the contribution to the decay will be included in a full NNLO calculation of top quark pair production and decay.

In the second part of the work, we examined how knowledge of higher order corrections is applied when comparing theory with experiment and looked in particular at processes involving heavy quarks. For the first time, we were able to use the NNLO corrections to top pair production for different values of the top mass along with ATLAS and CMS data to perform a simultaneous extraction of $m_t$ and $\alpha_s$. These are two extremely important parameters in the Standard Model which are among the least well constrained. Using 8 TeV measurements from ATLAS and CMS, we found that differential top data adds considerable constraining power to the parameter space beyond that which is possible using the total $t\bar{t}$ cross section alone. We considered four different spectra ($p_T^t$, $M_{t\bar{t}}$, $y_t$ and $y_{t\bar{t}}$) from the two experiments both separately and in combination. Based on a goodness-of-fit criterion and having combined distributions from the two experiments, we obtained the result $\alpha_s(M_Z) = 0.1164^{+0.0054}_{-0.0058}$ and $m_t = 173.3^{+0.9}_{-0.8}$. This was compatible with earlier extractions using top data and also with the PDG world averages to within error, although the quality of the experimental data means that such results should be treated with some caution.

Finally, we considered processes with two heavy quarks in the initial state and how they may be described using 4- and 5-flavour schemes. We were able to show that potentially large logarithms in the 4F scheme which might spoil convergence are, for most cases, actually modest in size for scales of interest. We estimated the effects of resummation of these logarithms into a heavy quark PDF in the 5F scheme. We then took the particular case of the production of a Higgs boson in association with two bottom quarks and studied its phenomenology at the LHC and a future collider. Motivated by arguments of compatibility between the schemes, we suggested renormalisation and factorisation scales at which calculations should be performed which are of order $M_H/3$.

## 7.2. Outlook

There is much work to be done on the subject of NNLO corrections. The lack of any complete 2-loop, five-point amplitude is perhaps one of the most important problems in the near future and the result will be highly relevant to experiment. To achieve this, the IBP equations for the non-planar topologies need to be solved for all scalar integrals appearing in squared amplitudes—we believe that, using the technology we have developed, this will be possible soon. Though this may be the main obstacle to a result that can be used for phenomenology, it is not the only one. The master integrals for the non-planar technologies are at present unknown and must be calculated, something that may be achieved with the help of IBP solutions through the method of differential equations. A suitable numerical implementation of the full amplitude must also be designed. Given the size of the analytic expressions involved and the complexity of the numerical evaluation of the masters, perhaps the most sensible approach to this problem would be the construction of a grid of points from which intermediate values can be interpolated. We leave this issue to future work.

The copious production of top quarks at the LHC has resulted in it being dubbed a 'top factory'. Knowledge of top quark production and decay is therefore vital to the experiments. The real-virtual contributions to the decay presented in this thesis are a necessary ingredient in a full NNLO calculation currently being implemented in the program `stripper`[134]. The final result, retaining full spin information, will be the first calculation of its kind and prove vital to many studies. An example of such an application might be an extraction of $\alpha_s$ and $m_t$ similar to the approach taken in Chapter 5, but using fully differential measurements of the leptons produced rather than reconstructed stable tops as done here. This may result in a more reliable determination of the parameters and sidestep current gaps in our understanding of the experimental uncertainties associated with the top reconstruction.

The High-Luminosity upgrade of the LHC promises unprecedented statistics, with an anticipated 3 ab$^{-1}$ of delivered integrated luminosity by the end of its lifetime. This has the potential to uncover BSM physics in scenarios where the relative statistical uncertainty in measurement can be very small, either by direct observation of new final states or detection of deviations from SM predictions in known final states. It will also allow the measurement of fundamental SM parameters with increasing precision, including the strong coupling and the top quark mass, considered in this thesis. The further future also beckons. Future hadron colliders promise higher energies of order 100 TeV and the opportunity to access heavy exotics, while lepton collider proposals such as the ILC would provide cleaner environments for precision measurements. It is clear from the discussion in this thesis, however, that in order to profit from such experimental advances we must advance progress in higher order calculations to match the level of experimental sophistication. We await with anticipation developments to come which may unveil more of Nature's mysteries.

# Appendix A.

# Colour matrices

## A.1. Useful identities

Here we present further identities involving the colour matrices $T^a$ and the structure constants $f^{abc}$ which can prove useful when evaluating Feynman diagrams.

$$f^{abc} = -2i\mathrm{Tr}(T^a[T^b, T^c]) \tag{A.1}$$

$$d^{abc} = 2\mathrm{Tr}(T^aT^b, T^c) \tag{A.2}$$

$$f^{acd}d^{bcd} = 0 \tag{A.3}$$

$$f^{ade}f^{efc}f^{dbf} = -\frac{N}{2}f^{abc} \tag{A.4}$$

$$T^aT^b = \frac{1}{2}\left(\frac{1}{N}\delta_{ab} + (d^{abc} + if^{abc})T^c\right) \tag{A.5}$$

$$T^b_{kj}T^b_{il}T^a_{lk} = -\frac{1}{2N}T^a_{ij} \tag{A.6}$$

$$\mathrm{Tr}(T^aT^bT^c) = \frac{1}{4}(d^{abc} + if^{abc}) \tag{A.7}$$

$$\mathrm{Tr}(T^aT^bT^aT^c) = -\frac{1}{4N}\delta^{bc} \tag{A.8}$$

## A.2. Proof of the Fierz identity

The $SU(N)$ generators $T^a$ form the complete set of traceless Hermitian $N$x$N$ matrices, i.e. we can write any such matrix $U$ as

$$U = c_1\mathbb{1} + c_aT^a \tag{A.9}$$

for arbitrary coefficients $c_1, c_a$. We can extract these coefficients:

$$c_1 = \frac{1}{N}\mathrm{Tr}(U),\ c_a = 2\mathrm{Tr}(T^aU) \tag{A.10}$$

and so write

$$U_{ij} = \frac{1}{N}\mathrm{Tr}(U)\delta_{ij} + 2T^a_{kl}U_{lk}T^a_{ij} \tag{A.11}$$

from which we see

$$U_{lk}\left(\frac{1}{N}\delta_{kl}\delta_{ij} + 2T^a_{ij}T^a_{kl} - \delta_{il}\delta_{jk}\right) = 0. \tag{A.12}$$

Thus we have the Fierz identity

$$T^a_{ij}T^a_{kl} = \frac{1}{2}\left(\delta_{il}\delta_{jk} - \frac{1}{N}\delta_{ij}\delta_{kl}\right) \tag{A.13}$$

# Appendix B.

# Master integrals for $q\bar{q} \to Q\bar{Q}g$ at 2-loop order

Using the definitions of the topologies given in Chapter 3, we find 113 masters in $B_1$, 75 in $B_2$, 62 in $C_1$, 28 in $C_2$ and 10 in $C_3$. We list these below.

$$
\begin{aligned}
B_1 : \{ & B(0,1,1,0,0,1,0,0,0,0,0), B(0,1,0,1,0,1,0,0,0,0,0), B(0,0,0,1,1,1,0,0,0,0,0),\\
& B(1,1,0,0,0,0,1,0,0,0,0), B(0,1,1,0,0,0,1,0,0,0,0), B(0,0,1,0,1,0,1,0,0,0,0),\\
& B(0,0,0,1,1,0,1,0,0,0,0), B(1,0,0,0,0,1,0,1,0,0,0), B(0,0,1,0,0,1,0,1,0,0,0),\\
& B(1,0,0,0,0,0,1,1,0,0,0), B(1,1,0,1,0,1,0,0,0,0,0), B(1,1,0,1,0,0,1,0,0,0,0),\\
& B(1,0,0,1,1,0,1,0,0,0,0), B(1,0,0,1,1,0,0,1,0,0,0), B(1,0,0,1,0,1,0,1,0,0,0),\\
& B(0,0,1,0,1,1,0,1,0,0,0), B(0,0,0,1,1,1,0,1,0,0,0), B(1,0,0,0,1,0,1,1,0,0,0),\\
& B(0,0,1,0,1,0,1,1,0,0,0), B(1,1,1,1,0,1,0,0,0,0,0), B(0,1,1,1,1,1,0,0,0,0,0),\\
& B(1,1,1,1,0,0,1,0,0,0,0), B(1,1,1,0,1,0,1,0,0,0,0), B(1,1,0,1,1,0,1,0,0,0,0),\\
& B(1,1,-1,1,1,0,1,0,0,0,0), B(1,0,1,1,1,0,1,0,0,0,0), B(0,1,1,1,1,0,1,0,0,0,0),\\
& B(1,1,1,0,0,1,1,0,0,0,0), B(1,1,0,1,0,1,1,0,0,0,0), B(1,1,-1,1,0,1,1,0,0,0,0),\\
& B(0,1,1,1,0,1,1,0,0,0,0), B(0,1,1,0,1,1,1,0,0,0,0), B(0,1,0,1,1,1,1,0,0,0,0),\\
& B(0,0,1,1,1,1,1,0,0,0,0), B(1,1,1,0,0,1,0,1,0,0,0), B(1,1,0,1,0,1,0,1,0,0,0),\\
& B(1,1,-1,1,0,1,0,1,0,0,0), B(1,0,1,1,0,1,0,1,0,0,0), B(0,1,1,1,0,1,0,1,0,0,0),\\
& B(0,1,1,0,1,1,0,1,0,0,0), B(1,0,0,1,1,1,0,1,0,0,0), B(0,1,0,1,1,1,0,1,0,0,0),\\
& B(0,0,1,1,1,1,0,1,0,0,0), B(-1,0,1,1,1,1,0,1,0,0,0), B(1,1,1,0,0,0,1,1,0,0,0),\\
& B(1,1,0,0,1,0,1,1,0,0,0), B(1,0,1,0,1,0,1,1,0,0,0), B(1,-1,1,0,1,0,1,1,0,0,0),\\
& B(0,1,1,0,1,0,1,1,0,0,0), B(1,0,0,1,1,0,1,1,0,0,0), B(1,1,0,0,0,1,1,1,0,0,0),
\end{aligned}
$$

$$B(1,0,1,0,0,1,1,1,0,0,0), B(0,1,1,0,0,1,1,1,0,0,0), B(0,0,1,0,1,1,1,1,0,0,0),$$
$$B(-1,0,1,0,1,1,1,1,0,0,0), B(1,1,1,1,1,0,1,0,0,0,0), B(1,1,1,1,1,-1,1,0,0,0,0),$$
$$B(1,1,1,1,0,1,1,0,0,0,0), B(1,1,1,1,-1,1,1,0,0,0,0), B(1,1,0,1,1,1,1,0,0,0,0),$$
$$B(1,1,-1,1,1,1,1,0,0,0,0), B(0,1,1,1,1,1,1,0,0,0,0), B(1,1,1,1,0,1,0,1,0,0,0),$$
$$B(1,1,1,1,-1,1,0,1,0,0,0), B(1,1,0,1,1,1,0,1,0,0,0), B(1,0,1,1,1,1,0,1,0,0,0),$$
$$B(0,1,1,1,1,1,0,1,0,0,0), B(-1,1,1,1,1,1,0,1,0,0,0), B(1,1,1,0,1,0,1,1,0,0,0),$$
$$B(1,1,1,-1,1,0,1,1,0,0,0), B(1,1,0,1,1,0,1,1,0,0,0), B(1,0,1,1,1,0,1,1,0,0,0),$$
$$B(1,1,1,0,0,1,1,1,0,0,0), B(1,1,0,1,0,1,1,1,0,0,0), B(1,1,-1,1,0,1,1,1,0,0,0),$$
$$B(1,0,1,0,1,1,1,1,0,0,0), B(1,-1,1,0,1,1,1,1,0,0,0), B(0,1,1,0,1,1,1,1,0,0,0),$$
$$B(-1,1,1,0,1,1,1,1,0,0,0), B(1,0,0,1,1,1,1,1,0,0,0), B(0,0,1,1,1,1,1,1,0,0,0),$$
$$B(-1,0,1,1,1,1,1,1,0,0,0), B(1,1,1,1,1,1,1,0,0,0,0), B(1,1,1,1,1,1,1,-1,0,0,0),$$
$$B(1,1,1,1,1,1,1,0,-1,0,0), B(1,1,1,1,1,1,0,1,0,0,0), B(1,1,1,1,1,1,-1,1,0,0,0),$$
$$B(1,1,1,1,1,1,0,1,-1,0,0), B(1,1,1,1,1,0,1,1,0,0,0), B(1,1,1,1,1,-1,1,1,0,0,0),$$
$$B(1,1,1,1,1,0,1,1,-1,0,0), B(1,1,1,1,0,1,1,1,0,0,0), B(1,1,1,1,-1,1,1,1,0,0,0),$$
$$B(1,1,1,1,0,1,1,1,-1,0,0), B(1,1,1,0,1,1,1,1,0,0,0), B(1,1,1,-1,1,1,1,1,0,0,0),$$
$$B(1,1,1,0,1,1,1,1,-1,0,0), B(1,1,0,1,1,1,1,1,0,0,0), B(1,1,-1,1,1,1,1,1,0,0,0),$$
$$B(1,0,1,1,1,1,1,1,0,0,0), B(1,-1,1,1,1,1,1,1,0,0,0), B(0,1,1,1,1,1,1,1,0,0,0),$$
$$B(-1,1,1,1,1,1,1,1,0,0,0), B(0,1,1,1,1,1,1,1,-1,0,0), B(1,1,1,1,1,1,1,1,0,0,0),$$
$$B(1,1,1,1,1,1,1,1,-1,0,0), B(1,1,1,1,1,1,1,1,0,-1,0), B(1,1,1,1,1,1,1,1,0,0,-1),$$
$$B(1,1,1,1,1,1,1,1,-2,0,0), B(1,1,1,1,1,1,1,1,-1,-1,0), B(1,1,1,1,1,1,1,1,-1,0,-1),$$
$$B(1,1,1,1,1,1,1,1,0,-2,0), B(1,1,1,1,1,1,1,1,0,-1,-1)\}$$

$$B_2 : \{B(0,1,1,0,0,1,0,0,0,0,0), B(0,1,0,1,0,1,0,0,0,0,0), B(1,1,0,0,0,0,1,0,0,0,0),$$
$$B(0,1,1,0,0,0,1,0,0,0,0), B(1,0,0,0,0,1,0,1,0,0,0), B(0,0,1,0,0,1,0,1,0,0,0),$$
$$B(1,0,0,0,0,0,1,1,0,0,0), B(0,0,0,0,0,0,1,1,0,0,1), B(1,1,0,1,0,1,0,0,0,0,0),$$
$$B(1,1,0,1,0,0,1,0,0,0,0), B(1,0,0,1,0,1,0,1,0,0,0), B(0,1,1,0,0,0,1,0,0,0,1),$$
$$B(0,1,0,1,0,0,1,0,0,0,1), B(0,0,1,0,0,1,0,1,0,0,1), B(0,0,0,1,0,1,0,1,0,0,1),$$
$$B(0,0,1,0,0,0,1,1,0,0,1), B(1,1,1,1,0,1,0,0,0,0,0), B(1,1,1,1,0,0,1,0,0,0,0),$$
$$B(1,1,1,0,0,1,1,0,0,0,0), B(1,1,0,1,0,1,1,0,0,0,0), B(1,1,-1,1,0,1,1,0,0,0,0),$$
$$B(0,1,1,1,0,1,1,0,0,0,0), B(1,1,1,0,0,1,0,1,0,0,0), B(1,1,0,1,0,1,0,1,0,0,0),$$
$$B(1,1,-1,1,0,1,0,1,0,0,0), B(1,0,1,1,0,1,0,1,0,0,0), B(0,1,1,1,0,1,0,1,0,0,0),$$

$$
\begin{aligned}
&B(1,1,1,0,0,0,1,1,0,0,0), B(1,1,0,0,0,1,1,1,0,0,0), B(1,0,1,0,0,1,1,1,0,0,0),\\
&B(0,1,1,0,0,1,1,1,0,0,0), B(1,1,1,0,0,0,1,0,0,0,1), B(1,1,0,1,0,0,1,0,0,0,1),\\
&B(0,1,1,1,0,0,1,0,0,0,1), B(1,0,1,0,0,1,0,1,0,0,1), B(1,0,0,1,0,1,0,1,0,0,1),\\
&B(0,0,1,1,0,1,0,1,0,0,1), B(1,1,0,0,0,0,1,1,0,0,1), B(1,0,1,0,0,0,1,1,0,0,1),\\
&B(0,1,1,0,0,0,1,1,0,0,1), B(-1,1,1,0,0,0,1,1,0,0,1), B(1,0,0,0,0,1,1,1,0,0,1),\\
&B(0,0,1,0,0,1,1,1,0,0,1), B(-1,0,1,0,0,1,1,1,0,0,1), B(1,1,1,1,0,1,1,0,0,0,0),\\
&B(1,1,1,1,-1,1,1,0,0,0,0), B(1,1,1,1,0,1,0,1,0,0,0), B(1,1,1,1,-1,1,0,1,0,0,0),\\
&B(1,1,1,0,0,1,1,1,0,0,0), B(1,1,0,1,0,1,1,1,0,0,0), B(1,1,-1,1,0,1,1,1,0,0,0),\\
&B(1,1,1,1,0,0,1,0,0,0,1), B(1,1,1,1,0,-1,1,0,0,0,1), B(1,0,1,1,0,1,0,1,0,0,1),\\
&B(1,-1,1,1,0,1,0,1,0,0,1), B(1,1,1,0,0,0,1,1,0,0,1), B(1,1,1,-1,0,0,1,1,0,0,1),\\
&B(1,0,1,0,0,1,1,1,0,0,1), B(1,-1,1,0,0,1,1,1,0,0,1), B(0,1,1,0,0,1,1,1,0,0,1),\\
&B(-1,1,1,0,0,1,1,1,0,0,1), B(0,1,0,1,0,1,1,1,0,0,1), B(1,1,1,1,0,1,1,1,0,0,0),\\
&B(1,1,1,1,-1,1,1,1,0,0,0), B(1,1,1,1,0,1,1,1,-1,0,0), B(1,1,1,0,0,1,1,1,0,0,1),\\
&B(1,1,1,-1,0,1,1,1,0,0,1), B(1,1,1,0,-1,1,1,1,0,0,1), B(1,1,0,1,0,1,1,1,0,0,1),\\
&B(1,1,-1,1,0,1,1,1,0,0,1), B(0,1,1,1,0,1,1,1,0,0,1), B(-1,1,1,1,0,1,1,1,0,0,1),\\
&B(1,1,1,1,0,1,1,1,0,0,1), B(1,1,1,1,-1,1,1,1,0,0,1), B(1,1,1,1,-2,1,1,1,0,0,1)\}
\end{aligned}
$$

$$
\begin{aligned}
C_1 : \{&C(0,0,0,1,0,1,0,0,1,0,0), C(0,0,0,1,1,0,0,0,1,0,0), C(0,0,0,1,1,0,0,1,0,0,0),\\
&C(0,1,0,1,0,0,0,1,0,0,0), C(0,1,1,1,0,0,0,0,0,0,0), C(1,0,0,1,0,1,0,0,0,0,0),\\
&C(0,0,1,1,1,0,0,0,1,0,0), C(0,1,0,1,0,1,0,0,1,0,0), C(0,1,0,1,0,1,0,1,0,0,0),\\
&C(0,1,1,0,0,1,0,0,1,0,0), C(0,1,1,1,0,0,0,0,1,0,0), C(1,0,0,1,0,1,0,1,0,0,0),\\
&C(1,0,0,1,1,0,0,1,0,0,0), C(1,0,1,1,1,0,0,0,0,0,0), C(1,1,0,0,0,1,0,1,0,0,0),\\
&C(1,1,1,0,0,1,0,0,0,0,0), C(0,0,0,1,1,1,0,1,1,0,0), C(0,0,1,1,1,0,0,1,1,0,0),\\
&C(0,1,0,1,0,1,0,1,1,0,0), C(0,1,0,1,1,0,0,1,1,0,0), C(0,1,0,1,1,1,0,0,1,0,0),\\
&C(0,1,0,1,1,1,0,1,0,0,0), C(0,1,1,1,0,0,0,1,1,0,0), C(0,1,1,1,0,1,0,0,1,0,0),\\
&C(0,1,1,1,1,0,0,0,1,0,0), C(0,1,1,1,1,0,0,1,0,0,0), C(1,0,0,1,0,1,0,1,1,0,0),\\
&C(1,0,0,1,1,0,0,1,1,0,0), C(1,0,0,1,1,1,0,0,1,0,0), C(1,0,0,1,1,1,0,1,0,0,0),\\
&C(1,0,1,1,1,0,0,0,1,0,0), C(1,0,1,1,1,0,0,1,0,0,0), C(1,1,0,1,0,1,0,1,0,0,0),\\
&C(-1,1,0,1,0,1,0,1,1,0,0), C(-1,1,1,1,1,0,0,0,1,0,0), C(1,-1,0,1,1,1,0,1,0,0,0),\\
&C(0,1,0,1,1,1,0,1,1,0,0), C(0,1,1,1,0,1,0,1,1,0,0), C(0,1,1,1,1,0,0,1,1,0,0),\\
&C(0,1,1,1,1,1,0,0,1,0,0), C(1,0,0,1,1,1,0,1,1,0,0), C(1,0,1,1,1,0,0,1,1,0,0),
\end{aligned}
$$

$$C(1,1,0,1,0,1,0,1,1,0,0), C(1,1,0,1,1,1,0,1,0,0,0), C(1,1,1,0,0,1,0,1,1,0,0),$$
$$C(-1,1,0,1,1,1,0,1,1,0,0), C(-1,1,1,1,1,0,0,1,1,0,0), C(1,-1,0,1,1,1,0,1,1,0,0),$$
$$C(1,-1,1,1,1,0,0,1,1,0,0), C(0,1,1,1,1,1,0,1,1,0,0), C(1,1,0,1,1,1,0,1,1,0,0),$$
$$C(1,1,1,1,1,1,0,0,1,0,0), C(1,1,1,1,1,1,0,1,0,0,0), C(-1,1,1,1,1,1,0,1,1,0,0),$$
$$C(0,1,1,1,1,1,-1,1,1,0,0), C(1,1,-1,1,1,1,0,1,1,0,0), C(1,1,0,1,1,1,-1,1,1,0,0),$$
$$C(1,1,1,1,1,1,-1,0,1,0,0), C(1,1,1,1,1,1,-1,1,0,0,0), C(1,1,1,1,1,1,0,1,1,0,0),$$
$$C(1,1,1,1,1,1,-1,1,1,0,0), C(1,1,1,1,1,1,0,1,1,0,-1)\}$$

$$C_2 : \{C(0,1,1,1,0,0,0,0,0,0,0), C(0,1,0,1,0,0,0,1,0,0,0), C(0,0,0,1,1,0,0,1,0,0,0),$$
$$C(0,0,0,1,1,0,0,0,1,0,0), C(1,0,1,1,1,0,0,0,0,0,0), C(1,0,0,1,1,0,0,1,0,0,0),$$
$$C(0,1,1,1,0,0,0,0,1,0,0), C(0,0,1,1,1,0,0,0,1,0,0), C(0,1,0,1,0,0,0,1,0,0,1),$$
$$C(0,1,0,1,0,0,0,0,1,0,1), C(1,0,1,1,1,0,0,1,0,0,0), C(0,1,1,1,1,0,0,1,0,0,0),$$
$$C(1,0,1,1,1,0,0,0,1,0,0), C(0,1,1,1,1,0,0,0,1,0,0), C(-1,1,1,1,1,0,0,0,1,0,0),$$
$$C(0,1,1,1,0,0,0,1,1,0,0), C(1,0,0,1,1,0,0,1,1,0,0), C(0,1,0,1,1,0,0,1,1,0,0),$$
$$C(0,0,1,1,1,0,0,1,1,0,0), C(0,1,1,1,0,0,0,1,0,0,1), C(0,1,1,1,0,0,0,0,1,0,1),$$
$$C(0,1,0,1,0,0,0,1,1,0,1), C(1,0,1,1,1,0,0,1,1,0,0), C(1,-1,1,1,1,0,0,1,1,0,0),$$
$$C(0,1,1,1,1,0,0,1,1,0,0), C(-1,1,1,1,1,0,0,1,1,0,0), C(0,1,1,1,0,0,0,1,1,0,1),$$
$$C(-1,1,1,1,0,0,0,1,1,0,1)\}$$

$$C_3 : \{C(1,0,0,1,0,1,0,0,0,0,0), C(0,0,0,1,1,0,0,1,0,0,0), C(1,0,1,1,1,0,0,0,0,0,0),$$
$$C(1,0,0,1,1,0,0,1,0,0,0), C(1,0,0,1,0,1,0,1,0,0,0), C(0,0,0,1,0,1,0,1,0,0,1),$$
$$C(1,0,1,1,1,0,0,1,0,0,0), C(1,0,0,1,1,1,0,1,0,0,0), C(1,-1,0,1,1,1,0,1,0,0,0),$$
$$C(1,0,0,1,0,1,0,1,0,0,1)\}$$

# Appendix C.

# Top mass systematics and the lower edge of $M_{t\bar{t}}$

Experimental measurements of the cross section and of distributions depend on the value of the top quark mass. This dependence enters because inclusive measurements of stable top quarks are actually extrapolations from measurements of top quark decay products in fiducial regions, and these extrapolations are derived from simulated Monte Carlo data which explicitly carries a dependence on $m_t$[1]. Typically, this 'experimental response' to the input value of $m_t$ is known to constitute a $\sim 1-2\%$ effect at the level of the inclusive cross section. For generic bins of a kinematic distribution this is a similarly small effect. However, there are kinematic regions where these effects could be significantly larger and are perhaps currently an underestimated systematic effect. This is particularly the case for the lowest bin in $M_{t\bar{t}}$. The reason for this is that for simulated stable top (parton) data, the lowest possible value of $M_{t\bar{t}}$ is $2m_t$ and therefore when using simulated data that uses $m_t = 172.5$ GeV, one can never obtain values $M_{t\bar{t}} < 345$ GeV.

Considering the case of the $M_{t\bar{t}}$ distribution, if the lowest value of $M_{t\bar{t}}$ considered is 345 GeV (as is done here, see Table 5.1), then for values of the top quark mass $m_t < 172.5$ GeV, the theoretical predictions will contain events with kinematics such that $M_{t\bar{t}} < 345$ GeV. In our extractions above, and consistent with what is done by the two experiments, these 'underflow' events are not included. Since this lowest bin of $M_{t\bar{t}}$ is the most sensitive to the value of $m_t$, this issue deserves a careful examination. If, in the extrapolation to stable top quarks, some underflow events do leak into the first bin of $M_{t\bar{t}}$, then this could have serious effects on the extraction of $m_t$ from this distribution. In order to study this further, we have considered how our extractions react to the addition of underflow events to this lowest bin. For the discrete set of values of $m_t$ considered, it is only the predictions for $m_t = 169.0$ GeV and $m_t = 171.0$ GeV that contain underflow events, but nevertheless, these do alter the fits of the theory predictions.

[1] The default choice of input $m_t$ used by both ATLAS and CMS is $m_t = 172.5$ GeV.

| | ATLAS | | | | | |
|---|---|---|---|---|---|---|
| | CT14 | | NNPDF30 | | NNPDF31 | |
| | $\alpha_s$ | $\chi^2_{\text{min}}$ | $\alpha_s$ | $\chi^2_{\text{min}}$ | $\alpha_s$ | $\chi^2_{\text{min}}$ |
| $M_{t\bar{t}}$ | $0.1174^{+0.0049}_{-0.0075}$ | 1.24 | $0.1195^{+0.0063}_{-0.0061}$ | 1.67 | $0.1199^{+0.0054}_{-0.0064}$ | 3.21 |
| $M^u_{t\bar{t}}$ | $0.1175^{+0.0049}_{-0.0077}$ | 1.23 | $0.1196^{+0.0063}_{-0.0061}$ | 1.67 | $0.1199^{+0.0054}_{-0.0064}$ | 3.27 |

| | CMS | | | | | |
|---|---|---|---|---|---|---|
| | CT14 | | NNPDF30 | | NNPDF31 | |
| | $\alpha_s$ | $\chi^2_{\text{min}}$ | $\alpha_s$ | $\chi^2_{\text{min}}$ | $\alpha_s$ | $\chi^2_{\text{min}}$ |
| $M_{t\bar{t}}$ | $0.1146^{+0.0035}_{-0.0036}$ | 9.84 | $0.1100^{+0.0052}_{-0.0081}$ | 7.03 | $0.1160^{+0.0053}_{-0.0138}$ | 6.10 |
| $M^u_{t\bar{t}}$ | $0.1146^{+0.0035}_{-0.0036}$ | 9.82 | $0.1100^{+0.0052}_{-0.0081}$ | 7.01 | $0.1160^{+0.0053}_{-0.0139}$ | 6.08 |

**Table C.1.:** Tabulated values of best-fit $\alpha_s$ (with uncertainties) and associated $\chi^2_{\text{min}}$ from extractions of $\alpha_s$ using ATLAS (upper table) and CMS (lower table) measurements of the normalized $M_{t\bar{t}}$ distribution and the total cross section. The case with and without underflow events is considered. Results are shown for three different PDF sets and $m_t$ has been set to the world average value of 173.3 GeV. The cells highlighted in gray correspond to extractions that satisfy the conditions of Equations (5.13, 5.14).

In Table C.1 we show the extraction of $\alpha_s$ for a fixed value of the top mass, $m_t = 173.3$ GeV, using either the invariant mass distribution, $M_{t\bar{t}}$, or the invariant mass distribution supplemented with underflow events, $M^u_{t\bar{t}}$. In Table C.2 we show the corresponding table for the extraction of $m_t$ for $\alpha_s = 0.118$. In the extraction of $\alpha_s$, for all PDF sets and for both experiments it is clear that the addition of the underflow events only slightly increases the value of $\chi^2_{\text{min}}$ and barely affects the best-fit value and the associated uncertainties. In the extraction of $m_t$, when using data from either experiment, the addition of underflow events again has a very small effect on $\chi^2_{\text{min}}$. Interestingly however, in the case of CMS, the best-fit value of $m_t$ is shifted upwards by 0.3 GeV in all cases and the associated uncertainties decrease.

We have also investigated the effects of the addition of underflow events in simultaneous extractions of $\alpha_s$ and $m_t$, the results of which are tabulated for each experiment in Table C.3. Reflecting the results of the individual extractions, we see that adding the underflow events has only a very small effect on the extractions when using ATLAS data. This is not the case when using the CMS data, where we now observe significant increases in both the best-fit values of $\alpha_s$ and $m_t$, with $m_t$ in particular being shifted by almost 1 $\sigma$ for all PDFs. Although these shifts are consistent within uncertainties, it is however interesting to note that they push the extracted values upwards towards the respective world averages.

| | ATLAS | | | | | |
|---|---|---|---|---|---|---|
| | CT14 | | NNPDF30 | | NNPDF31 | |
| | $m_t$ | $\chi^2_{\rm min}$ | $m_t$ | $\chi^2_{\rm min}$ | $m_t$ | $\chi^2_{\rm min}$ |
| $M_{t\bar{t}}$ | $173.3^{+0.6}_{-0.5}$ | 1.24 | $173.3^{+0.6}_{-0.5}$ | 1.72 | $173.5^{+0.6}_{-0.5}$ | 3.21 |
| $M^u_{t\bar{t}}$ | $173.3^{+0.5}_{-0.5}$ | 1.23 | $173.3^{+0.5}_{-0.5}$ | 1.74 | $173.5^{+0.5}_{-0.5}$ | 3.24 |

| | ATLAS | | | | | |
|---|---|---|---|---|---|---|
| | CT14 | | NNPDF30 | | NNPDF31 | |
| | $m_t$ | $\chi^2_{\rm min}$ | $m_t$ | $\chi^2_{\rm min}$ | $m_t$ | $\chi^2_{\rm min}$ |
| $M_{t\bar{t}}$ | $170.5^{+1.7}_{-1.8}$ | 7.61 | $170.6^{+1.8}_{-1.8}$ | 6.23 | $170.9^{+1.8}_{-1.8}$ | 4.04 |
| $M^u_{t\bar{t}}$ | $170.8^{+1.4}_{-1.4}$ | 7.63 | $170.9^{+1.4}_{-1.4}$ | 6.25 | $171.2^{+1.4}_{-1.4}$ | 4.09 |

**Table C.2.:** Tabulated values of best-fit $m_t$ (with uncertainties) and associated $\chi^2_{\rm min}$ from extractions of $m_t$ using ATLAS (upper table) and CMS (lower table) measurements of the normalized $M_{t\bar{t}}$ distribution and the total cross section. The case with and without underflow events is considered. Results are shown for three different PDF sets and $\alpha_s$ has been set to the world average value of 0.118. The cells highlighted in gray correspond to extractions that satisfy the conditions of Equations (5.13, 5.14).

| | ATLAS | | | | | | | | |
|---|---|---|---|---|---|---|---|---|---|
| | CT14 | | | NNPDF30 | | | NNPDF31 | | |
| | $\alpha_s$ | $m_t$ | $\chi^2_{\rm min}$ | $\alpha_s$ | $m_t$ | $\chi^2_{\rm min}$ | $\alpha_s$ | $m_t$ | $\chi^2_{\rm min}$ |
| $M_{t\bar{t}}$ | $0.1174^{+0.0050}_{-0.0080}$ | $173.2^{+0.6}_{-0.5}$ | 1.23 | $0.1195^{+0.0063}_{-0.0061}$ | $173.3^{+0.6}_{-0.5}$ | 1.67 | $0.1203^{+0.0055}_{-0.0062}$ | $173.5^{+0.6}_{-0.5}$ | 3.06 |
| $M^u_{t\bar{t}}$ | $0.1175^{+0.0049}_{-0.0080}$ | $173.3^{+0.5}_{-0.5}$ | 1.23 | $0.1196^{+0.0063}_{-0.0061}$ | $173.3^{+0.5}_{-0.5}$ | 1.67 | $0.1203^{+0.0055}_{-0.0062}$ | $173.5^{+0.5}_{-0.5}$ | 3.09 |

| | ATLAS | | | | | | | | |
|---|---|---|---|---|---|---|---|---|---|
| | CT14 | | | NNPDF30 | | | NNPDF31 | | |
| | $\alpha_s$ | $m_t$ | $\chi^2_{\rm min}$ | $\alpha_s$ | $m_t$ | $\chi^2_{\rm min}$ | $\alpha_s$ | $m_t$ | $\chi^2_{\rm min}$ |
| $M_{t\bar{t}}$ | $0.1109^{+0.0036}_{-0.0029}$ | $168.7^{+2.0}_{-2.3}$ | 4.77 | $0.1054^{+0.0055}_{-0.0053}$ | $168.8^{+2.2}_{-2.5}$ | 2.15 | $0.1101^{+0.0054}_{-0.0081}$ | $169.2^{+2.0}_{-2.3}$ | 2.60 |
| $M^u_{t\bar{t}}$ | $0.1120^{+0.0038}_{-0.0032}$ | $170.0^{+1.5}_{-1.5}$ | 5.37 | $0.1067^{+0.0058}_{-0.0056}$ | $170.2^{+1.5}_{-1.5}$ | 2.85 | $0.1126^{+0.0050}_{-0.0097}$ | $170.6^{+1.5}_{-1.5}$ | 3.11 |

**Table C.3.:** Tabulated values of best-fit $\alpha_s$ and $m_t$ (with uncertainties) and associated $\chi^2_{\rm min}$ from simultaneous extractions of $m_t$ and $\alpha_s$ using ATLAS (upper table) and CMS (lower table) measurements of the normalized $M_{t\bar{t}}$ distribution and the total cross section. The case with and without underflow events is considered. Results are shown for three different PDF sets. The cells highlighted in gray correspond to extractions that satisfy the conditions of Equations (5.13, 5.14).

Perhaps the most surprising feature of the numbers presented in this section is the fact that the change in the extractions is not the same for both experiments. This may point to a different treatment by the two experiments of underflow events, either in the direct measurements themselves or in the process of the extrapolation of direct measurements to stable tops. Given that we have found that different treatments of underflow events can have a significant effect on extractions of $m_t$ and $\alpha_s$, and this aspect of the measurement of $M_{t\bar{t}}$ deserves careful consideration in future measurements.

# Appendix D.

# Cross section for the process $gg \to b\bar{b}H$ in the collinear limit

In this Appendix we illustrate in some detail the calculation of the cross section for the partonic process

$$g(p_1) + g(p_2) \to b(k_1) + \bar{b}(k_2) + H(k) \tag{D.1}$$

in the limit of collinear emission of $b$ quarks. We choose, as independent kinematic invariants,

$$\begin{aligned}
\hat{s} &= (p_1 + p_2)^2 = 2p_1p_2 && \text{(D.2)}\\
t_1 &= (p_1 - k_1)^2 = -2p_1k_1 + m_b^2 && \text{(D.3)}\\
t_2 &= (p_2 - k_2)^2 = -2p_2k_2 + m_b^2 && \text{(D.4)}\\
s_1 &= (k_1 + k)^2 = 2k_1k + m_b^2 + M_H^2 && \text{(D.5)}\\
s_2 &= (k_2 + k)^2 = 2k_2k + m_b^2 + M_H^2. && \text{(D.6)}
\end{aligned}$$

The remaining invariants

$$\begin{aligned}
u_1 &= (p_1 - k_2)^2 = -2p_1k_2 + m_b^2 && \text{(D.7)}\\
u_2 &= (p_2 - k_1)^2 = -2p_2k_1 + m_b^2 && \text{(D.8)}\\
s_{12} &= (k_1 + k_2)^2 = 2k_1k_2 + 2m_b^2 && \text{(D.9)}\\
t &= (p_1 - k)^2 - M_H^2 = -2kp_1 && \text{(D.10)}\\
u &= (p_2 - k)^2 - M_H^2 = -2kp_2 && \text{(D.11)}
\end{aligned}$$

are related to the independent invariants by

$$u_1 = s_1 - \hat{s} - t_2 + m_b^2 \tag{D.12}$$

$$u_2 = s_2 - \hat{s} - t_1 + m_b^2 \tag{D.13}$$

$$t = -s_1 + t_2 - t_1 + m_b^2 \tag{D.14}$$

$$u = -s_2 + t_1 - t_2 + m_b^2 \tag{D.15}$$

$$s_{12} = \hat{s} - s_1 - s_2 + M_H^2 + 2m_b^2. \tag{D.16}$$

The leading-order Feynman diagrams are shown in Figure D.1. The squared invariant amplitude (averaged over initial state summed over final state spin and color variables) has the general structure

$$|\mathcal{M}|^2 = \frac{G(s, s_1, s_2, t_1, t_2)}{(t_1 - m_b^2)^2(t_2 - m_b^2)^2(u_1 - m_b^2)^2(u_2 - m_b^2)^2}. \tag{D.17}$$

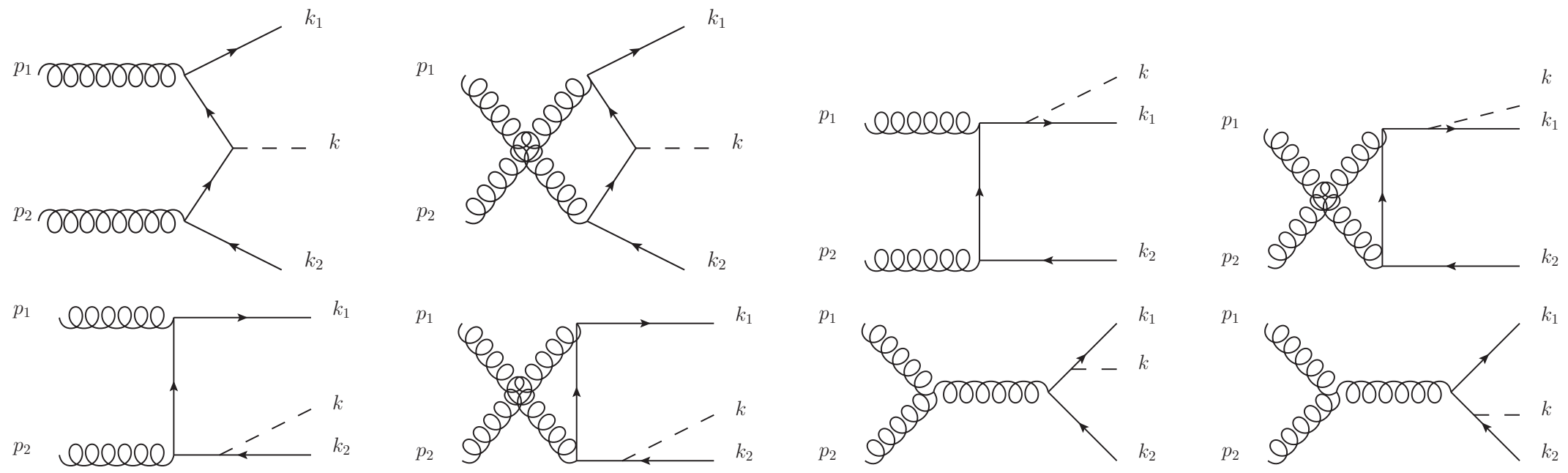

**Figure D.1.:** Leading order diagrams for $gg \to b\bar{b}H$.

The function $G(s, s_1, s_2, t_1, t_2)$ is a polynomial in $t_1, t_2$. It can be shown on general grounds[165,166] that each double pole is suppressed by a factor of $m_b^2$. Furthermore, it is well known that collinear singularities do not arise in interference terms among different amplitudes. Thus,

$$|\mathcal{M}|^2 = \frac{G_t}{(t_1 - m_b^2)(t_2 - m_b^2)} + \frac{G_u}{(u_1 - m_b^2)(u_2 - m_b^2)} + |\mathcal{M}|^2_{\text{reg}} \tag{D.18}$$

where the term $|\mathcal{M}|^2_{\text{reg}}$ does not give rise to collinear singularities in the limit $m_b = 0$. An explicit calculation gives

$$G_t = G_u = \frac{32\alpha_s^2\pi^2 m_b^2 G_F M_H^2 \sqrt{2}}{3} \frac{P_{qg}(z_1)}{z_1} \frac{P_{qg}(z_2)}{z_2}, \tag{D.19}$$

where

$$z_1 = \frac{M_H^2}{s_1}; \qquad z_2 = \frac{M_H^2}{s_2} \tag{D.20}$$

and $P_{qg}(z)$ is defined in Equation 6.11.

The 3-body phase-space invariant measure

$$\begin{aligned} & d\phi_3(p_1,p_2;k_1,k_2,k) \\ & = \frac{d^3k_1}{(2\pi)^3 2k_1^0}\frac{d^3k_2}{(2\pi)^3 2k_2^0}\frac{d^3k}{(2\pi)^3 2k^0}(2\pi)^4\delta(p_1+p_2-k_1-k_2-k) \end{aligned} \tag{D.21}$$

can be factorised as

$$d\phi_3(p_1,p_2;k_1,k_2,k) = \frac{dt_1}{2\pi}\frac{dt_2}{2\pi}d\phi_2(p_1;k_1,q_1)d\phi_2(p_2;k_2,q_2)d\phi_1(q_1,q_2;k), \tag{D.22}$$

where

$$q_1^2 = t_1; \qquad q_2^2 = t_2. \tag{D.23}$$

We now compute each factor explicitly. We have

$$\begin{aligned} d\phi_2(p_1;k_1,q_1) &= \frac{d^3k_1}{(2\pi)^3 2k_1^0}\frac{d^3q_1}{(2\pi)^3 2q_1^0}(2\pi)^4\delta(p_1-k_1-q_1) \\ &= \frac{1}{16\pi^2}\frac{|\vec{k}_1|^2 d|\vec{k}_1| d\cos\theta_1 d\phi_1}{k_1^0 q_1^0}\delta(p_1^0-k_1^0-q_1^0) \end{aligned} \tag{D.24}$$

where

$$k_1^0 = \sqrt{|\vec{k}_1|^2+m_b^2} \tag{D.25}$$

$$q_1^0 = \sqrt{|\vec{p}_1|^2+|\vec{k}_1|^2-2|\vec{p}_1||\vec{k}_1|\cos\theta_1+t_1} \tag{D.26}$$

We may now integrate over $\cos\theta_1$ using the delta function

$$\delta(p_1^0-k_1^0-q_1^0) = \frac{q_1^0}{|\vec{p}_1||\vec{k}_1|}\delta(\cos\theta_1-\cos\bar{\theta}_1) \tag{D.27}$$

with $\bar{\theta}_1$ a solution of

$$p_1^0 - \sqrt{|\vec{k}_1|^2+m_b^2} - \sqrt{|\vec{p}_1|^2+|\vec{k}_1|^2-2|\vec{p}_1||\vec{k}_1|\cos\bar{\theta}_1+t_1} = 0. \tag{D.28}$$

This gives

$$d\phi_2(p_1;k_1,q_1) = \frac{1}{16\pi^2}\frac{|\vec{k}_1|d|\vec{k}_1|d\varphi_1}{k_1^0|\vec{p}_1|}; \qquad d\phi_2(p_2;k_2,q_2) = \frac{1}{16\pi^2}\frac{|\vec{k}_2|d|\vec{k}_2|d\varphi_2}{k_2^0|\vec{p}_2|} \tag{D.29}$$

and therefore

$$d\phi_3(p_1,p_2;k_1,k_2,k) = \frac{1}{1024\pi^6}dt_1 dt_2 \frac{|\vec{k}_1|d|\vec{k}_1|d\varphi_1}{k_1^0|\vec{p}_1|}\frac{|\vec{k}_2|d|\vec{k}_2|d\varphi_2}{k_2^0|\vec{p}_2|}d\phi_1(q_1,q_2;k). \tag{D.30}$$

It will be convenient to adopt the centre-of-mass frame, where

$$p_1 = \frac{\sqrt{\hat{s}}}{2}(1,0,0,1), \qquad p_2 = \frac{\sqrt{\hat{s}}}{2}(1,0,0,-1) \tag{D.31}$$

In this frame

$$s_1 = (k+k_1)^2 = (p_1+p_2-k_2)^2 = \hat{s} + m_b^2 - 2\sqrt{\hat{s}}\sqrt{|\vec{k}_2|^2 + m_b^2} \tag{D.32}$$

$$s_2 = (k+k_2)^2 = (p_1+p_2-k_1)^2 = \hat{s} + m_b^2 - 2\sqrt{\hat{s}}\sqrt{|\vec{k}_1|^2 + m_b^2} \tag{D.33}$$

and therefore

$$\frac{|\vec{k}_1|d|\vec{k}_1|}{k_1^0|\vec{p}_1|}\frac{|\vec{k}_2|d|\vec{k}_2|}{k_2^0|\vec{p}_2|} = \frac{ds_1}{\hat{s}}\frac{ds_2}{\hat{s}}. \tag{D.34}$$

Furthermore, we may use the invariance of the cross section upon rotations about the $z$ axis to replace

$$d\varphi_1 d\varphi_2 \to 2\pi d\varphi; \qquad \varphi = \varphi_1 - \varphi_2. \tag{D.35}$$

Finally,

$$d\phi_1(q_1,q_2;k) = 2\pi\delta\left((q_1+q_2)^2 - M_H^2\right), \tag{D.36}$$

and therefore

$$d\phi_3(p_1,p_2;k_1,k_2,k) = \frac{1}{256\pi^4\hat{s}^2}ds_1 ds_2 dt_1 dt_2\, d\varphi\delta\left((q_1+q_2)^2 - M_H^2\right). \tag{D.37}$$

It is a tedious, but straightforward, task to show that, upon integration over the azimuth $\varphi$ using the delta function, this expression is the same as the one given in [167] for the three-body phase-space measure in terms of four invariants.

The two invariants $u_1, u_2$ are related to independent invariants through Equations (D.12,D.13), which can be written

$$u_1 - m_b^2 = -(t_2 - a_2) \tag{D.38}$$

$$u_2 - m_b^2 = -(t_1 - a_1) \tag{D.39}$$

where we have defined

$$a_1 = s_2 - \hat{s}; \qquad a_2 = s_1 - \hat{s}. \tag{D.40}$$

The bounds for $t_1$ are easily obtained. In the centre-of-mass frame we have

$$t_1 = \frac{1}{2}\left[a_1 + m_b^2 - \cos\bar{\theta}_1\sqrt{(a_1 + m_b^2)^2 - 4m_b^2(a_1 + \hat{s})}\right] \tag{D.41}$$

$$t_2 = \frac{1}{2}\left[a_2 + m_b^2 + \cos\bar{\theta}_2\sqrt{(a_2 + m_b^2)^2 - 4m_b^2(a_2 + \hat{s})}\right]. \tag{D.42}$$

The upper and lower bound are obtained for $\cos\bar{\theta}_1 = \pm 1$, $\cos\bar{\theta}_2 = \pm 1$. We get

$$t_1^- \le t_1 \le t_1^+; \qquad t_2^- \le t_2 \le t_2^+, \tag{D.43}$$

where

$$t_1^\pm = \frac{1}{2}\left[a_1 + m_b^2 \pm \sqrt{(a_1 + m_b^2)^2 - 4m_b^2(a_1 + \hat{s})}\right] \tag{D.44}$$

$$t_2^\pm = \frac{1}{2}\left[a_2 + m_b^2 \pm \sqrt{(a_2 + m_b^2)^2 - 4m_b^2(a_2 + \hat{s})}\right]. \tag{D.45}$$

For small $m_b^2$,

$$t_i^+ = m_b^2 + \frac{m_b^2\hat{s}}{a_i} + O(m^4); \qquad t_i^- = a_i - \frac{m_b^2\hat{s}}{a_i} + O(m^4); \qquad i = 1, 2. \tag{D.46}$$

All the ingredients to compute the total partonic cross section in the collinear limit are now available. In this limit, the relative azimuth $\phi$ between $b$ and $\bar{b}$ is irrelevant, and simply provides a factor of $2\pi$. Furthermore

$$\hat{s} = \frac{M_H^2}{z_1 z_2}; \qquad s_1 = \hat{s}z_2; \qquad s_2 = \hat{s}z_1 \tag{D.47}$$

and therefore

$$\frac{ds_1\, ds_2}{\hat{s}^2} = dz_1\, dz_2. \tag{D.48}$$

The integrals over $t_1, t_2$ are easily computed:

$$\int_{t_i^-}^{t_i^+} dt_i \, \frac{1}{t_i - m_b^2} = \log \frac{a_1^2}{m_b^2 \hat{s}} + O(1) = \log \frac{M_H^2}{m_b^2} \frac{(1-z_i)^2}{z_1 z_2} \tag{D.49}$$

$$\int_{t_i^-}^{t_i^+} dt_i \, \frac{1}{t_i - a_i} = -\log \frac{a_i^2}{m_b^2 \hat{s}} + O(1) = -\log \frac{M_H^2}{m_b^2} \frac{(1-z_i)^2}{z_1 z_2} + O(1). \tag{D.50}$$

Finally,

$$\delta\left((q_1+q_2)^2 - M_H^2\right) = \delta(z_1 z_2 \hat{s} - M_H^2). \tag{D.51}$$

We find

$$\begin{aligned}
\hat{\sigma}^{4\mathrm{F,coll}}(\hat{\tau}) &= \frac{1}{2\hat{s}} \int d\phi_3(p_1, p_2; k_1, k_2, k) \, G_u \left[ \frac{1}{(t_1 - m_b^2)(t_2 - m_b^2)} + \frac{1}{(t_1 - a_1)(t_2 - a_2)} \right] \\
&= \hat{\tau} \frac{\alpha_s^2}{4\pi^2} \frac{m_b^2}{M_H^2} \frac{G_F \pi}{3\sqrt{2}} 2 \int_0^1 dz_1 \int_0^1 dz_2 \, \delta(z_1 z_2 - \hat{\tau}) \\
&\times P_{qg}(z_1) \log \left[ \frac{M_H^2}{m_b^2} \frac{(1-z_1)^2}{\hat{\tau}} \right] P_{qg}(z_2) \log \left[ \frac{M_H^2}{m_b^2} \frac{(1-z_2)^2}{\hat{\tau}} \right].
\end{aligned} \tag{D.52}$$

# Colophon

This thesis was made in $\LaTeX\,2_\varepsilon$ using the "hepthesis" class [168].